\documentclass[%
onecolumn,
 amsmath,amssymb,
 aps,
longbibliography
]{revtex4-1}

\usepackage{graphicx}% Include figure files
\usepackage{dcolumn}% Align table columns on decimal point
\usepackage{bm}% bold math
\usepackage{subfigure}
\usepackage{soul}
\DeclareMathOperator\erf{erf}

\begin{document}

%\preprint{APS/123-QED}

\title{High-resolution Navier--Stokes simulations of \\
Richtmyer--Meshkov instability with re-shock}% Force line

\author{Man Long Wong}
 \email{wongml@stanford.edu}
 \affiliation{%
  Department of Aeronautics and Astronautics, Stanford University \\
  Stanford, CA 94305, USA
 }%

\author{Daniel Livescu}
 \email{livescu@lanl.gov}
 \affiliation{%
  CCS-2, Los Alamos National Laboratory \\
  Los Alamos, NM 87545, USA
 }%

\author{Sanjiva K. Lele}
 \email{lele@stanford.edu
 \\ \\ \\ \textit{\normalsize Second revision submitted to Physical Review Fluids}}
 \affiliation{%
  Department of Aeronautics and Astronautics \& \\
  Department of Mechanical Engineering, Stanford University \\
  Stanford, CA 94305, USA
 }%

\date{\today}% It is always \today, today,

\begin{abstract}
The interaction of a Mach 1.45 shock wave with a perturbed planar interface between sulphur hexafluoride and air is studied through high-resolution two-dimensional (2D) and three-dimensional (3D) shock-capturing adaptive mesh refinement simulations of multi-species Navier--Stokes equations. The sensitivities of time-dependent statistics on grid resolution for 2D and 3D simulations are evaluated to ensure the accuracy of the results. The numerical results are used to examine the differences between the development of 2D and 3D Richtmyer--Meshkov instability during two different stages: (1) initial growth of hydrodynamic instability from multi-mode perturbations after first shock and (2) transition to chaotic or turbulent state after re-shock. The effects of the Reynolds number on the mixing in 3D simulations are also studied through varying the transport coefficients.

\iffalse
\begin{description}
\item[Usage]
Secondary publications and information retrieval purposes.
\item[PACS numbers]
May be entered using the \verb+\pacs{#1}+ command.
\item[Structure]
You may use the \texttt{description} environment to structure your abstract;
use the optional argument of the \verb+\item+ command to give the category of each item. 
\end{description}
\fi

\end{abstract}

\pacs{Valid PACS appear here}% PACS, the Physics and Astronomy

\maketitle

%\tableofcontents

%%%%%%%%%%%%%%%%%%%%%%%%%%%%%%%%%%%%%%%%%%%%%%%%%%%%%%%%%%%%%%%%%%%%%%%%%%%%%%%%
%%%%%%%%%%%%%%%%%%%%%%%%%%%%%%%%%%%%%%%%%%%%%%%%%%%%%%%%%%%%%%%%%%%%%%%%%%%%%%%%
%%%%%%%%%%%%%%%%%%%%%%%%%%%%%%%%%%%%%%%%%%%%%%%%%%%%%%%%%%%%%%%%%%%%%%%%%%%%%%%%

\section{\label{sec:introduction} Introduction}

Richtmyer--Meshkov (RM) instability, or RMI~\cite{richtmyer1960taylor, meshkov1969instability}, is a fundamental hydrodynamic flow instability that occurs when a shock wave passes through a material interface between fluids of different densities. The instability is initiated by the deposition of vorticity at the interface due to the misalignment in the pressure and density gradients across the shock wave and material interface. Studying RMI is very important for understanding the dynamics of supernovae and other astrophysical explosions, mixing enhancement in supersonic/hypersonic combustion, and turbulence and mixing limitations for inertial confinement fusion~\cite{brouillette2002richtmyer, zhou2017arayleigh, zhou2017brayleigh}.

The study of RMI through physical experiments is challenging because the flows are transient, the initial conditions of the interface are difficult to be characterized, and uncertainty-quantified measurements are hard to be obtained. The initial perturbation grows nonlinearly soon after the passage of the shock through the interface. The growth of the mixing region is known to be largely dependent on the characteristics of initial perturbation~\cite{thornber2010influence, schilling2010high, thornber2011growth, hahn2011richtmyer, grinstein2011simulations, gowardhan2011numerical, gowardhan2011bipolar, balasubramanian2012experimental} and hence a precise control of the initial conditions is required from experiments. If the Reynolds number is sufficiently high, small-scale turbulent features will develop beyond the mixing transition. Characterizing the turbulent mixing requires high-resolution temporal and spatial measurements of both the density and velocity fields. The simultaneous planar laser-induced fluorescence (PLIF) and particle image velocimetry (PIV) measurements have been used to collect density and velocity data from shock-induced mixing problems in many previous studies~\cite{balasubramanian2012experimental, balakumar2012turbulent, orlicz2015mixing, mohaghar2017evaluation, reese2018simultaneous}. However, the method is only limited to the measurements of the fields in a two-dimensional plane and data of other important quantities such as pressure and temperature are still out of reach. As a consequence, numerical simulations are commonly used as a complement to study the turbulent mixing induced by RMI.

As this moment, direct numerical simulation (DNS) of turbulent mixing from RMI is still computationally too expensive. However, many high-fidelity simulations and large eddy simulations (LES's) have been conducted in the past to study the development of turbulent mixing from RMI. \citet{thornber2010influence} showed the strong influence on the growth of instability from perturbations with different bandwidths using shock-capturing simulations. \citet{hill2006large} performed LES's of RMI with re-shock using hybrid shock-capturing/bandwidth optimized centre-difference scheme with stretched-vortex subgrid-scale model in an adaptive mesh refinement framework and their results showed good agreement with experimental data. \citet{lombardini2011atwood} later used the same numerical method to examine the Atwood number dependence of RMI, followed by a Mach number dependence study~\cite{lombardini2012transition}. The uncertainty of turbulence statistics in under-resolved simulations of RMI was later explored by \citet{tritschler2014richtmyer} for different numerical methods. While all studies mentioned above focused on the baroclinic instability in planar geometry, there are also some studies of the instabilities with inclined~\cite{subramaniam2017turbulence} and spherical interfaces~\cite{lombardini2014turbulent1,lombardini2014turbulent2}.

Although typical applications of shock-driven turbulent mixing require the multi-component governing equations to be solved in three-dimensional (3D) domains, one may still desire to conduct two-dimensional (2D) simulations to characterize the dynamics in the nonlinear pre-turbulent regime and take advantage of the lower requirement on computational resources. Before the onset of fully developed turbulence and especially during the initial growth of the shock-induced instability, there may be similarities in the development of the instability between 2D and 3D configurations. \citet{olson2014comparison} compared the 2D and 3D simulations of RMI using the artificial fluid LES method but the mixing statistics obtained from their highest resolution simulations are still very grid-sensitive. \citet{thornber2015numerical} also studied the 2D RMI with up to 64 realizations of inviscid simulations to reduce the statistical variability of their results but they did not study the effect of re-shock on the 2D instability. In this paper, we have conducted a large number of realizations of high-resolution Navier--Stokes simulations of 2D RMI with re-shock to investigate the difference between 2D and 3D RMI induced mixing.

In many previous RMI studies~\cite{thornber2010influence, schilling2010high, grinstein2011simulations, gowardhan2011numerical}, the hydrodynamic equations were solved without the molecular viscous and diffusive terms. The argument for not including these terms is that the physical dissipation or diffusion can be represented by numerical dissipation or diffusion, by assuming that there is only forward cascade of energy from large scale to small scale features~\cite{grinstein2007implicit, garnier2009large}. While this might be relevant during the decay of the fully turbulent mixing region, the molecular transport terms could have a strong effect on the initial growth of mixing layer before mixing transition occurs. In this work, we have chosen the initial conditions for the 3D simulations such that the flows inside the mixing layers have not transitioned to turbulence yet before re-shock. By varying the transport coefficients, multiple cases with effectively different turbulent Reynolds numbers have been conducted to study the diffusive and viscous effects on the shock-driven mixing before and after re-shock.

%%%%%%%%%%%%%%%%%%%%%%%%%%%%%%%%%%%%%%%%%%%%%%%%%%%%%%%%%%%%%%%%%%%%%%%%%%%%%%%%
%%%%%%%%%%%%%%%%%%%%%%%%%%%%%%%%%%%%%%%%%%%%%%%%%%%%%%%%%%%%%%%%%%%%%%%%%%%%%%%%
%%%%%%%%%%%%%%%%%%%%%%%%%%%%%%%%%%%%%%%%%%%%%%%%%%%%%%%%%%%%%%%%%%%%%%%%%%%%%%%%

\section{\label{sec:governing_equations} Governing equations}

The equations solved in this study are the conservative multi-component Navier--Stokes equations:
\begin{align}
	\frac{\partial \rho Y_i}{\partial t} + \nabla \cdot \left( \rho \bm{u} Y_i \right) &= - \nabla \cdot \bm{J_i}, \label{eq:species_continuity_eqn} \\
    \frac{\partial \rho \bm{u}}{\partial t} + \nabla \cdot \left( \rho \bm{uu} + p \bm{\delta} \right) &= \nabla \cdot \bm{\tau}, \label{eq:mixture_momentum_eqn} \\
    \frac{\partial E}{\partial t} + \nabla \cdot \left[ \left( E + p \right) \bm{u} \right] &= \nabla \cdot \left( \bm{\tau} \cdot \bm{u} - \bm{q_c} - \bm{q_d} \right), \label{eq:mixture_energy_eqn}
\end{align}
\noindent where $\rho$, $\bm{u}=[u,v,w]^{T}=[u_1,u_2,u_3]^{T}$, $p$, and $E$ are the density, velocity vector, pressure, and total energy per unit volume of the fluid mixture, respectively. $Y_i$ is the mass fraction of species $i \in \left[ 1,2,...,N \right]$, with $N$ the total number of species. All $Y_i$'s sum up to one by definition. $\bm{J_i}$ is diffusive mass flux for species $i$. $\bm{\tau}$, $\bm{q_c}$, and $\bm{q_d}$ are viscous stress tensor, conductive heat flux, and inter-species diffusional enthalpy flux, respectively. $\bm{\delta}$ is the identity tensor.

The mixture is assumed to be ideal and calorically perfect, with:
\begin{align}
    E = \rho \left( e + \frac{1}{2} \bm{u} \cdot \bm{u} \right), \\
    p = \left( \gamma - 1 \right) \rho e, \quad  e = c_v T,
\end{align}
\noindent where $e$ and $T$ are respectively specific internal energy and temperature of the mixture. $\gamma$ and $c_v$ are the ratio of specific heats and specific heat at constant volume of the mixture.

The multi-component diffusive mass flux of species $i$ is given by \citet{hirschfelder1954molecular}:
\begin{equation}
    \bm{J_i} = \rho \frac{M_i}{M^2} \sum^{N}_{j=1} M_j \tilde{D}_{ij} \nabla X_j,
    \label{eqn:multicomponent_diffusive_flux}
\end{equation}

\noindent where $M_i$ and $X_i$ are respectively the molecular weight and mole fraction of species $i$. $M$ is the molecular weight of the mixture and $\tilde{D}_{ij}$ is the $mn$th element of the matrix of ordinary multi-component diffusion coefficients $\tilde{\bm{D}}$. The mole fraction of species $i$ is given by:
\begin{equation}
    X_i = \frac{M}{M_i} Y_i.
\end{equation}
\noindent Baro-diffusion is neglected in this study since the shock Mach number is relatively small~\cite{livescu2013} and after the passage of the shock the flow is close to incompressible (see below). The multi-component diffusive mass flux is reduced to the Fick's law for binary mixture:
\begin{equation}
	\bm{J_i} = -\rho D_i \nabla Y_i,\ i=1, 2,
\end{equation}
\noindent where $D_1 = D_2$ is the binary diffusion coefficient.

The viscous stress tensor $\bm{\tau}$ for a Newtonian mixture is:
\begin{equation}
	\bm{\tau} = 2 \mu \bm{S} + \left( \mu_v - \frac{2}{3} \mu \right) \bm{\delta} \left( \nabla \cdot \mathbf{u} \right),
\end{equation}
\noindent where $\mu$ and $\mu_v$ are the shear viscosity and bulk viscosity respectively of the mixture. $\bm{S}$ is the strain-rate tensor given by:
\begin{equation}
    \bm{S} = \frac{1}{2} \left[ \nabla \bm{u} + \left( \nabla \bm{u} \right) ^{T} \right].
\end{equation}

The conductive flux and the inter-species diffusional enthalpy flux~\cite{williams2018combustion} are given by:
\begin{align}
	\bm{q_c} = - \kappa \nabla T, \\
	\bm{q_d} = \sum^{N}_{i=1} h_i \bm{J_i},
\end{align}
\noindent where $\kappa$ is the thermal conductivity of the mixture. $h_i$ is the specific enthalpy of species $i$:
\begin{equation}
    h_i = c_{p,i} T,
\end{equation}
\noindent where $c_{p,i}$ is the specific heat at constant volume of species $i$.

The equations and mixing rules for the fluid properties $\gamma$, $c_v$, $c_{p,i}$,  $\mu$, $\mu_v$, $\kappa$, and $D_i$ are given in appendices~\ref{sec:appendix_TC} and \ref{sec:appendix_mixing_rules}.

%%%%%%%%%%%%%%%%%%%%%%%%%%%%%%%%%%%%%%%%%%%%%%%%%%%%%%%%%%%%%%%%%%%%%%%%%%%%%%%%
%%%%%%%%%%%%%%%%%%%%%%%%%%%%%%%%%%%%%%%%%%%%%%%%%%%%%%%%%%%%%%%%%%%%%%%%%%%%%%%%
%%%%%%%%%%%%%%%%%%%%%%%%%%%%%%%%%%%%%%%%%%%%%%%%%%%%%%%%%%%%%%%%%%%%%%%%%%%%%%%%

\section{\label{sec:numerical_methods} Numerical methods}

Simulations were performed using the in-house Hydrodynamics Adaptive Mesh Refinement Simulator (HAMeRS). The parallelization of the code and all the construction, management, and storage of cells are facilitated by the Structured Adaptive Mesh Refinement Application Infrastructure (SAMRAI) library~\cite{gunney2016advances, gunney2006parallel, hornung2006managing, hornung2002managing, wissink2001large} from Lawrence Livermore National Laboratory (LLNL). Explicit form of sixth order localized dissipation weighted compact nonlinear scheme (WCNS)~\cite{wong2017high}, which is designed to add sufficient dissipation around shocks and regions with large gradients for numerical stability and minimum dissipation in smooth regions to capture vortical features, is used for the hyperbolic part of the governing equations. Explicit sixth order finite differences are used to compute the derivatives of diffusive and viscous fluxes in non-conservative form. A third order total variation diminishing Runge--Kutta (RK-TVD) scheme~\cite{shu1989efficient} is adopted for time integration. Various sensors are employed to identify regions of interest for adaptive mesh refinement (AMR). These include gradient sensor on pressure field and wavelet sensor~\cite{wong2016multiresolution} on density field to detect shock waves and mixing regions, respectively. An additional sensor based on mass fractions is also used to assist the detection of mixing regions. Further details on the numerical methods and the sensors for mesh refinement are given in appendices~\ref{sec:spatial_numerics} and \ref{sec:AMR_sensors}, respectively.

%%%%%%%%%%%%%%%%%%%%%%%%%%%%%%%%%%%%%%%%%%%%%%%%%%%%%%%%%%%%%%%%%%%%%%%%%%%%%%%%
%%%%%%%%%%%%%%%%%%%%%%%%%%%%%%%%%%%%%%%%%%%%%%%%%%%%%%%%%%%%%%%%%%%%%%%%%%%%%%%%
%%%%%%%%%%%%%%%%%%%%%%%%%%%%%%%%%%%%%%%%%%%%%%%%%%%%%%%%%%%%%%%%%%%%%%%%%%%%%%%%

\section{\label{sec:numerical_setup} Numerical setup}

\begin{figure}[!ht]
\centering
\subfigure[$\ $2D configuration]{%
\includegraphics[width=0.5\textwidth]{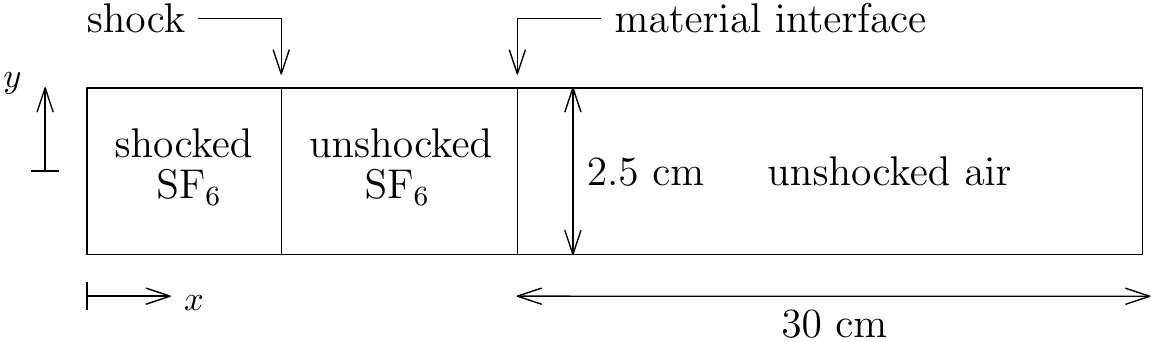} \label{fig:2D_configuration}}
\subfigure[$\ $3D configuration]{%
\includegraphics[width=0.5\textwidth]{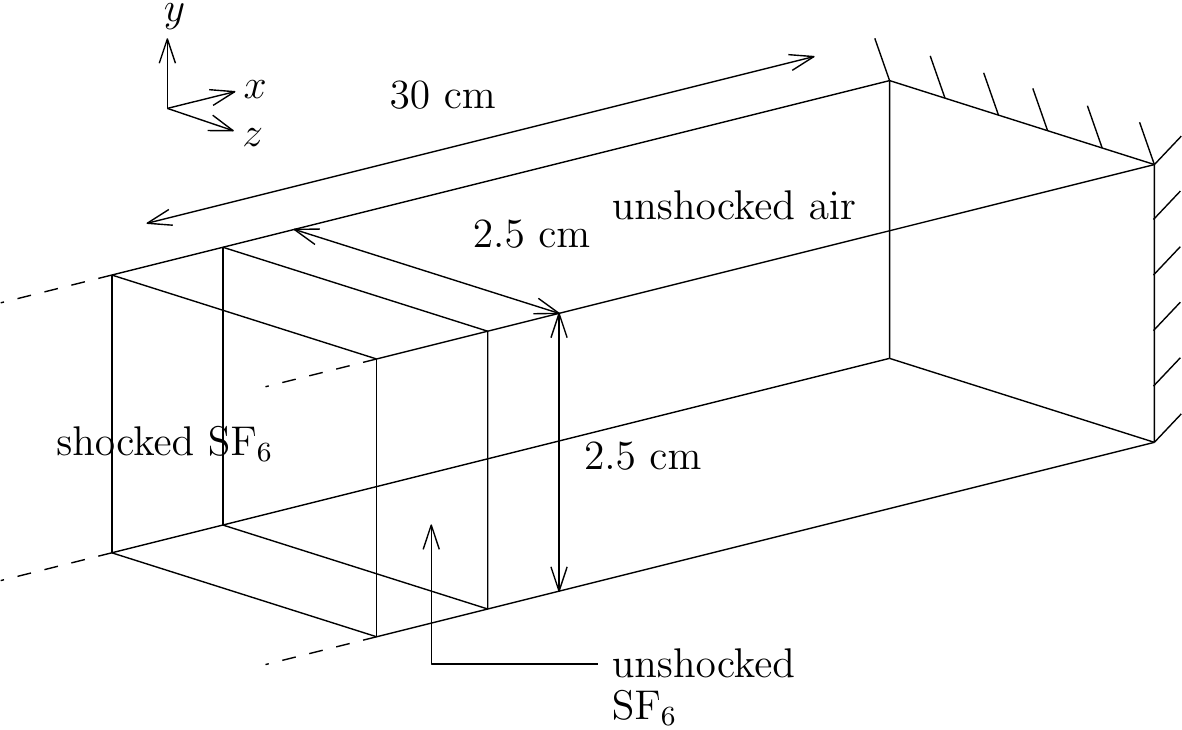} \label{fig:3D_configuration}}
\caption{Schematic diagrams of initial flow fields and computational domains for the 2D and 3D simulations.}
\label{fig:2D_3D_configuration}
\end{figure}

In our 2D numerical set-up, the height of the domain in the transverse direction is chosen to be $2.5\ {\mathrm{cm}}$. To facilitate the comparison between 2D and 3D results in later sections, the 3D numerical set-up is chosen to be the extension of the 2D set-up with a transverse sectional area of $2.5\ {\mathrm{cm}} \times 2.5\ {\mathrm{cm}}$. In both 2D and 3D simulations, a shock wave with Mach number $Ma = 1.45$ is launched initially in a sulphur hexafluoride ($\mathrm{SF_6}$) region and interacts with a diffuse material interface between ideal gases $\mathrm{SF_6}$ and air. The computational domain and initial conditions for the 3D simulations are shown in figure~\ref{fig:3D_configuration}. Boundaries are periodic in the transverse directions and reflective boundary conditions are applied at the end wall. The length of the numerical shock tube is chosen to be large enough ($50\ \mathrm{cm}$) such that no features leave the domain at the open-sided boundary during the simulations. The 2D configuration (figure~\ref{fig:2D_configuration}) is basically a cross-sectional view of the 3D set-up.

The pre-shocked conditions correspond to $T=298\ {\mathrm{K}}$ and $p=101325\ {\mathrm{Pa}}$. The fluids are initially at rest in the simulation reference frame. Table~\ref{tab:IC} shows the initial conditions in different portions of the domain. The initial Atwood number, $At = ( \rho_{\mathrm{SF_6}} - \rho_{\mathrm{air}} ) / ( \rho_{\mathrm{SF_6}} + \rho_{\mathrm{air}} )$, across the interface is 0.68. The density, pressure, temperature, and Mach number relations across the initial shock in $\mathrm{SF_6}$ in the shock reference frame are given by the Rankine-Hugoniot jump conditions:

\begin{align}
    \frac{\rho^{\prime}}{\rho} &= \frac{ \left( \frac{\gamma_{\mathrm{SF_6}} + 1}{2} \right) Ma^2 }{ 1 + \left( \frac{\gamma_{\mathrm{SF_6}} - 1}{2} \right) Ma^2 }, \\
    \frac{p^{\prime}}{p} &= \frac{\frac{2\gamma_{\mathrm{SF_6}}}{\gamma_{\mathrm{SF_6}}-1} Ma^2 - 1}{\frac{\gamma_{\mathrm{SF_6}}+1}{\gamma_{\mathrm{SF_6}}-1}}, \\
    \frac{T^{\prime}}{T} &= \left( \frac{\gamma_{\mathrm{SF_6}} Ma^2 - \left( \frac{\gamma_{\mathrm{SF_6}}-1}{2} \right)}{\frac{\gamma_{\mathrm{SF_6}}+1}{2}} \right) \left( \frac{ 1 + \left( \frac{\gamma_{\mathrm{SF_6}} - 1}{2} \right) Ma^2 }{ \left( \frac{\gamma_{\mathrm{SF_6}} + 1}{2} \right) Ma^2 } \right), \\
    {Ma^{\prime}}^2 &= \left( \frac{1 + \left( \frac{\gamma_{\mathrm{SF_6}} - 1}{2} \right) Ma^2 }{ \gamma_{\mathrm{SF_6}} Ma^2 - \left( \frac{ \gamma_{\mathrm{SF_6}} - 1 }{2} \right) } \right),
\end{align}

\noindent where quantities without primes are pre-shock quantities and those with primes are post-shock quantities.

\begin{table}[!ht]
\caption{\label{tab:IC}%
Initial conditions of the post-shock state and the pre-shock states of the light- and heavy-fluid sides.}
\begin{ruledtabular}
\begin{tabular}{ c c c c  }
 Quantity & Post-shock $\mathrm{SF_6}$ & Pre-shock $\mathrm{SF_6}$ & Air \\ 
 \hline
 $\rho\ ({\mathrm{kg\ m^{-3}}})$ & 11.97082 & 5.972866 & 1.145601 \\
 $p\ ({\mathrm{Pa}})$ & 218005.4 & 101325.0 & 101325.0 \\ 
 $T\ ({\mathrm{K}})$ & 319.9084 & 298.0 & 298.0 \\
 $u\ ({\mathrm{m\ s^{-1}}})$ & 98.93441 & 0.0 & 0.0 \\
\end{tabular}
\end{ruledtabular}
\end{table}

\begin{figure}[!ht]
\centering
\includegraphics[width=0.4\textwidth]{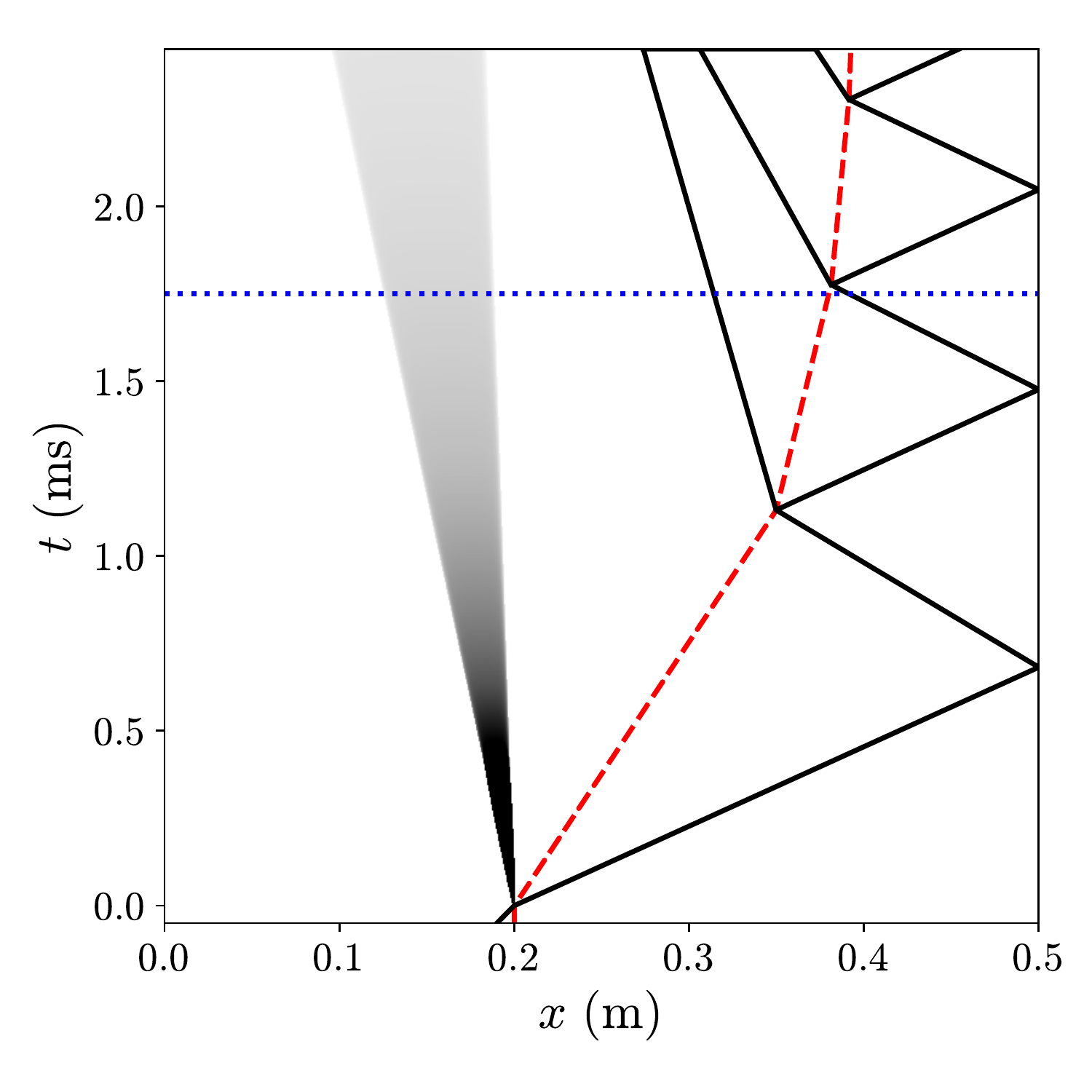}
\caption{$x$-$t$ diagram showing the propagation of material interface, shock waves, and rarefraction. Red dashed line: material interface; black lines: shock waves; gray region: rarefraction. The blue dotted line indicates the end time of all simulations.}
\label{fig:x_t_diagram}
\end{figure}

Figure~\ref{fig:x_t_diagram} shows the space-time ($x$-$t$) diagram for different features in a one-dimensional (1D) flow representation. $t=0$ is set as the instance of the first arrival of the shock at the interface. The initial position of the incident shock is chosen such that the first shock-interface interaction occurs at $0.05\ {\mathrm{ms}}$ after the start of simulation. Since the shock traverses the interface from the heavy fluid side, a shock is transmitted to the light fluid side and a rarefaction wave is reflected back to the heavy fluid side. The shock wave also accelerates the interface towards the end wall. After hitting the end wall, the shock is reflected back and interacts with the interface again at $t \approx 1.15\ {\mathrm{ms}}$. Since the reflected shock passes from the light fluid side during the re-shock of interface, a transmitted shock and a reflected shock are generated. The reflected shock will eventually lead to a second re-shock. The simulations are stopped at $t = 1.75\ {\mathrm{ms}}$, which is the time just before second re-shock, as the grid resolution requirements become too large to accurately capture this flow stage.

In order to trigger the flow instability, perturbations are seeded on the material interfaces. Multi-mode perturbations for 2D and 3D simulations are given by the following two equations:

\begin{equation}
S(y) = A \sum_m \cos \left( \frac{2\pi m}{L_{y}} y + \phi_m \right),
\label{eqn:2D_perturbation}
\end{equation}

\begin{equation}
S(y,z) = A \sum_{m} \cos \left( \frac{2\pi m}{L_{yz}} y + \phi_m \right)\cos \left( \frac{2\pi m}{L_{yz}} z + \psi_m \right),
\label{eqn:3D_perturbation}
\end{equation}

\noindent where $L_{y} = L_{yz} = 2.5\ {\mathrm{cm}}$. In both 2D and 3D simulations, there are 11 perturbation modes with wavenumber $m$ between 20 and 30 in each transverse direction. Constant amplitude $A=\sqrt{2} \times 0.01\ {\mathrm{mm}}$ is used for each mode and random phase shifts $\phi_m$ and $\psi_m$ between 0 and $2\pi$ are added to each mode to prevent adding up of harmonics. The perturbation for the 3D case is a combination of ``egg-crate" modes, which are mode forms commonly used to model perturbations created by wire meshes in experiments~\cite{vetter1995experiments, poggi1998velocity, prasad2000late}. However, instead of using a single mode, we have chosen a narrowband of ``egg-crate" modes with different phases to account for possible variation in wire mesh spacing in a physical experiment and to facilitate symmetry breaking. Another approach commonly used in previous studies~\cite{hill2006large, grinstein2011simulations, tritschler2014richtmyer} to break symmetry is to superpose a small random perturbation on a single dominant ``egg-crate" mode. Note that all 3D simulations presented in this work have the exact perturbation with same set of relative phases.

A finite thickness is prescribed for the initial interface between $\mathrm{SF_6}$ and air. The partial density of each gas, $\rho Y_i$, is smoothed across the material interface initially with the following formulation:
\begin{align}
	f_{sm} &= \frac{1}{2} \left(1 + \erf\left(\frac{x - L_i - S}{\epsilon_i}\right) \right), \\
    \left( \rho Y_i \right)_{sm} &= \left( \rho Y_i \right)_{L}\left(1 - f_{sm}\right) + \left(\rho Y_i \right)_{R}f_{sm},
\end{align}

\noindent where $L_i$ is the initial location of the material interface. $\epsilon_i$ is the characteristic thickness of the initial material interface and is set to $1\ \mathrm{mm}$. Therefore, the initial thickness has the same order of magnitude as the wavelengths of the perturbation modes. In the RMI simulations by~\citet{lombardini2012transition} and~\citet{tritschler2014richtmyer}, the initial interface thicknesses are also comparable to the wavelengths of the dominant perturbation modes. Note that the relatively large initial diffuse interface thickness will lower the early time growth rate compared to the sharp interface case~\cite{duff1962effects, brouillette1994experiments}, but the study of this effect is not the focus of this work.

To reduce the statistical variability associated with the very few modes present at the 1D interface, all statistical results for the 2D simulations are obtained by ensemble averaging over 24 realizations with different random phase shifts $\phi_m$ in the initial perturbations. We have found that 24 realizations are sufficient for the statistical convergence of quantities of interests. The details are discussed in appendix~\ref{sec:2D_stat_convergence}.

%%%%%%%%%%%%%%%%%%%%%%%%%%%%%%%%%%%%%%%%%%%%%%%%%%%%%%%%%%%%%%%%%%%%%%%%%%%%%%%%
%%%%%%%%%%%%%%%%%%%%%%%%%%%%%%%%%%%%%%%%%%%%%%%%%%%%%%%%%%%%%%%%%%%%%%%%%%%%%%%%
%%%%%%%%%%%%%%%%%%%%%%%%%%%%%%%%%%%%%%%%%%%%%%%%%%%%%%%%%%%%%%%%%%%%%%%%%%%%%%%%

\section{\label{sec:characteristic scales_2D_3D} Characteristic scales of two-dimensional and three-dimensional problems}

In this section, the characteristic length scales, time scales, and growth rates of the 2D and 3D problems are determined from the initial conditions and the linear theory.

The growth rate $\dot{\eta}_{\mathrm{imp}}^{2D}$ for a 2D small single-mode perturbation with initial amplitude $\eta_0^{2D}$ from an impulsive acceleration on a discontinuous interface obtained from linear theory by Richtmyer~\cite{richtmyer1960taylor} is given by:
\begin{equation}
    \dot{\eta}_{\mathrm{imp}}^{2D} = \frac{2\pi}{\lambda^{2D}} U_i At \eta_0^{2D}, \label{eq:2D_imp_growth_rate}
\end{equation}
\noindent where $\lambda^{2D}$ is the wavelength of the perturbation and $U_i$ is the change in interface velocity induced by the incident shock wave. It can be extended to the growth rate of a 3D single-mode interface as:
\begin{equation}
    \dot{\eta}_{\mathrm{imp}}^{3D} = \sqrt{2} \frac{2\pi}{\lambda^{3D}} U_i At \eta_0^{3D}, \label{eq:3D_imp_growth_rate}
\end{equation}

\noindent where $\lambda^{3D}=\lambda_y^{3D}=\lambda_z^{3D}$ is the wavelength of the perturbation in either $y$ or $z$ directions.

Although material interfaces are smoothed initially in the numerical setup, the growth rates from linear theory still become relevant after the initial diffusive growth, as the initial characteristic thicknesses of all cases in this work are the same. In this work, the equivalent single-mode amplitudes $\eta_0^{2D}$ and $\eta_0^{3D}$ are estimated as the standard deviations of the initial perturbations.

The equivalent single-mode wavelengths, $\lambda^{2D}$ and $\lambda^{3D}$, can be estimated from the integral length scales. Since the initial perturbations are periodic with only few modes, the integral length scales are computed from the energy spectra of the perturbations, instead of the autocorrelations. This is necessary, as the autocorrelation of a periodic signal is itself periodic and does not approach zero when based on only few modes, no matter how large the computational domain is. The integral length scales for the 2D and 3D cases, $l^{2D}$ and $l^{3D}$, are computed as:
\begin{equation}
    l^{2D}/l^{3D} = \frac{2\pi}{\int^{\infty}_{0} E(k) \, dk} \int^{\infty}_{0} \frac{E(k)}{k} \, dk,
\end{equation}

\noindent where $E(k)$ is the energy spectrum of the perturbation that depends on angular wavenumber $k$. Note that $E(k)$ is the radial energy spectrum in the $yz$ plane and $k$ is the radial angular wavenumber for the 3D case. The wavelengths $\lambda^{2D}$ and $\lambda^{3D}$ are then estimated as:
\begin{equation}
    \lambda^{2D} = l^{2D}, \quad \lambda^{3D} = \sqrt{2} \, l^{3D}.
\end{equation}

The characteristic time scales can be obtained from the impulsive growth rates as:
\begin{equation}
    \tau^{2D} = \frac{\lambda^{2D}}{\dot{\eta}_{\mathrm{imp}}^{2D}},
    \quad \tau^{3D} = \frac{\lambda^{3D}}{\dot{\eta}_{\mathrm{imp}}^{3D}}.
\end{equation}

Table~\ref{tab:characteristic scales} shows the standard deviations, wavelengths, growth rates from impulsive theory, and characteristic time scales from the initial conditions of the 2D and 3D cases. It can be seen from the table that the growth rates and characteristic time scales for both 2D and 3D cases are virtually identical with the initial conditions chosen. $\eta_0$ values for both cases are smaller than 5\% of the corresponding $\lambda$ value. Therefore, it is expected that the perturbations grow linearly after first shock. Note that in many previous simulations, including those by~\citet{lombardini2012transition} and~\citet{tritschler2014richtmyer}, the initial amplitudes of the perturbations are more than 10\% of the wavelengths and hence the perturbations are already in the nonlinear regime initially.

A method that can take the mode compression and coupling effects into account to estimate initial growth rates was proposed by~\citet{weber2013growth}. The 2D and 3D growth rates for discontinuous interfaces estimated with the corresponding method are, respectively, 10.3 $\mathrm{m\ s^{-1}}$ and 10.0 $\mathrm{m\ s^{-1}}$.  The 2D and 3D initial growth rates measured from the mixing widths (defined in section~\ref{sec:grid_sensitivity}) computed from the simulations are 0.86 $\mathrm{m\ s^{-1}}$ and 0.82 $\mathrm{m\ s^{-1}}$, respectively. The essentially identical initial growth rates for the 2D and 3D problems estimated with different methods show that the 2D and 3D initial conditions are designed appropriately for fair comparison between the two problems. The growth rates measured from the simulation results are much smaller than the ones given by the impulsive theory or method by~\citet{weber2013growth} for discontinuous interfaces due to the large initial diffuse interface thicknesses. For simplicity, we have chosen the impulsive growth rates given by equations~\eqref{eq:2D_imp_growth_rate} and \eqref{eq:3D_imp_growth_rate} as the characteristic growth rates to normalize physical quantities in this work.

\begin{table}[!ht]
\caption{\label{tab:characteristic scales}%
Initial characteristic scales for the 2D and 3D cases.}
\begin{ruledtabular}
\begin{tabular}{ c c c c c }
 Case & $\eta_0\ (\mathrm{mm})$ & $\lambda\ (\mathrm{mm})$ & $\dot{\eta}_{\mathrm{imp}}\ (\mathrm{m\ s^{-1}})$ & $\tau\ (\mathrm{ms})$ \\
 \hline
 2D & 0.0331 & 1.02 & 18.4 & 0.0552 \\
 3D & 0.0234 & 1.02 & 18.4 & 0.0551 \\
\end{tabular}
\end{ruledtabular}
\end{table}

%\begin{table}[!ht]
%\caption{\label{tab:characteristic scales}%
%Initial characteristic scales for two-dimensional and three-dimensional cases.}
%\begin{ruledtabular}
%\begin{tabular}{ c c c }
% Case & 2D & 3D \\ 
% \hline
% $\eta_0\ (\mathrm{mm})$                  & 0.0331 & 0.0234 \\
% $\lambda\ (\mathrm{mm})$                 & 1.02   & 1.02   \\ 
% $\dot{\eta}_{\mathrm{imp}}\ (\mathrm{m\ s^{-1}})$ & 18.4   & 18.4   \\
% $\tau\ (\mathrm{ms})$                    & 0.0552 & 0.0551 \\
%\end{tabular}
%\end{ruledtabular}
%\end{table}

%%%%%%%%%%%%%%%%%%%%%%%%%%%%%%%%%%%%%%%%%%%%%%%%%%%%%%%%%%%%%%%%%%%%%%%%%%%%%%%%
%%%%%%%%%%%%%%%%%%%%%%%%%%%%%%%%%%%%%%%%%%%%%%%%%%%%%%%%%%%%%%%%%%%%%%%%%%%%%%%%
%%%%%%%%%%%%%%%%%%%%%%%%%%%%%%%%%%%%%%%%%%%%%%%%%%%%%%%%%%%%%%%%%%%%%%%%%%%%%%%%

\section{\label{sec:grid_sensitivity} Grid sensitivity analysis}

\begin{figure*}[!ht]
\centering
\subfigure[$\ t=1.10\ \mathrm{ms}$]{%
\includegraphics[width=0.45\textwidth]{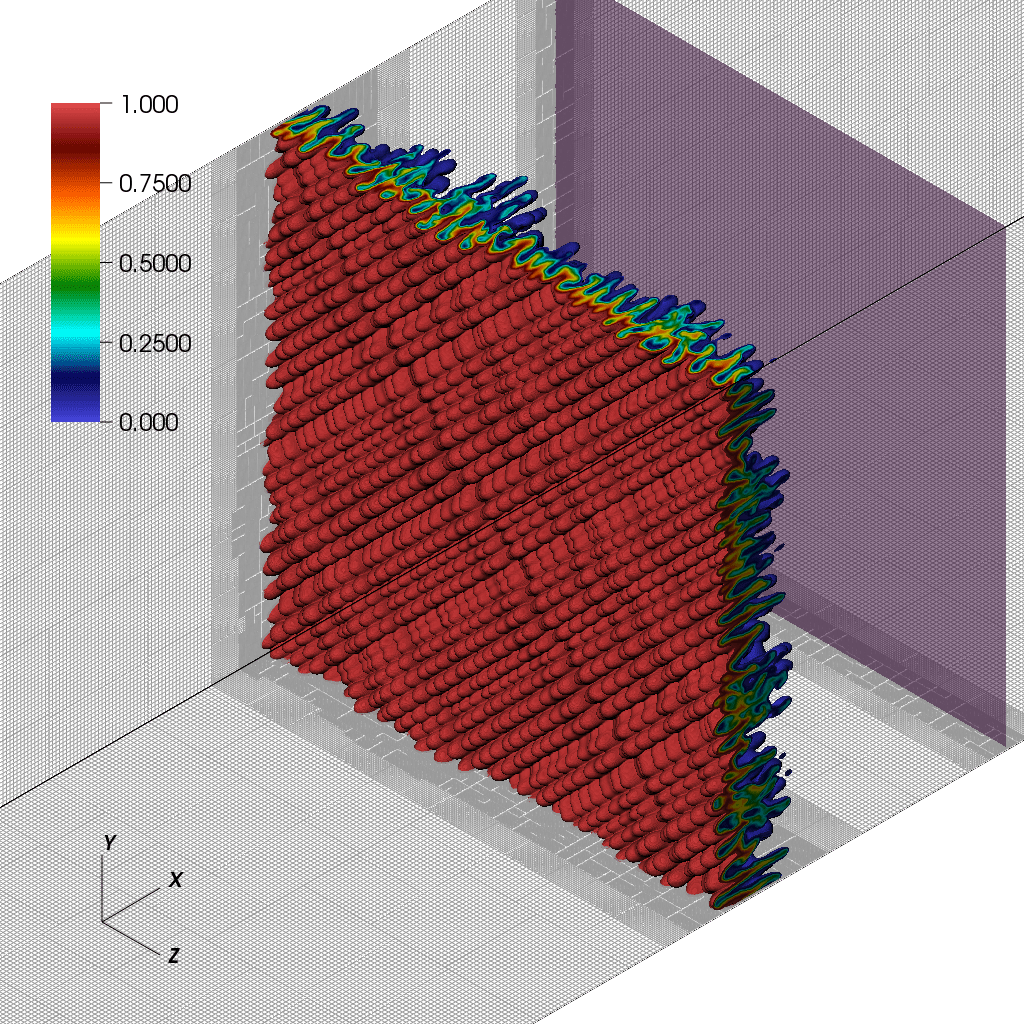}}
\subfigure[$\ t=1.75\ \mathrm{ms}$]{%
\includegraphics[width=0.45\textwidth]{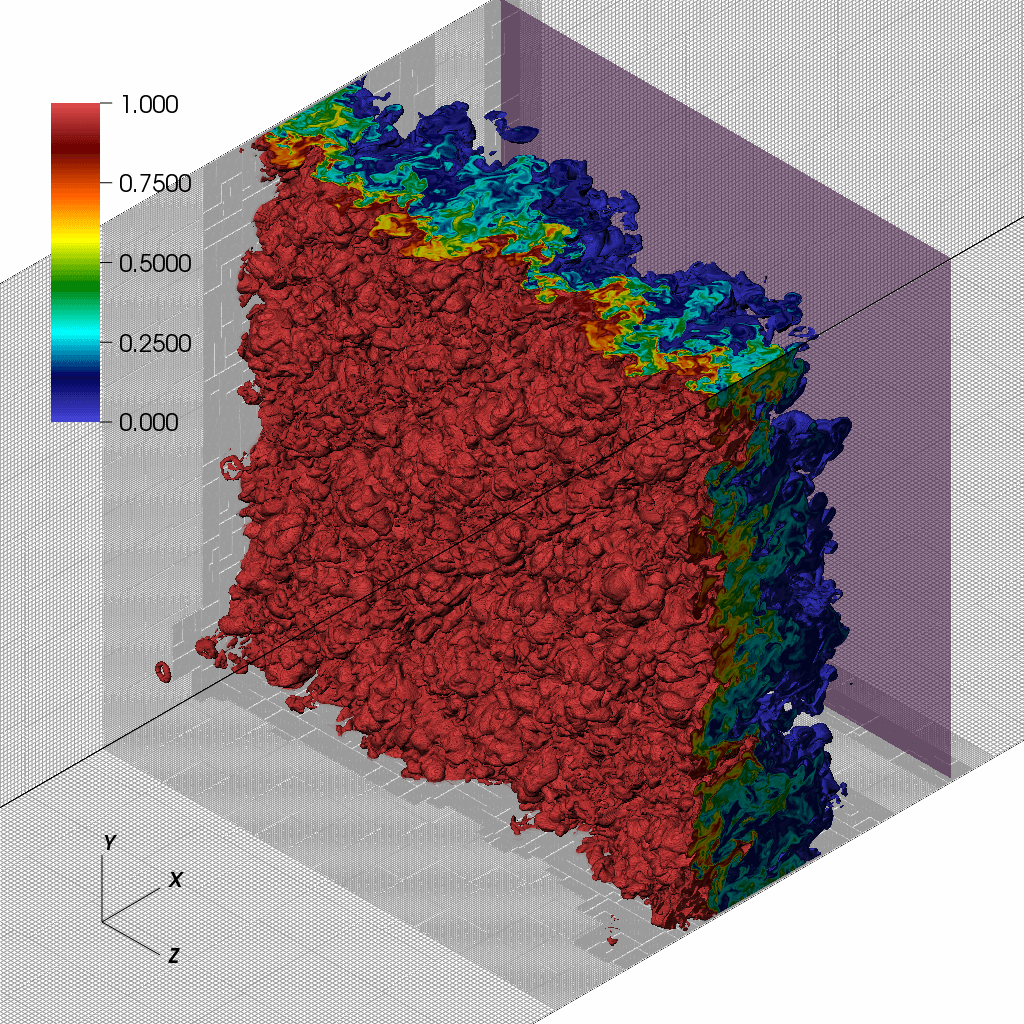}}
\caption{Isovolumes of the $\mathrm{SF_6}$ mole fraction, $X_{\mathrm{SF_6}}$, at different times for the 3D problem with grid D. The colorbars indicate the value of $X_{\mathrm{SF_6}}$. The transparent magenta planes represent the locations of shocks. The AMR grid hierarchy is shown on the side walls of the domain.}
\label{fig:3D_plot}
\end{figure*}

To understand the sensitivities of numerical results to the grid spacing, a grid sensitivity study is conducted for both 2D and 3D problems. Tables~\ref{tab:2D_grids} and \ref{tab:3D_grids} show the grids used in the grid sensitivity analysis. In all grid settings, there are three levels of meshes in total, with two levels of mesh refinement. The refinement ratios in each direction are respectively 1:2 and 1:4 from base level to second level and from second level to the finest level. For the 2D problem, the base grids have number of grid points in the transverse direction varying from 128 points (grid D) to 1024 points (grid G). The highest resolution grid has grid spacing of $3.05\ \mathrm{\mu m}$ at the finest level. 24 realizations are conducted on each grid to reduce the statistical variability of the data. As for the 3D problem, the base grids have number of grid points in the transverse directions varying from 32 points (grid B) to 128 points (grid D). The highest resolution grid has grid spacing of $24.4\ \mathrm{\mu m}$ at the finest level. Figure~\ref{fig:3D_plot} shows the visualizations of the mixing layer with the shock waves and the AMR grid for the 3D problem just before re-shock and at the end of simulation after re-shock with grid D settings.

\begin{table}[!ht]
\caption{\label{tab:2D_grids}%
Different grids used for the 2D problem. Three levels of grids, with 1:8 overall refinement ratio in each direction, are used in all cases.}
\begin{ruledtabular}
\begin{tabular}{ c c c c }
Grid & Base grid resolution & Refinement ratios & Finest grid spacing ($\mathrm{\mu m}$) \\ 
\hline
D & $2560  \times 128$  & 1:2, 1:4 & 24.4 \\
E & $5120  \times 256$  & 1:2, 1:4 & 12.2 \\
F & $10240 \times 512$  & 1:2, 1:4 & 6.10 \\
G & $20480 \times 1024$ & 1:2, 1:4 & 3.05 \\
\end{tabular}
\end{ruledtabular}
\end{table}

\begin{table}[!ht]
\caption{\label{tab:3D_grids}%
Different grids used for the 3D problem. Three levels of grids with 1:8 overall refinement ratio in each direction are used in all cases.}
\begin{ruledtabular}
\begin{tabular}{ c c c c }
Grid & Base grid resolution & Refinement ratios & Finest grid spacing ($\mathrm{\mu m}$) \\ 
\hline
\iffalse
A & $320  \times 16  \times 16$  & 1:2, 1:4 & 195.3 \\
\fi
B & $640  \times 32  \times 32$  & 1:2, 1:4 & 97.7  \\
C & $1280 \times 64  \times 64$  & 1:2, 1:4 & 48.8  \\
D & $2560 \times 128 \times 128$ & 1:2, 1:4 & 24.4  \\
\end{tabular}
\end{ruledtabular}
\end{table}

The grid sensitivities of different statistical quantities are examined through comparison of their temporal evolution corresponding to different grids. The quantities analyzed include the mixing width, $W$, mixedness, $\Theta$, integrated turbulent kinetic energy, TKE, integrated scalar dissipation rate, $\chi$, and integrated enstrophy, $\Omega$. The mixing width and mixedness are defined as:
\begin{align}
    W &= \int 4 \bar{X}_{\mathrm{SF_6}} \left( 1 -  \bar{X}_{\mathrm{SF_6}} \right) dx, \label{eqn:mixing_width_definition} \\
    \Theta &= \frac{\int \overline{ X_{\mathrm{SF_6}} \left(1 - X_{\mathrm{SF_6}} \right) } \, dx}{\int \bar{X}_{\mathrm{SF_6}} \left( 1 -  \bar{X}_{\mathrm{SF_6}} \right) dx}, \label{eqn:mixedness_definition}
\end{align}
where $\bar{ \cdot }$ is the mean of the respective quantity in the homogeneous directions ($y$ direction for 2D and $yz$ plane for 3D). The mean is also computed using an additional averaging over the 24 realizations (ensemble averaging) for the 2D case.

The mathematical formulations for TKE, scalar dissipation rate, and enstrophy are given by:
\begin{align}
    \mathrm{TKE} &= \frac{1}{2} \rho u_i^{\prime\prime}u_i^{\prime\prime}, \\
    \chi &= D_{\mathrm{SF_6}} \nabla Y_{\mathrm{SF6}} \cdot \nabla Y_{\mathrm{SF6}}, \\
    \Omega &= \rho \bm{\omega} \cdot \bm{\omega},
\end{align}
\noindent where $u_i^{\prime\prime} = u_i - \tilde{u}_i$ and $\bm{\omega} = \nabla \times \bm{u}$ is the vorticity. The Favre (density-weighted) mean of the velocity is defined by $\tilde{u}_i = \overline{\rho u_i} / \bar{\rho} $. The integrated quantities are calculated over the full domain.

\begin{figure*}[!ht]
\centering
\subfigure[$\ $Mixing width]{%
\includegraphics[width=0.4\textwidth]{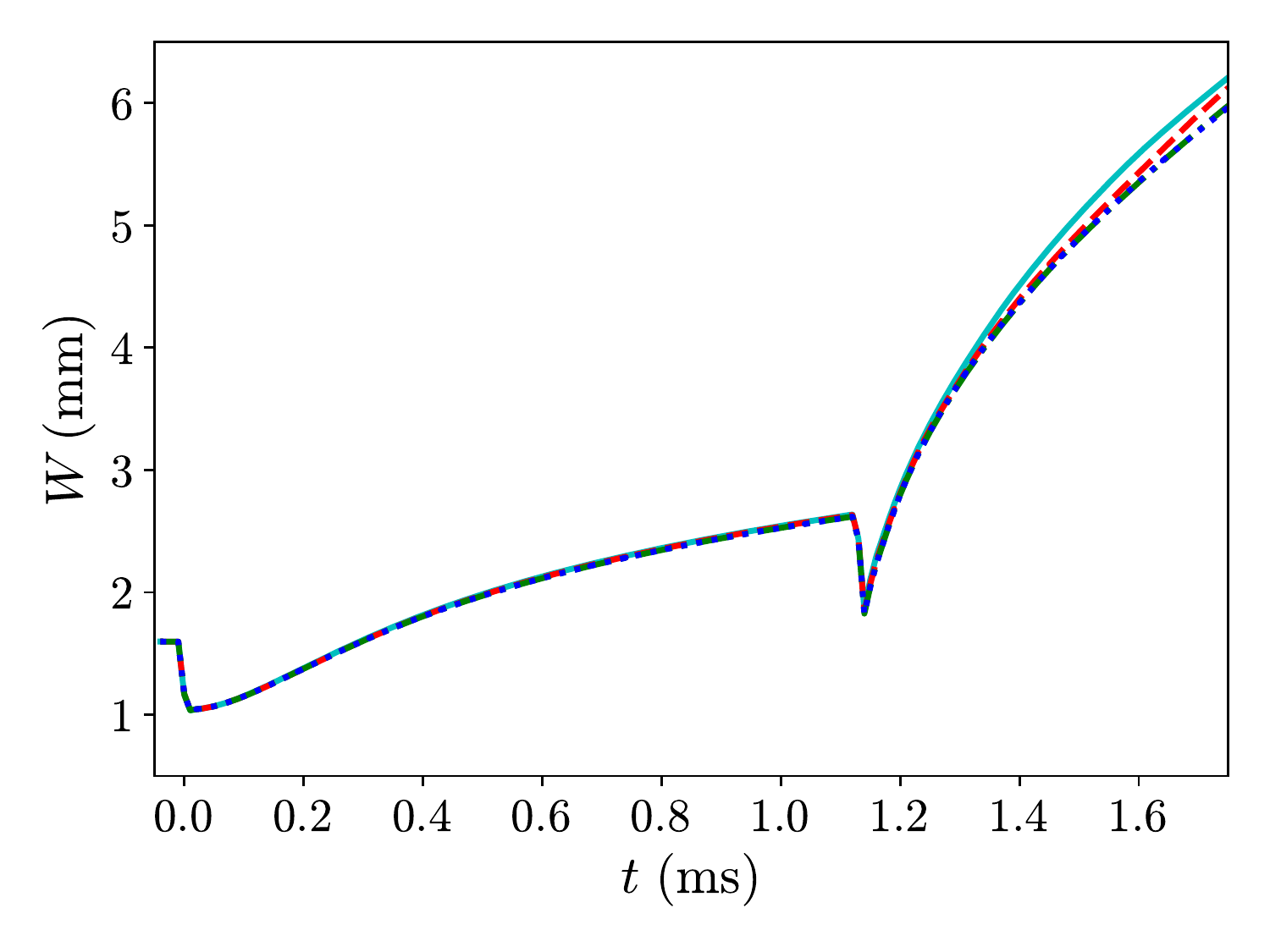}}
\subfigure[$\ $Mixedness]{%
\includegraphics[width=0.4\textwidth]{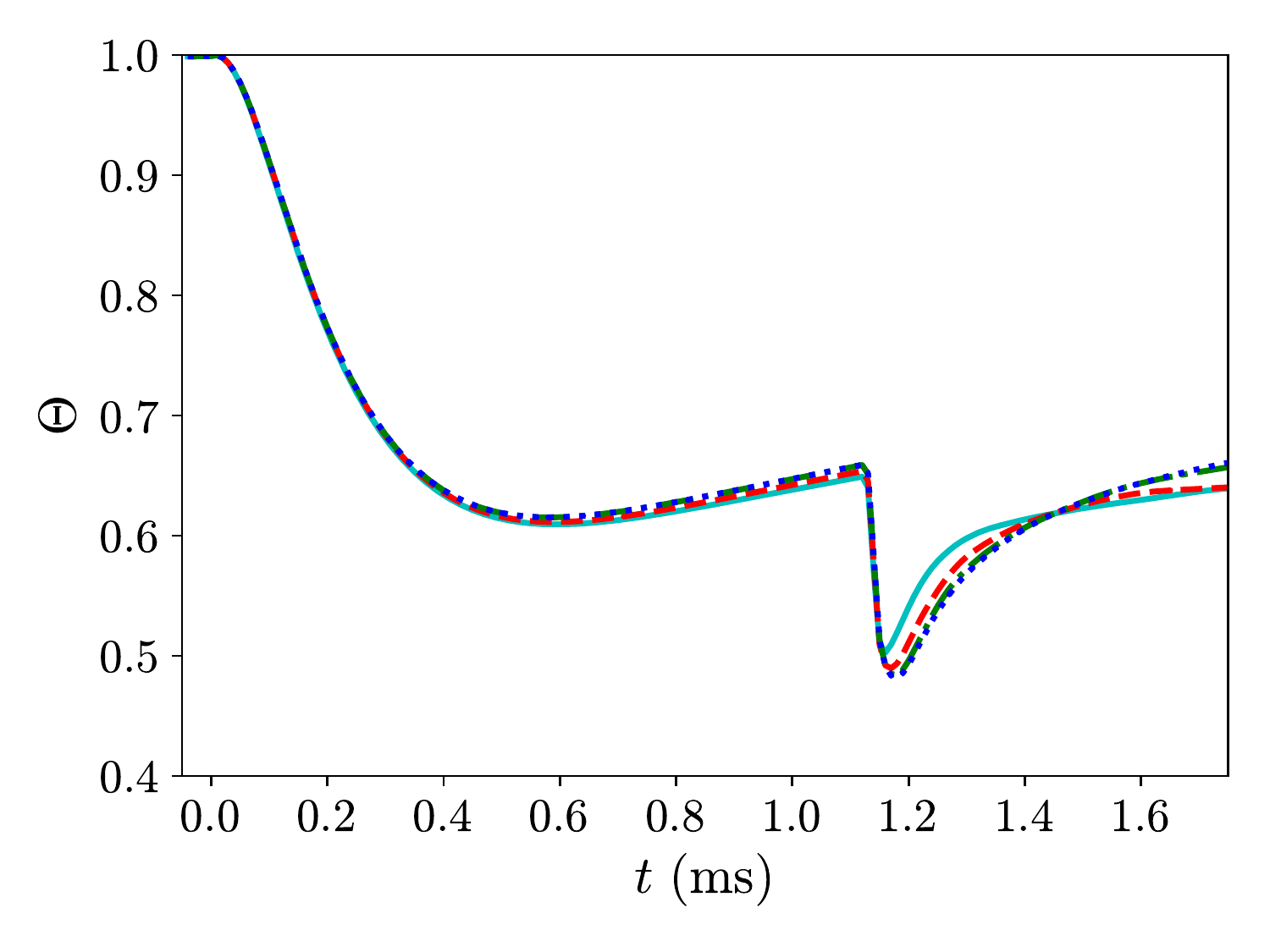}}
\subfigure[$\ $Integrated $\mathrm{TKE}$]{%
\includegraphics[width=0.4\textwidth]{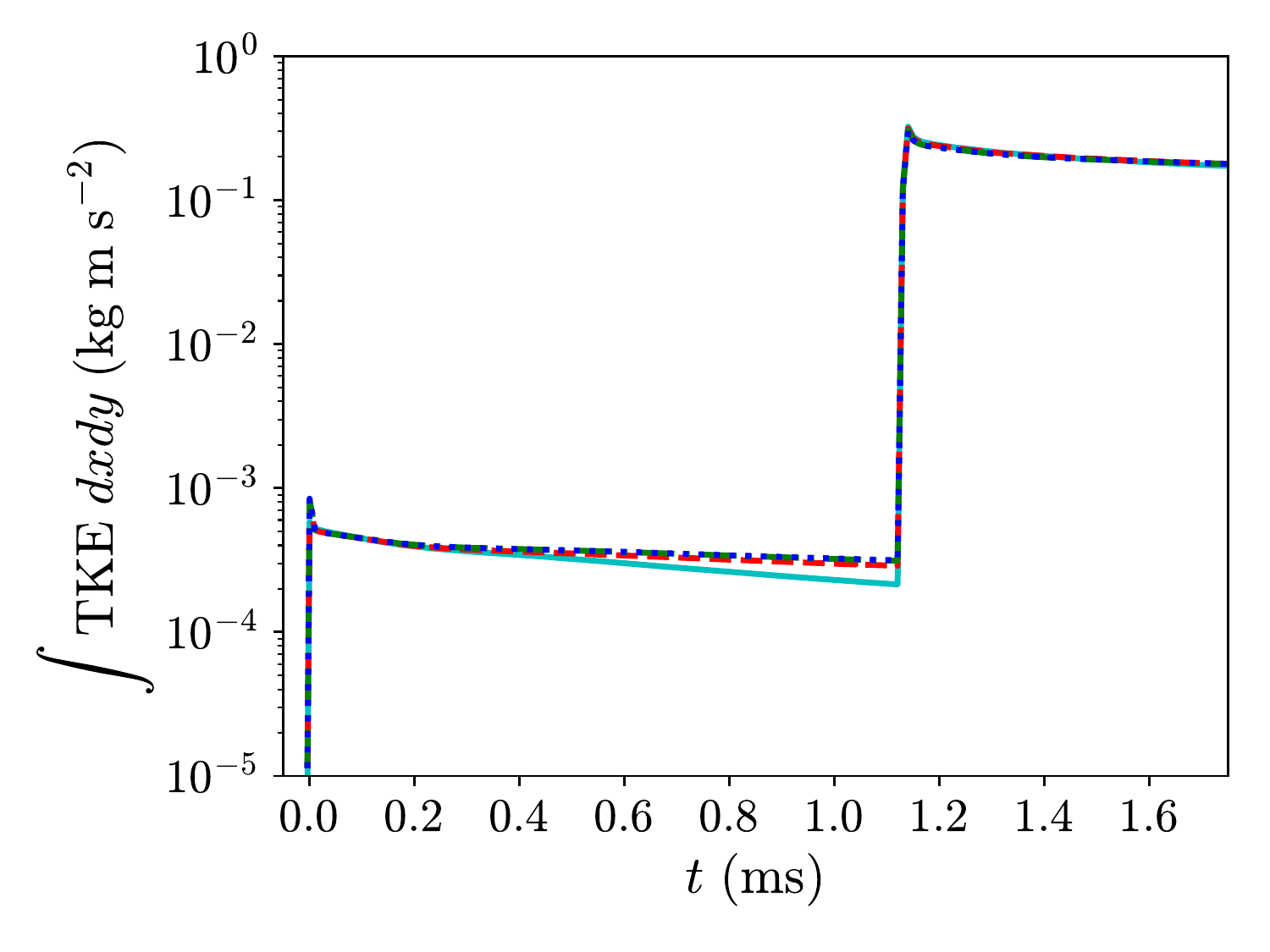}}
\subfigure[$\ $Integrated $\mathrm{TKE}$ (after re-shock)]{%
\includegraphics[width=0.45\textwidth]{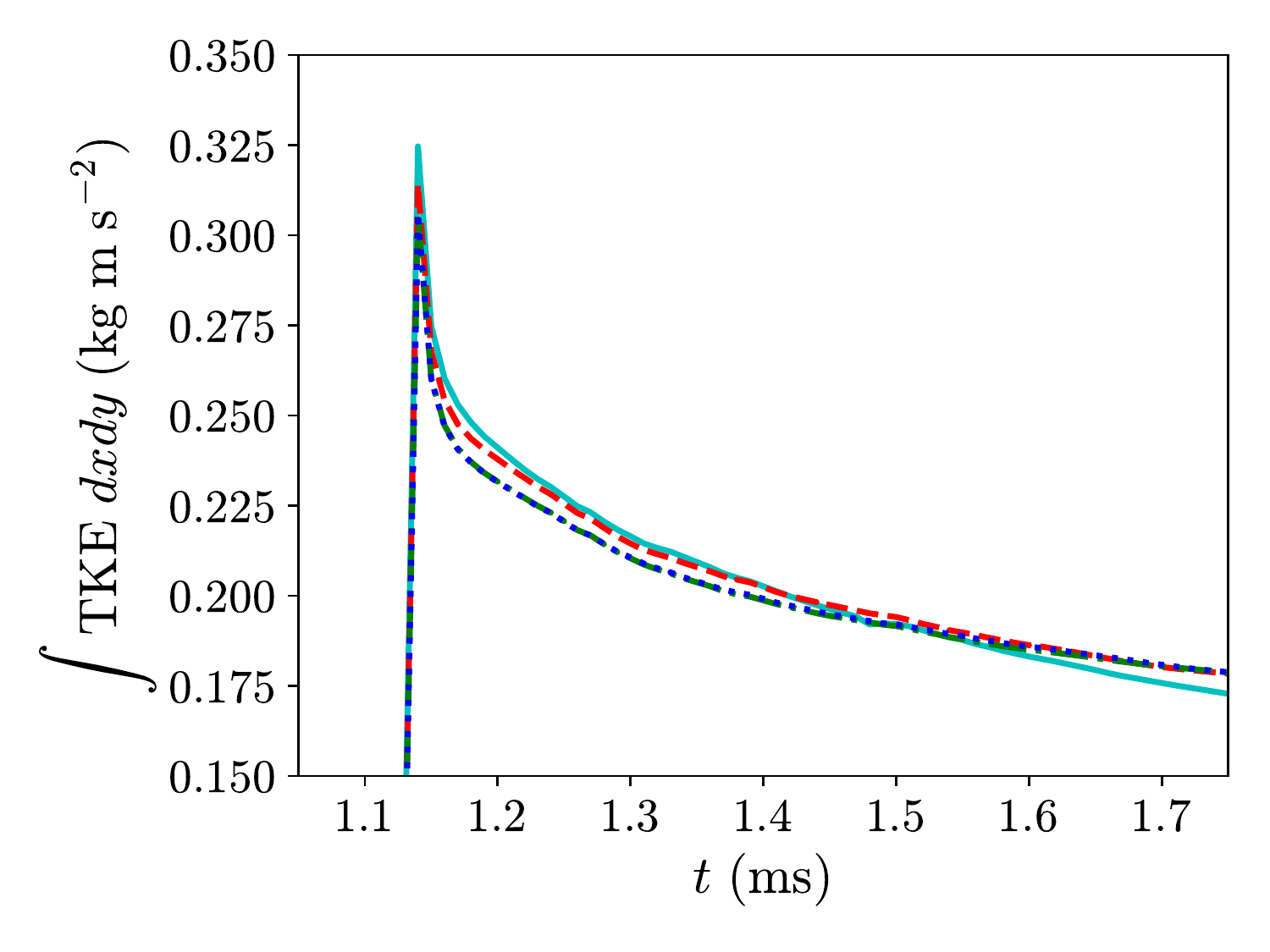}}
\subfigure[$\ $Integrated scalar dissipation rate]{%
\includegraphics[width=0.4\textwidth]{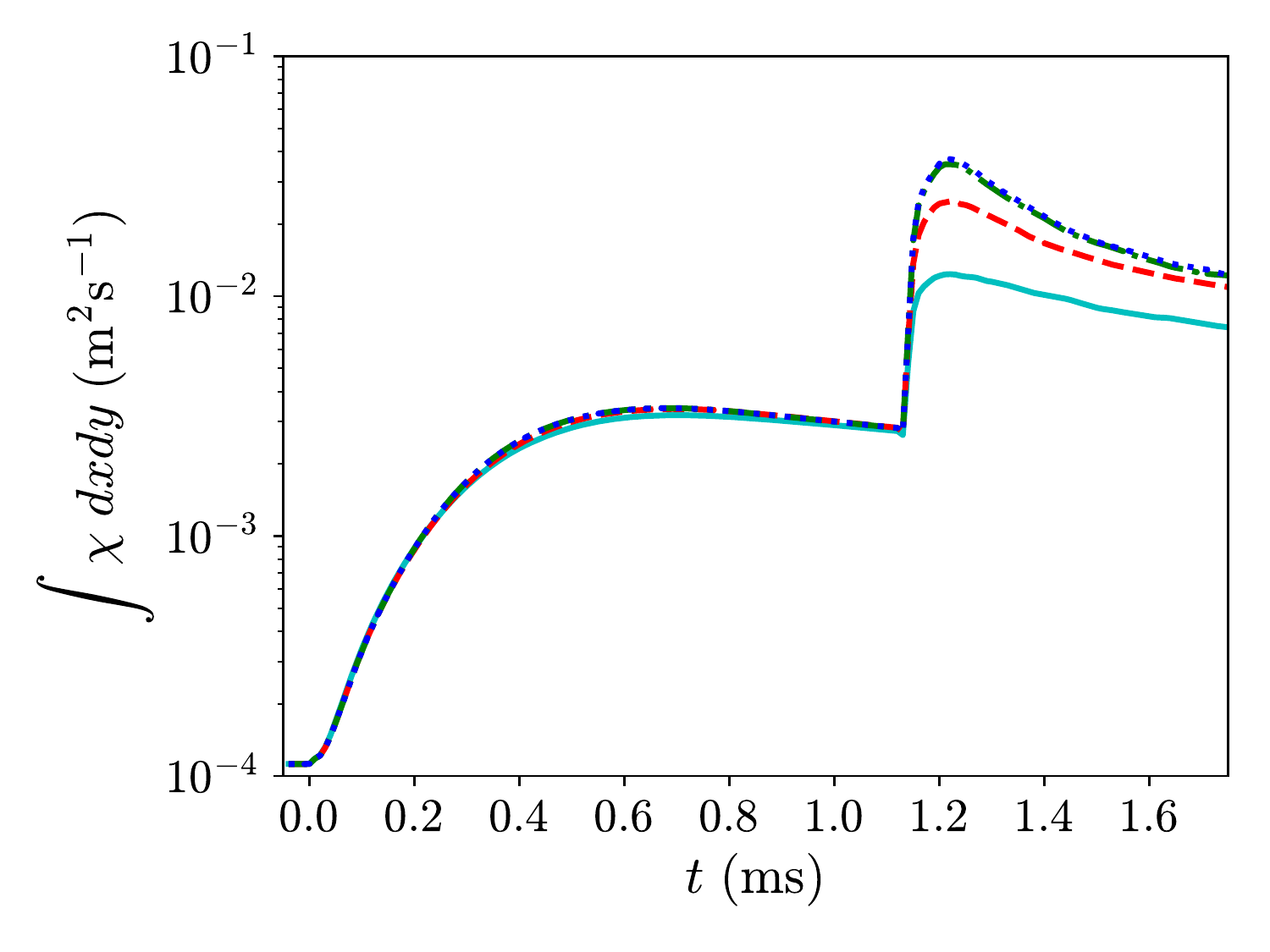}}
\subfigure[$\ $Integrated enstrophy]{%
\includegraphics[width=0.4\textwidth]{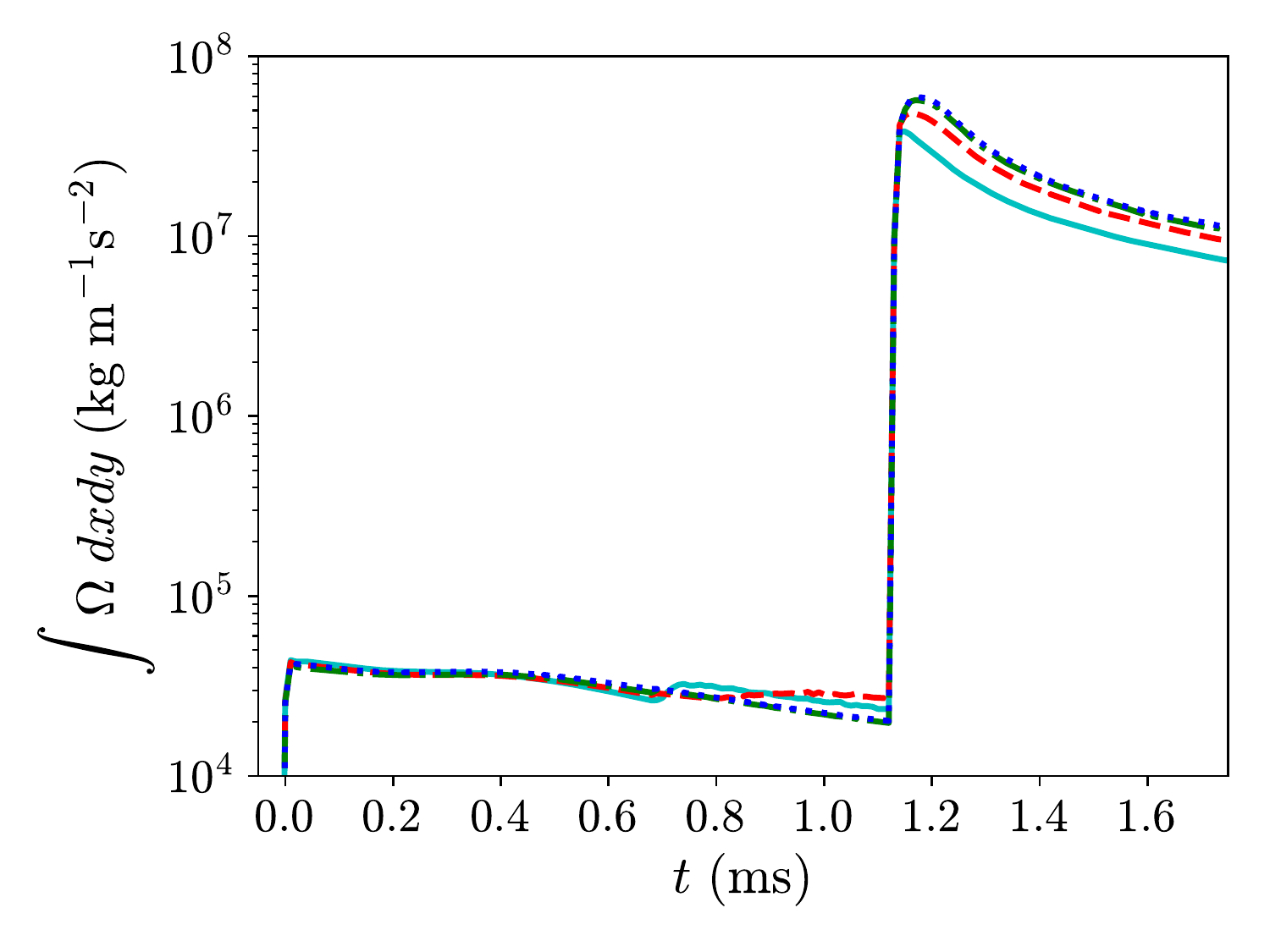}}
\caption{Grid sensitivities of different statistical quantities over time for the 2D problem. Cyan solid line: grid D; red dashed line: grid E; green dash-dotted line: grid F; blue dotted line: grid G.}
\label{fig:2D_grid_sensitivity}
\end{figure*}

\begin{figure*}[!ht]
\centering
\subfigure[$\ $Mixing width]{%
\includegraphics[width=0.4\textwidth]{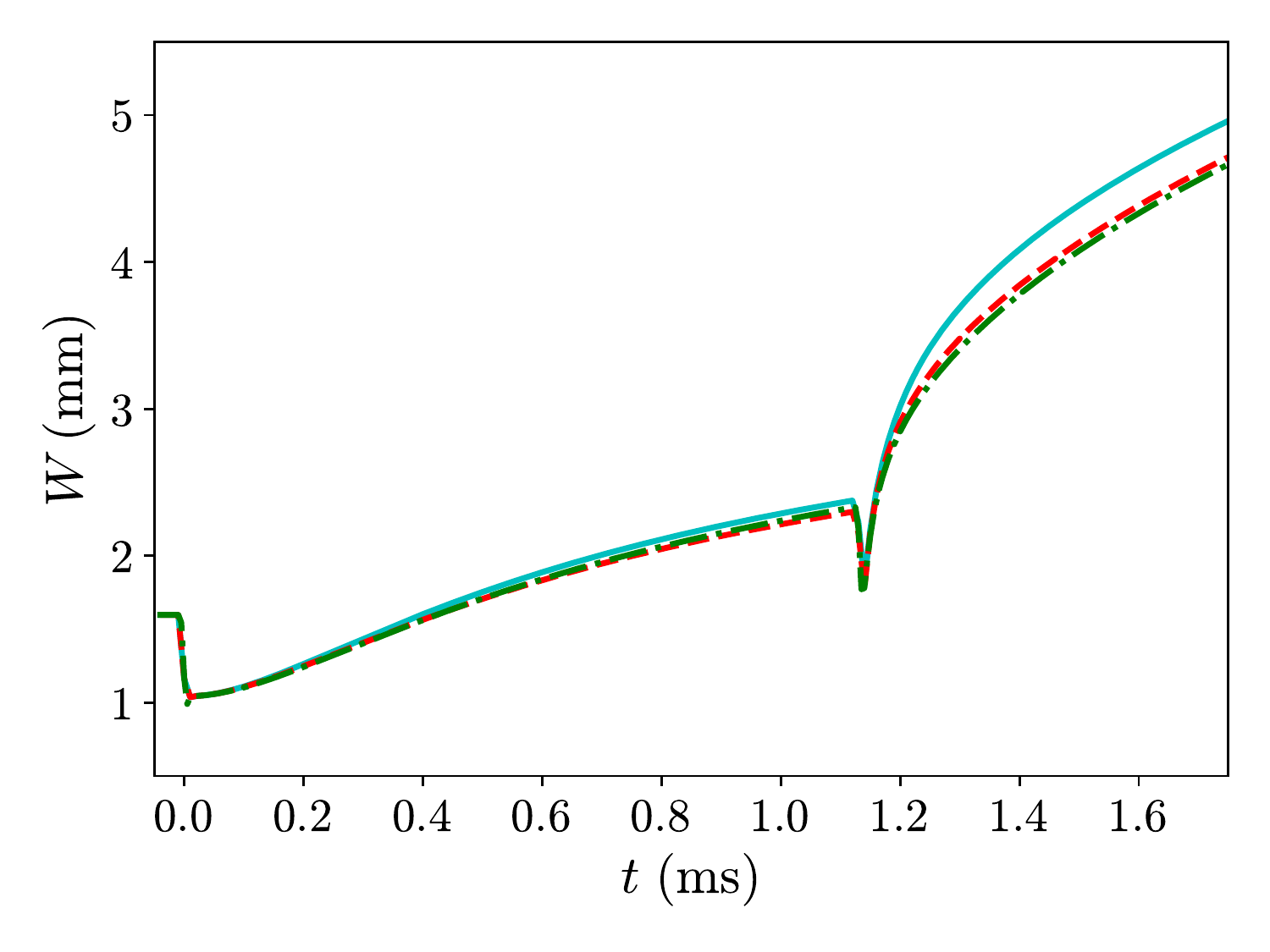}}
\subfigure[$\ $Mixedness]{%
\includegraphics[width=0.4\textwidth]{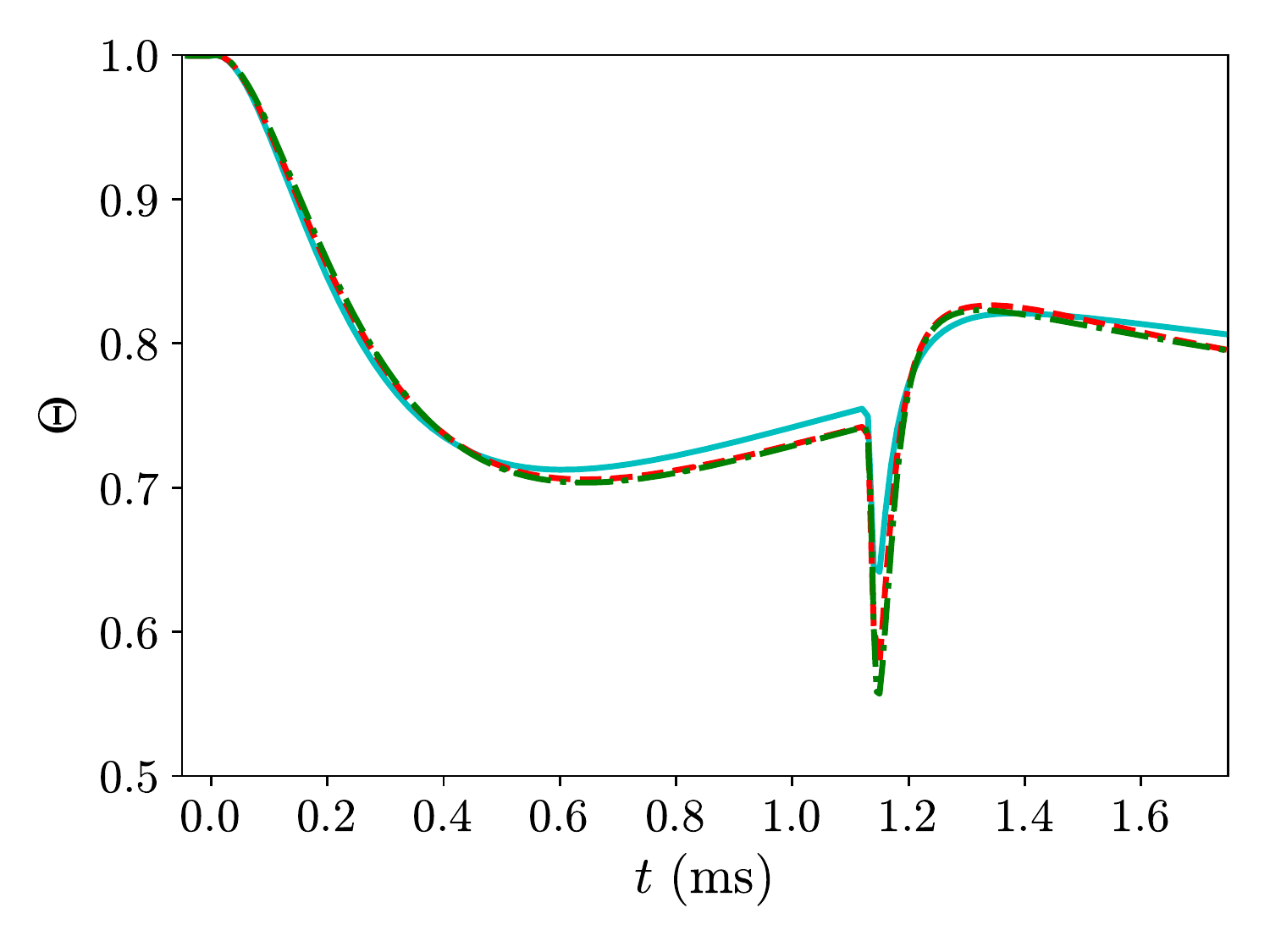}}
\subfigure[$\ $Integrated $\mathrm{TKE}$]{%
\includegraphics[width=0.4\textwidth]{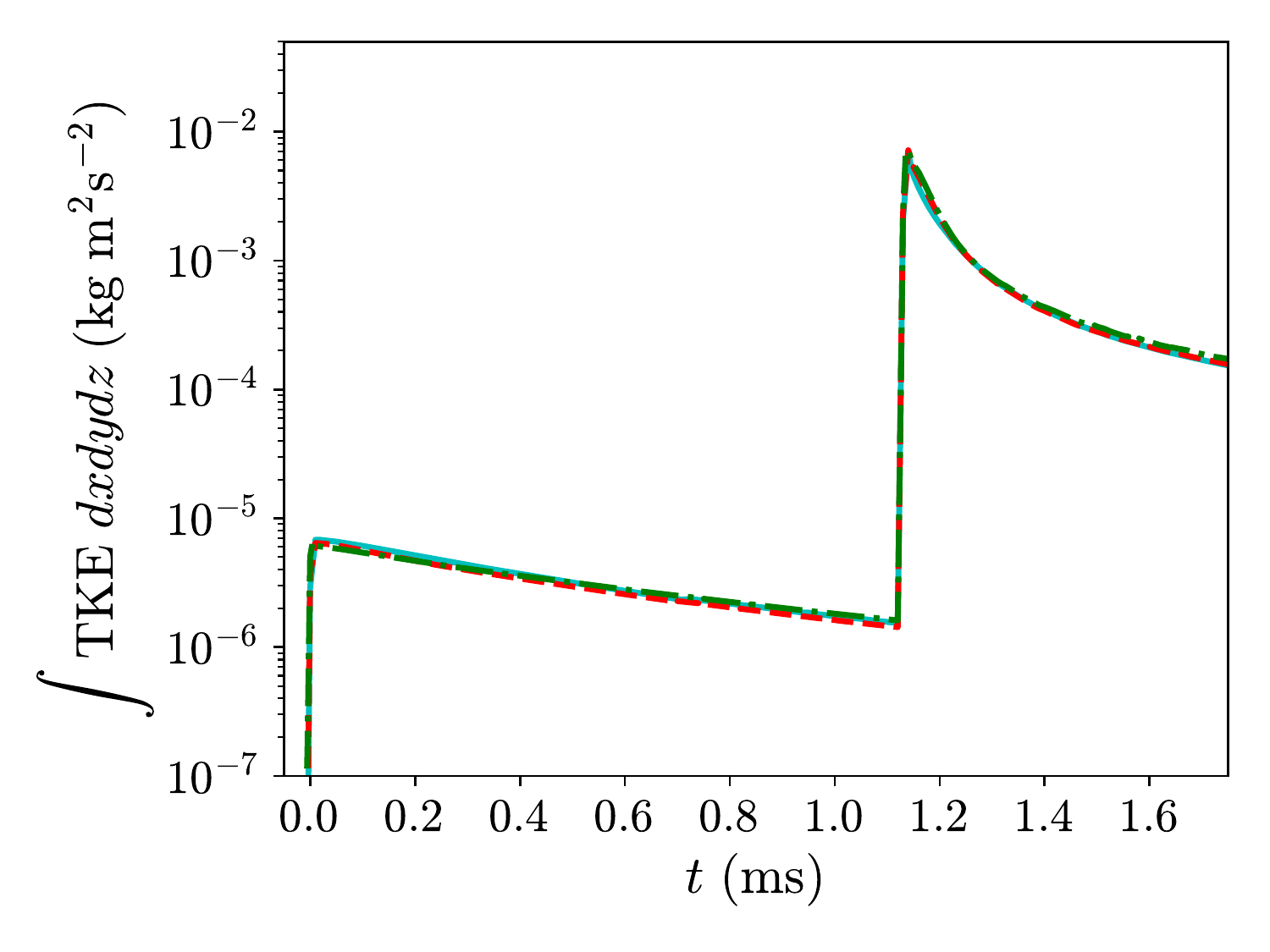}}
\subfigure[$\ $Integrated $\mathrm{TKE}$ (after re-shock)]{%
\includegraphics[width=0.45\textwidth]{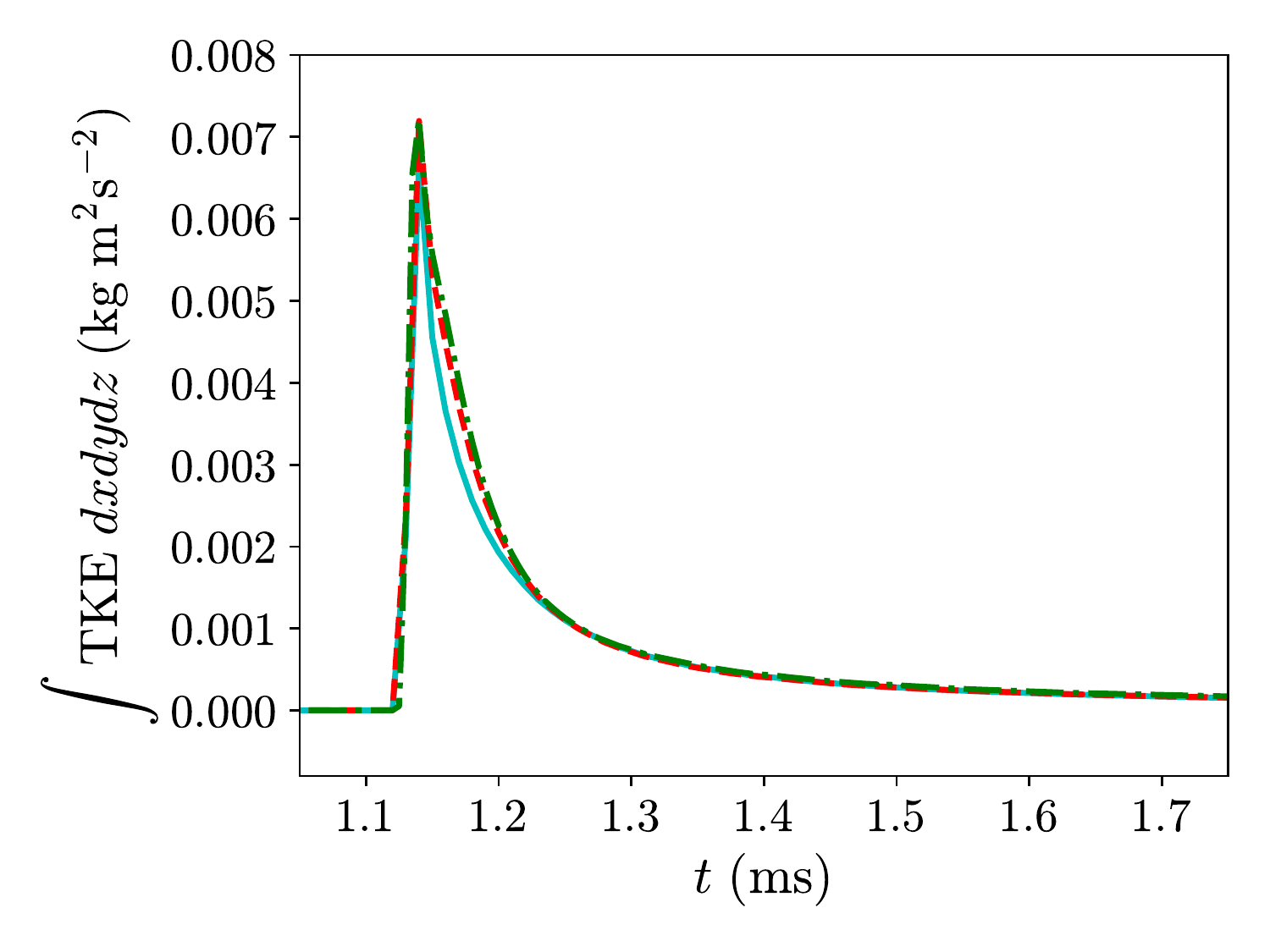}}
\subfigure[$\ $Integrated scalar dissipation rate]{%
\includegraphics[width=0.4\textwidth]{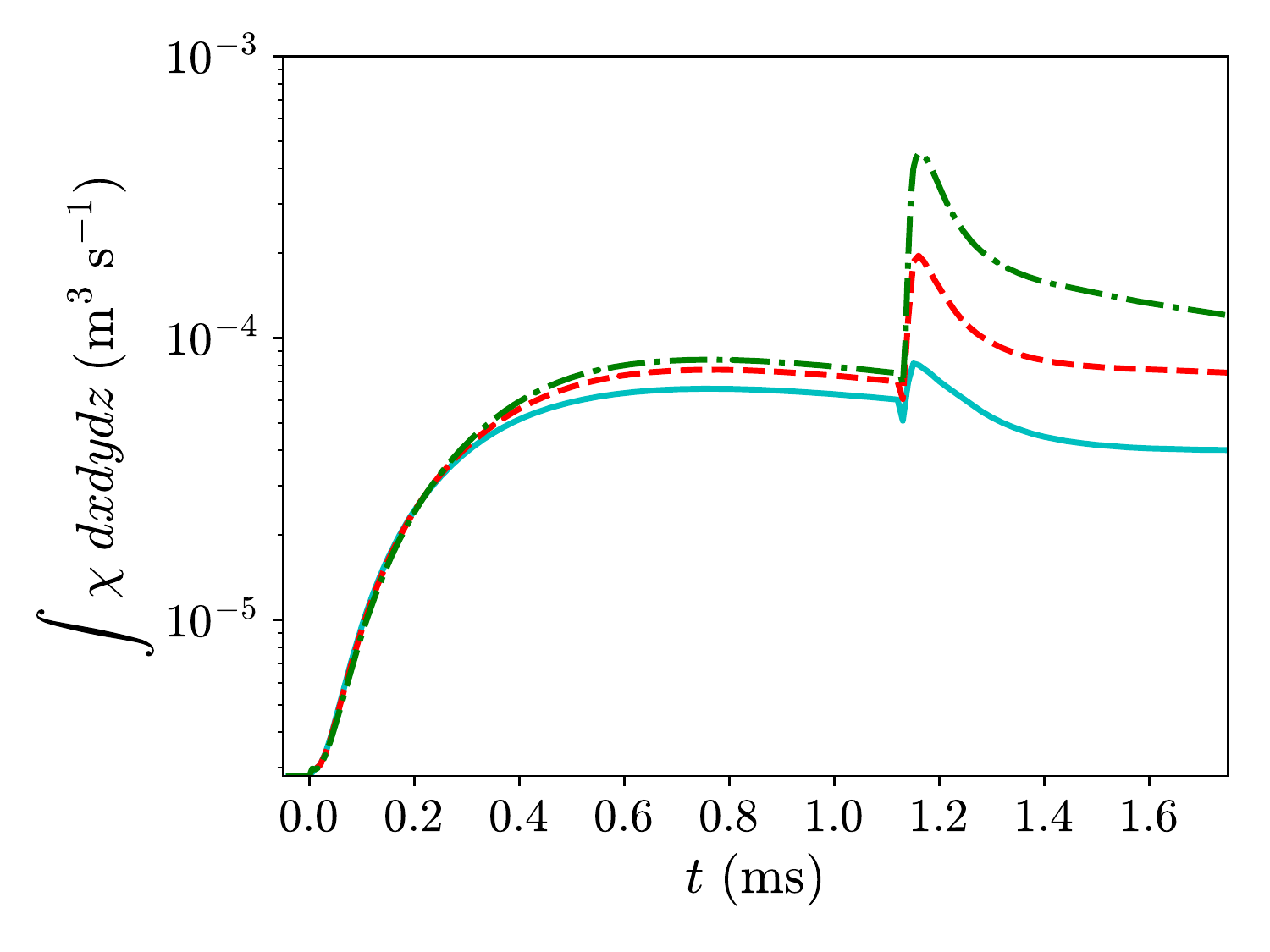}}
\subfigure[$\ $Integrated enstrophy]{%
\includegraphics[width=0.4\textwidth]{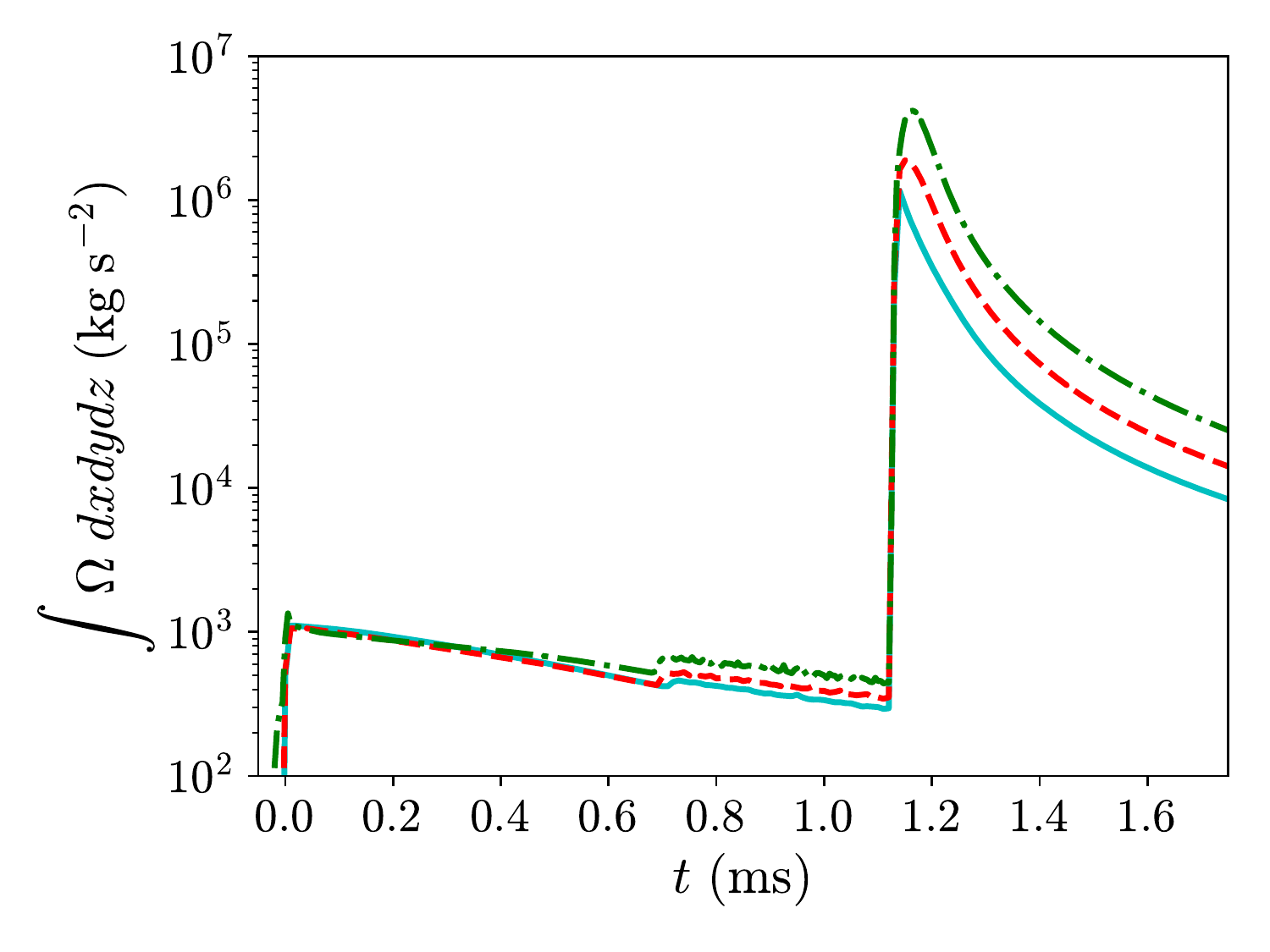}}
\caption{Grid sensitivities of different statistical quantities over time for the 3D problem. Cyan solid line: grid B; red dashed line: grid C; green dash-dotted line: grid D.}
\label{fig:3D_grid_sensitivity}
\end{figure*}

Figures~\ref{fig:2D_grid_sensitivity} and \ref{fig:3D_grid_sensitivity} show the time evolution of various statistical quantities computed on different grids for the 2D and 3D cases, respectively. The mixing width defines the large scale characteristic length of the mixing layer. The growth of this quantity is largely dominated by the entrainment of the fluids through convective motions. The mixedness is commonly used as an estimation of the amount of fluids that is molecularly mixed within the mixing layer~\cite{youngs1994numerical, lombardini2012transition, tritschler2014richtmyer, mohaghar2017evaluation, reese2018simultaneous}. It is generally interpreted as the ratio of the amount of fluids molecularly mixed to the amount of fluids entrained by convection. By comparing the results obtained on grids F and G from figure~\ref{fig:2D_grid_sensitivity}, it can be seen that both mixing width and mixedness for the 2D case are fully grid converged. Similarly, for the 3D case, it can be seen from figure~\ref{fig:3D_grid_sensitivity} that the grid also has very little effect on the mixing width and mixedness at the highest grid resolution level (grid D).

In the flow problem being studied, kinetic energy is deposited at the material interface when the shock passes through the interface and TKE is generated subsequently. TKE is largely dominated by the low wavenumber eddies or large scale features. After the passage of the shock, TKE decays as it is converted to internal energy through viscous dissipation. Grid convergence of the domain-integrated TKE is obtained at the highest resolutions considered for both 2D and 3D problems, as seen in figures~\ref{fig:2D_grid_sensitivity} and \ref{fig:3D_grid_sensitivity}.

While mixing width and TKE are dominated by large scale features, scalar dissipation rate and enstrophy are mainly associated with high wavenumber features closer to Batchelor and Kolmogorov scales. \citet{tritschler2014richtmyer} noticed that peaks of the scalar dissipation rate and enstrophy spectra shift to higher wavenumbers and magnitudes when smaller scales are resolved with finer grid in the grid sensitivity analysis of their RMI simulations. Therefore, the grid convergence requirements for these two quantities are expected to be stricter than mixing width and TKE. From figure~\ref{fig:2D_grid_sensitivity}, we can see that grid convergence for both domain-integrated scalar dissipation rate and enstrophy is still achieved for the 2D problem, but only at the finest grid. For the 3D problem, grid convergence of both integrated quantities is only obtained for a short duration after first shock (until $t \approx 0.25 \ \mathrm{ms}$) with the finest grid. Since the quantities that depend on high wavenumber features are not grid-converged for the 3D case, the simulations performed are still far from DNS level, especially immediately after re-shock. Moreover, we estimate that the incident and transmitted shock thicknesses are at $O(10)$ and $O(1)$ microns respectively. Even with the ultra-high grid resolution setting (grid G) for the 2D case, it is believed that the transmitted shock is still far from well-resolved and the solutions around the shock are regularized by the shock-capturing scheme.

%{\color{blue} % SKL commented out 08/27 % In the implicit LES community, it is believed that one may use a numerically dissipative scheme as a substitute of an explicit sub-grid scale model by assuming that the scheme has the same dissipative effects as sub-grid scales on resolved scales~\cite{grinstein2007implicit, garnier2009large}. While we cannot prevent one from classifying the 3D simulations conducted in this work to be of implicit LES type since numerical dissipation is introduced in general during the simulations, one has to be cautious on whether the numerical dissipation is truly a good representation of the sub-grid scale effects for the flows being studied. In fact, sub-grid scale effects in variable-density flows are still being studied in recent years~\cite{subramaniam2017turbulence, sidharth2018subgrid} and require further research with DNS and experimental data. We also have to clarify that the regularization effects or truncation errors of the scheme have never been designed to model the sub-grid scale effects. Therefore, in order to examine the numerical effects on the statistics, an extensive grid convergence analysis is provided in this section to show that most statistics of interests that will be discussed in later section of this work are reasonably captured by the 3D simulations.}
% SKL's rewording 08/27
An assumption of implicit LES (ILES) is that one may use a numerically dissipative scheme as a substitute for an explicit sub-grid scale (SGS) model, {\it i.e.} assuming that the numerical scheme has the same dissipative effects as sub-grid scales have on the resolved scales~\cite{grinstein2007implicit, garnier2009large}. While the present simulations do not use an explicit SGS
model and thus share this aspect of ILES, we do not assume that the numerical scheme used provides a good model for the SGS effects in the flows we study.
We recognize that SGS effects in variable-density flows are subject of active recent studies~\cite{candler2014baroclinic, subramaniam2017turbulence, sidharth2018subgrid} and require further research using DNS and experimental data. We also stress that the regularization effects or truncation errors of the scheme used in the present work were {\it not} designed to model the SGS effects. Therefore, in order to examine the numerical effects on the turbulence statistics, an extensive grid convergence analysis is provided in this section to show that most statistics of interest discussed in later sections are reasonably captured by the 3D simulations.
%Despite the application of numerical regularization on the solutions, the simulations are also not of implicit LES type since the numerical dissipation introduced from the regularization is not designed to model the subgrid scale effects. However, through the grid convergence analysis discussed in this section, we have shown that the statistics that are weakly dependent on small scale features are fairly well captured.

%%%%%%%%%%%%%%%%%%%%%%%%%%%%%%%%%%%%%%%%%%%%%%%%%%%%%%%%%%%%%%%%%%%%%%%%%%%%%%%%
%%%%%%%%%%%%%%%%%%%%%%%%%%%%%%%%%%%%%%%%%%%%%%%%%%%%%%%%%%%%%%%%%%%%%%%%%%%%%%%%
%%%%%%%%%%%%%%%%%%%%%%%%%%%%%%%%%%%%%%%%%%%%%%%%%%%%%%%%%%%%%%%%%%%%%%%%%%%%%%%%

\section{\label{sec:reduced_reynolds_number_simulations} Set-up for three-dimensional simulations with reduced Reynolds numbers}

In order to study the effect of the Reynolds number on the development of 3D RMI, the 3D problem presented in the previous sections is repeated with the same initial conditions and the grid D settings, but with increased transport coefficients for each species ($D_i$, $\mu_i$, $\mu_{v,i}$, and $\kappa_i$). The transport coefficients are increased in the simulations by artificially multiplying the values computed using physical models given in appendix~\ref{sec:appendix_TC} with constant factors. Two cases, with transport coefficients increased by factors of 2 and 4, are simulated.

\begin{figure*}[!ht]
\centering
\subfigure[$\ $Reynolds number based on mixing width]{%
\includegraphics[width = 0.45\textwidth]{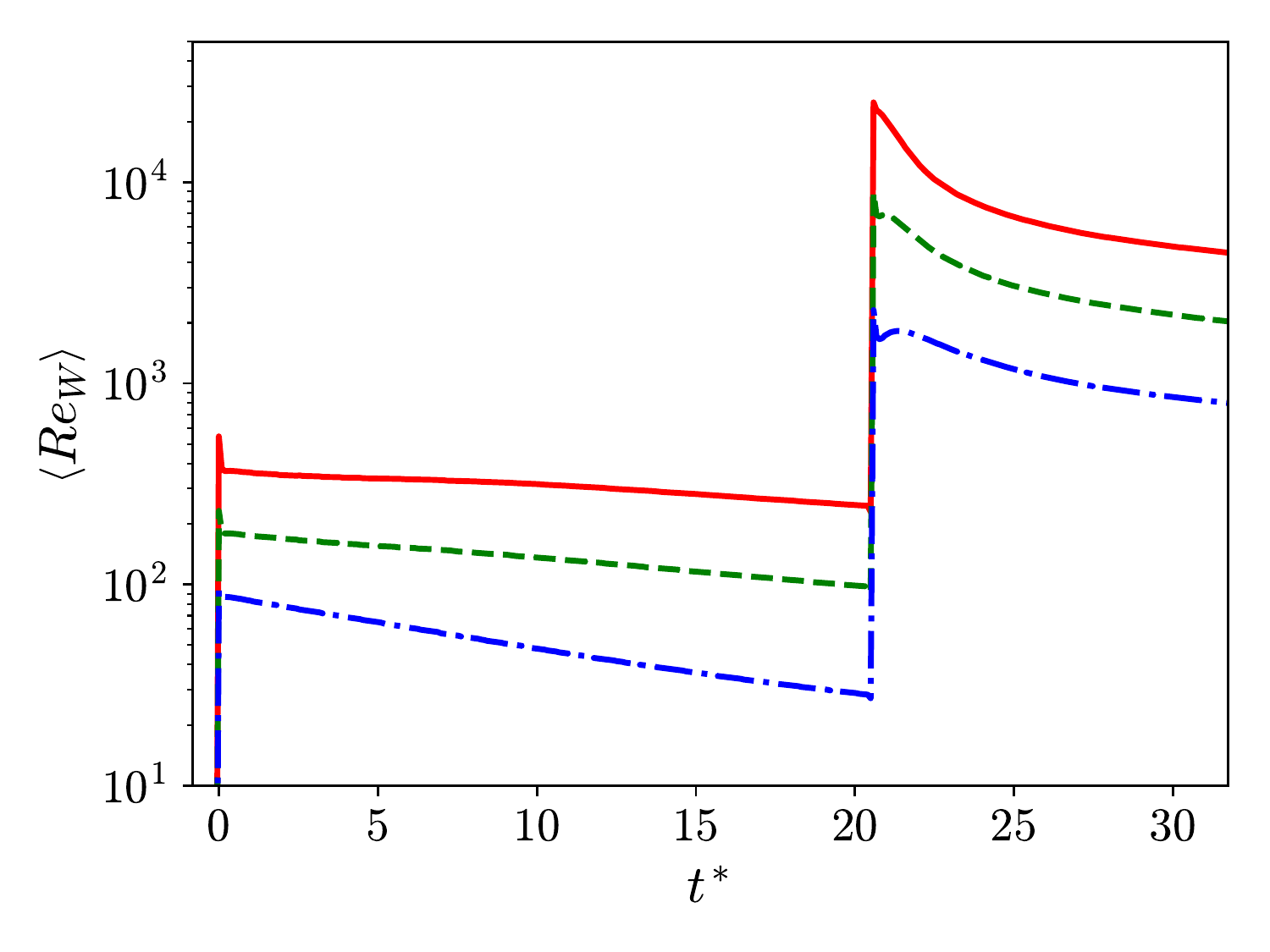}}
\subfigure[$\ $Reynolds number based on integral length scale]{%
\includegraphics[width = 0.45\textwidth]{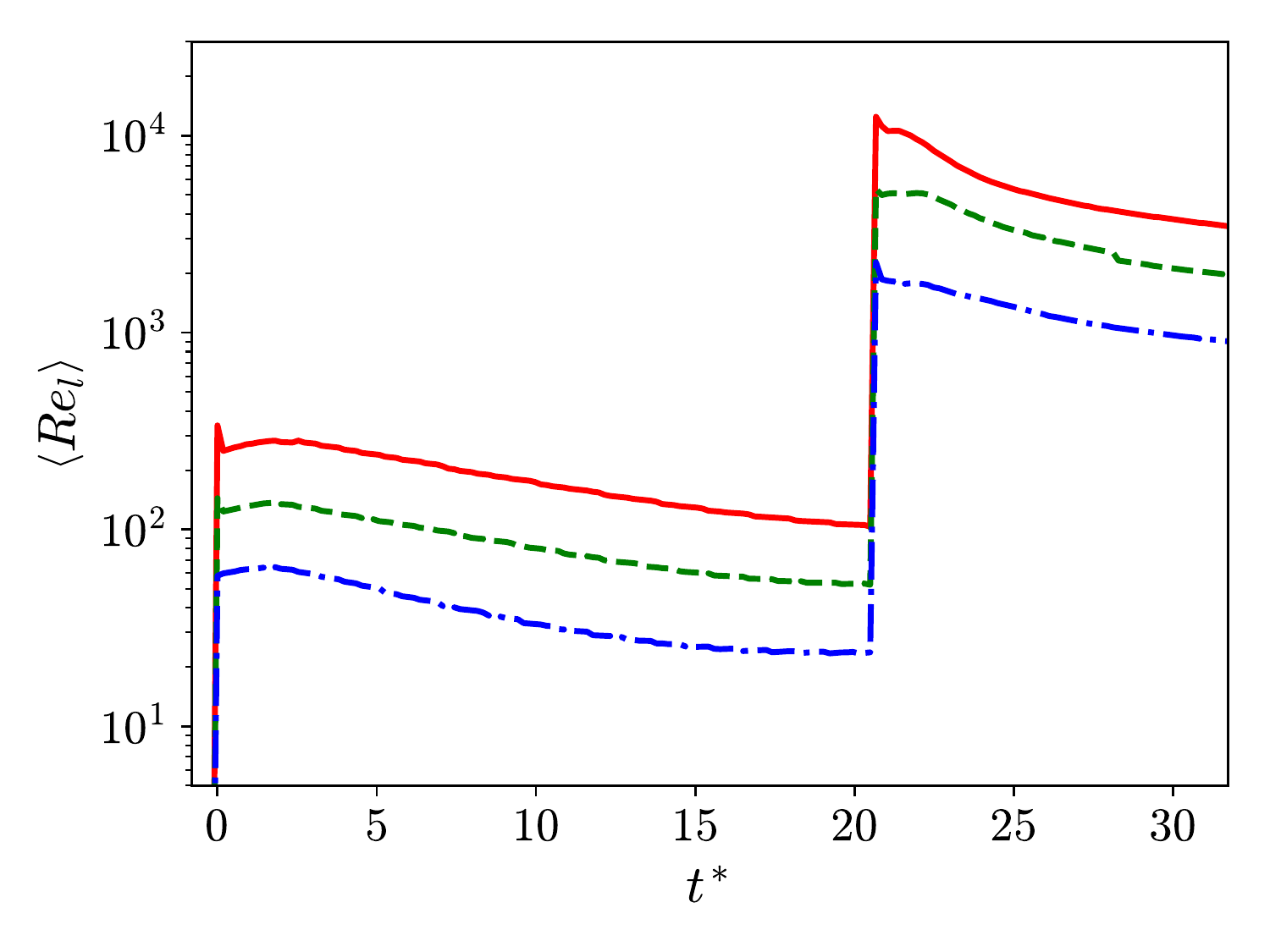}}
\subfigure[$\ $Schmidt number]{%
\includegraphics[width = 0.45\textwidth]{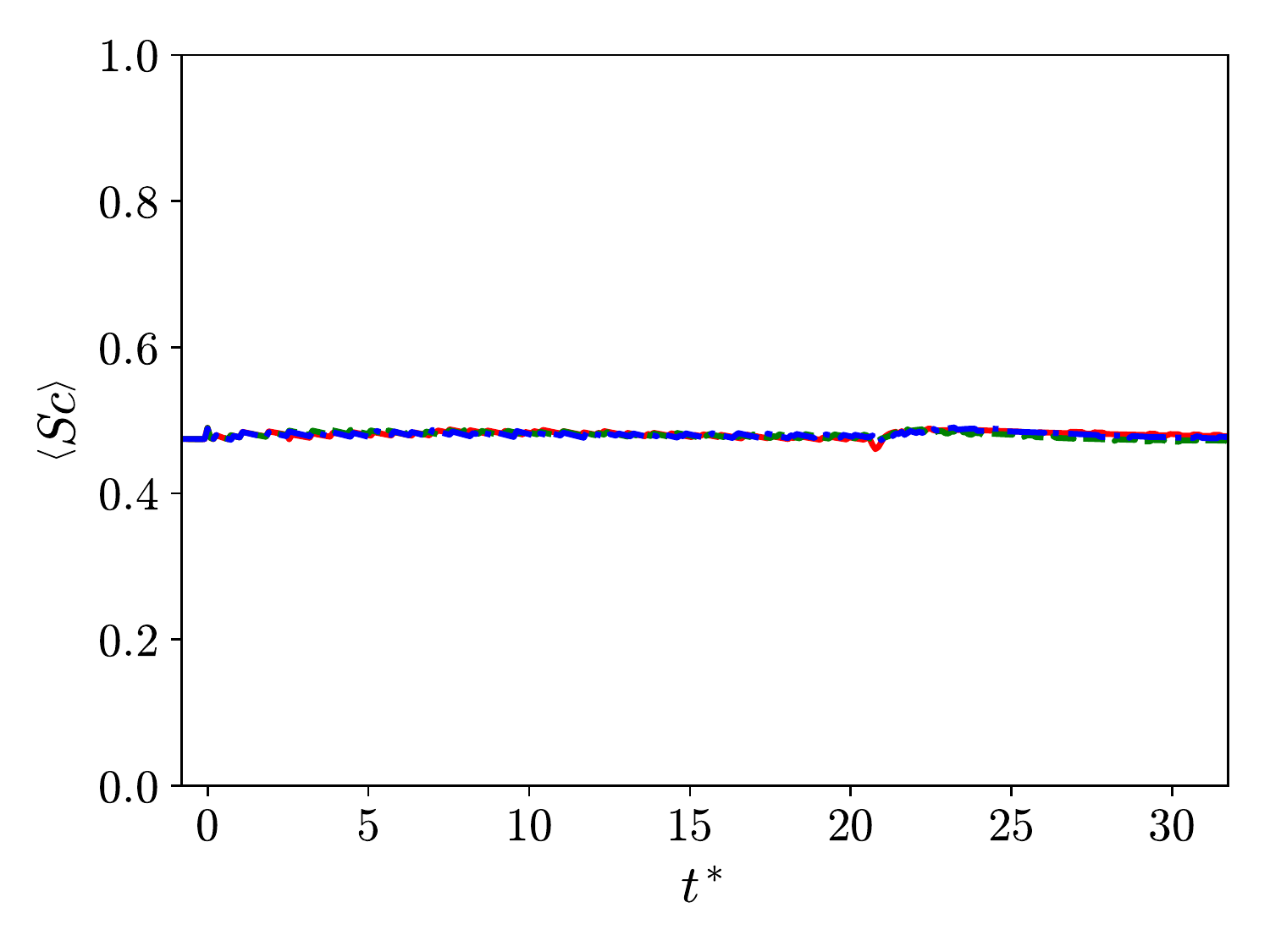}}
\subfigure[$\ $Prandtl number]{%
\includegraphics[width = 0.45\textwidth]{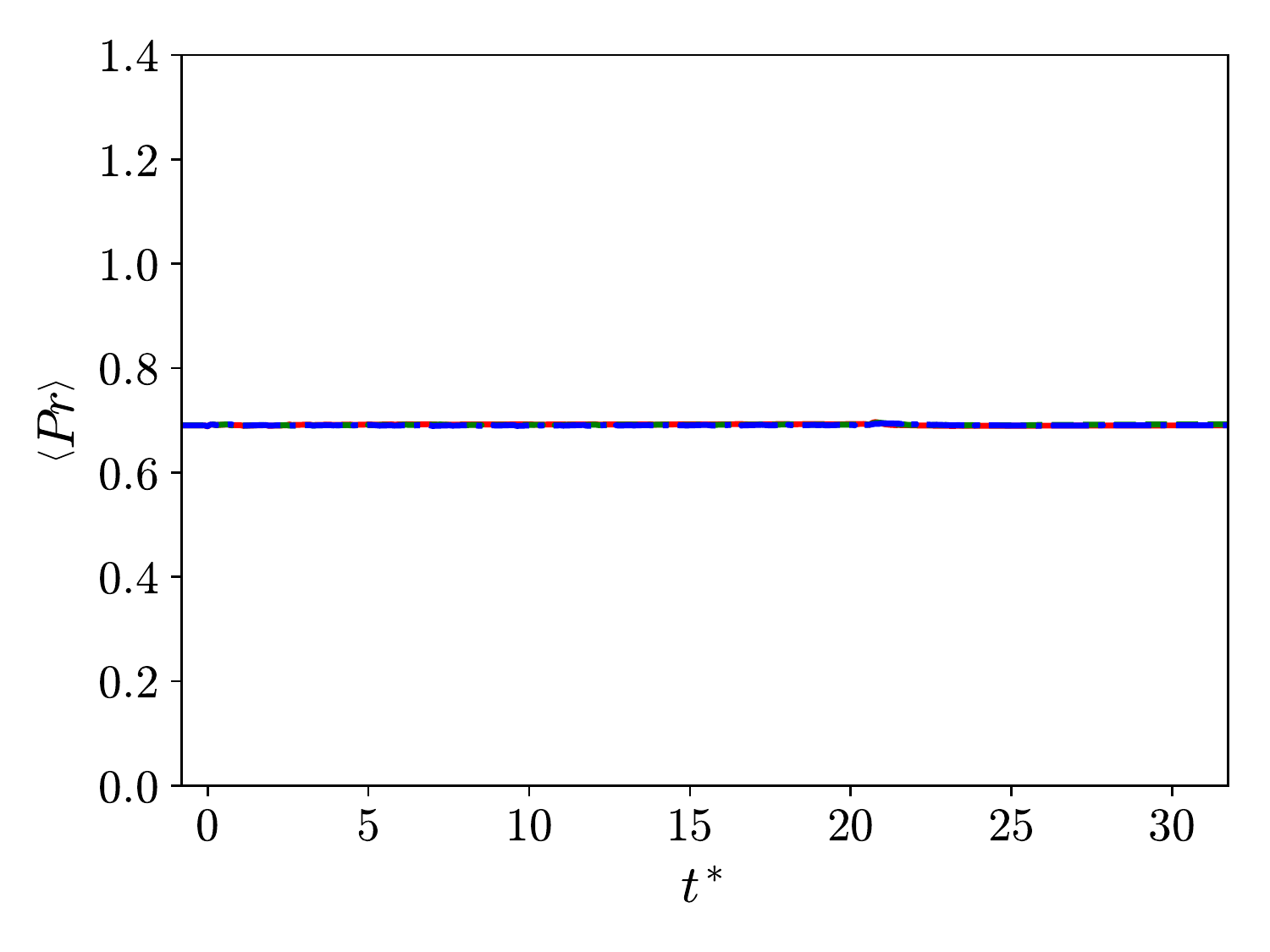}}
\caption{Comparison of the time evolution of different dimensionless numbers between the 3D cases with different transport coefficients. Red solid line: physical transport coefficients; green dashed line: $2 \times$physical transport coefficients; blue dash-dotted line: $4 \times$physical transport coefficients.}
\label{fig:Reynolds_dimensionless_number}
\end{figure*}

Figure~\ref{fig:Reynolds_dimensionless_number} shows the time evolution of the means of the Reynolds numbers based on the mixing width, $W$, and mean integral length scale of $\mathrm{SF_6}$ mole fraction, $\left< l_{X_\mathrm{SF_6}} \right>$, ($\left< Re_W \right>$ and $\left< Re_l \right>$), Schmidt number ($\left< Sc \right>$), and Prandtl number ($\left< Pr \right>$). The mean denoted by $\left< \cdot \right>$ corresponds to an additional averaging over the central part of the mixing layer of the quantities computed over the lines or planes. $Re_W$, $Re_l$, $Sc$, and $Pr$ are defined as:
\begin{align}
    Re_W &= \frac{ \bar{\rho} u_{\mathrm{rms}} W }{\bar{\mu}}, \\
    Re_l &= \frac{ \bar{\rho} u_{\mathrm{rms}} \left< l_{X_\mathrm{SF_6}} \right> }{\bar{\mu}}, \\
    Sc &= \frac{\bar{\mu}}{\bar{\rho} \bar{D}_{\mathrm{SF_6}}}, \\
    Pr &= \frac{\bar{c}_p \bar{\mu}}{\bar{\kappa}},
\end{align}

\noindent where $u_{\mathrm{rms}}=\sqrt{ \overline{ u^{\prime\prime}_i u^{\prime\prime}_i } /3}$. The central part of the mixing layer is defined as the regions where cross-sectional planes satisfy:
\begin{equation}
    4 \bar{X}_{\mathrm{SF_6}} \left( 1 - \bar{X}_{\mathrm{SF_6}} \right) \geq 0.9.
\end{equation}

\noindent The nondimensional time, $t^{*}$, is defined by:
\begin{equation}
    t^{*} = \frac{t}{\tau_c},
\end{equation}
\noindent where $\tau_c = \tau^{3D}$. Figure~\ref{fig:Reynolds_dimensionless_number} shows that increasing the transport coefficients mostly affects the Reynolds numbers, while other non-dimensional quantities such as Schmidt and Prandtl numbers within the central part of mixing layer remain almost unchanged over time. The Reynolds number values are close when calculated based on the mixing width or integral length scale since these length scales have similar orders of magnitude. The simulations with reduced Reynolds numbers are further discussed in the next section.

%%%%%%%%%%%%%%%%%%%%%%%%%%%%%%%%%%%%%%%%%%%%%%%%%%%%%%%%%%%%%%%%%%%%%%%%%%%%%%%%
%%%%%%%%%%%%%%%%%%%%%%%%%%%%%%%%%%%%%%%%%%%%%%%%%%%%%%%%%%%%%%%%%%%%%%%%%%%%%%%%
%%%%%%%%%%%%%%%%%%%%%%%%%%%%%%%%%%%%%%%%%%%%%%%%%%%%%%%%%%%%%%%%%%%%%%%%%%%%%%%%

\section{Comparison and analysis of two-dimensional and three-dimensional cases}

%%%%%%%%%%%%%%%%%%%%%%%%%%%%%%%%%%%%%%%%%%%%%%%%%%%%%%%%%%%%%%%%%%%%%%%%%%%%%%%%
%%%%%%%%%%%%%%%%%%%%%%%%%%%%%%%%%%%%%%%%%%%%%%%%%%%%%%%%%%%%%%%%%%%%%%%%%%%%%%%%
%%%%%%%%%%%%%%%%%%%%%%%%%%%%%%%%%%%%%%%%%%%%%%%%%%%%%%%%%%%%%%%%%%%%%%%%%%%%%%%%

This section contains the main results of the paper. First, the levels  of compressibility and non-Boussinesq effects are assessed using the turbulent Mach number and an effective Atwood number. Then, some features of the mixing are discussed using the mean and variance profiles of the mole fraction as well as its probability density function. This is then followed by a discussion of the turbulent kinetic energy and the anisotropy of Reynolds normal stresses. Finally, the time evolution of the spectra of the mole fraction and energy defined based on the momentum with density-weighted fluctuation of the velocity after re-shock is examined. All results for the 2D case in this section are obtained by ensemble averaging over 24 realizations computed with grid G, while those for the 3D cases are obtained with grid D. It is found that 24 realizations are sufficient to achieve statistical convergence for most of the quantities of interest for the 2D case in this work (as discussed in appendix~\ref{sec:2D_stat_convergence}).

\subsection{Flow compressibility and effective Atwood number}

Several aspects of compressibility of a flow can be measured from the turbulent Mach number, $Ma_{t}$, which is defined as:
\begin{equation}
    Ma_{t} = \frac{ \sqrt{ \overline{ u^{\prime\prime}_i u^{\prime\prime}_i } }}{\bar{c}},
\end{equation}
\noindent where $c=\sqrt{\gamma p / \rho}$ is the speed of sound.

\begin{figure*}[!ht]
\centering
\includegraphics[width = 0.45\textwidth]{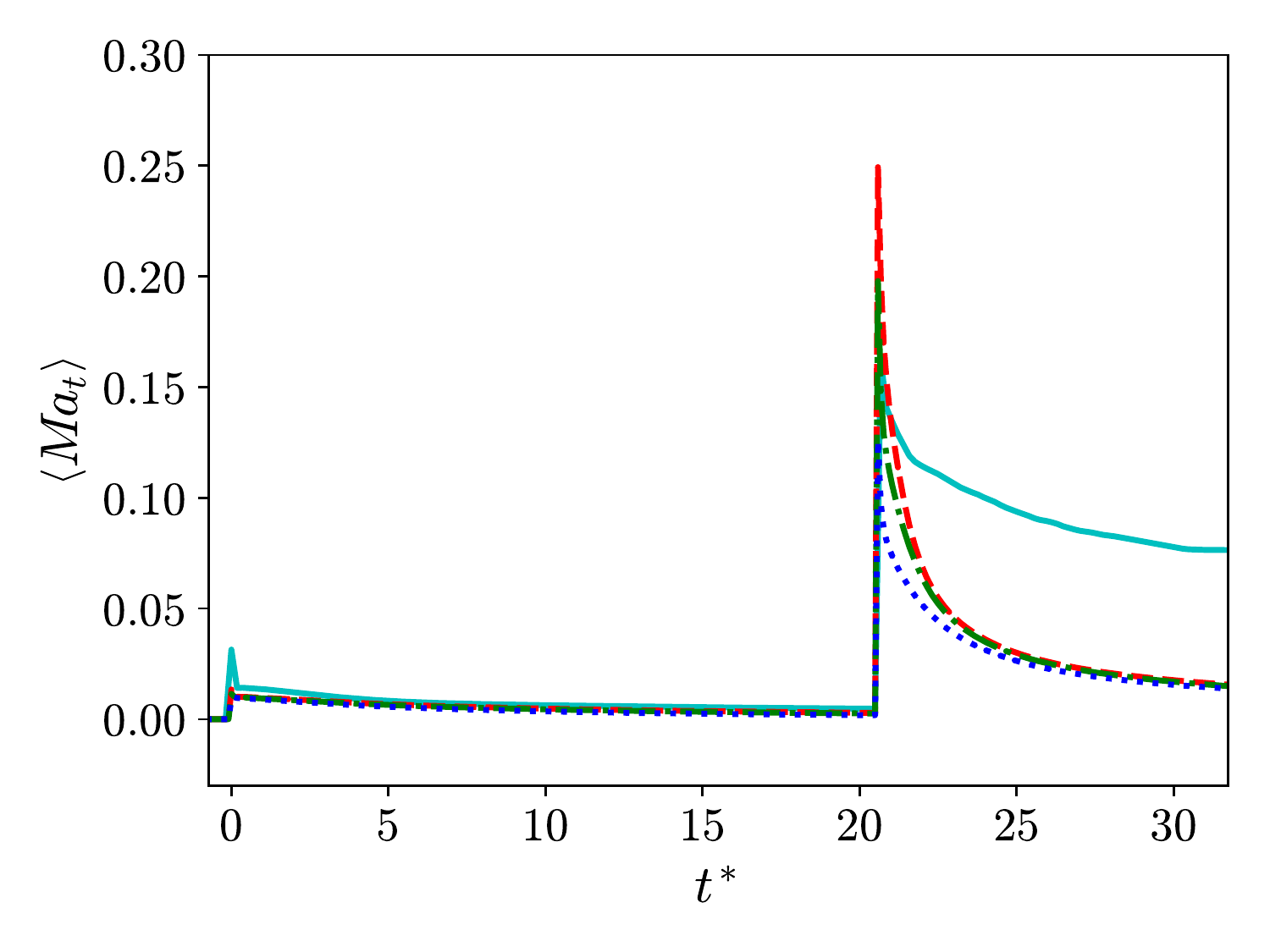}
\caption{Comparison of the time evolution of means of turbulent Mach number, $\left< Ma_{t} \right>$, in the central part of mixing layer between the 2D and 3D problems. Cyan solid line: 2D with physical transport coefficients; red dashed line: 3D with physical transport coefficients; green dash-dotted line: 3D with $2 \times$physical transport coefficients; blue dotted line: 3D with $4 \times$physical transport coefficients.}
\label{fig:2D_3D_transport_coeffs_Ma_t}
\end{figure*}

Figure~\ref{fig:2D_3D_transport_coeffs_Ma_t} shows the time evolution of turbulent Mach number within the central part of the mixing layer for the 2D and 3D cases with different transport coefficients. The turbulent Mach number is very close to zero for all cases before re-shock. After re-shock, the turbulent Mach number for all 3D cases decays quickly to zero asymptotically after a jump. The turbulent Mach number for the 2D case is larger than that for the 3D cases because velocity fluctuations remain larger in the 2D case after re-shock. This is further discussed in section~\ref{sec:TKE}. Overall though, all simulations are only weakly compressible as $\left< Ma_{t} \right>$ for each case is always smaller than $0.3$. 

\begin{figure*}[!ht]
\centering
\subfigure[$\ t^{*}=0.9$, 2D]{%
\includegraphics[height = 0.33\textwidth]{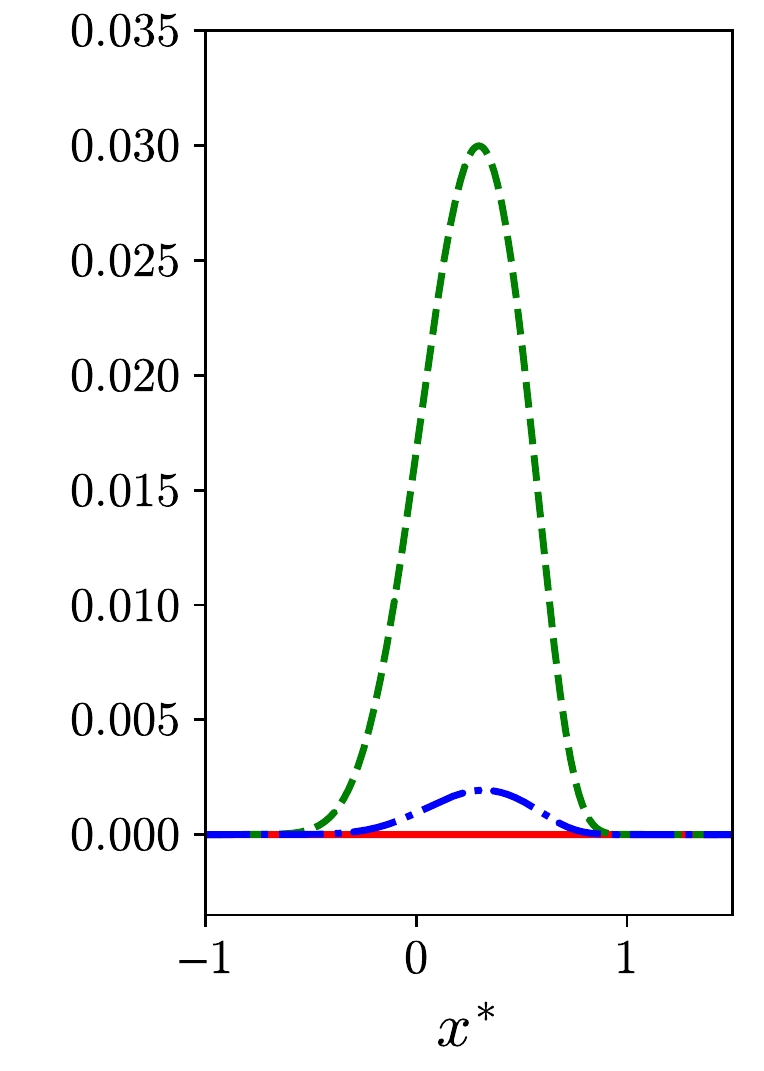}}
\subfigure[$\ t^{*}=0.9$, 3D]{%
\includegraphics[height = 0.33\textwidth]{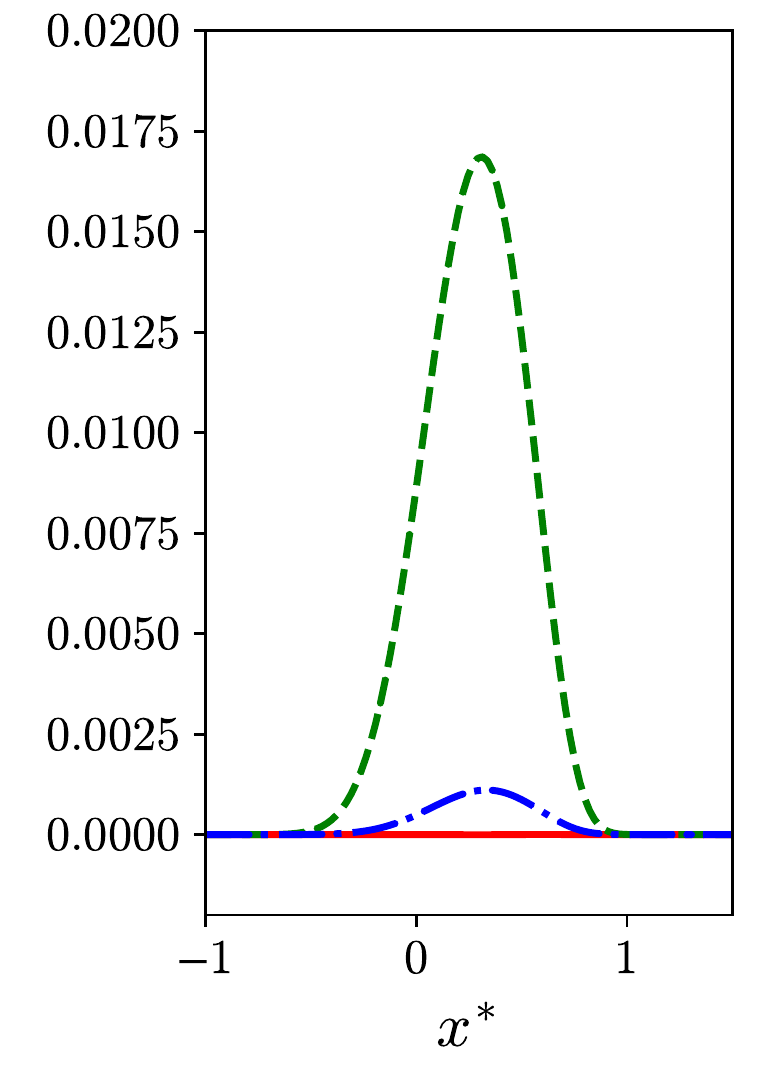}}
\subfigure[$\ t^{*}=22.6$, 2D]{%
\includegraphics[height = 0.33\textwidth]{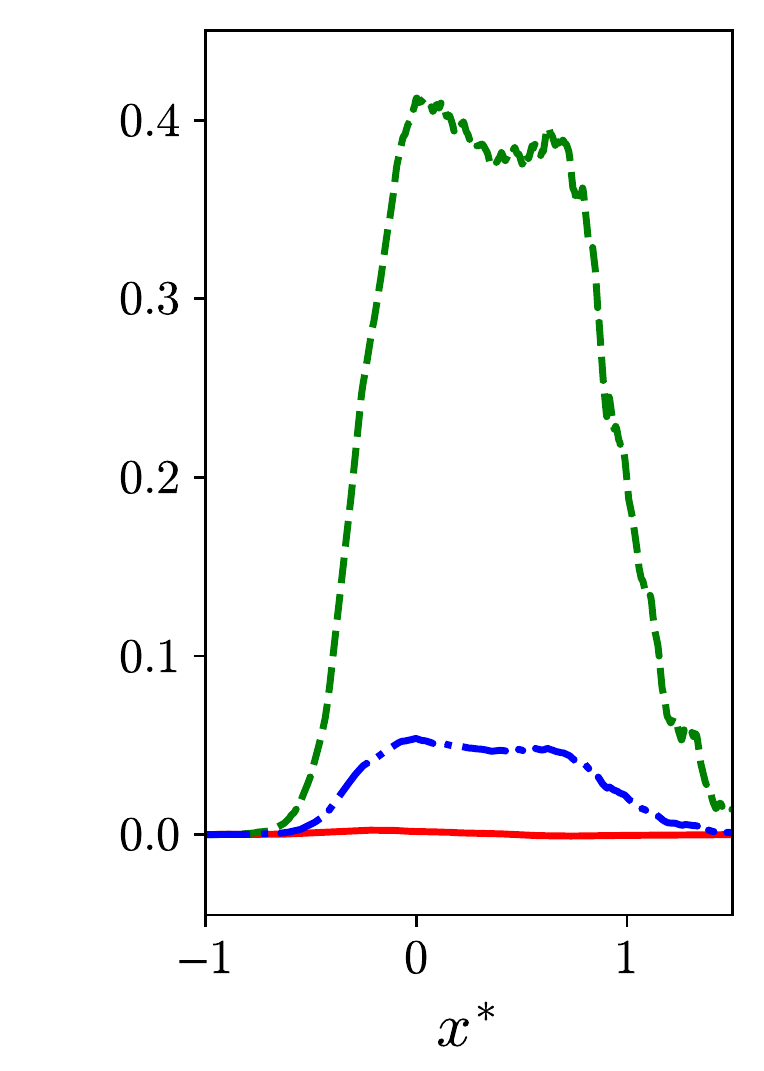}}
\subfigure[$\ t^{*}=22.6$, 3D]{%
\includegraphics[height = 0.33\textwidth]{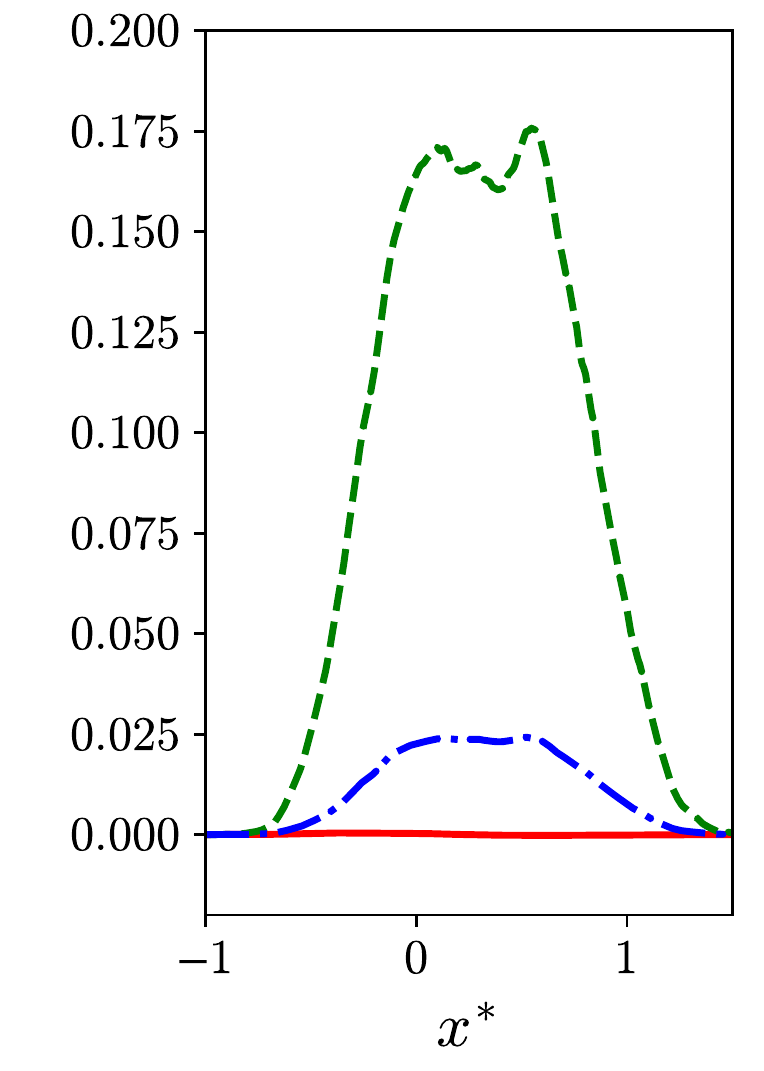}}
\caption{Profiles of normalized covariances of mixture density with other quantities just after first shock ($t^{*}=0.9$) and just after re-shock ($t^{*}=22.6$) for the 2D and 3D problems with physical transport coefficients. Red solid line: $\overline{\rho^{\prime}p^{\prime}}/(\bar{\rho}\bar{p})$; green dashed line: $\overline{\rho^{\prime}M^{\prime}}/(\bar{\rho}\bar{M})$; blue dash-dotted line: $\overline{\rho^{\prime}(1/T)^{\prime}}/(\bar{\rho}\overline{(1/T)})$.}
\label{fig:2D_vs_3D_compressibility_profiles}
\end{figure*}

Another measure of the compressibility effects is the magnitudes of thermodynamic fluctuations. From the ideal equation of state, the mixture density, $\rho$, depends on pressure, $p$, mixture molecular weight, $M$, and temperature, $T$, through $\rho = p M / (R_u T)$. Figure~\ref{fig:2D_vs_3D_compressibility_profiles} shows the normalized covariances of $\rho$ with $p$, $M$, and $1/T$ across the normalized position $x^{*}$ just after first shock and re-shock from the 2D and 3D simulations with physical transport coefficients. The normalized position is defined as:
\begin{equation}
    x^{*}(x,t) = \frac{x - x_i(t)}{W(t)},
\end{equation}
\noindent where $x_i$ is the location of the interface from the $x$-$t$ diagram (figure~\ref{fig:x_t_diagram}). It is shown in figure~\ref{fig:2D_vs_3D_compressibility_profiles} that, for both cases, the covariance of the mixture density and mixture molecular weight is the largest among the three covariances at the two times just after impulsive accelerations considered. Compared to this covariance, the covariance of the mixture density and reciprocal of temperature is much smaller and that of the mixture density and pressure is virtually zero, which means the mixing layer is very weakly compressible.

\begin{figure*}[!ht]
\centering
\subfigure[$\ $Before re-shock, 2D]{%
\includegraphics[height = 0.33\textwidth]{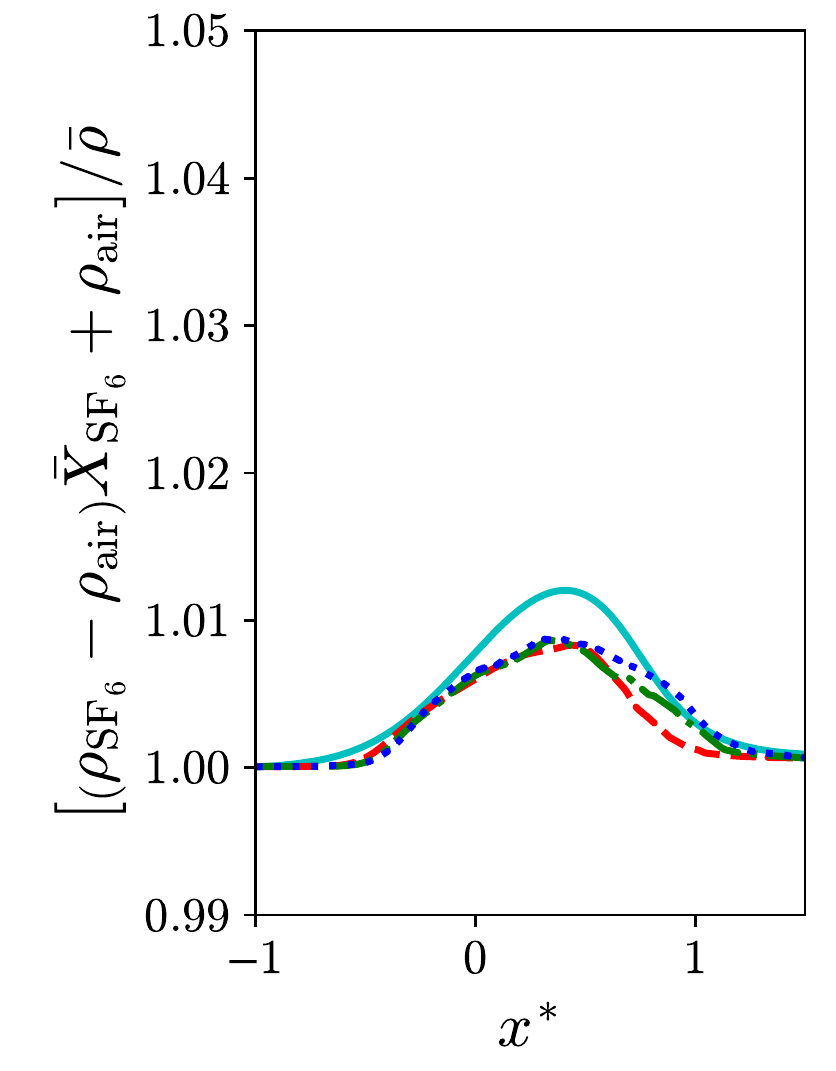}}
\subfigure[$\ $Before re-shock, 3D]{%
\includegraphics[height = 0.33\textwidth]{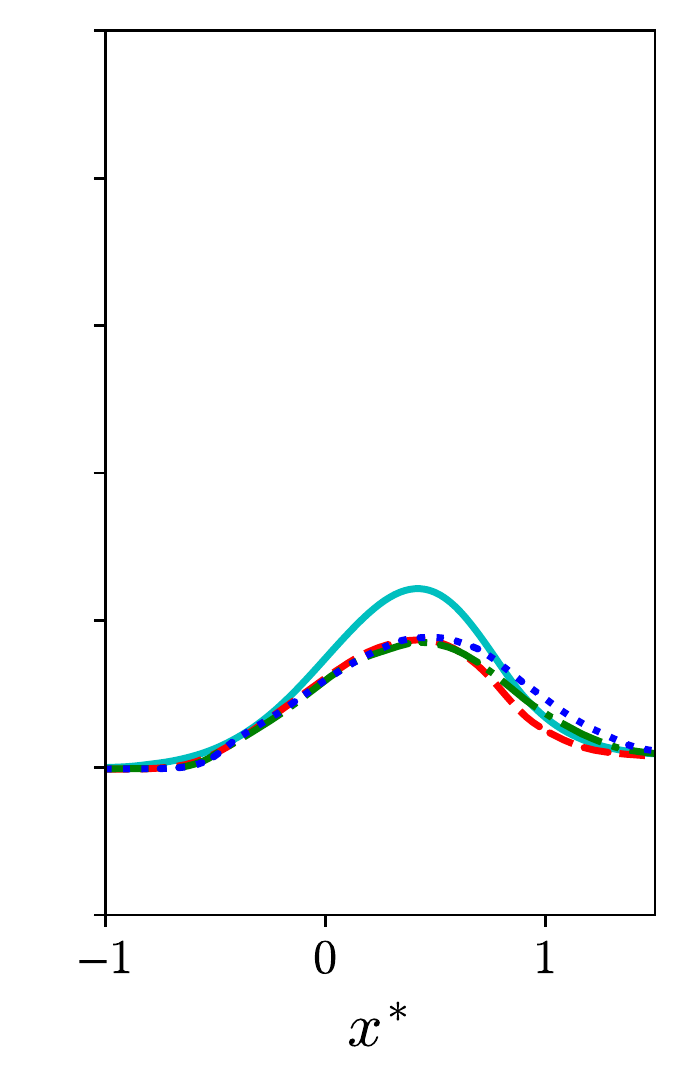}}
\subfigure[$\ $After re-shock, 2D]{%
\includegraphics[height = 0.33\textwidth]{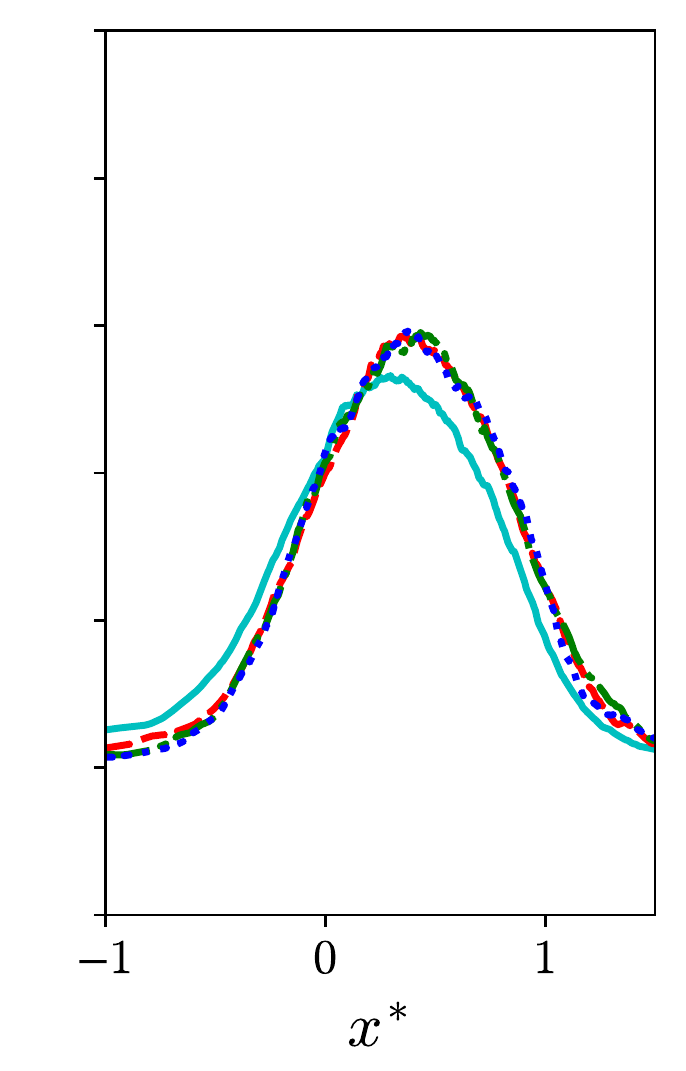}}
\subfigure[$\ $After re-shock, 3D]{%
\includegraphics[height = 0.33\textwidth]{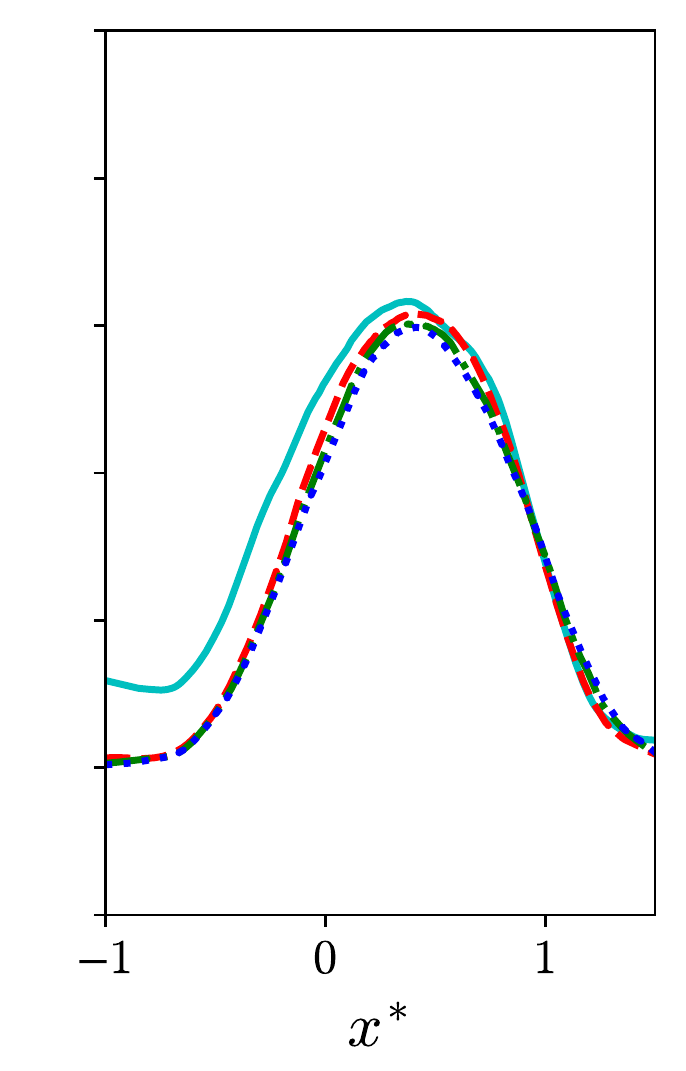}}
\caption{Profiles of the ratio between the reconstructed density using the incompressible assumption and true density at different times for the 2D and 3D problems with physical transport coefficients. Before re-shock: $t^{*}=0.9$ (cyan solid line); $t^{*}=7.5$ (red dashed line); $t^{*}=14.1$ (green dash-dotted line); $t^{*}=20.7$ (blue dotted line). After re-shock: $t^{*}=22.6$ (cyan solid line); $t^{*}=26.3$ (red dashed line); $t^{*}=30.1$ (green dash-dotted line); $t^{*}=32.9$ (blue dotted line).}
\label{fig:2D_vs_3D_incompressibility_profiles}
\end{figure*}

As a consequence of the nearly incompressible flow behavior, from the ideal equation of state, density mainly depends on the composition of the flow through the mixture molecular weight, $M$, as the quantity $\rho R_u/M = p/T$, is quasi-uniform across the mixing regions, where $R_u$ is the universal gas constant. The quasi-uniform behavior of the ratio $p/T$ is further discussed in appendix~\ref{sec:p_over_T}. Note that if $p/T$ is uniform in the flow, the variable-density Navier--Stokes equations can be derived as the infinite speed of sound limit of the fully compressible Navier--Stokes equations with multi-species transport~\cite{livescu2013}. Further, if the ratio $p/T$ is strictly uniform, one can also easily derive a linear relationship between mole fraction and density as:

\begin{equation}
    X_{\mathrm{SF_6}} = \frac{\rho - \rho_{\mathrm{air}}}{\rho_{\mathrm{SF_6}} - \rho_{\mathrm{air}}}, \label{eq:mole_fraction_density_relationship}
\end{equation}
or
\begin{equation}
    \rho = \left( \rho_{\mathrm{SF_6}} - \rho_{\mathrm{air}} \right) X_{\mathrm{SF_6}} + \rho_{\mathrm{air}}, \label{eq:mole_fraction_density_relationship_2}
\end{equation}

\noindent where $\rho_{\mathrm{air}}$ and $\rho_{\mathrm{SF_6}}$ are time-dependent densities of air and $\mathrm{SF_6}$, respectively, on either side of the material interface from the solutions of the 1D flow representation. $\rho_{\mathrm{air}}$ and $\rho_{\mathrm{SF_6}}$ only change at first shock and re-shock. Figure~\ref{fig:2D_vs_3D_incompressibility_profiles} compares the ratios between the density mean profiles reconstructed from the mole fraction mean profiles using equation~\eqref{eq:mole_fraction_density_relationship_2} and the true density mean profiles for the 2D and 3D cases with physical transport coefficients at different times. The density ratios have less than 5\% deviation from one at all times for both 2D and 3D cases, as the flow is quasi-incompressible and the ratio $p/T$ does not vary much inside the mixing regions. The quasi-linear relationship between mole fraction and density suggests that mole fraction field can be viewed as normalized density field within the mixing region.

To quantify the non-Boussinesq effects inside the mixing layer, we can study the effective Atwood number, $At_e$, within the central part of the mixing layer, which is defined by \citet{cook2004mixing} as:
\begin{equation}
    At_e = \left< \frac{\sqrt{ \overline{{\rho^{\prime}}^{2}} }}{\bar{\rho}} \right>. \label{eq:At_e_density}
\end{equation}

\noindent The Boussinesq approximation is invalid if $At_e \geq 0.05$. Since the flows are nearly incompressible, the effective Atwood number is in fact also related to the mole fraction through equation~\eqref{eq:mole_fraction_density_relationship_2}:
\begin{equation}
    At_e \approx \left< \frac{\sqrt{ \overline{{X_{\mathrm{SF_6}}^{\prime}}^{2}} }}{\bar{X}_{\mathrm{SF_6}} + \rho_{\mathrm{air}}/\left( \rho_{\mathrm{SF_6}} - \rho_{\mathrm{air}} \right)} \right>.
\end{equation}

\begin{figure*}[!ht]
\centering
\includegraphics[width = 0.45\textwidth]{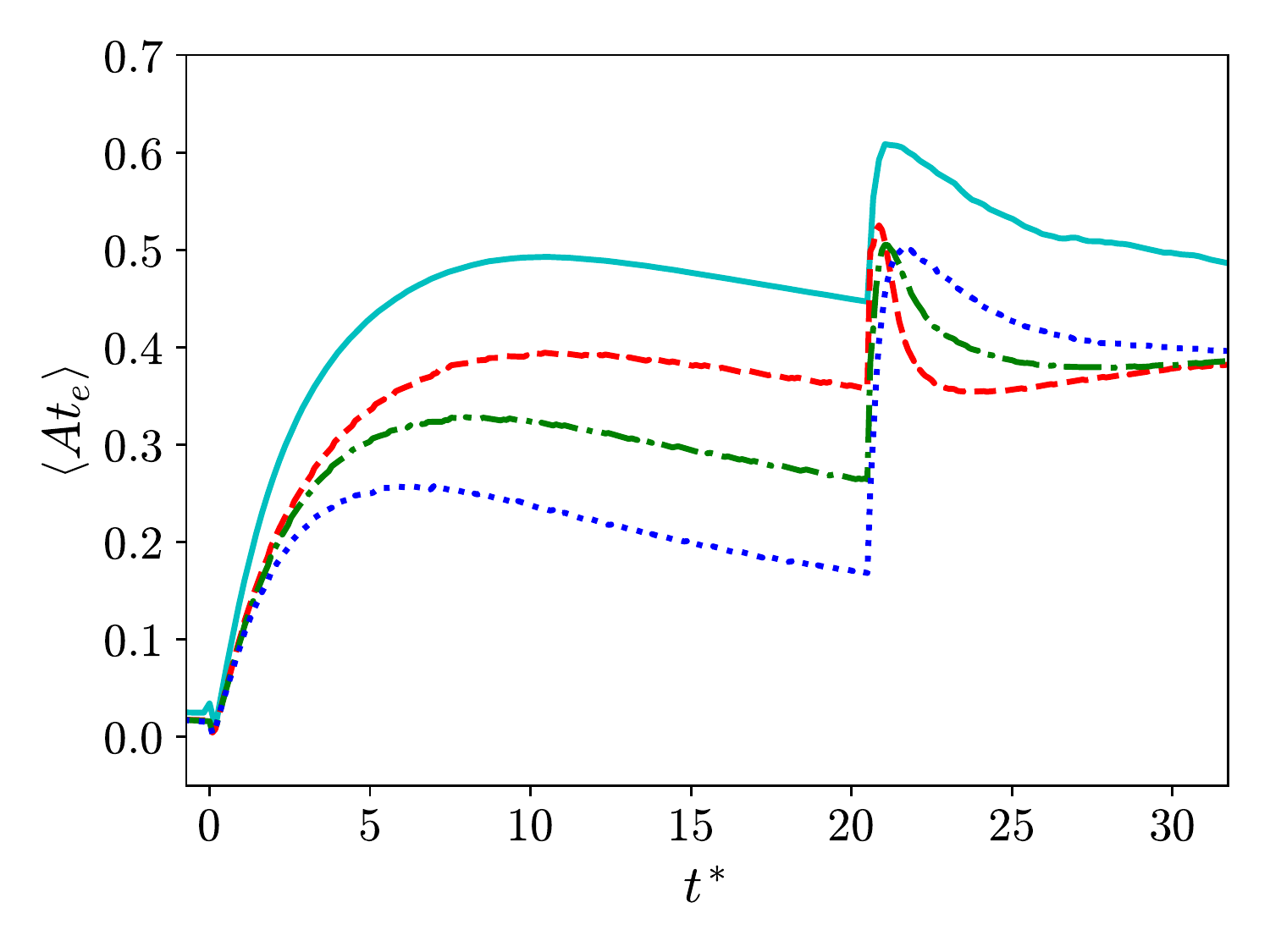}
\caption{Comparison of the time evolution of means of effective Atwood number, $\left< At_{e} \right>$, in the central part of mixing layer between the 2D and 3D problems. Cyan solid line: 2D with physical transport coefficients; red dashed line: 3D with physical transport coefficients; green dash-dotted line: 3D with $2 \times$physical transport coefficients; blue dotted line: 3D with $4 \times$physical transport coefficients.}
\label{fig:2D_3D_transport_coeffs_At_e}
\end{figure*}

Figure~\ref{fig:2D_3D_transport_coeffs_At_e} shows the effective Atwood number computed with equation~\eqref{eq:At_e_density} against time. Before first shock, $At_e$ is small and very different from the Atwood number of a discontinuous interface (0.68), since the material interface is smoothed initially. However, after first shock, $At_e$ increases rapidly as the interface becomes sharper, followed by gradual decrease due to molecular diffusion. After re-shock, there is a jump in $At_e$ due to further interface intensification. $At_e$ seems to plateau at late times in all cases. The flows in all cases are very non-Boussinesq after the first shock. In general, the mixing layer is more non-Boussinesq in the 2D configuration than in the 3D configuration, as $At_e$ is always larger for the 2D case.

%%%%%%%%%%%%%%%%%%%%%%%%%%%%%%%%%%%%%%%%%%%%%%%%%%%%%%%%%%%%%%%%%%%%%%%%%%%%%%%%
%%%%%%%%%%%%%%%%%%%%%%%%%%%%%%%%%%%%%%%%%%%%%%%%%%%%%%%%%%%%%%%%%%%%%%%%%%%%%%%%
%%%%%%%%%%%%%%%%%%%%%%%%%%%%%%%%%%%%%%%%%%%%%%%%%%%%%%%%%%%%%%%%%%%%%%%%%%%%%%%%

\subsection{\label{sec:mole_fraction_visualization} Visualization of mole fraction fields}

\begin{figure*}[!ht]
\centering
\subfigure[$\ t^{*}=7.25 \ \left( t=0.40\ \mathrm{ms} \right)$]{%
\includegraphics[trim={0 0 1.75cm 0},clip,height=0.2\textwidth]{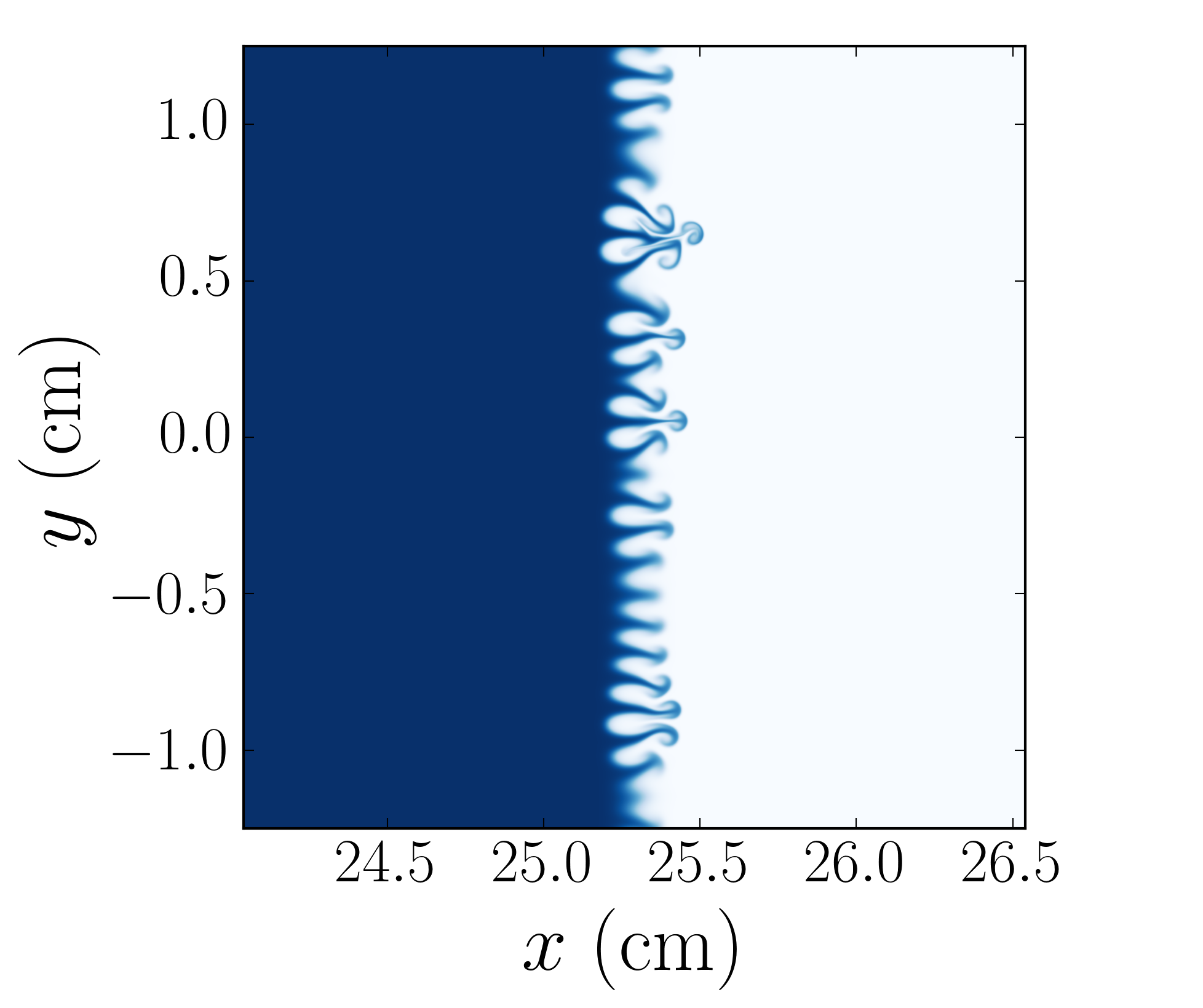}}
\subfigure[$\ t^{*}=19.9 \ \left( t=1.10\ \mathrm{ms} \right)$]{%
\includegraphics[trim={0 0 1.75cm 0},clip,height=0.2\textwidth]{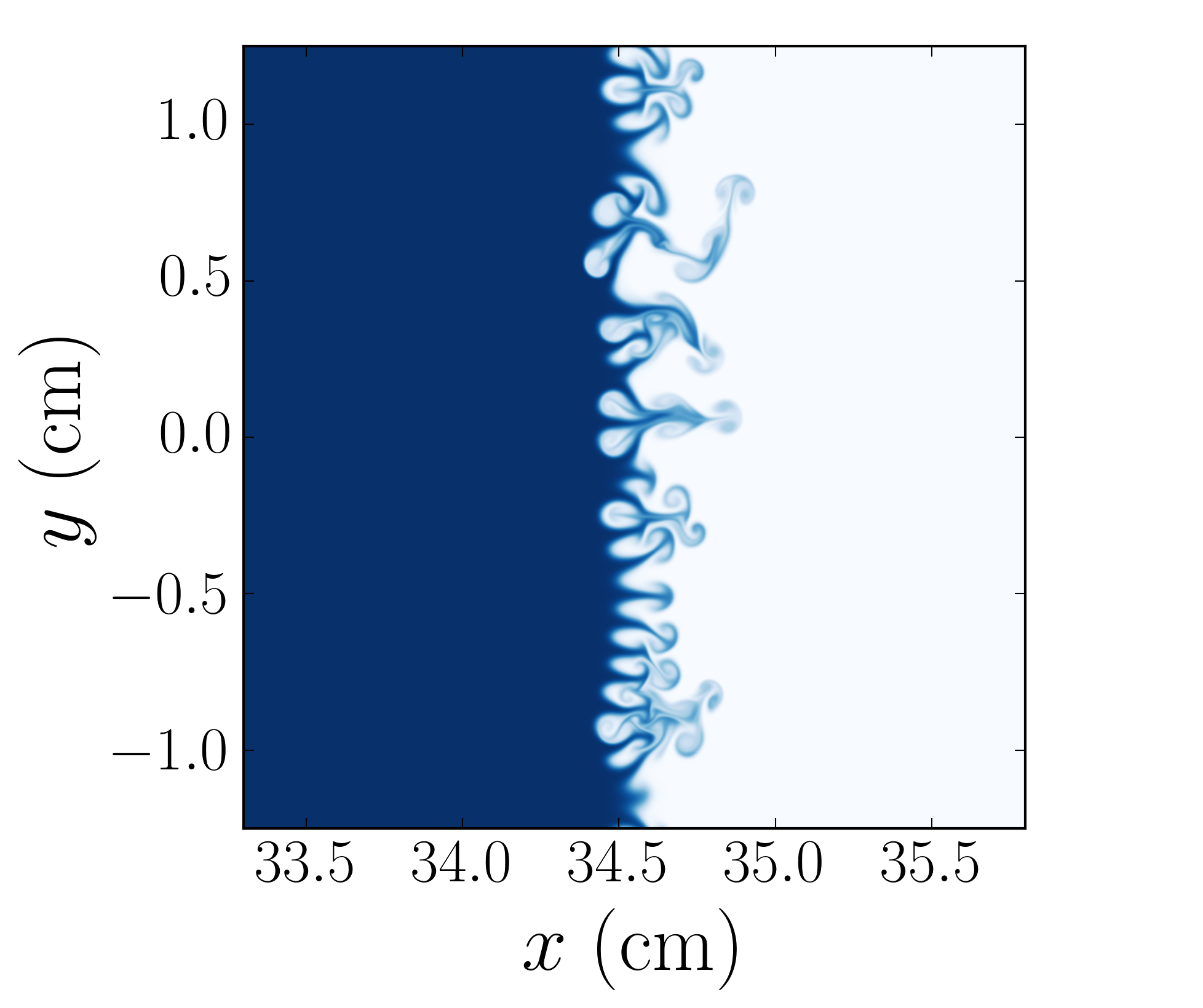}}
\subfigure[$\ t^{*}=21.8 \ \left( t=1.20\ \mathrm{ms} \right)$]{%
\includegraphics[trim={0 0 1.75cm 0},clip,height=0.2\textwidth]{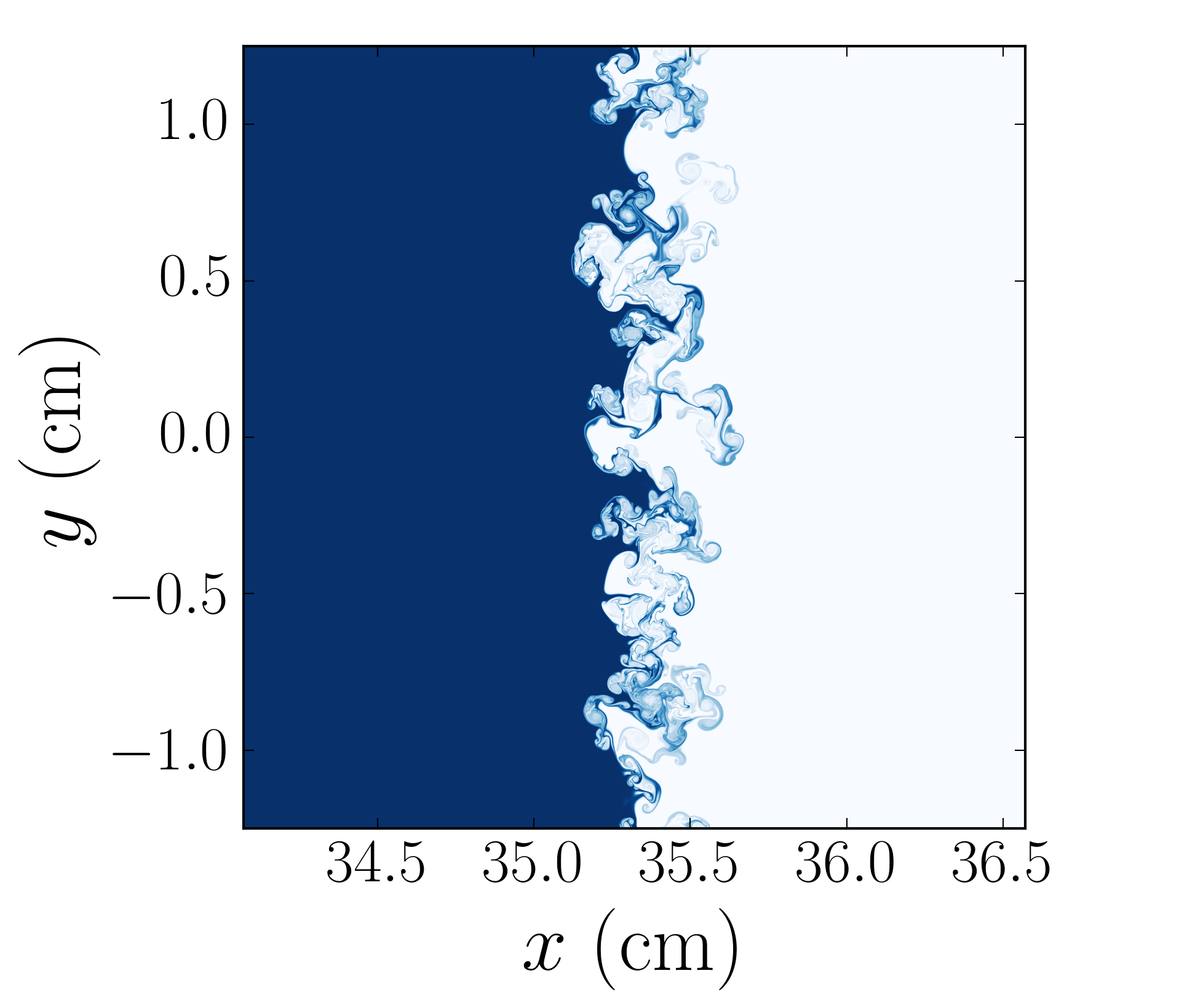}}
\subfigure[$\ t^{*}=31.7 \ \left( t=1.75\ \mathrm{ms} \right)$]{%
\includegraphics[height=0.2\textwidth]{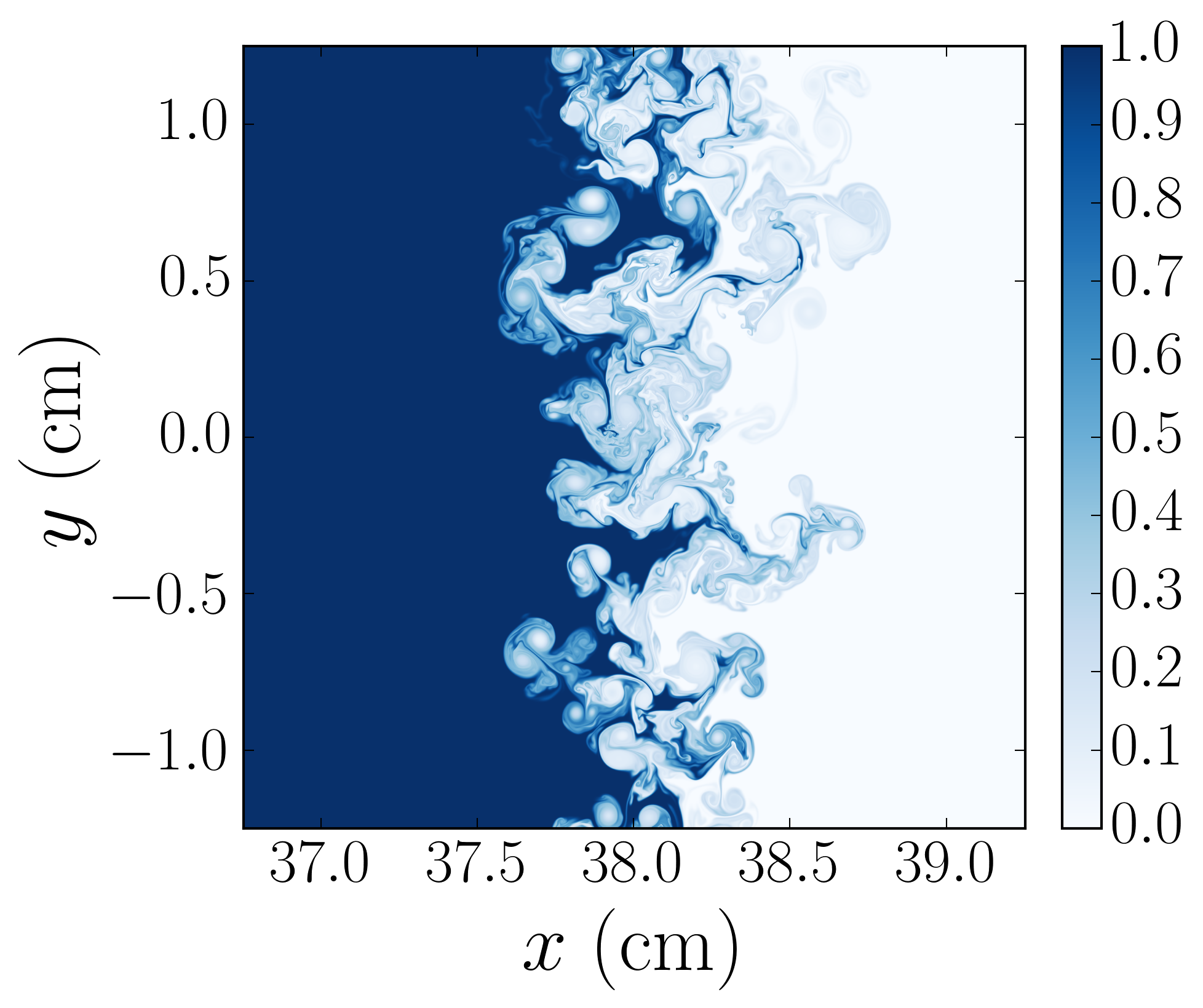}}
\caption{$\mathrm{SF_6}$ mole fraction fields, $X_{\mathrm{SF_6}}$, at different times from one of the realizations for the 2D problem computed with grid G.}
\label{fig:mole_fraction_2D}

\centering
\subfigure[$\ t^{*}=7.25 \ \left( t=0.40\ \mathrm{ms} \right)$]{%
\includegraphics[trim={0 0 1.75cm 0},clip,height=0.2\textwidth]{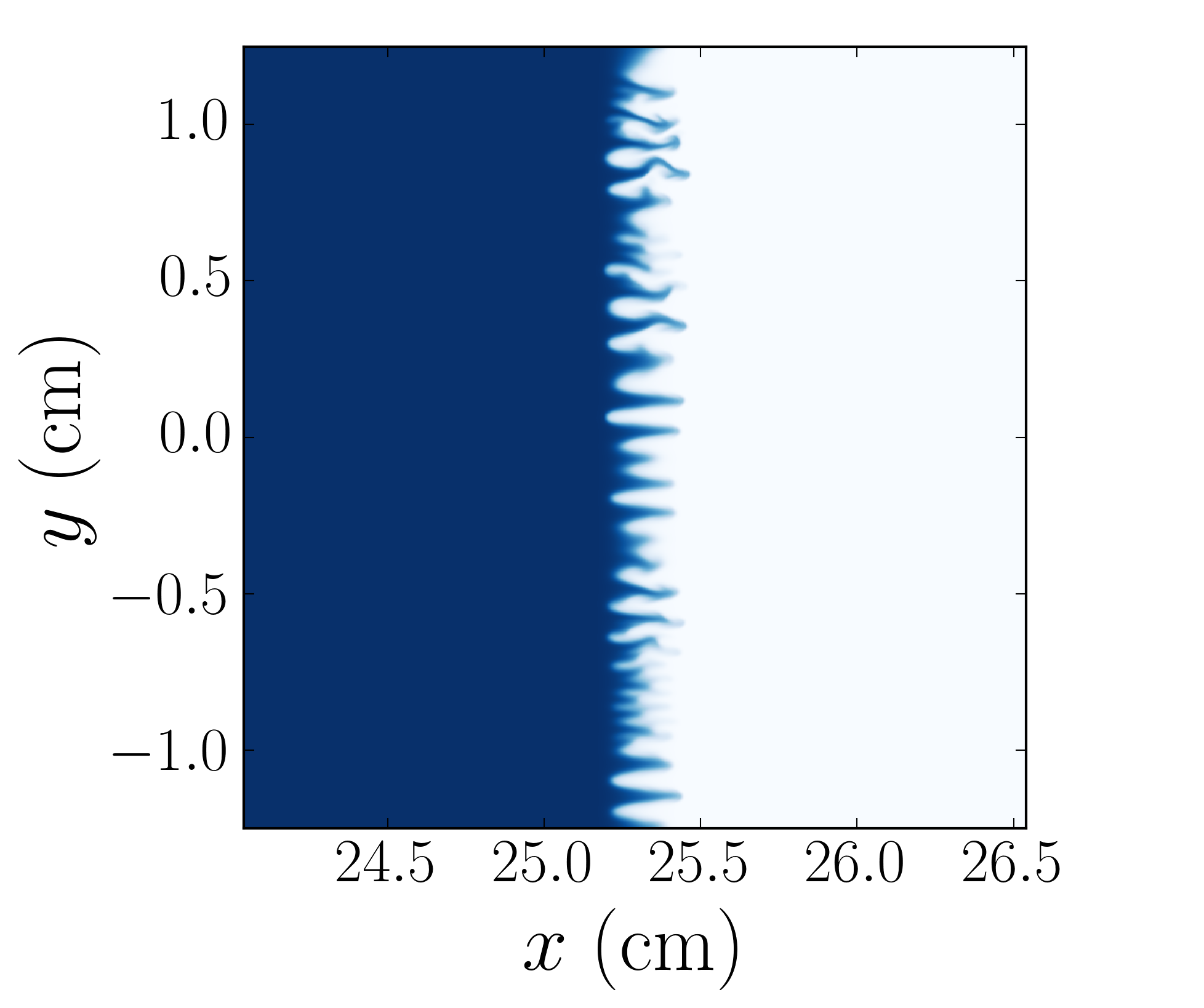}}
\subfigure[$\ t^{*}=19.9 \ \left( t=1.10\ \mathrm{ms} \right)$]{%
\includegraphics[trim={0 0 1.75cm 0},clip,height=0.2\textwidth]{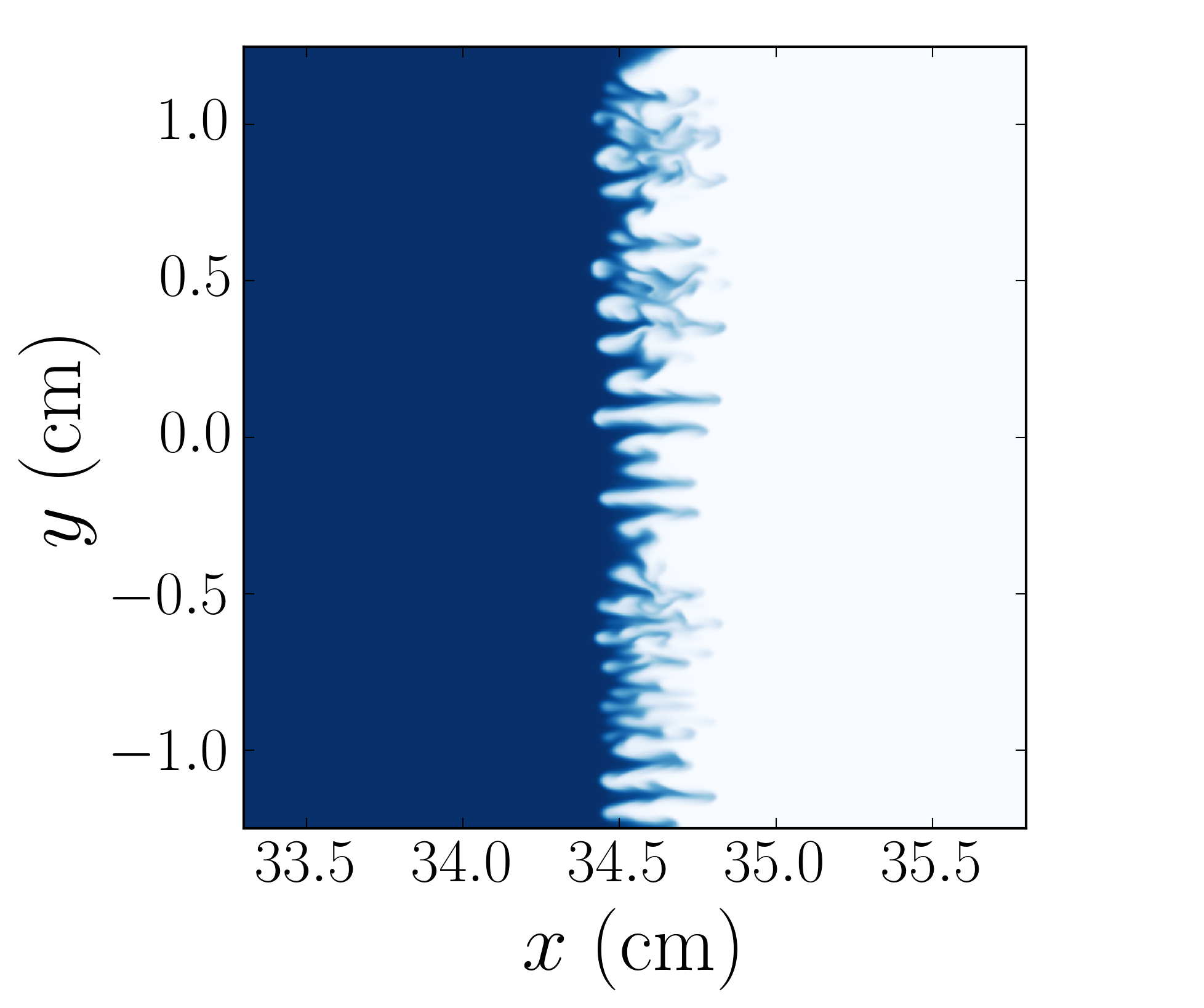}}
\subfigure[$\ t^{*}=21.8 \ \left( t=1.20\ \mathrm{ms} \right)$]{%
\includegraphics[trim={0 0 1.75cm 0},clip,height=0.2\textwidth]{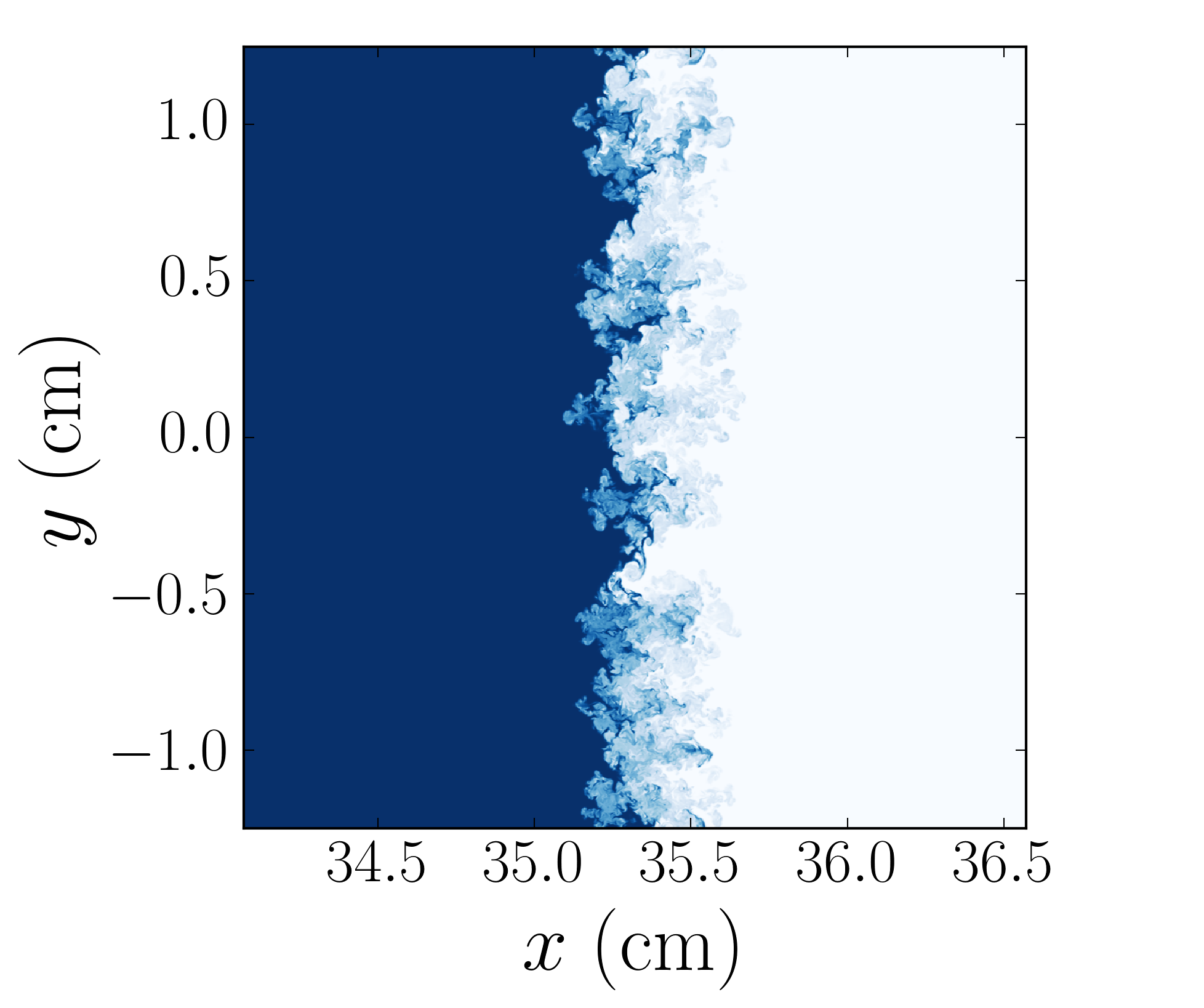}}
\subfigure[$\ t^{*}=31.7 \ \left( t=1.75\ \mathrm{ms} \right)$]{%
\includegraphics[height=0.2\textwidth]{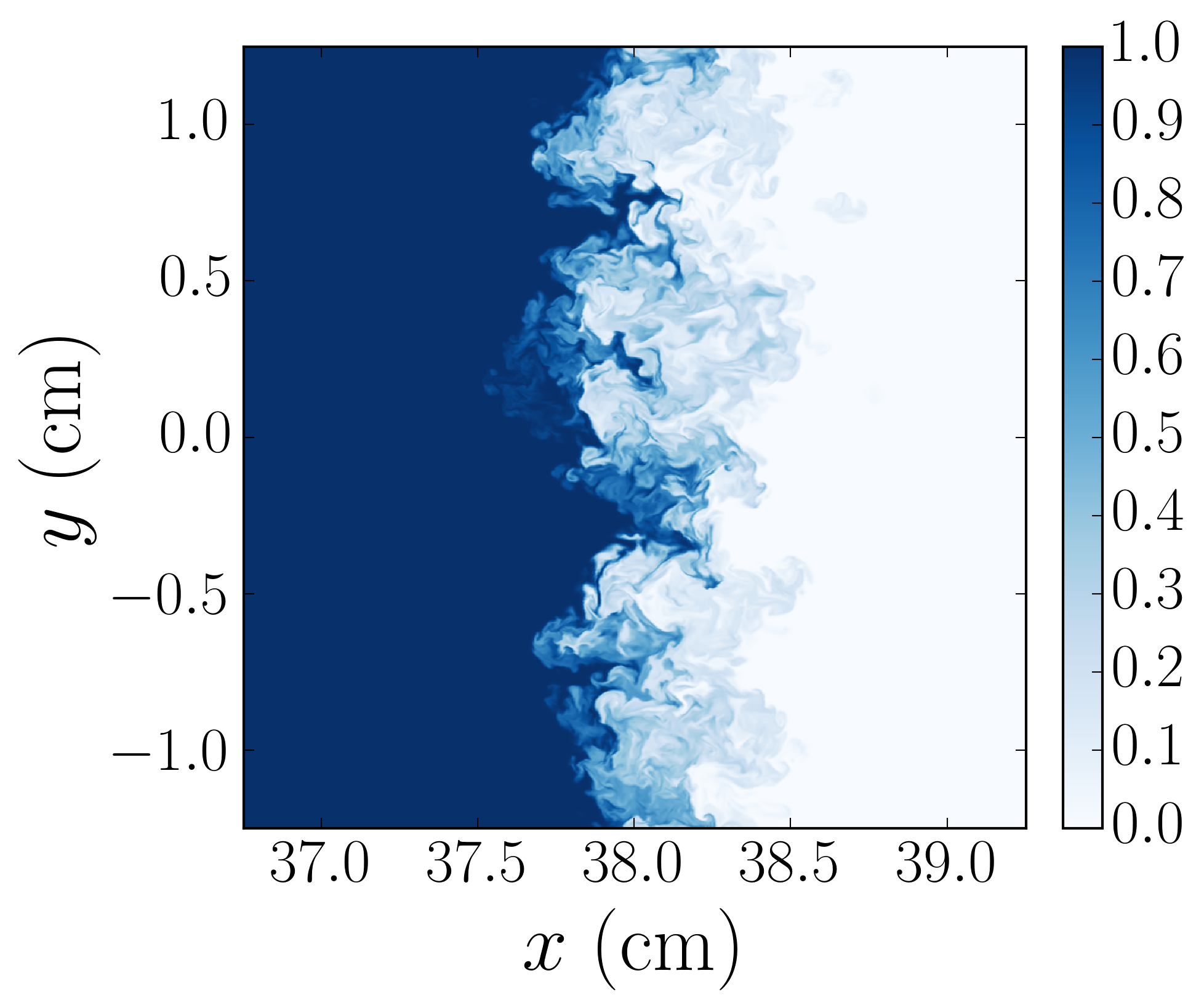}}
\caption{$\mathrm{SF_6}$ mole fraction fields, $X_{\mathrm{SF_6}}$, in $xy$ plane at $z=0$, at different times for the 3D problem with physical transport coefficients computed with grid D.}
\label{fig:mole_fraction_3D}

\centering
\subfigure[$\ t^{*}=7.25 \ \left( t=0.40\ \mathrm{ms} \right)$]{%
\includegraphics[trim={0 0 1.75cm 0},clip,height=0.2\textwidth]{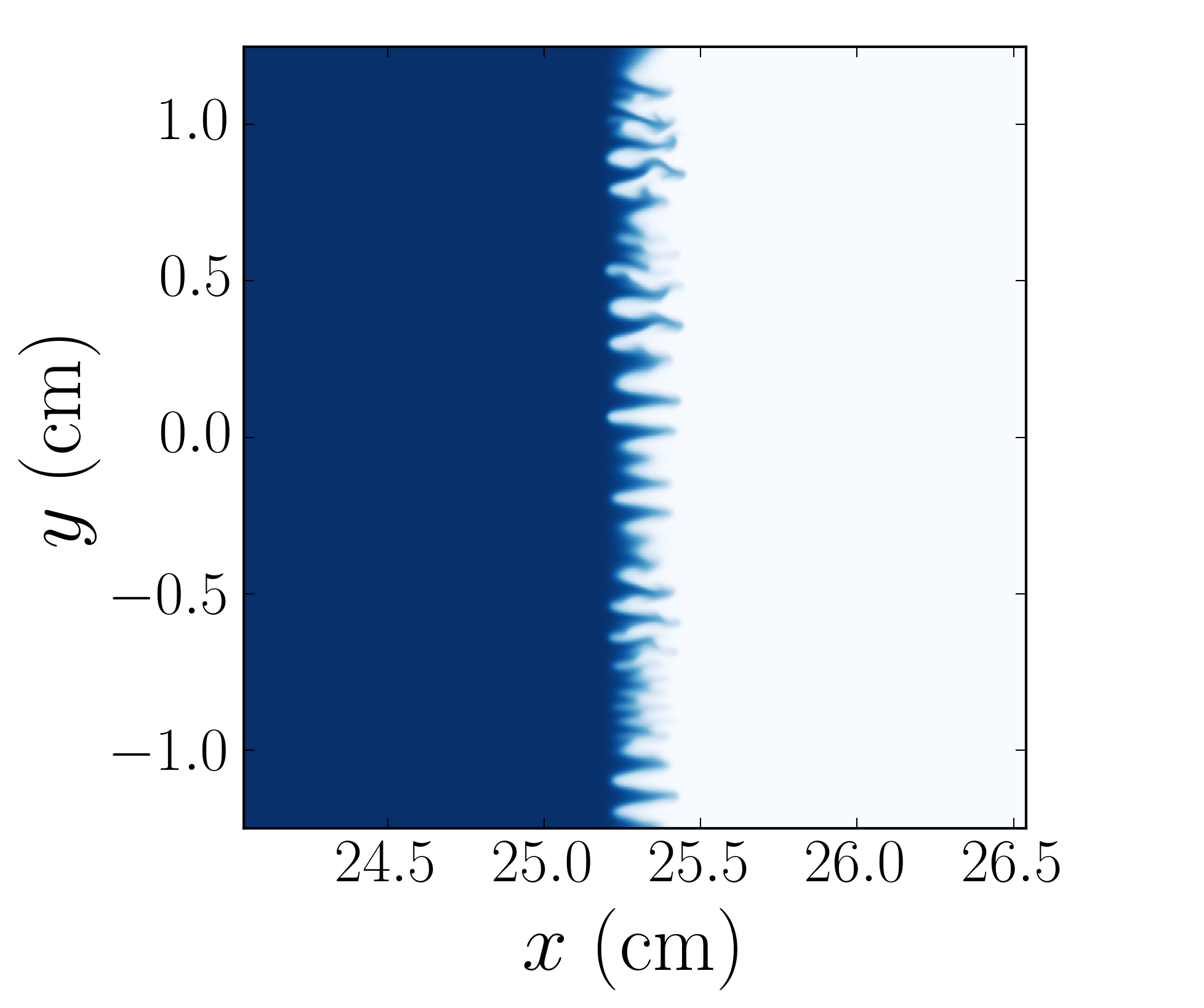}}
\subfigure[$\ t^{*}=19.9 \ \left( t=1.10\ \mathrm{ms} \right)$]{%
\includegraphics[trim={0 0 1.75cm 0},clip,height=0.2\textwidth]{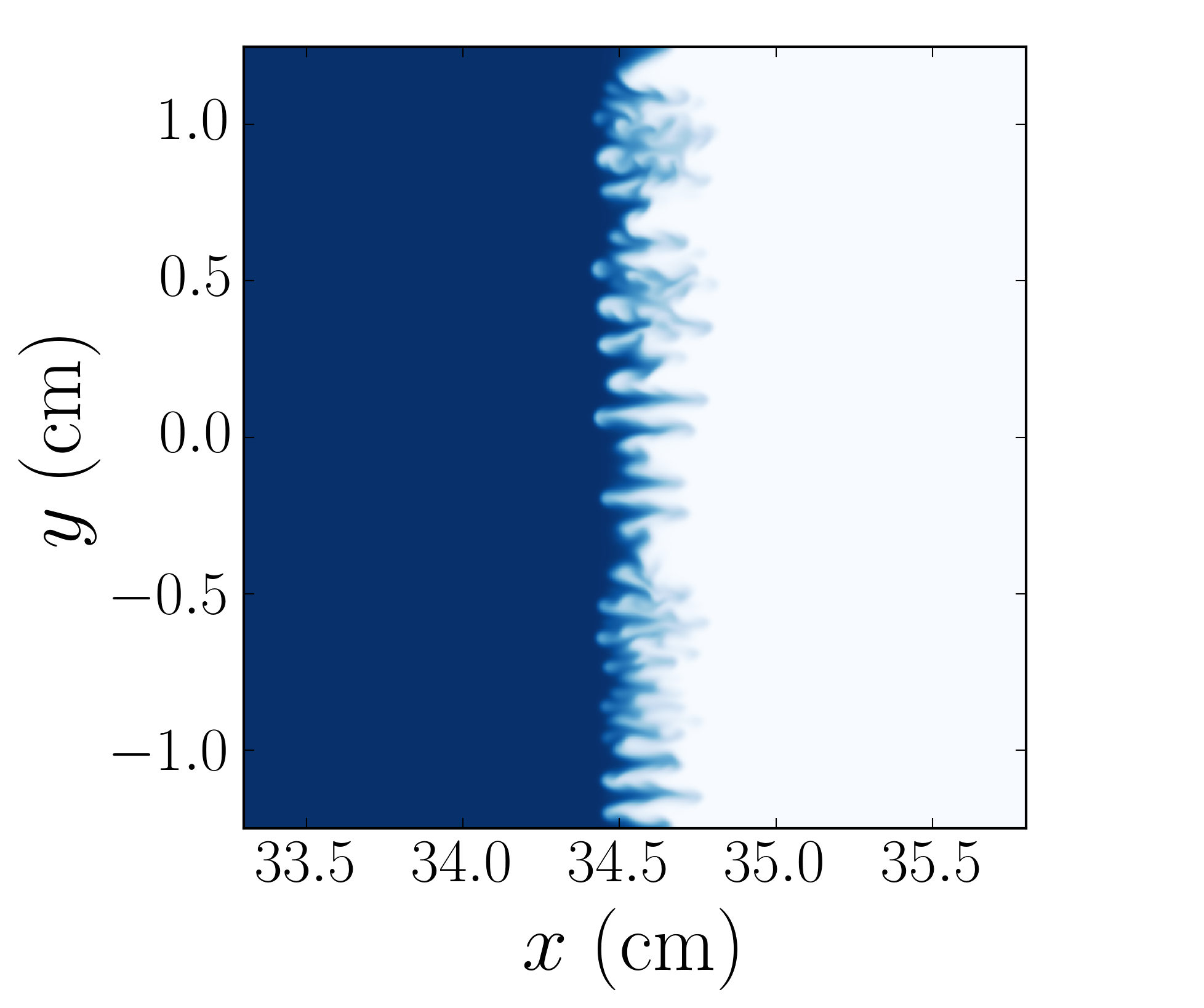}}
\subfigure[$\ t^{*}=21.8 \ \left( t=1.20\ \mathrm{ms} \right)$]{%
\includegraphics[trim={0 0 1.75cm 0},clip,height=0.2\textwidth]{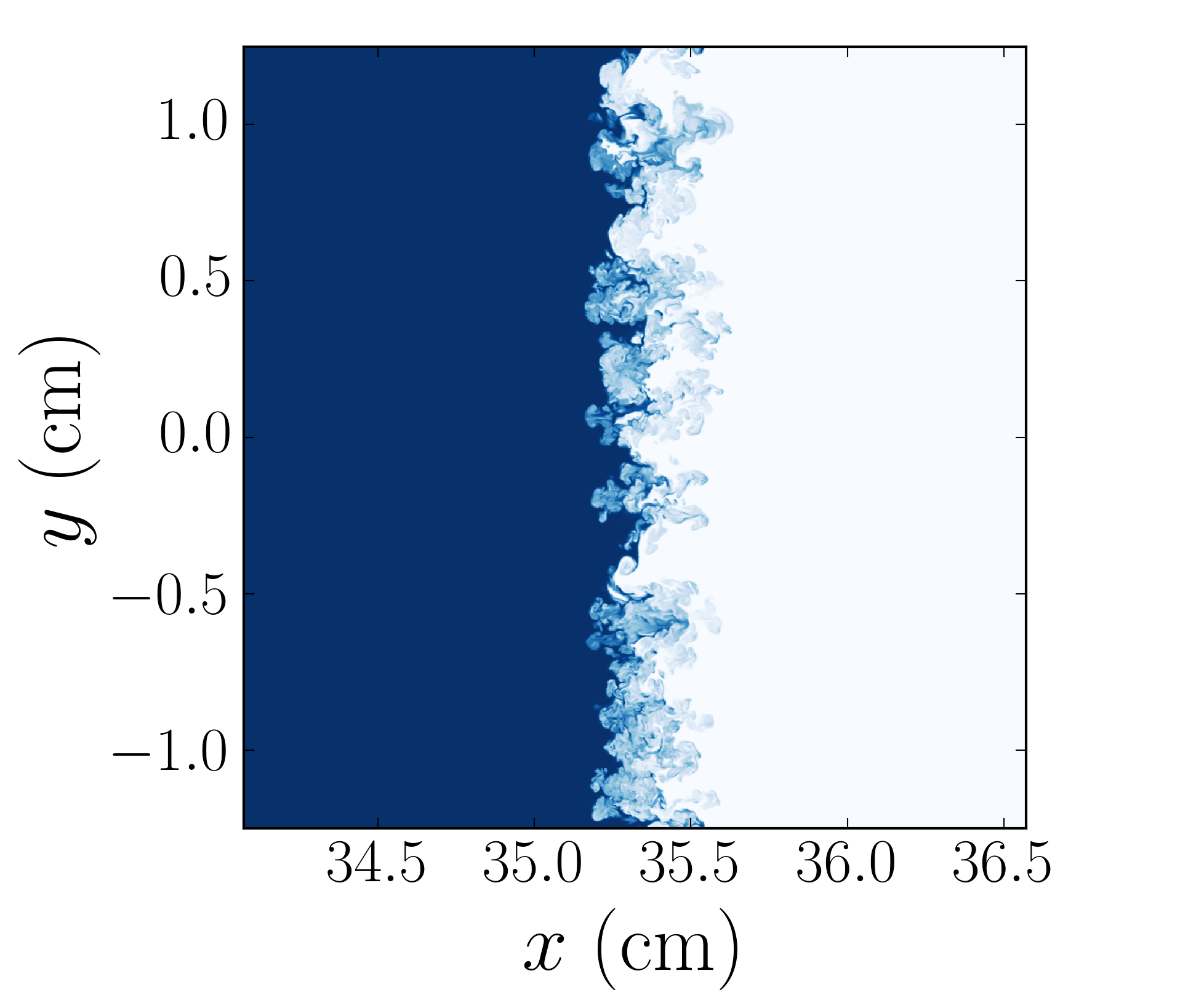}}
\subfigure[$\ t^{*}=31.7 \ \left( t=1.75\ \mathrm{ms} \right)$]{%
\includegraphics[height=0.2\textwidth]{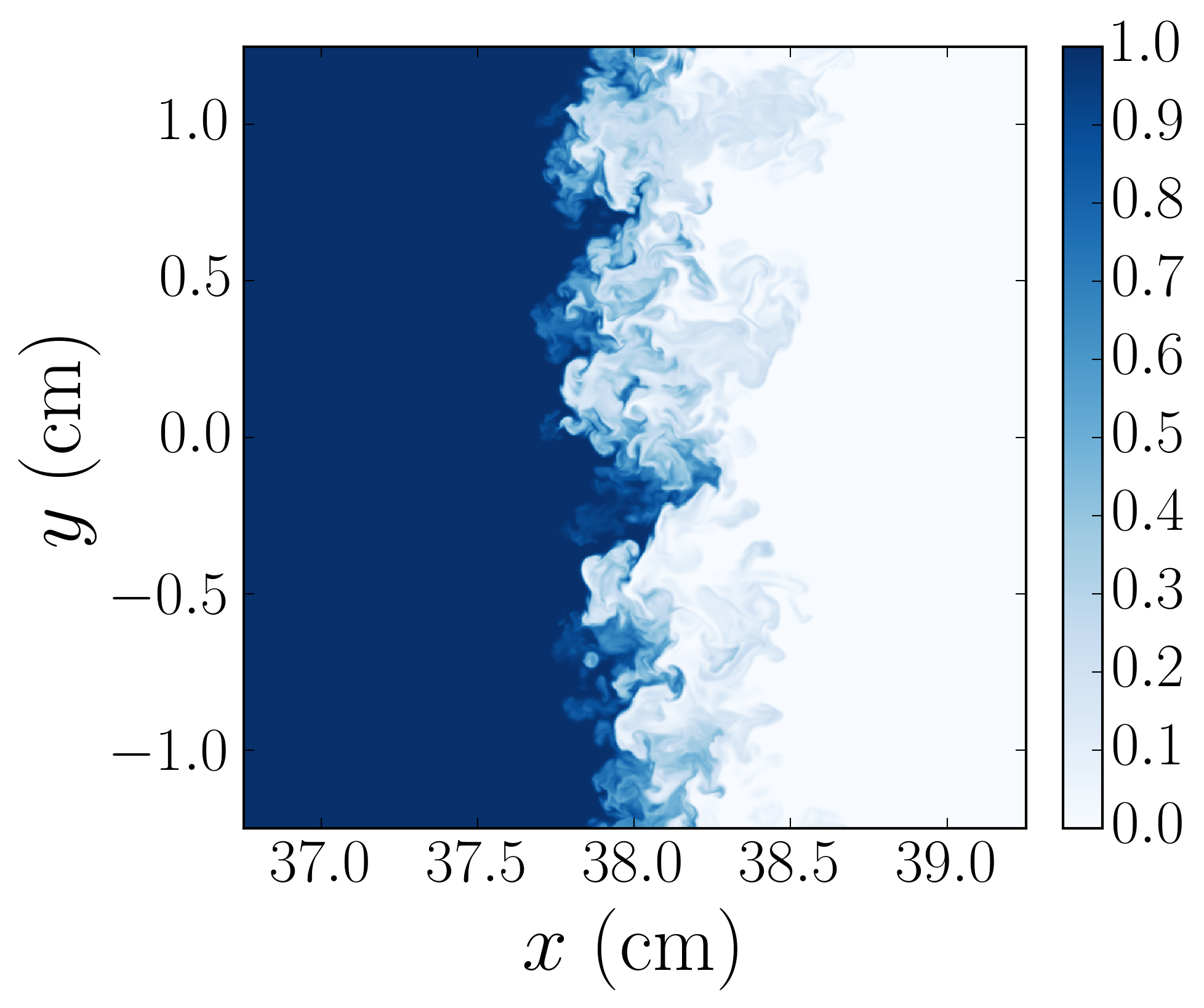}}
\caption{$\mathrm{SF_6}$ mole fraction fields, $X_{\mathrm{SF_6}}$, in $xy$ plane at $z=0$, at different times for the 3D problem with $2 \times$physical transport coefficients computed with grid D.}
\label{fig:mole_fraction_3D_ITC_2}

\centering
\subfigure[$\ t^{*}=7.25 \ \left( t=0.40\ \mathrm{ms} \right)$]{%
\includegraphics[trim={0 0 1.75cm 0},clip,height=0.2\textwidth]{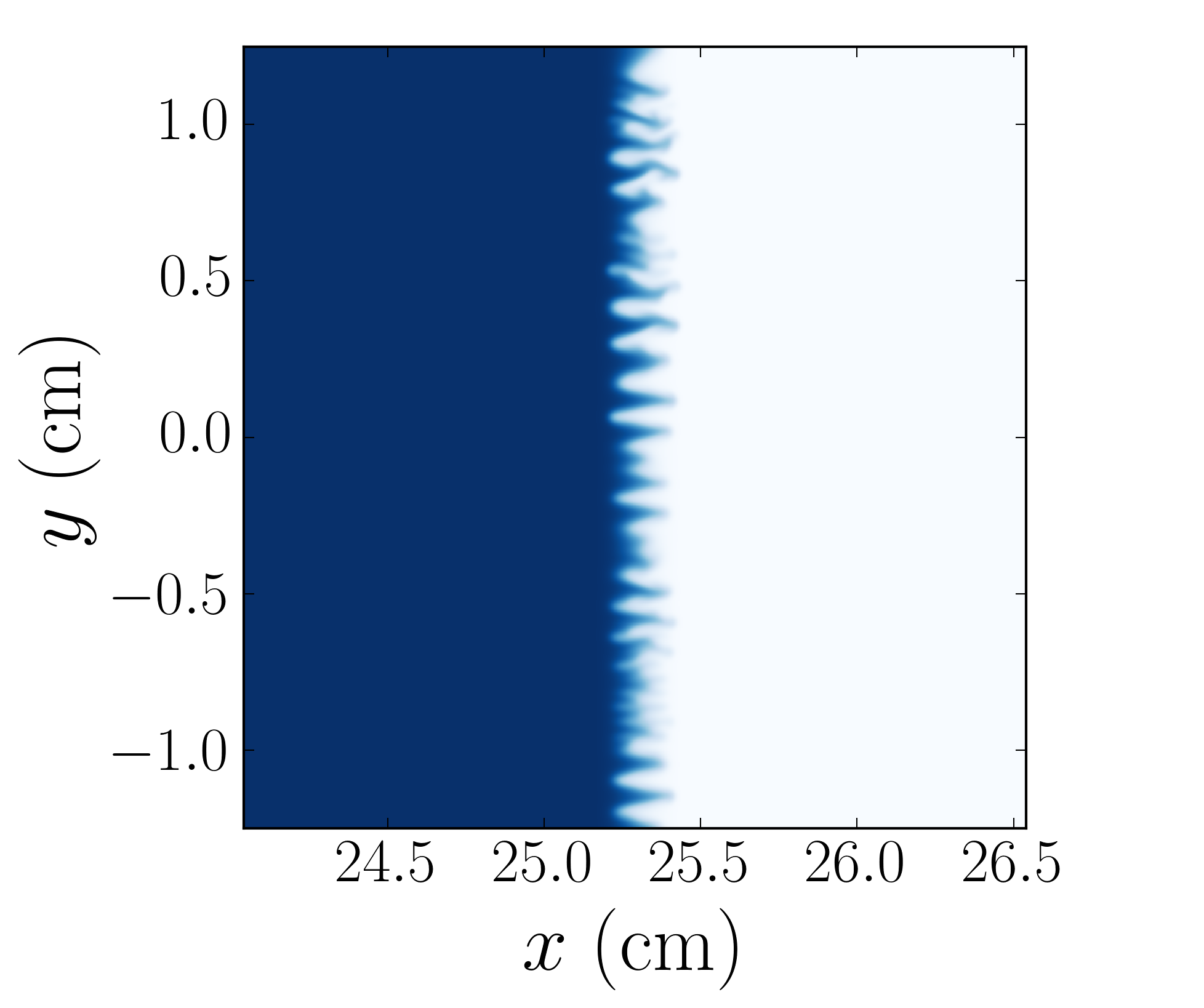}}
\subfigure[$\ t^{*}=19.9 \ \left( t=1.10\ \mathrm{ms} \right)$]{%
\includegraphics[trim={0 0 1.75cm 0},clip,height=0.2\textwidth]{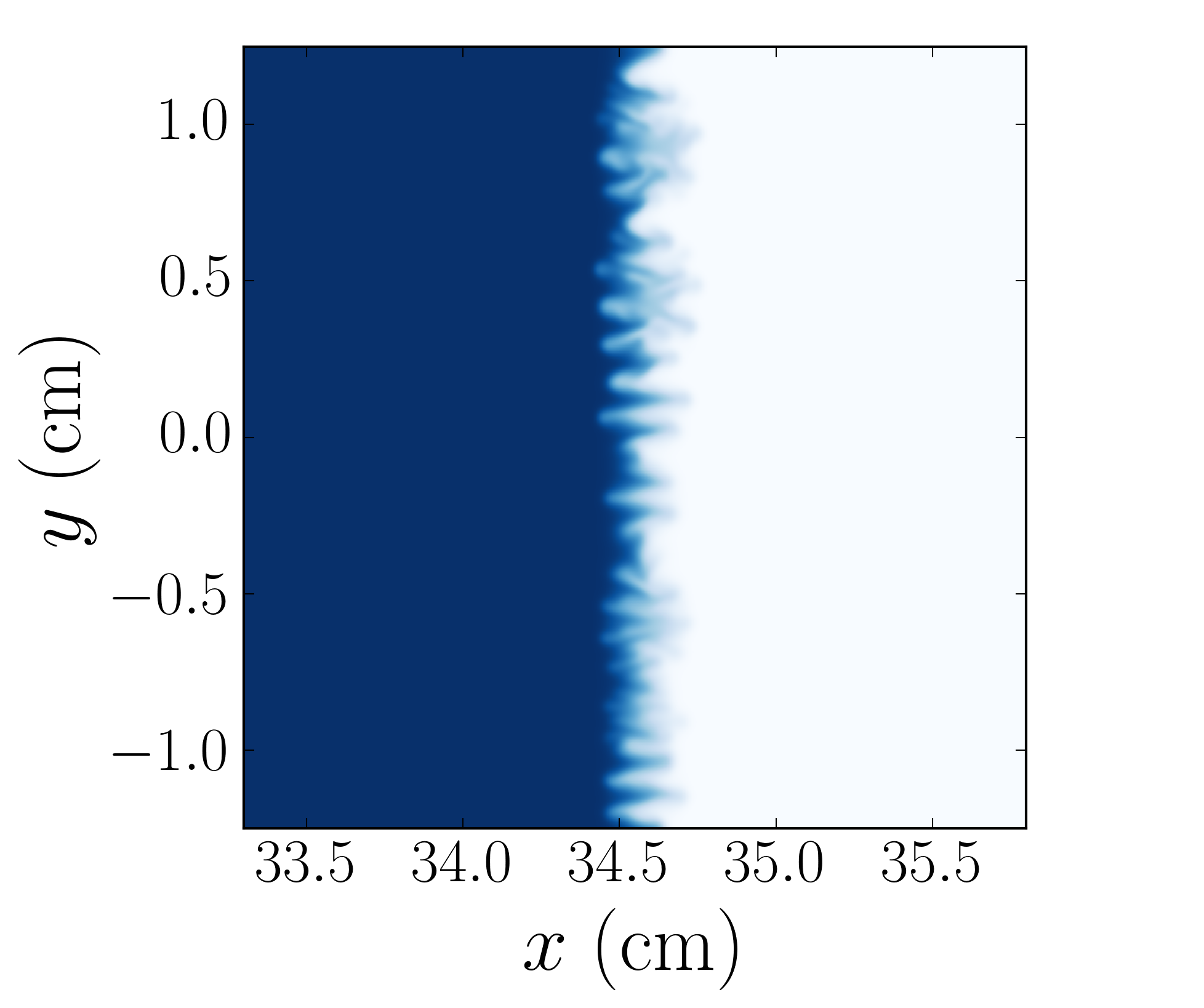}}
\subfigure[$\ t^{*}=21.8 \ \left( t=1.20\ \mathrm{ms} \right)$]{%
\includegraphics[trim={0 0 1.75cm 0},clip,height=0.2\textwidth]{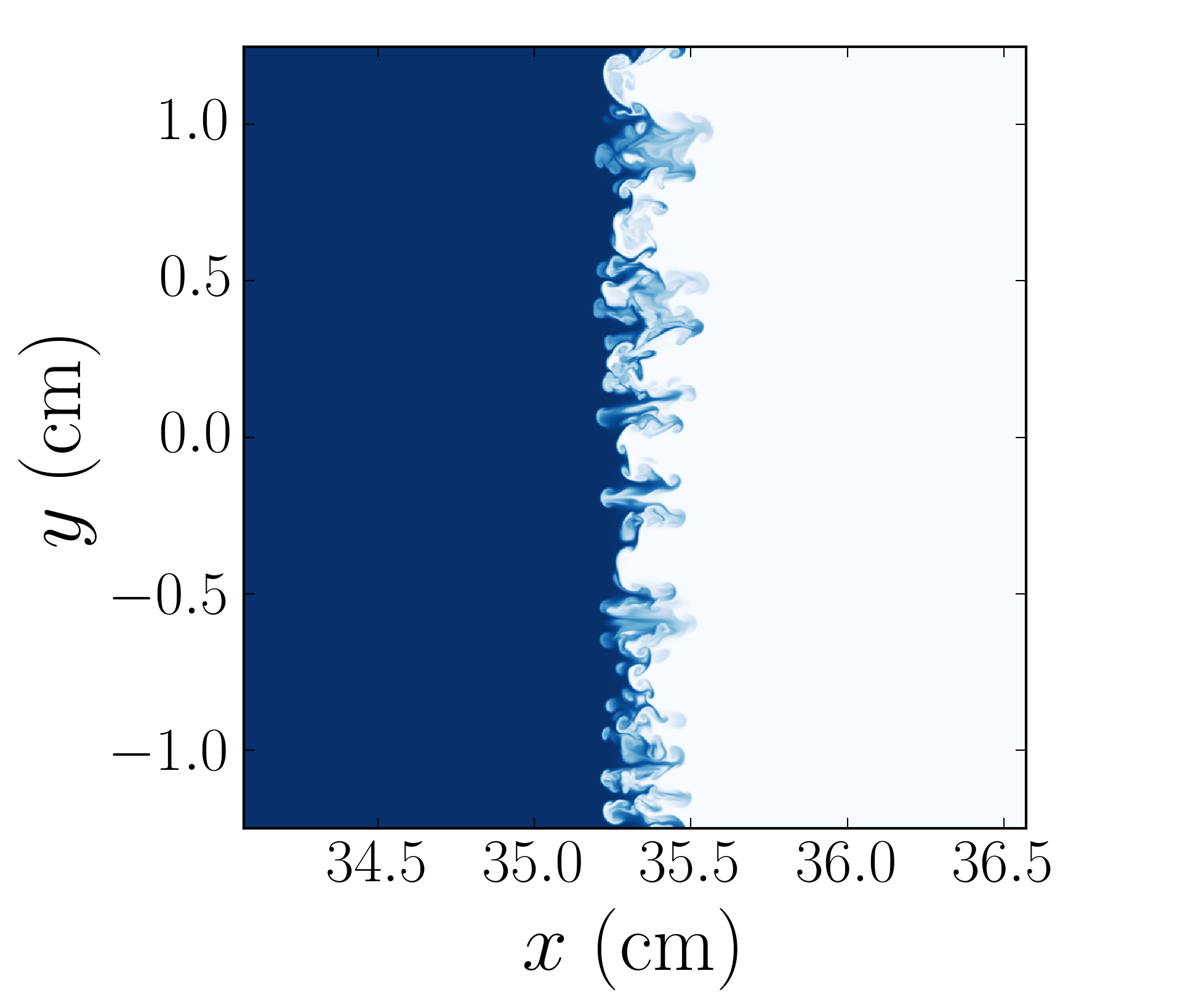}}
\subfigure[$\ t^{*}=31.7 \ \left( t=1.75\ \mathrm{ms} \right)$]{%
\includegraphics[height=0.2\textwidth]{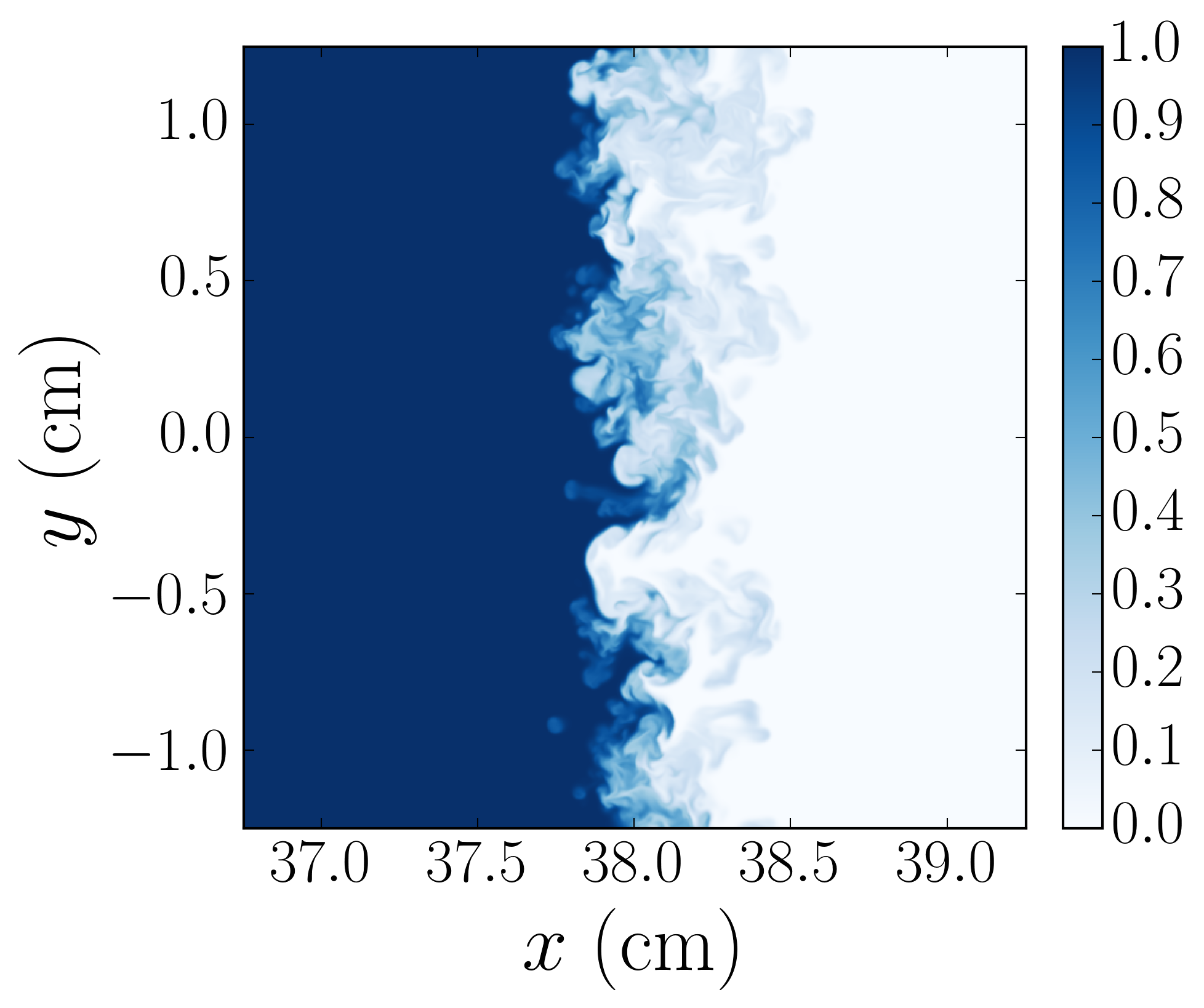}}
\caption{$\mathrm{SF_6}$ mole fraction fields, $X_{\mathrm{SF_6}}$, in $xy$ plane at $z=0$, at different times for the 3D problem with $4 \times$physical transport coefficients computed with grid D.}
\label{fig:mole_fraction_3D_ITC_4}
\end{figure*}

Figures~\ref{fig:mole_fraction_2D} and \ref{fig:mole_fraction_3D} show the $\mathrm{SF_{6}}$ mole fraction fields for the 2D and 3D problems at different normalized times $t^{*}$. The development of the bubble and spike structures is very similar at early times after first shock for both cases. As time evolves, the differences in the development of the instability become more distinguishable. In the 2D case, many mushroom structures, which are caused by the roll-up of the interface due to baroclinic torque can be observed. In the 3D case, the roll-up of the interface is less prominent. Instead, the bubble and spike  grow into long structures at the moment just before re-shock. Right after re-shock, at $t^{*} \approx 21.5$, there is an immediate enhancement in mixing for both cases. However, the distinction between the 2D and 3D cases remains clear. In the 3D case, particular structures can no longer be identified, which is indicative of the mixing transition. On the contrary, distinct mushroom dipole structures can still be observed for the 2D case, although the flow field becomes much more chaotic than the state just before re-shock.

Figures~\ref{fig:mole_fraction_3D}, \ref{fig:mole_fraction_3D_ITC_2}, and \ref{fig:mole_fraction_3D_ITC_4} compare the $\mathrm{SF_{6}}$ mole fraction fields from the 3D simulations with different transport coefficients. Before re-shock, features are smeared out for cases with increased transport coefficients, as the viscous and diffusive effects are acting on faster time scales relative to those of the inviscid linear and nonlinear instability effects. After re-shock, the mixing transitions are delayed for the reduced Reynolds number cases. This is most clear for the case with 4 times increase in transport coefficients. Right after re-shock, at $t^{*}=22.6$, features maintaining coherence can still be observed. Nevertheless, based on the visual inspection of the figures, turbulent mixing is observed at the end times for all simulations, indicating that mixing transitions occur in all cases.

%%%%%%%%%%%%%%%%%%%%%%%%%%%%%%%%%%%%%%%%%%%%%%%%%%%%%%%%%%%%%%%%%%%%%%%%%%%%%%%%
%%%%%%%%%%%%%%%%%%%%%%%%%%%%%%%%%%%%%%%%%%%%%%%%%%%%%%%%%%%%%%%%%%%%%%%%%%%%%%%%
%%%%%%%%%%%%%%%%%%%%%%%%%%%%%%%%%%%%%%%%%%%%%%%%%%%%%%%%%%%%%%%%%%%%%%%%%%%%%%%%

\subsection{Mole fraction profiles and mixing widths}

\begin{figure*}[!ht]
\centering
\subfigure[$\ $Before re-shock, 2D]{%
\includegraphics[height = 0.33\textwidth]{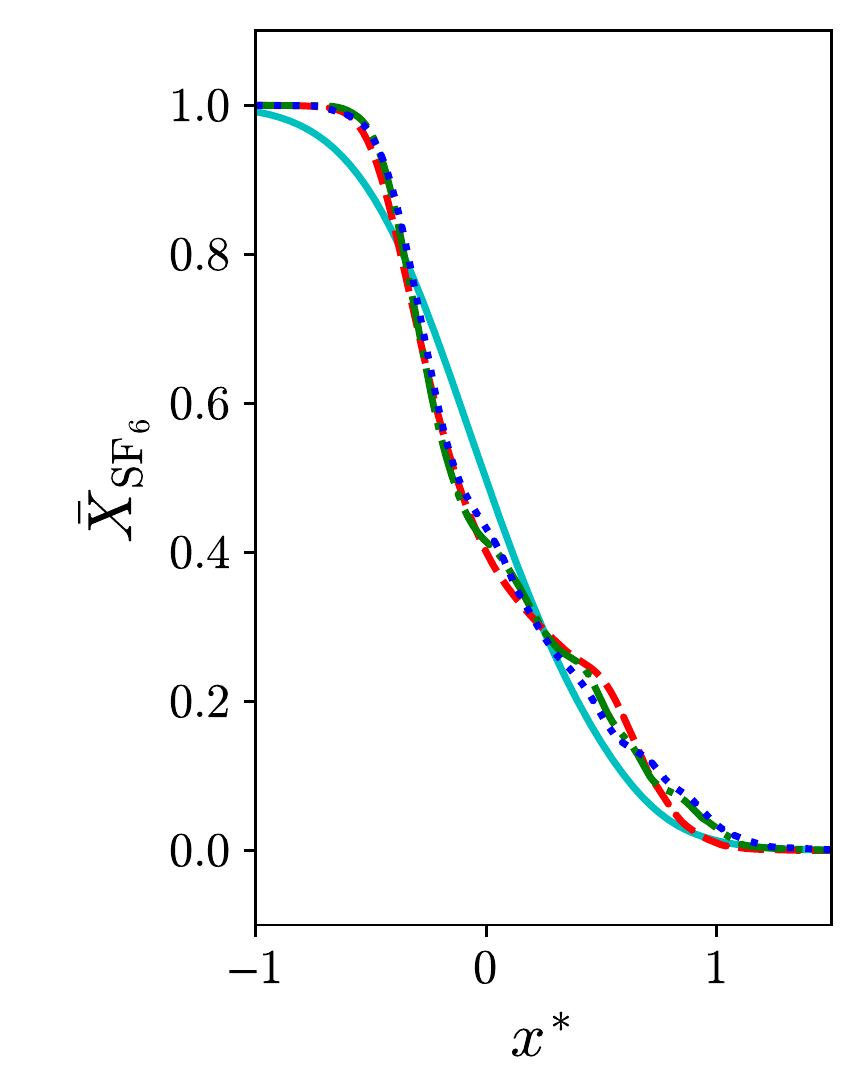}}
\subfigure[$\ $Before re-shock, 3D]{%
\includegraphics[height = 0.33\textwidth]{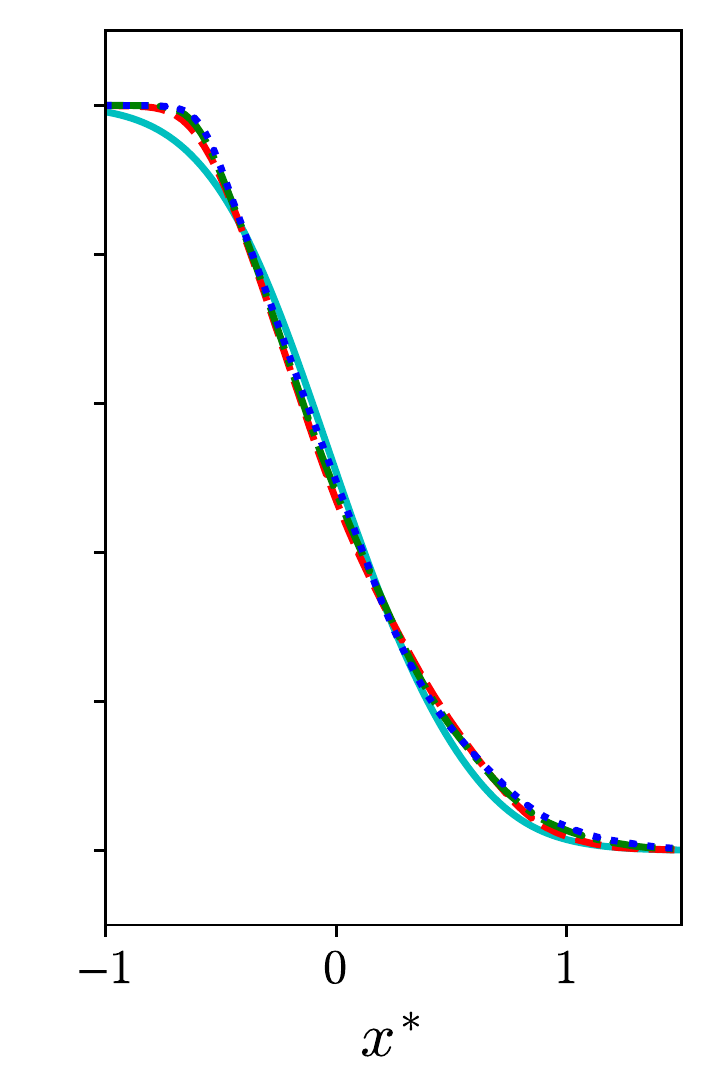}}
\subfigure[$\ $After re-shock, 2D]{%
\includegraphics[height = 0.33\textwidth]{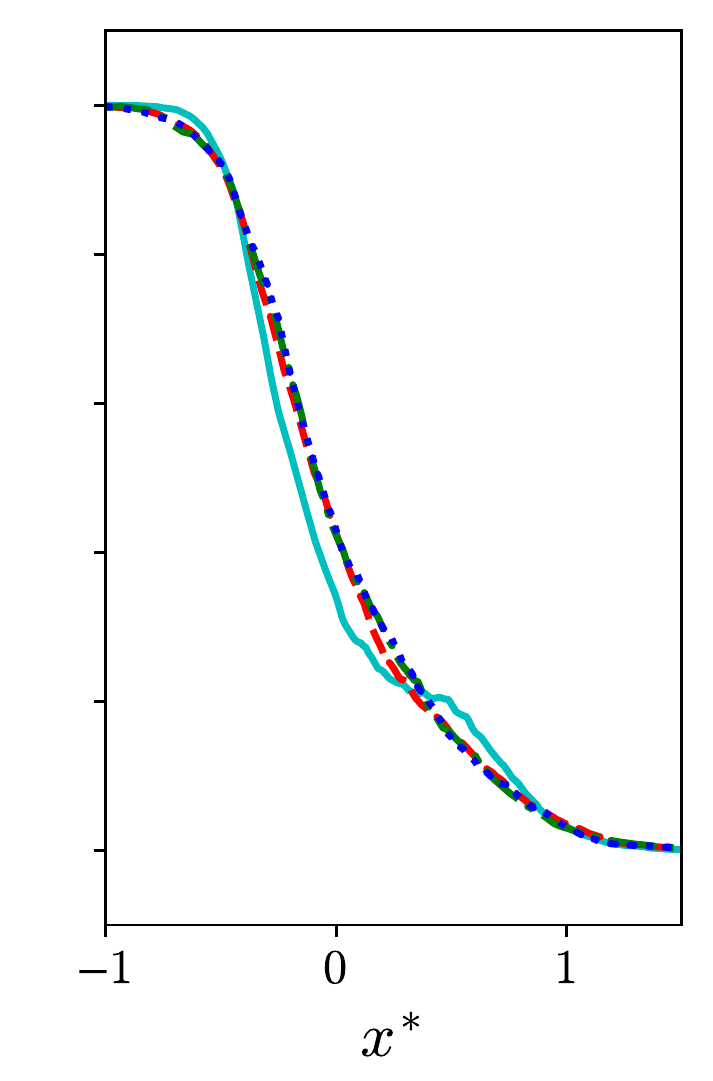}}
\subfigure[$\ $After re-shock, 3D]{%
\includegraphics[height = 0.33\textwidth]{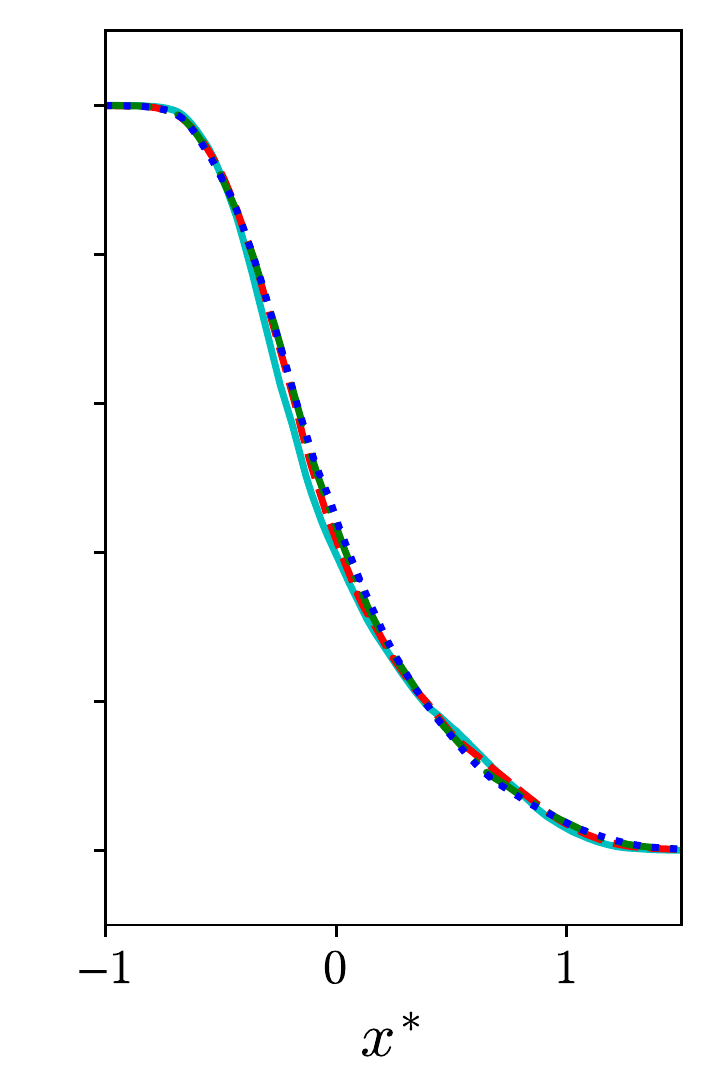}}
\caption{Profiles of $\mathrm{SF_6}$ mole fraction, $\bar{X}_{\mathrm{SF_6}}$, at different times for the 2D and 3D problems with physical transport coefficients. Before re-shock: $t^{*}=0.91$ (cyan solid line); $t^{*}=7.25$ (red dashed line); $t^{*}=13.6$ (green dash-dotted line); $t^{*}=19.9$ (blue dotted line). After re-shock: $t^{*}=21.8$ (cyan solid line); $t^{*}=25.4$ (red dashed line); $t^{*}=29.0$ (green dash-dotted line); $t^{*}=31.7$ (blue dotted line).}
\label{fig:2D_vs_3D_mole_fraction_profiles}
\end{figure*}

Figure~\ref{fig:2D_vs_3D_mole_fraction_profiles} shows the mean profiles of $\mathrm{SF_6}$ mole fraction, $\bar{X}_{\mathrm{SF_6}}$, from the 2D and 3D simulations with physical transport coefficients at different times. The mean mole fraction profiles collapse quite well at late times after both first shock and re-shock for both 2D and 3D cases. The profiles are asymmetric, as the spikes penetrate more than the bubbles. Similar findings were also observed for the density profiles in the planar Rayleigh--Taylor (RT) instability, or RTI, studied by \citet{livescu2009high,livescu2010npv}. It is also interesting to see that the normalized density profiles at late times become quite indistinguishable for both 2D and 3D cases.

\begin{figure*}[!ht]
\centering
\subfigure[$\ $Global time evolution]{%
\includegraphics[width = 0.45\textwidth]{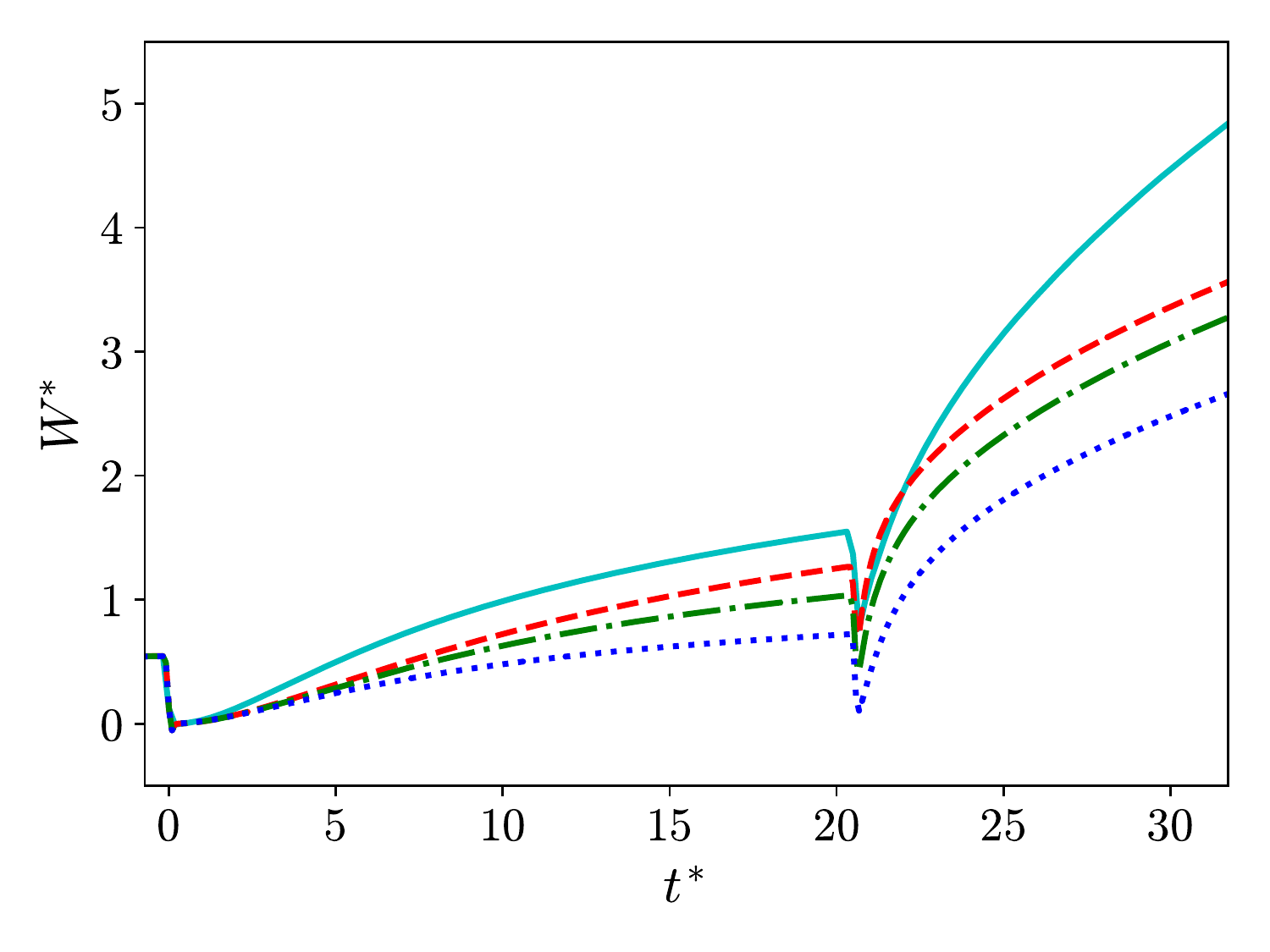}
\label{fig:2D_3D_transport_coeffs_mixing_widths}}
\subfigure[$\ $After re-shock]{%
\includegraphics[width = 0.45\textwidth]{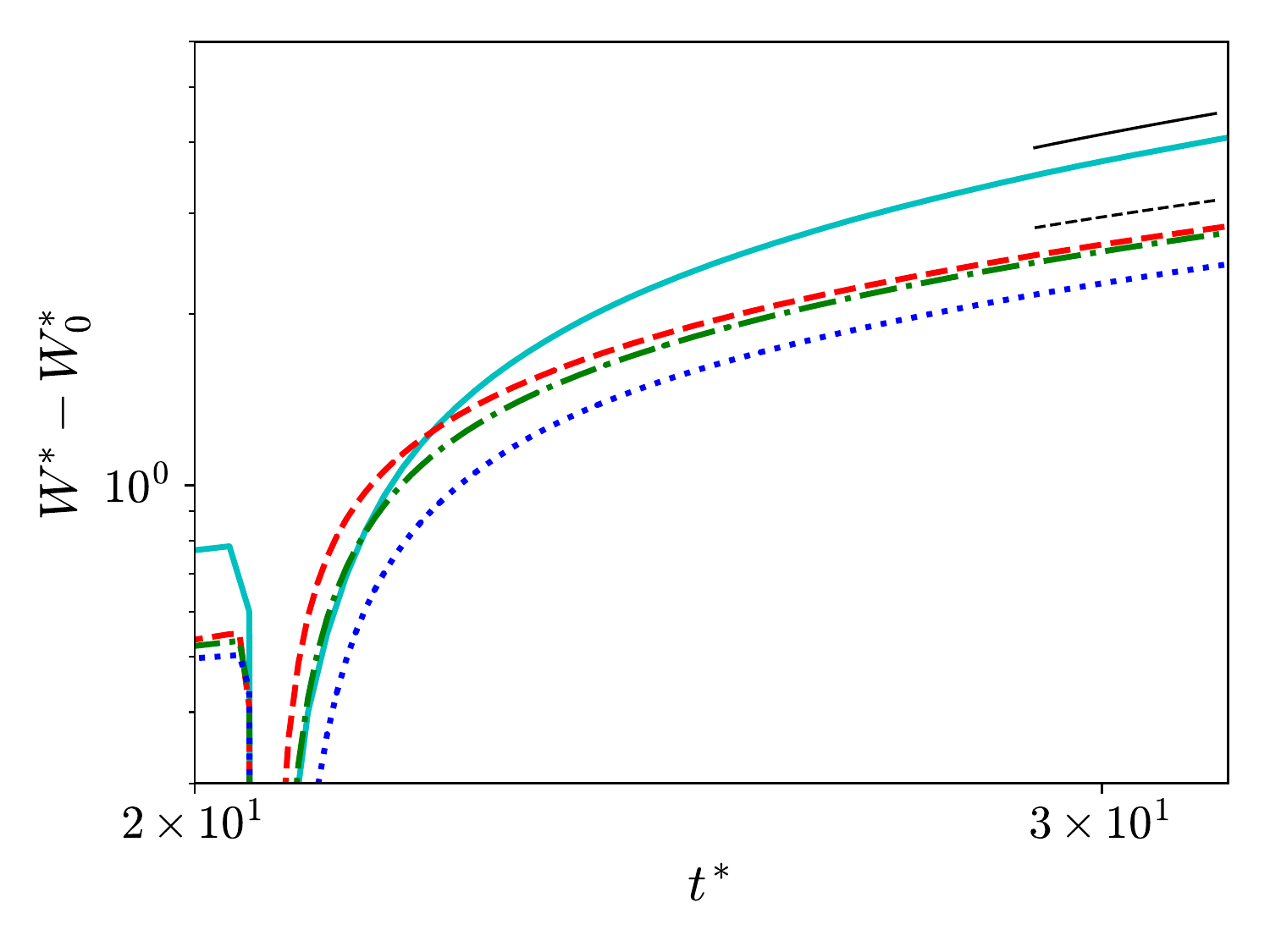}
\label{fig:2D_3D_transport_coeffs_mixing_widths_late}}
\caption{Comparison of the time evolution of mixing widths between the 2D and 3D problems. Cyan solid line: 2D with physical transport coefficients; red dashed line: 3D with physical transport coefficients; green dash-dotted line: 3D with $2 \times$physical transport coefficients; blue dotted line: 3D with $4 \times$physical transport coefficients. The black thin solid and dashed lines indicate the scalings, $\sim \left( t^{*}-t^{*}_0 \right)^{0.55}$ and $\sim \left( t^{*}-t^{*}_0 \right)^{0.44}$ respectively, where $t^{*}_0 = 20.5$ is the re-shock time.}
\label{fig:2D_3D_transport_coeffs_mixing_widths_combined}
\end{figure*}

The development of the mixing widths for the 2D and 3D cases is compared in figure~\ref{fig:2D_3D_transport_coeffs_mixing_widths}. The mixing width is normalized by $\dot{\eta}_{\mathrm{imp}}$ and $\tau_c$:
\begin{equation}
    W^{*} = \frac{W - \left. W \right| _{t=0}}{\dot{\eta}_{\mathrm{imp}} \tau_c}.
\end{equation}

With physical transport coefficients, $W^{*}$ initially grows at a slightly faster rate in the 2D case compared to the 3D case after first shock, but the growth rates become similar before re-shock. After re-shock, the 2D mixing width grows at a much higher rate than the 3D mixing width until the end of simulations. Comparing the 3D cases with different transport coefficients on the same figure, it can be seen that although the growths of the normalized mixing widths are very similar just after first shock, they become smaller with decreasing Reynolds number until re-shock. After re-shock, the growth rates that are different initially become very similar near the end of simulations among all 3D cases and this can be seen from figure~\ref{fig:2D_3D_transport_coeffs_mixing_widths_late}.

In many previous works of RMI, the relation between the turbulent mixing layer width and time was commonly studied through the scaling law: $W^{*} \sim \left( t^{*} - t^{*}_{0} \right)^{\theta}$, where $t^{*}_{0}$ is the virtual time origin that is normally chosen as the time when the shock traverses the interface. With that scaling, a wide range of asymptotic values for the exponent $\theta$ between 0.25 to 0.67 \cite{dimonte1995richtmyer, dimonte2000density, zhou2001scaling, lombardini2012transition, tritschler2014richtmyer, thornber2017late} was suggested from theoretical analysis or observed from experiments and simulations. Here, we are mainly interested in how the mixing widths grow at late times after re-shock. It is found that at late times after re-shock, there are reasonable fits for both the 2D and 3D mixing widths using a modified scaling law: $W^{*} - W^{*}_{0} \sim \left( t^{*} - t^{*}_{0} \right)^{\theta}$, where $W^{*}_{0}$ is the mixing width at re-shock time and $t^{*}_{0}$ is the re-shock time. This scaling law, originally proposed by~\citet{weber2013growth}, leads to better convergence for $\theta$ over time for both 2D and 3D cases compared to another scaling law mentioned above. We believe both scaling laws should agree with each other at very late times, when $W^{*} \gg W^{*}_{0}$. However, this cannot be verified for our cases here as our simulations are constrained by the second re-shock. As an additional consistency check on the results, similar values are obtained using a nonlinear fit with $W^{*}_{0}$ and $t^{*}_{0}$ as unknowns, as well as estimating the scalings from derivatives of the mixing widths, without knowledge of $W^{*}_{0}$ and $t^{*}_{0}$~\cite{thornber2010influence}. The $\theta$ values are estimated to be $0.55$ and $0.44$, respectively, for the 2D and 3D cases with physical transport coefficients. The $\theta$ value for the 2D case falls between $\theta \approx 0.48$ and $\theta \approx 0.63$ observed from the 2D single-shocked RMI simulations with narrowband and broadband initial conditions, respectively by~\citet{thornber2015numerical}. \citet{tritschler2014richtmyer} reported $\theta \approx 0.29$ at late times after re-shock from their 3D RMI simulations with different numerical methods. In the paper by~\citet{thornber2011growth}, the authors proposed that there is an intermediate stage in time where $\theta = 0.4$ before $\theta$ becomes 0.26 at the latest time after re-shock and this may explain the discrepancy between the values found for $\theta$. The comparison of mixing width time evolution with 2D and 3D scaling laws after re-shock is shown in figure~\ref{fig:2D_3D_transport_coeffs_mixing_widths_late}.

%%%%%%%%%%%%%%%%%%%%%%%%%%%%%%%%%%%%%%%%%%%%%%%%%%%%%%%%%%%%%%%%%%%%%%%%%%%%%%%%
%%%%%%%%%%%%%%%%%%%%%%%%%%%%%%%%%%%%%%%%%%%%%%%%%%%%%%%%%%%%%%%%%%%%%%%%%%%%%%%%
%%%%%%%%%%%%%%%%%%%%%%%%%%%%%%%%%%%%%%%%%%%%%%%%%%%%%%%%%%%%%%%%%%%%%%%%%%%%%%%%

\subsection{Mixedness and mole fraction variance profiles}

\begin{figure*}[!ht]
\centering
\includegraphics[width = 0.5\textwidth]{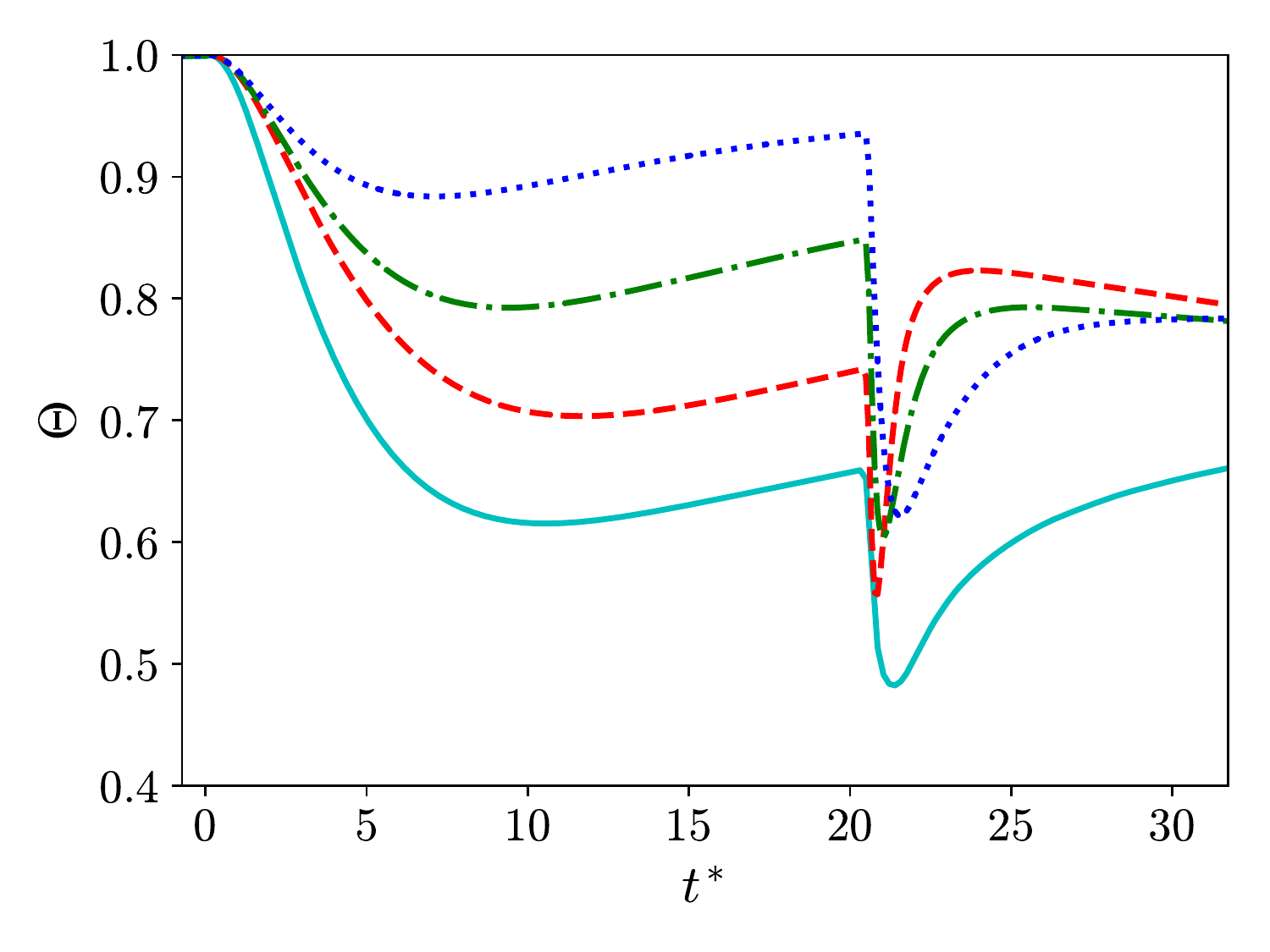}
\caption{Comparison of the time evolution of mixedness between the 2D and 3D problems. Cyan solid line: 2D with physical transport coefficients; red dashed line: 3D with physical transport coefficients; green dash-dotted line: 3D with $2 \times$physical transport coefficients; blue dotted line: 3D with $4 \times$physical transport coefficients.}
\label{fig:2D_3D_transport_coeffs_mixednesses}
\end{figure*}

The time evolution of mixedness is different for the 2D and 3D cases, as seen in figure~\ref{fig:2D_3D_transport_coeffs_mixednesses}. After first shock, from $t^{*}=0$ to $t^{*}=5$, the mixedness decreases for all cases, as the interface stretches due to rapid gradient intensification, while molecular diffusion is not fast enough to counteract this process. With physical transport coefficients, the 2D mixedness decreases at a faster rate than the 3D mixedness after first shock. At later times after first shock and before the re-shock, the mixedness of all cases increases again, as the rate of molecular mixing becomes larger than the rate of entraining pure fluids. Comparing the 3D cases before re-shock, it can be seen that mixedness increases with decreasing Reynolds number because of larger molecular diffusion inside the mixing region. After re-shock, there is a sudden decrease in mixedness in all cases as the interface stretches again due to gradient intensification. However, the mixedness recovers soon after the sudden reduction of mixedness in all cases. Comparing the 2D and 3D cases with physical transport coefficients, mixedness of the 3D case increases at a faster rate. This is likely associated with turbulent mixing in the 3D case, while transition to turbulence is absent in the 2D case. At late times after re-shock, mixedness for all 3D cases seems to approach to the same asymptotic value of around $0.8$. For the 2D case, the mixedness approaches a different value between $0.65$ and $0.7$. Similar asymptotic values of mixedness in 3D configurations were observed in previous studies. \citet{lombardini2012transition} and \citet{tritschler2014richtmyer} found that the mixedness in their simulations approached 0.85 asymptotically. \citet{mohaghar2017evaluation} also found an asymptotic value of 0.8 for mixedness in their experimental study. As for the 2D configuration, \citet{thornber2015numerical} found an asymptotic value of 0.63 for mixedness in their 2D single-shocked simulations with narrowband initial perturbations. The mixedness of the 2D case is smaller than that of the 3D case with physical transport coefficients at all times. This suggests that fluids are less molecularly mixed in the 2D configuration. The difference in mixedness between the 2D and 3D cases becomes larger after re-shock, which is probably due to lack of turbulence to enhance mixing in the 2D case.

Mixing can be further studied through the spatial profiles of mole fraction variance. Smaller variance indicates a larger extent of molecular mixing, as the fluid regions are more homogeneous. In fact, mixedness can be related to the spatial profile of mole fraction variance, by rewritting equation~\eqref{eqn:mixedness_definition} as:

\begin{equation}
    \Theta = 1 - 4 \frac{\int \overline{ X^{\prime ^2} _{SF_6} } dx}{W} = 1 - 4 \int \overline{ X^{\prime ^2} _{SF_6} } dx^{*}.
\end{equation}

\begin{figure*}[!ht]
\centering
\subfigure[$\ $Before re-shock, 2D]{%
\includegraphics[height = 0.33\textwidth]{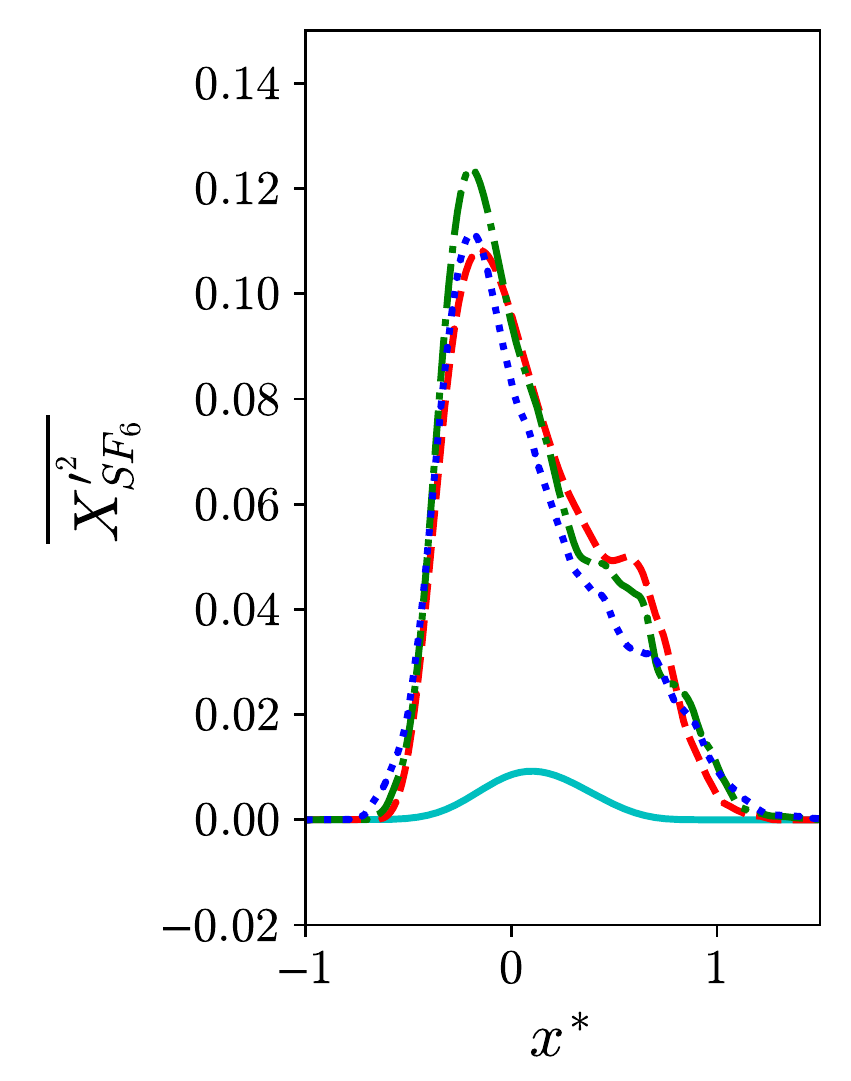}}
\subfigure[$\ $Before re-shock, 3D]{%
\includegraphics[height = 0.33\textwidth]{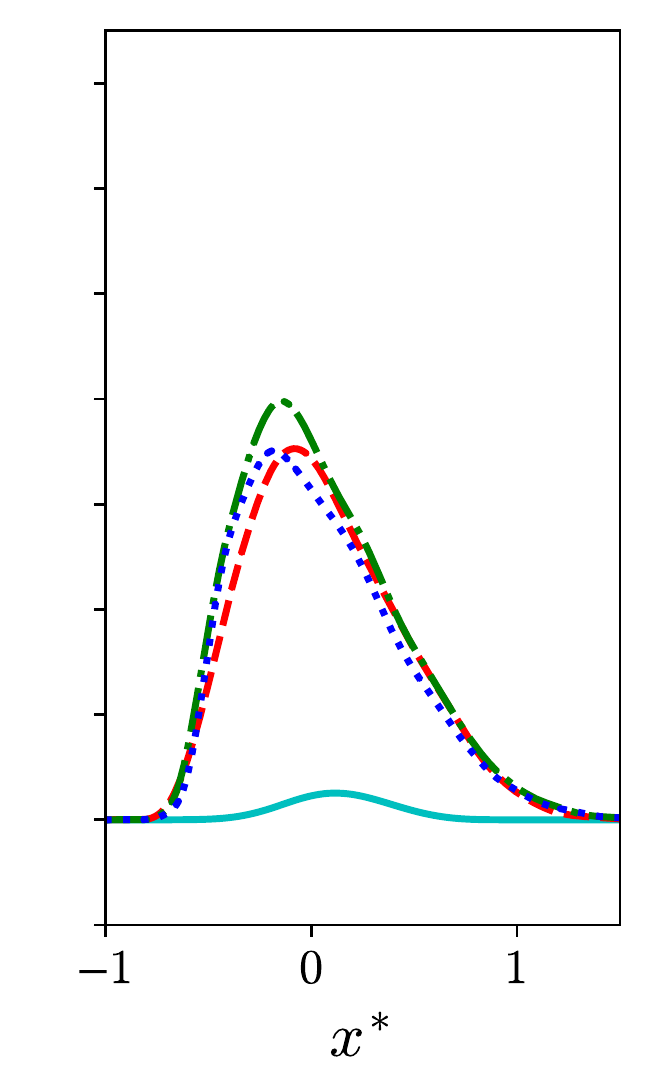}}
\subfigure[$\ $After re-shock, 2D]{%
\includegraphics[height = 0.33\textwidth]{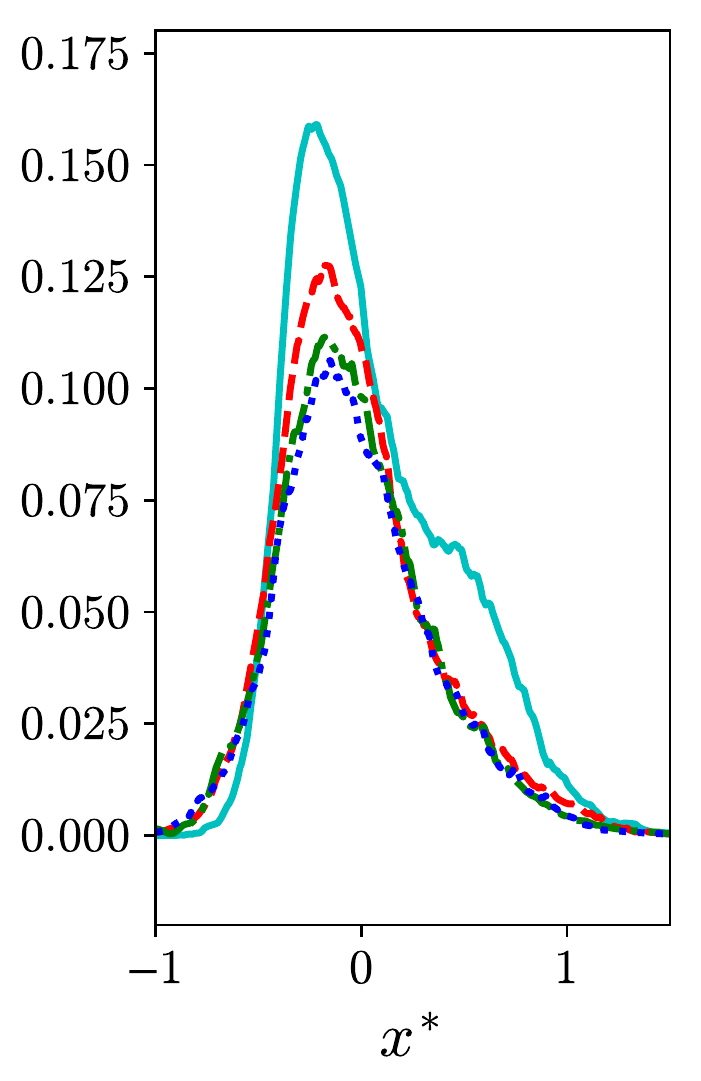}}
\subfigure[$\ $After re-shock, 3D]{%
\includegraphics[height = 0.33\textwidth]{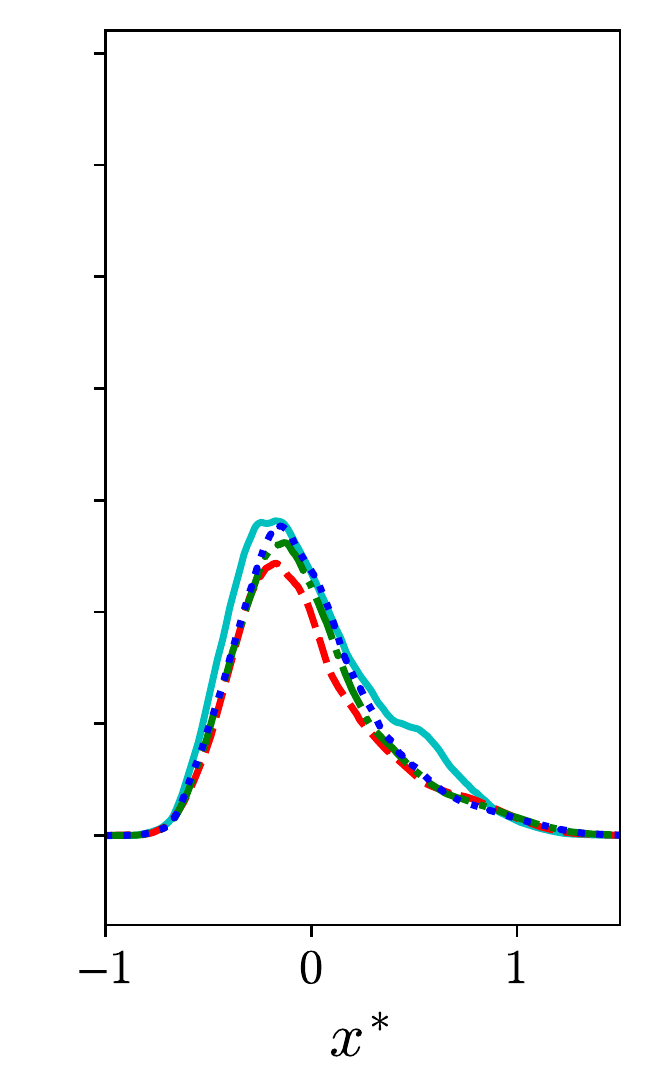}}
\caption{Profiles of $\mathrm{SF_6}$ mole fraction variance, $\overline{ X^{\prime ^2} _{SF_6} }$, at different times for the 2D and 3D problems with physical transport coefficients. Before re-shock: $t^{*}=0.91$ (cyan solid line); $t^{*}=7.25$ (red dashed line); $t^{*}=13.6$ (green dash-dotted line); $t^{*}=19.9$ (blue dotted line). After re-shock: $t^{*}=21.8$ (cyan solid line); $t^{*}=25.4$ (red dashed line); $t^{*}=29.0$ (green dash-dotted line); $t^{*}=31.7$ (blue dotted line).}
\label{fig:2D_vs_3D_mole_fraction_variance_profiles}
\end{figure*}

Figure~\ref{fig:2D_vs_3D_mole_fraction_variance_profiles} shows the profiles of $\mathrm{SF_6}$ mole fraction variance, $\overline{ X^{\prime ^2} _{SF_6} }$, across the normalized position $x^{*}$ from the 2D and 3D simulations with physical transport coefficients at different times. All profiles have peaks on the heavier fluid side, which indicates that fluids mix more slowly on the heavier fluid side for this variable-density flow. Similar asymmetries were also observed in other flows with strong variable-density effects~\cite{livescu2008variable,livescu2010npv,tian2017}, where they were associated with different inertia of the light and heavy fluid regions. Comparing 2D and 3D profiles at different times, the profile peaks are always larger in the 2D case and this is consistent with the smaller mixedness values for the 2D case. After re-shock, the profiles become more self-similar for both 2D and 3D cases at late times, consistent with the mixedness approaching asymptotic limits. This also suggests that a balanced state between entrainment of pure fluids at the edges of mixing layer and molecular mixing is approximately obtained in both configurations at late times. A similar balanced state was observed in freely decaying variable-density turbulence starting from isotropic turbulent fields by \citet{movahed2015mixing}.

%%%%%%%%%%%%%%%%%%%%%%%%%%%%%%%%%%%%%%%%%%%%%%%%%%%%%%%%%%%%%%%%%%%%%%%%%%%%%%%%
%%%%%%%%%%%%%%%%%%%%%%%%%%%%%%%%%%%%%%%%%%%%%%%%%%%%%%%%%%%%%%%%%%%%%%%%%%%%%%%%
%%%%%%%%%%%%%%%%%%%%%%%%%%%%%%%%%%%%%%%%%%%%%%%%%%%%%%%%%%%%%%%%%%%%%%%%%%%%%%%%

\subsection{Probability density functions of mole fraction}

The discrete probability density function (PDF) of a quantity $\phi$ bounded by $\phi_{\mathrm{min}}$ and $\phi_{\mathrm{max}}$ for the $k$th bin, where $k \in \left[1, 2,..., N_b \right]$, can be computed with:
\begin{equation}
    \mathrm{PDF} = \frac{N_k}{\left( \Delta \phi \right) N},
\end{equation}
where $N$ and $N_k$ are the total number of cells and the number of cells for the $k$th bin in the central part of mixing layer respectively. $\Delta \phi$ is given by:
\begin{equation}
    \Delta \phi = \frac{ \phi_{\mathrm{max}} - \phi_{\mathrm{min}} }{ N_b },
\end{equation}
\noindent where $N_b$ is the total number of bins.

\begin{figure*}[!ht]
\centering
\subfigure[$\ $2D with physical transport coefficients]{%
\includegraphics[width = 0.45\textwidth]{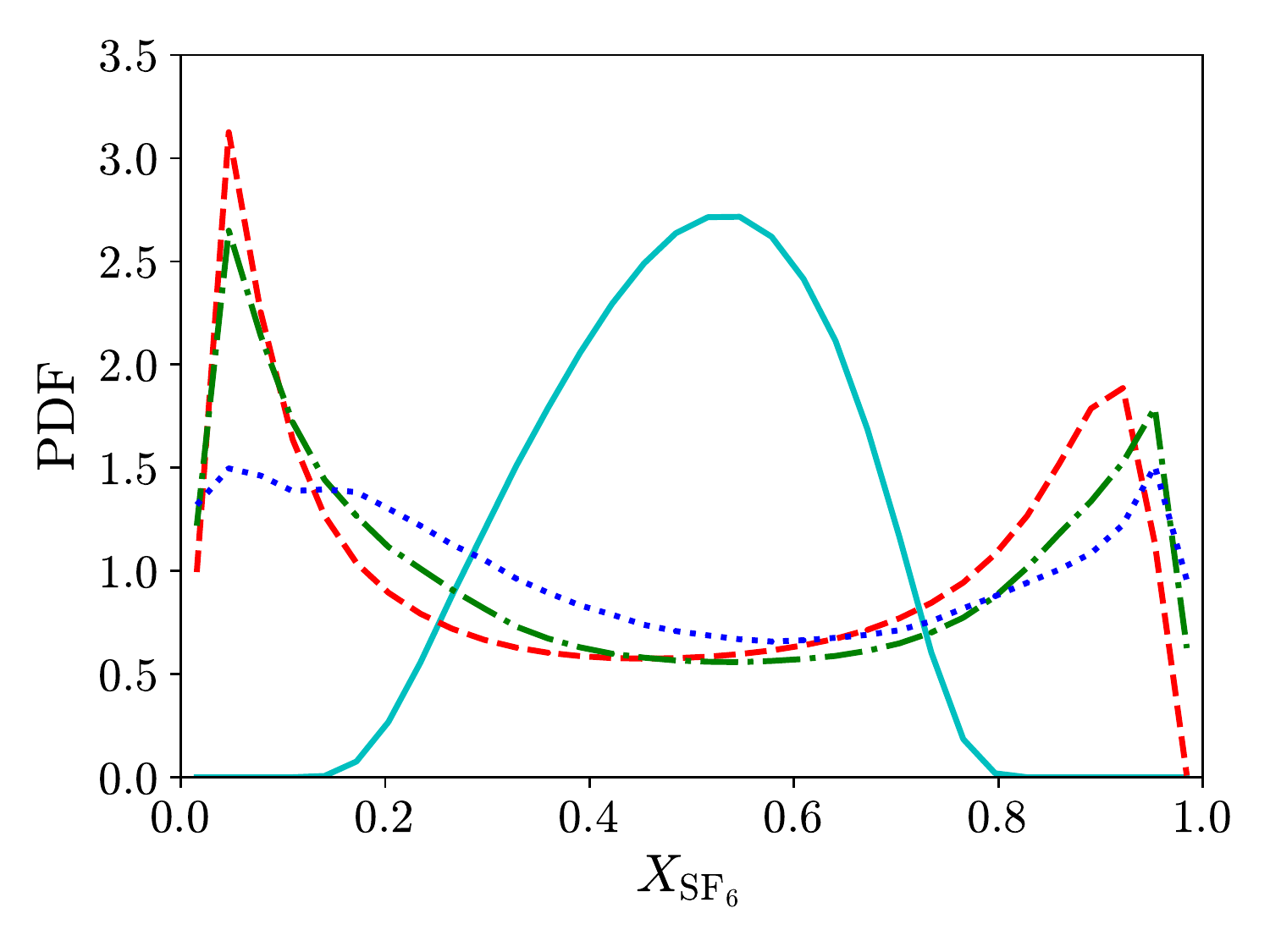}
\label{fig:pdf_physical_TC_before_reshock_2D}}
\subfigure[$\ $3D with physical transport coefficients]{%
\includegraphics[width = 0.45\textwidth]{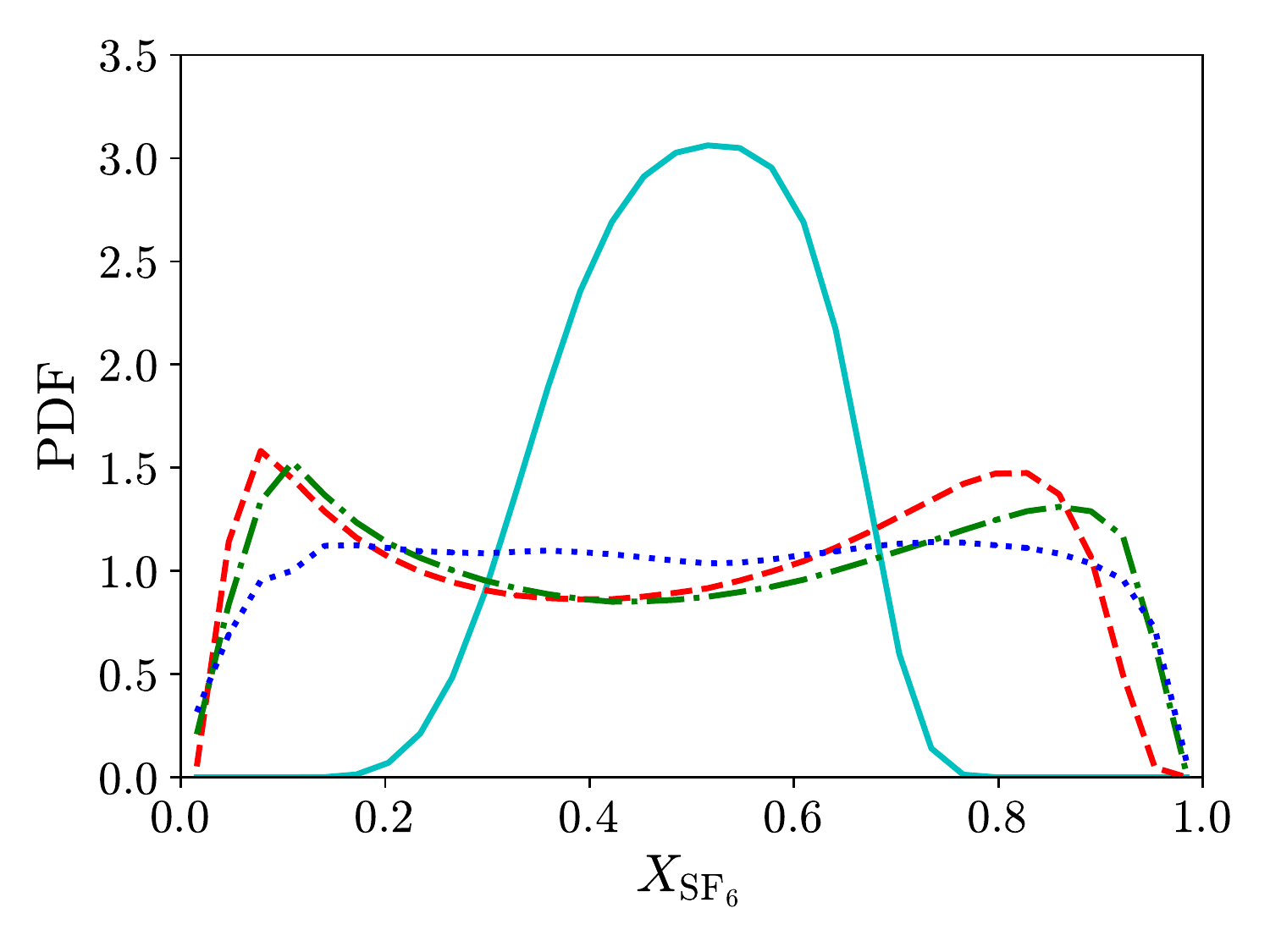}
\label{fig:pdf_physical_TC_before_reshock_3D}}

\subfigure[$\ $3D with $2 \times$physical transport coefficients]{%
\includegraphics[width = 0.45\textwidth]{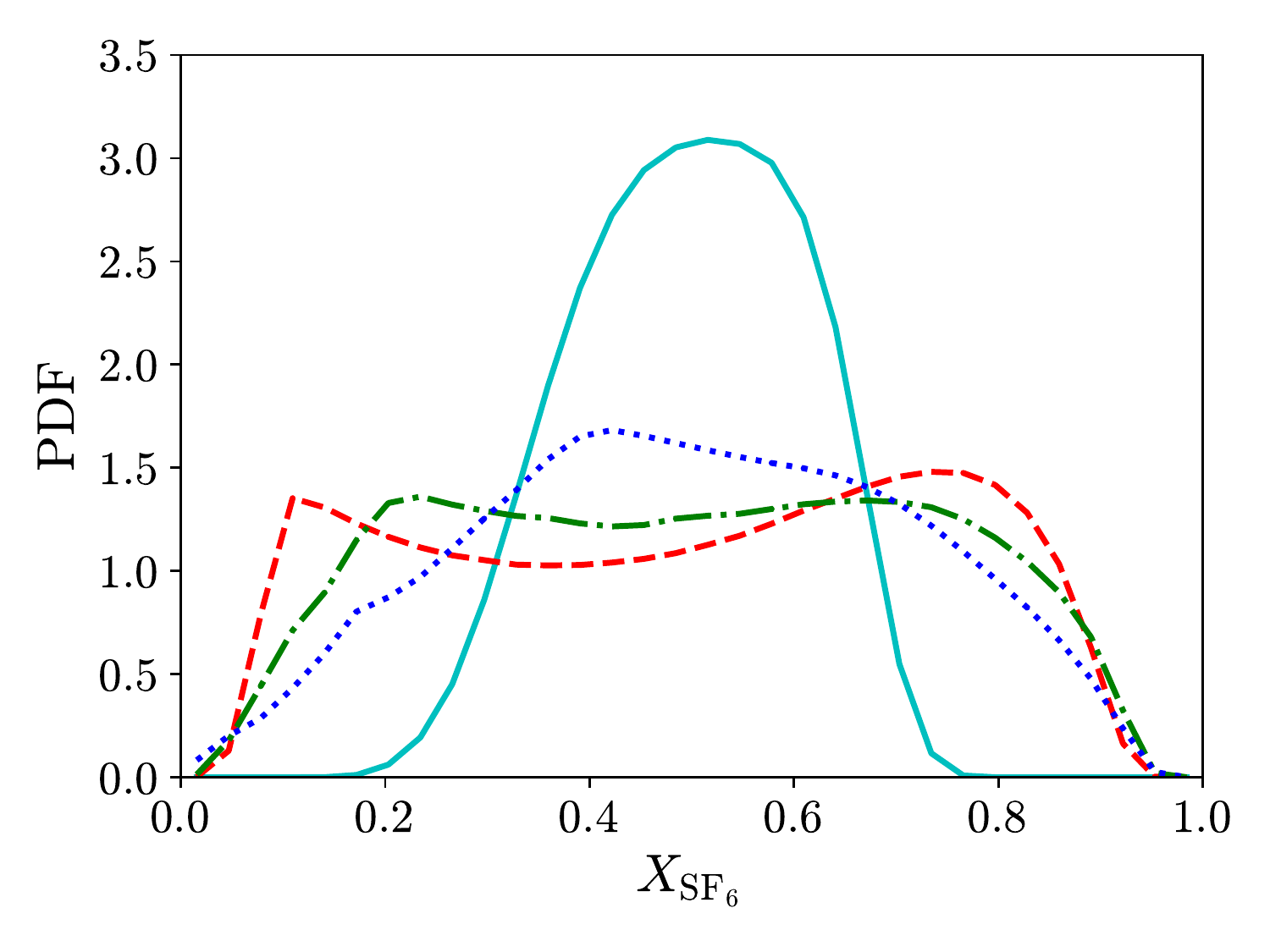}
\label{fig:2xpdf_physical_TC_before_reshock_3D}}
\subfigure[$\ $3D with $4 \times$physical transport coefficients]{%
\includegraphics[width = 0.45\textwidth]{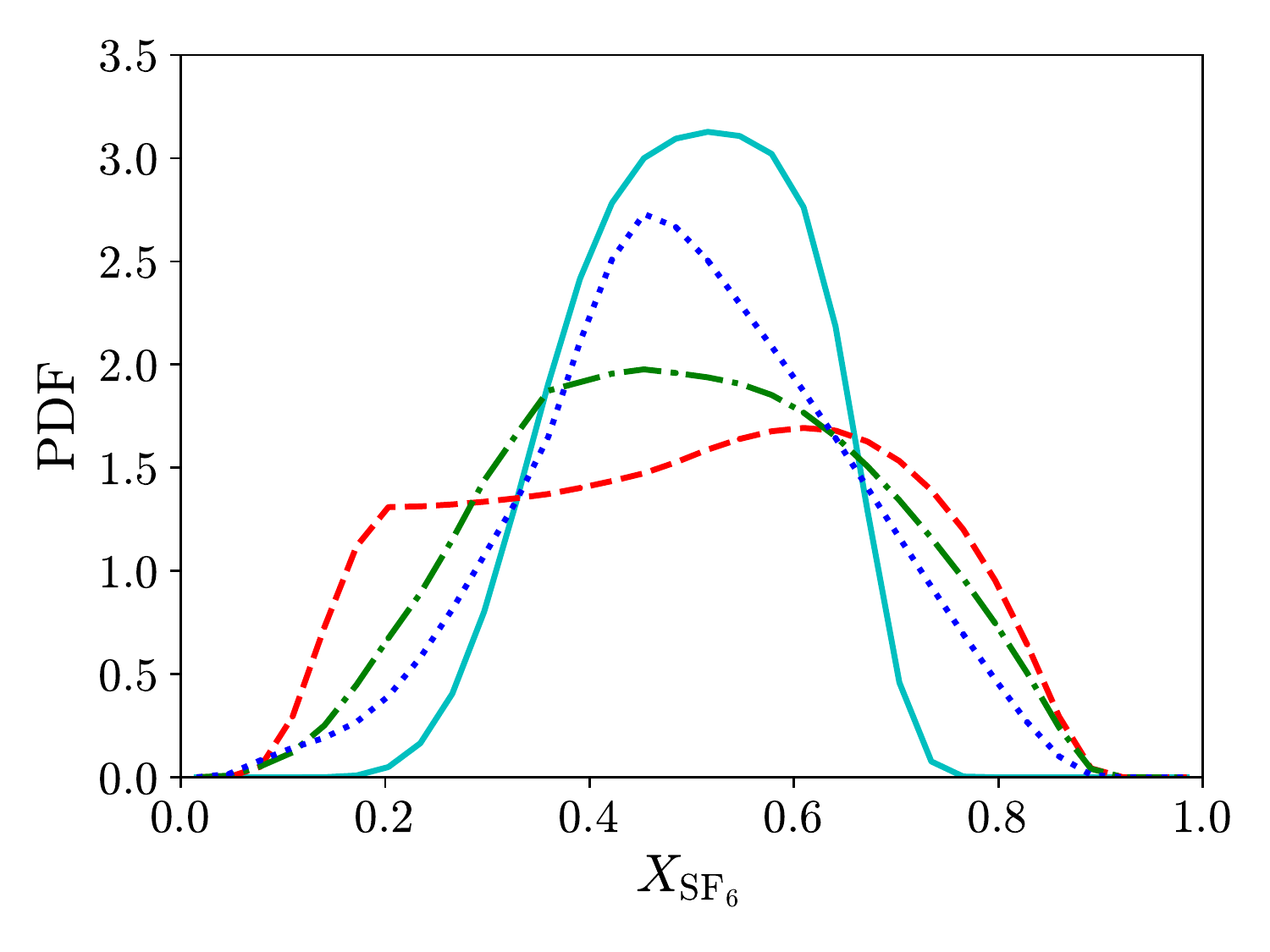}
\label{fig:4xpdf_physical_TC_before_reshock_3D}}
\caption{PDF's of $\mathrm{SF_6}$ mole fraction, $X_{\mathrm{SF_6}}$, for the 2D and 3D problems before re-shock. Cyan solid line: $t^{*}=0.91$; red dashed line: $t^{*}=7.25$; green dash-dotted line: $t^{*}=13.6$; blue dotted line: $t^{*}=19.9$.}
\label{fig:PDF_comparison_before_reshock}
\end{figure*}

\begin{figure*}[!ht]
\centering
\subfigure[$\ $2D with physical transport coefficients]{%
\includegraphics[width = 0.45\textwidth]{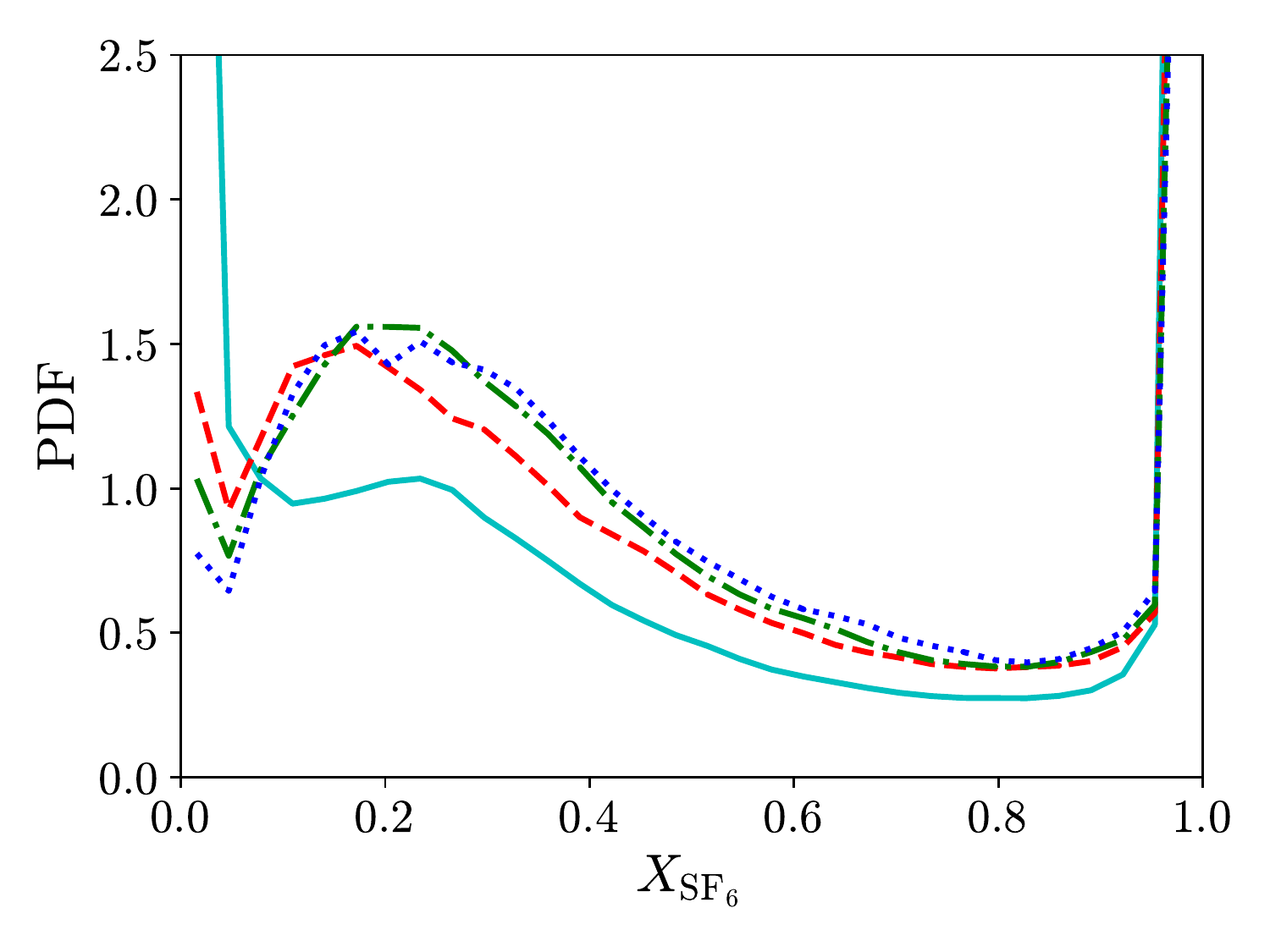}
\label{fig:pdf_physical_TC_after_reshock_2D}}
\subfigure[$\ $3D with physical transport coefficients]{%
\includegraphics[width = 0.45\textwidth]{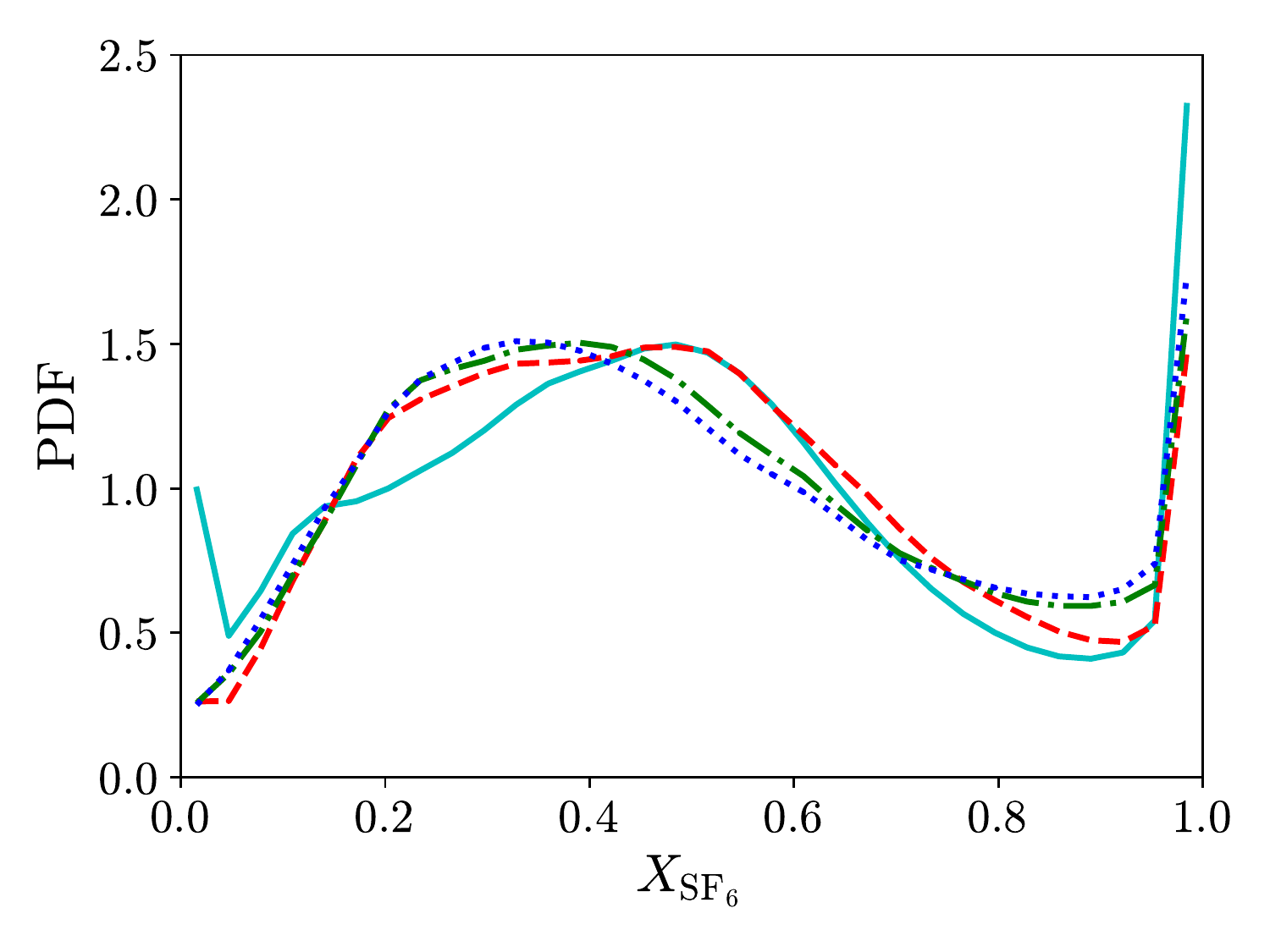}
\label{fig:pdf_physical_TC_after_reshock_3D}}

\subfigure[$\ $3D with $2 \times$physical transport coefficients]{%
\includegraphics[width = 0.45\textwidth]{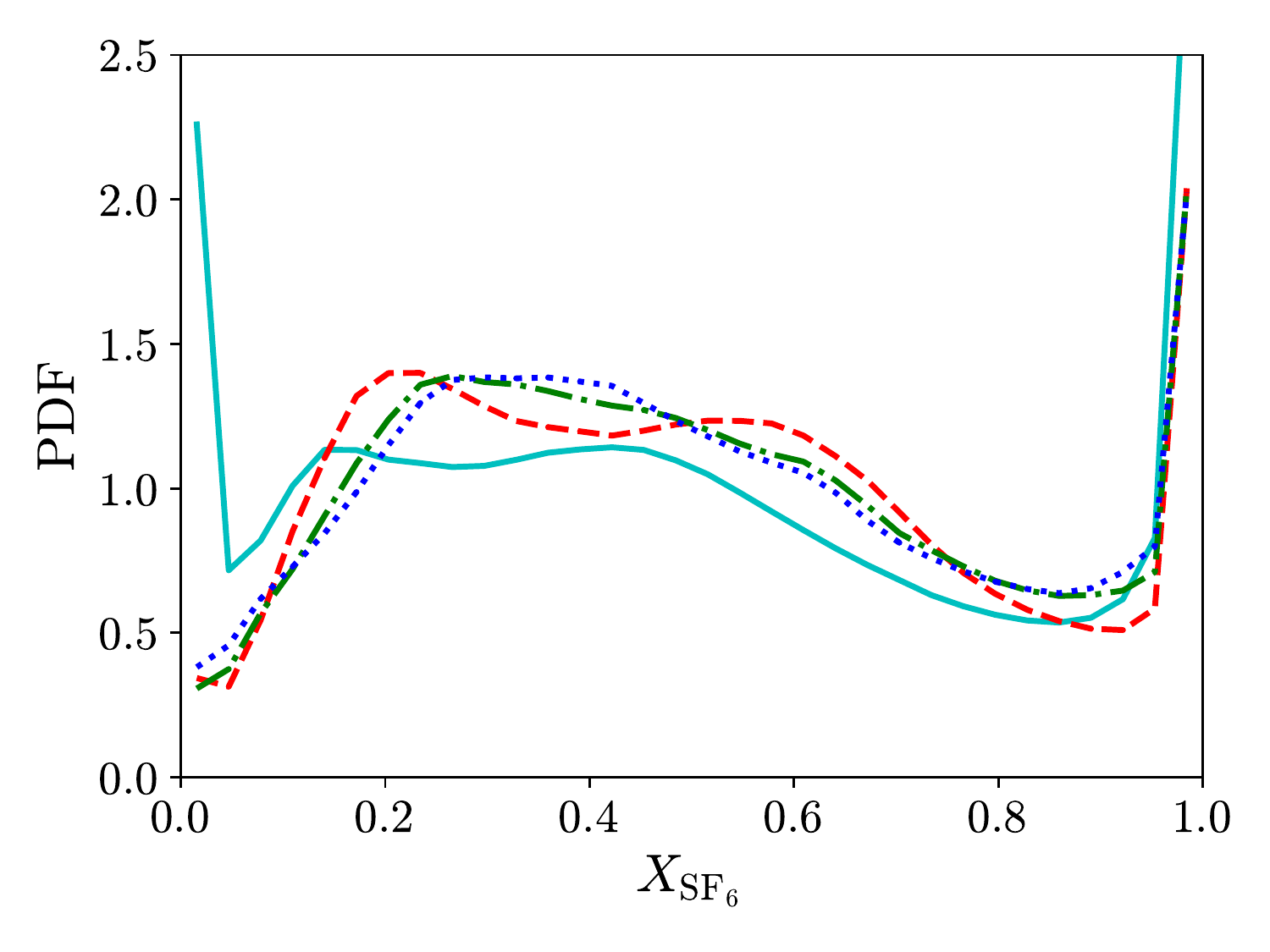}
\label{fig:2xpdf_physical_TC_after_reshock_3D}}
\subfigure[$\ $3D with $4 \times$physical transport coefficients]{%
\includegraphics[width = 0.45\textwidth]{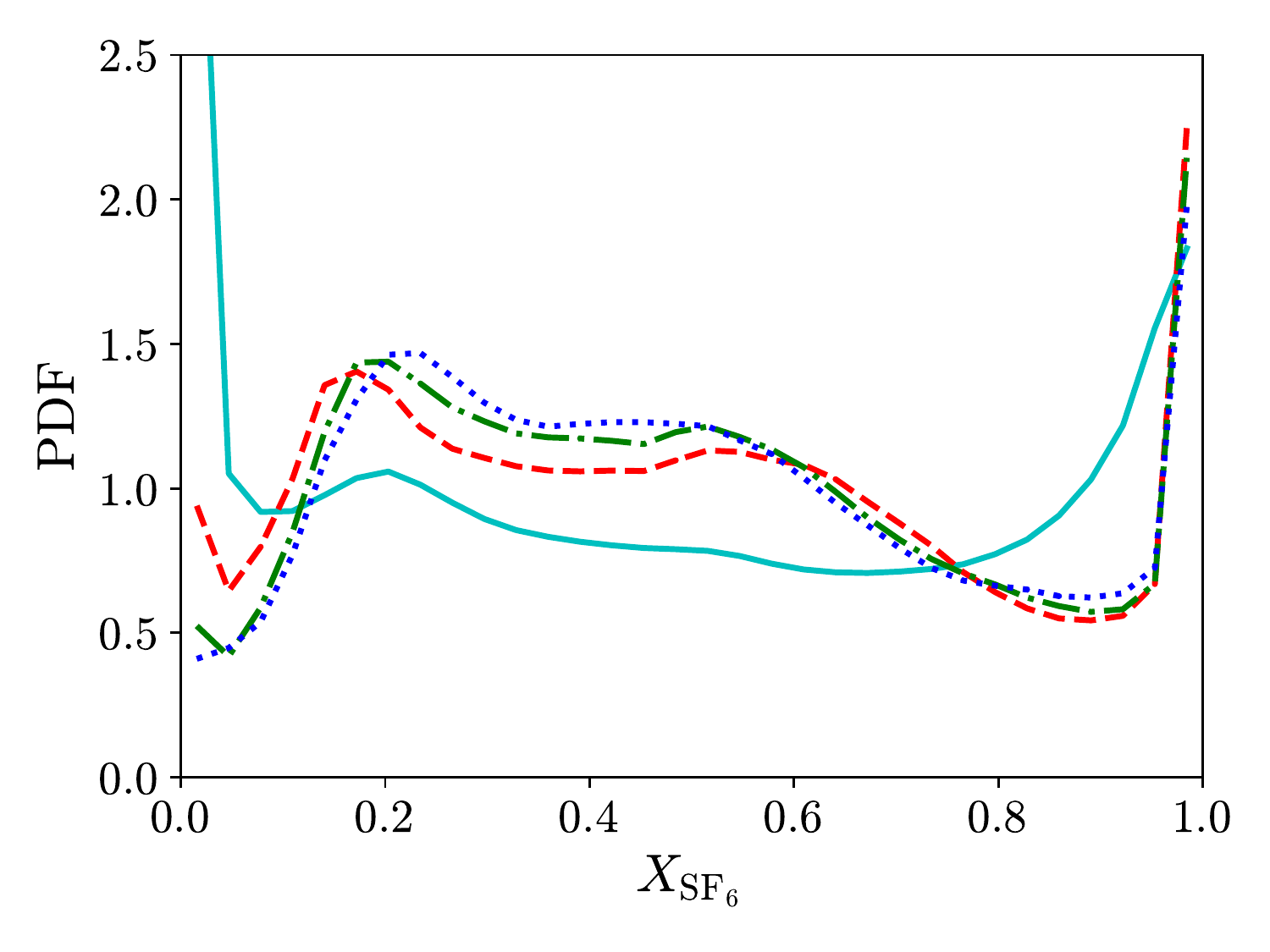}
\label{fig:4xpdf_physical_TC_after_reshock_3D}}
\caption{PDF's of $\mathrm{SF_6}$ mole fraction, $X_{\mathrm{SF_6}}$, for the 2D and 3D problems after re-shock. Cyan solid line: $t^{*}=21.8$; red dashed line: $t^{*}=25.4$; green dash-dotted line: $t^{*}=29.0$; blue dotted line: $t^{*}=31.7$.}
\label{fig:PDF_comparison_after_reshock}
\end{figure*}

The PDF's of $\mathrm{SF_6}$ mole fraction at different times for 2D and 3D cases are compared in figure~\ref{fig:PDF_comparison_before_reshock} before re-shock and figure~\ref{fig:PDF_comparison_after_reshock} after re-shock. For the 3D case with physical transport coefficients, it can be seen that the PDF has a quasi-Gaussian shape immediately after first shock resulted from the initial smooth profile of the material interface. However, as pure fluids are being entrained into the central part of mixing layer, two peaks at the pure fluid ends appear. As the instability evolves and the molecular effects start dominating the entrainment, the amplitude of each peak diminishes. With increased transport coefficients or reduced Reynolds number, the peaks at the pure fluid ends can no longer be clearly observed at late times after first shock. The evolution of the PDF for the 2D case is very similar to that for the 3D case with physical transport coefficients, but there seems to be more entrainment over time and the molecular diffusion effect is smaller. This is consistent with previous results that fluids in the 2D configuration are less mixed than those in 3D configuration before re-shock.

After re-shock, there is a common fundamental change in each mole fraction PDF. For each case, the PDF returns to a unimodal shape in the central region of $X_{\mathrm{SF_6}}$ values; however, this shape is strongly asymmetric, with the peak in the light fluid region and elongated tail in the heavy fluid region. Moreover, shortly after re-shock, each PDF has two additional peaks near the two pure fluid values. At later times, only the heavy fluid peak ($X_{\mathrm{SF_6}} \sim 1$) survives. The same form was also noticed for the density PDF of \citet{hill2006large} and mass fraction PDF of \citet{tritschler2014richtmyer} from 3D simulations. The similar shape of the PDF's indicates that each flow is composed of moderately mixed regions of lighter than average fluid and some pure regions of heavier fluid. At later times, the pure light fluid regions are significantly reduced but significant pure heavy fluid regions are still present. The same effect was also observed in other variable-density flows, such as homogeneous RTI by \citet{livescu2008variable}, classical RTI by \citet{livescu2009high,livescu2010npv}, variable-density round jet of \citet{gerashchenko2015density}, and variable-density shock-turbulence interaction by \citet{tian2017}. The central peak of the PDF shifts even more towards the lighter fluid side with smaller Reynolds number, which could be attributed to smaller baroclinic torque generated from smaller perturbation amplitude at re-shock. Thus, the subsequent reduced stirring makes it more difficult for the higher inertia and heavier fluid to mix.

%%%%%%%%%%%%%%%%%%%%%%%%%%%%%%%%%%%%%%%%%%%%%%%%%%%%%%%%%%%%%%%%%%%%%%%%%%%%%%%%
%%%%%%%%%%%%%%%%%%%%%%%%%%%%%%%%%%%%%%%%%%%%%%%%%%%%%%%%%%%%%%%%%%%%%%%%%%%%%%%%
%%%%%%%%%%%%%%%%%%%%%%%%%%%%%%%%%%%%%%%%%%%%%%%%%%%%%%%%%%%%%%%%%%%%%%%%%%%%%%%%

\subsection{\label{sec:TKE} Turbulent kinetic energy and Reynolds stress anisotropy}

\begin{figure*}[!ht]
\centering
\subfigure[$\ $Global time evolution]{%
\includegraphics[width = 0.45\textwidth]{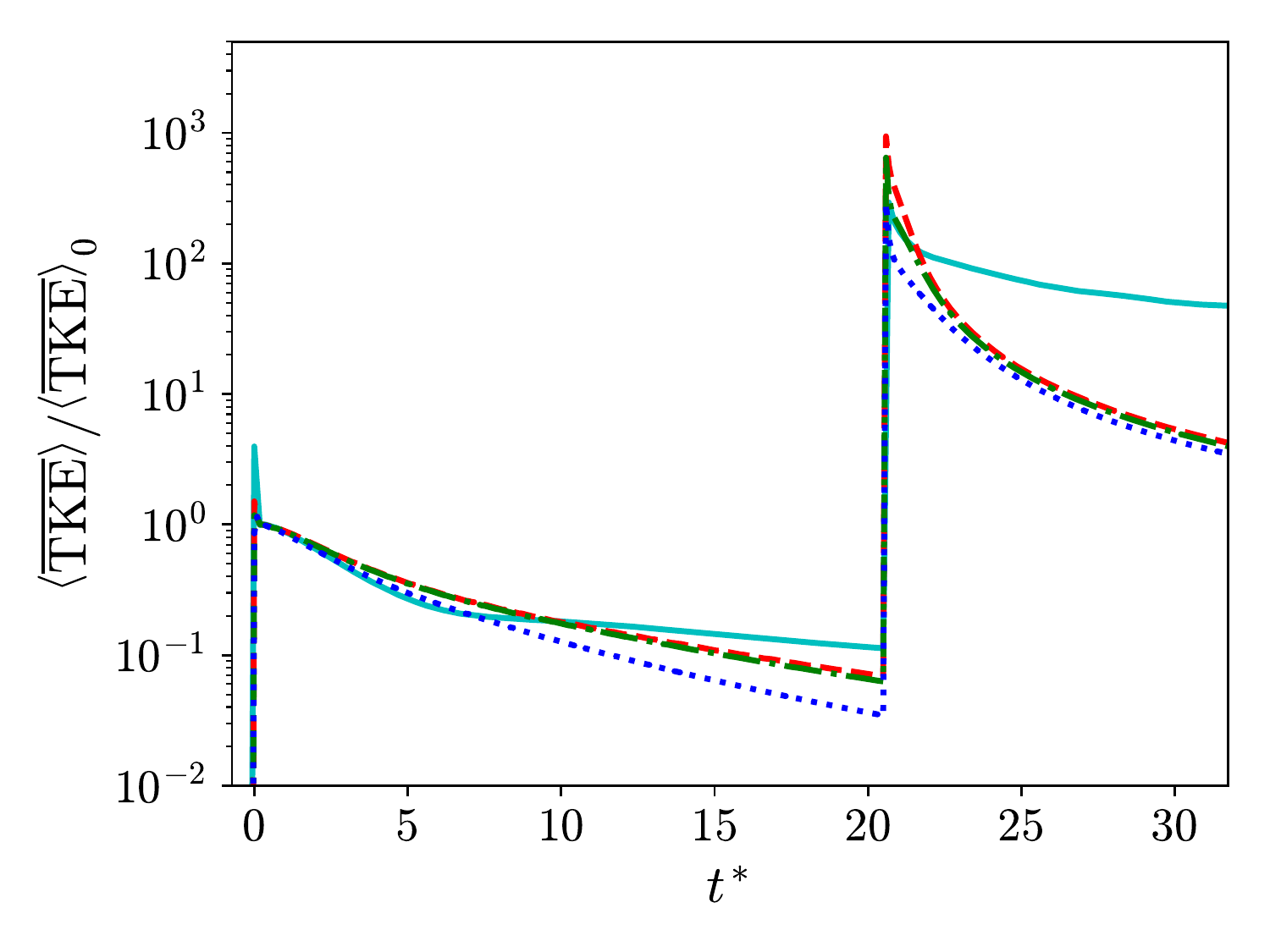}
\label{fig:2D_3D_transport_coeffs_TKE}}
\subfigure[$\ $After re-shock]{%
\includegraphics[width = 0.45\textwidth]{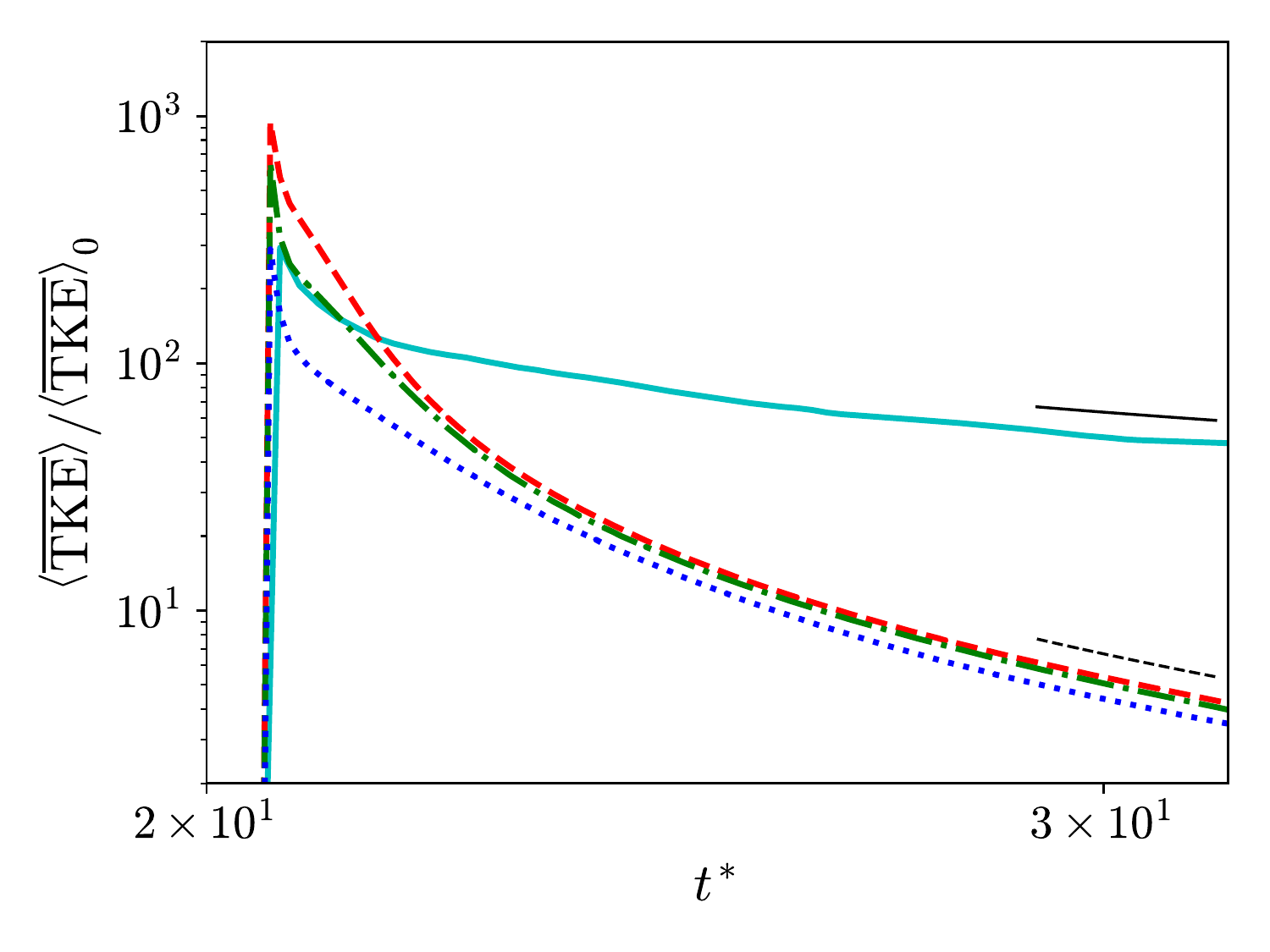}
\label{fig:3D_transport_coeffs_TKE_late}}
\caption{Comparison of the time evolution of the normalized means of turbulent kinetic energy, TKE, in the central part of mixing layer between the 2D and 3D problems. Mean TKE is normalized by its value at $t^{*}=0$, $\left< \overline{\mathrm{TKE}} \right>_0$. Cyan solid line: 2D with physical transport coefficients; red dashed line: 3D with physical transport coefficients; green dash-dotted line: 3D with $2 \times$physical transport coefficients; blue dotted line: 3D with $4 \times$physical transport coefficients. The black thin solid and dashed lines indicate the scalings, $\sim \left( t^{*}-t^{*}_0 \right)^{-0.5}$ and $\sim \left( t^{*}-t^{*}_0 \right)^{-1.4}$ respectively, where $t^{*}_0 = 20.5$ is the re-shock time.}
\label{fig:2D_3D_transport_coeffs_TKE_combined}
\end{figure*}

Figure~\ref{fig:2D_3D_transport_coeffs_TKE} compares the time evolution of mean TKE in the central part of the mixing layer, $\left< \overline{\mathrm{TKE}} \right>$, between the 2D and 3D cases. The mean TKE is normalized in each case such that it is equal to one at $t^{*}=0$ for better comparison between different cases. From figure~\ref{fig:2D_3D_transport_coeffs_TKE}, it can be seen that after first shock, there is a sudden jump in the TKE value for each case. Comparing the 2D and 3D cases with physical transport coefficients, the mean TKE decays after the jump, with similar decay rates for both cases until $t^{*} \approx 3$. At later times, TKE decays at a faster rate for the 3D case. This is likely associated with the presence of the vortex stretching mechanism in 3D domain, which enhances the breakdown of large scale features into smaller scales. After re-shock, the distinction in the decay rates between the 2D and 3D cases is much larger, with an even faster decay rate for the 3D case.

Comparing the 3D cases, after first shock, TKE decays more rapidly with smaller Reynolds number (i.e. with larger transport coefficients) and this effect holds until re-shock. After re-shock, although the decay rates are different among the cases initially, they become very similar and almost identical at late times. This is evident from figure~\ref{fig:3D_transport_coeffs_TKE_late} which only shows the temporal decay of the normalized mean TKE after re-shock for different cases. We have also examined the scaling law after re-shock for TKE: $\sim \left( t^{*} - t^{*}_{0} \right)^{-n}$ that was investigated in many previous studies~\cite{lombardini2012transition, tritschler2014richtmyer, thornber2016impact}. Similar to the study of scaling law for mixing widths, we have chosen $t^{*}_{0}$ to be the re-shock time. This choice was also verified using a nonlinear curve fitting with unknown $t^{*}_{0}$. At late times after re-shock, the values of $n$ for 2D and 3D cases with physical transport coefficients approach $0.5$ and $1.4$, respectively. The comparison of TKE time evolution with the 2D and 3D scaling laws after re-shock can be seen in figure~\ref{fig:3D_transport_coeffs_TKE_late}. The value of $n$ found for the 3D case is essentially as same as the value $n = 10/7 \approx 1.42$ observed by~\citet{tritschler2014richtmyer} after re-shock from their 3D simulations. \citet{lombardini2012transition} also reported that the value of $n$ was close to $10/7$ at late times of their single-shocked simulations. A TKE scaling with $n=10/7$ is identical to the TKE scaling of decaying turbulence of Batchelor-type~\cite{batchelor1956large}, in contrast to that of Saffman-type~\cite{saffman1967note,saffman1967large} with $n=6/5$.

\begin{figure*}[!ht]
\centering
\subfigure[$\ $Before re-shock, 2D]{%
\includegraphics[height = 0.33\textwidth]{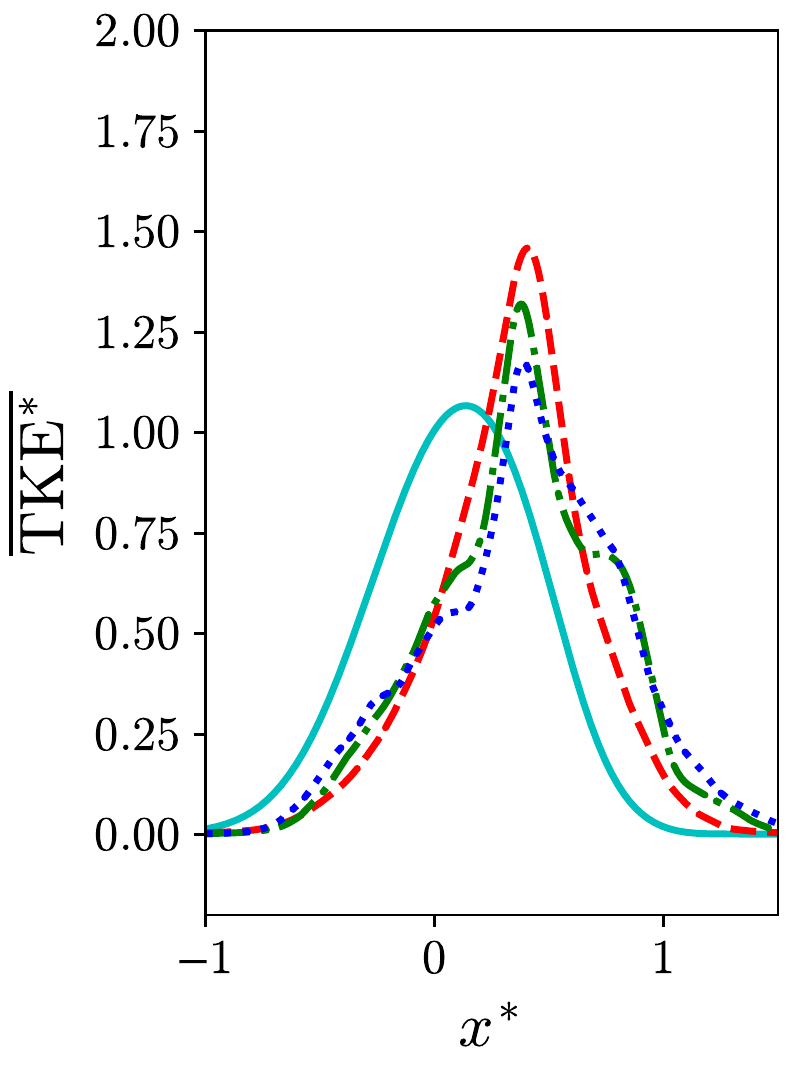}}
\subfigure[$\ $Before re-shock, 3D]{%
\includegraphics[height = 0.33\textwidth]{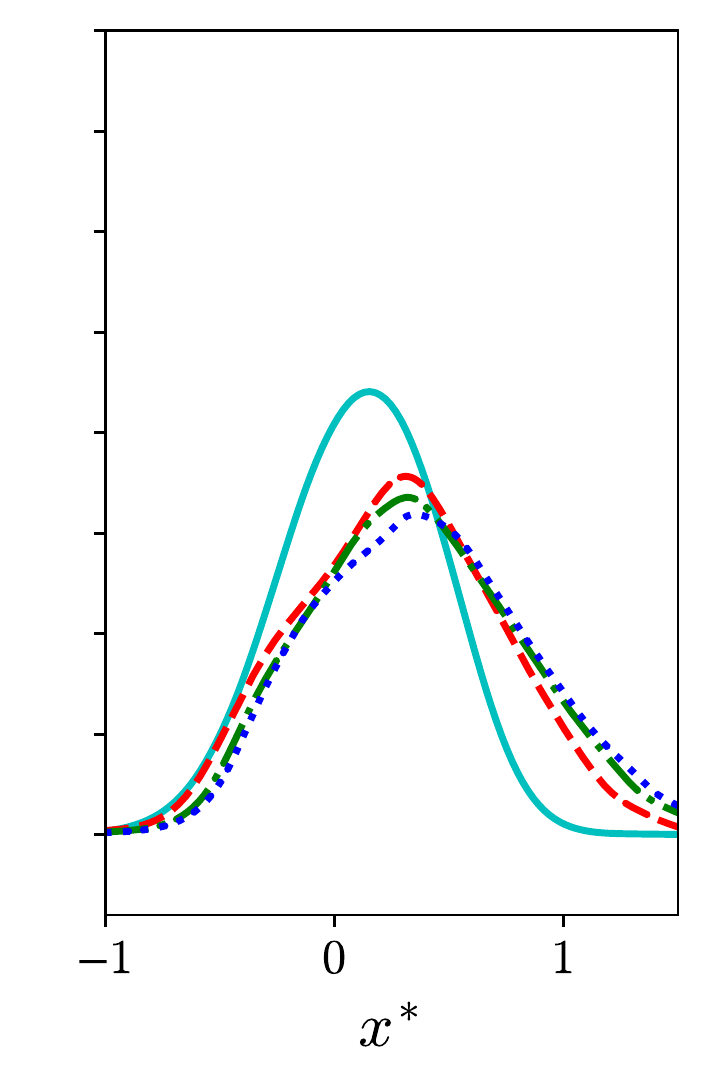}}
\subfigure[$\ $After re-shock, 2D]{%
\includegraphics[height = 0.33\textwidth]{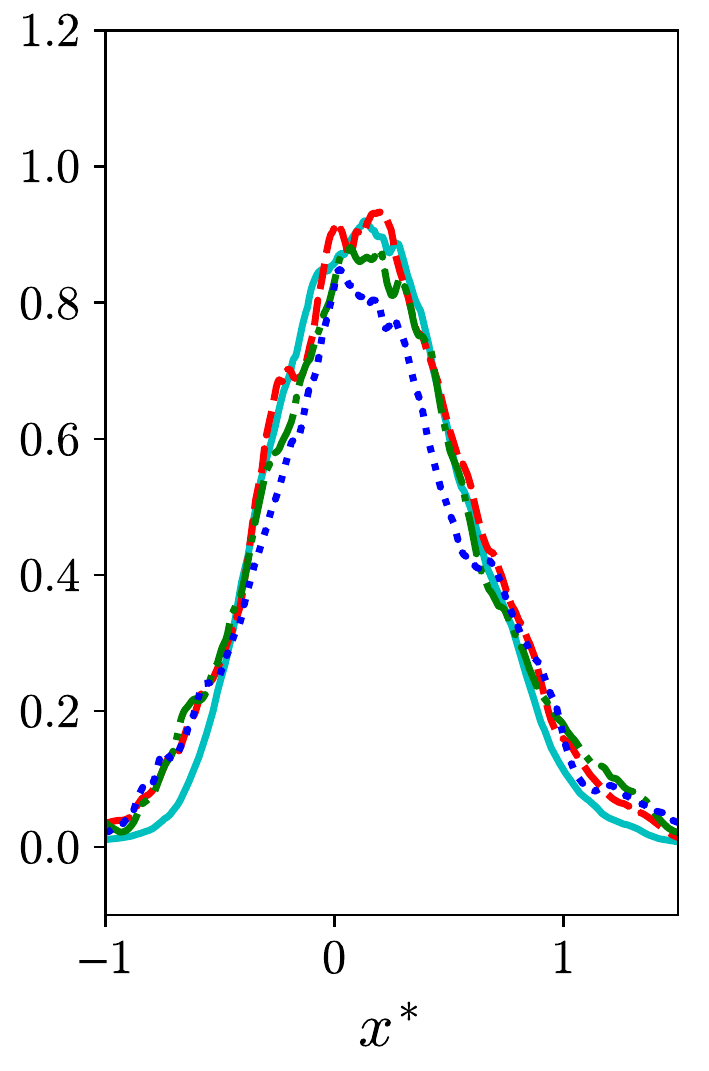}}
\subfigure[$\ $After re-shock, 3D]{%
\includegraphics[height = 0.33\textwidth]{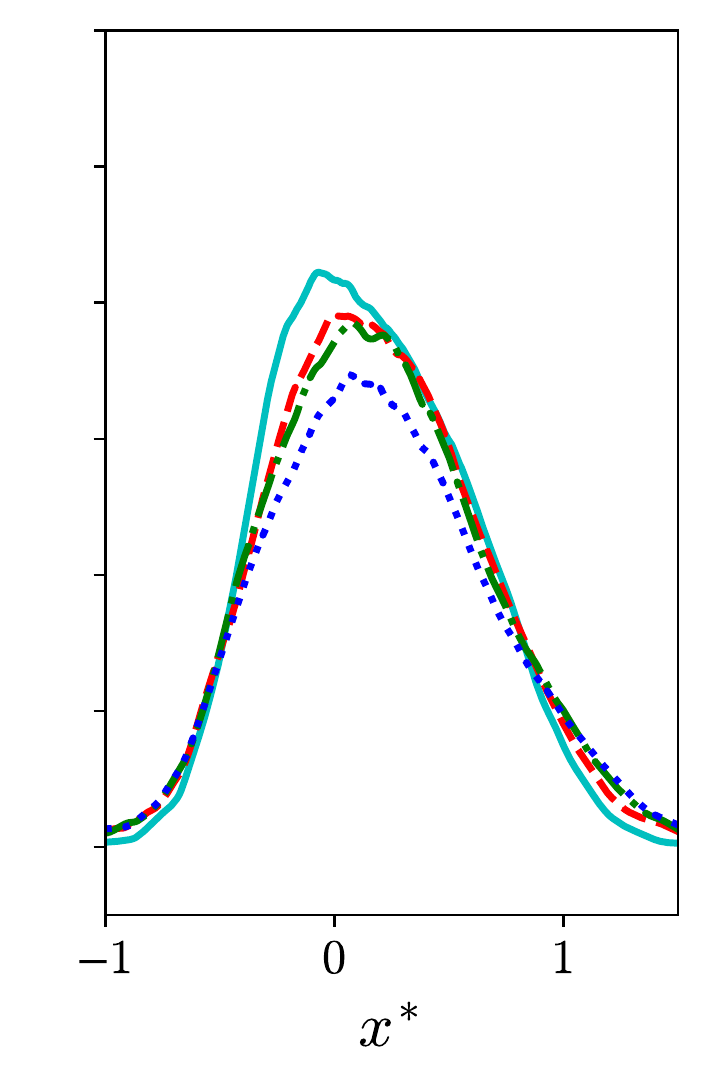}}
\caption{Profiles of normalized turbulent kinetic energy at different times for the 2D and 3D problems with physical transport coefficients. Before re-shock: $t^{*}=0.91$ (cyan solid line); $t^{*}=7.25$ (red dashed line); $t^{*}=13.6$ (green dash-dotted line); $t^{*}=19.9$ (blue dotted line). After re-shock: $t^{*}=21.8$ (cyan solid line); $t^{*}=25.4$ (red dashed line); $t^{*}=29.0$ (green dash-dotted line); $t^{*}=31.7$ (blue dotted line).}
\label{fig:2D_vs_3D_TKE_profiles}
\end{figure*}

Figure~\ref{fig:2D_vs_3D_TKE_profiles} compares the mean profiles of TKE at different times for the 2D and 3D cases with physical transport coefficients. Before averaging in the homogeneous directions to get the mean profiles, TKE is first normalized as:
\begin{equation}
    \mathrm{TKE}^{*} = \frac{\left( \mathrm{TKE} \right) W}{\int \overline{\mathrm{TKE}} \ dx}.
\end{equation}

Except at the very early times, the normalized TKE profiles approximately collapse when comparing the profiles at later times after both the first shock and re-shock, with only some small variation in the magnitude of the peak. Before re-shock, TKE peaks on the light fluid side ($x^*>0$) for both 2D and 3D cases, as mixing is more prominent on this side and  this is consistent with the mole fraction variance profiles. The TKE peak in the 2D case is higher than that in the 3D case, which means TKE is more localized towards light fluid side under the 2D configuration. However, after re-shock, the peaks of mean $\mathrm{TKE}^{*}$ for both cases move closer to the mid-line or mid-plane ($x^*=0$) of the mixing layer and the shapes of the profiles for the 2D and 3D problems look more similar.

Besides TKE, the Favre-averaged Reynolds stress anisotropy tensor, $b_{ij}$, is another statistical quantity of interest from the velocity field. The 2D and 3D anisotropy tensors, $b^{2D}_{ij}$ and $b^{3D}_{ij}$, are defined as:
\begin{align}
    b^{2D}_{ij} &= \frac{\tilde{R}_{ij}}{\tilde{R}_{kk}} - \frac{1}{2} \delta_{ij}, \\
    b^{3D}_{ij} &= \frac{\tilde{R}_{ij}}{\tilde{R}_{kk}} - \frac{1}{3} \delta_{ij},
\end{align}

\noindent where $\delta_{ij}$ is the Kronecker delta and $\tilde{R}_{ij}$ is the Favre-averaged Reynolds stress tensor given by:
\begin{equation}
    \tilde{R}_{ij} = \frac{\overline{\rho u_i^{\prime \prime} u_j^{\prime \prime}}}{\bar{\rho}}.
\end{equation}

$b^{2D}_{11}$ and $b^{3D}_{11}$ indicate the amount of TKE contributed from the streamwise component of Reynolds normal stresses. $b^{2D}_{11} = 1/2$ and $b^{3D}_{11} = 2/3$ correspond to having all TKE contributed from $\bar{\rho} \tilde{R}_{11}/2$ while $b^{2D}_{11} = -1/2$ and $b^{3D}_{11} = -1/3$ mean that there is no contribution to TKE from that component. The Reynolds normal stresses are isotropic if $b^{2D}_{11}=0$ in 2D flow or $b^{3D}_{11}=0$ in 3D flow.

\begin{figure*}[!ht]
\centering
\includegraphics[width = 0.45\textwidth]{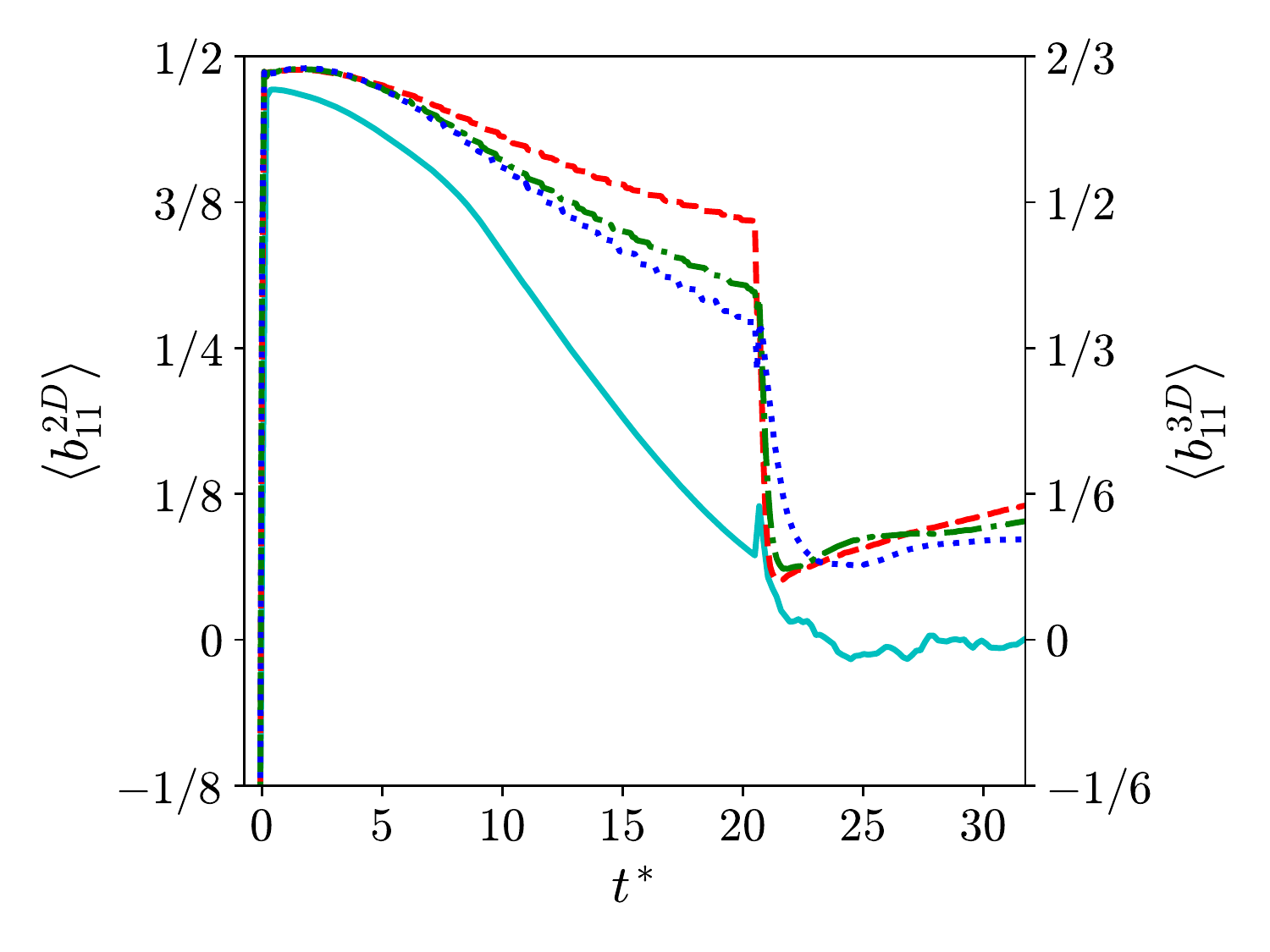}
\caption{Comparison of the time evolution of the means of Reynolds stress anisotropy component, $\left< b_{11} \right>$, in the central part of mixing layer between the 2D and 3D problems. Cyan solid line: 2D with physical transport coefficients; red dashed line: 3D with physical transport coefficients; green dash-dotted line: 3D with $2 \times$physical transport coefficients; blue dotted line: 3D with $4 \times$physical transport coefficients.}
\label{fig:2D_3D_transport_coeffs_anisotropy}
\end{figure*}

Figure~\ref{fig:2D_3D_transport_coeffs_anisotropy} shows the time evolution of mean $b_{11}$ in the central part of mixing layer for the 2D and 3D cases. In all cases, the means of $b_{11}$ almost attain their maximum values of 1/2 and 2/3 for the 2D and 3D cases, respectively, right after first shock. This is followed by a reduction of $\left<b_{11}\right>$ until just before re-shock. However, comparing cases with the physical transport coefficients, $\left<b_{11}\right>$ decreases more rapidly for the 2D case. Since the kinetic energy decays slower for the 2D case, the faster return towards isotropy for this case can be associated with a more efficient TKE redistribution among the Reynolds normal stresses. For quasi-incompressible flows, this redistribution is largely associated with the pressure-strain terms~\cite{livescu2002} in the Reynolds stress tensor transport equation. Thus, figure~\ref{fig:2D_3D_transport_coeffs_anisotropy} implies a stronger pressure-strain correlation for the 2D case as the flow evolves after the interaction with the shock.

At re-shock, there is a sudden decline in $\left<b_{11}\right>$ values for all cases. For the 2D case, $\left<b_{11}\right>$ reaches the isotropic value of zero rapidly after re-shock, while for the 3D cases, the TKE fields remain anisotropic until the end of simulations. \citet{thornber2015numerical} also noticed rapid isotropization of the Reynolds normal stresses from their narrowband 2D simulations compared to the 3D cases. Both results concerning the 3D cases by \citet{lombardini2012transition} and \citet{tritschler2014richtmyer} show slow isotropization of the Reynolds normal stresses over a long period of time and small non-zero asymptotic limits in anisotropy were observed. The Reynolds normal stresses in our 3D simulations still remain very anisotropic at late times compared to the asymptotic limit in~\citet{tritschler2014richtmyer} probably because of a shorter time duration from re-shock time to the end of simulation constrained by the arrival of the second re-shock. Comparing the time evolution of $\left<b_{11}\right>$ among different 3D cases, there is a slightly smaller anisotropy before re-shock at lower Reynolds numbers, as diffusive effects are more important before the flows become fully turbulent. After re-shock, all 3D cases have similar anisotropy.
 
\begin{figure*}[!ht]
\centering
\subfigure[$\ $Before re-shock, 2D]{%
\includegraphics[height = 0.33\textwidth]{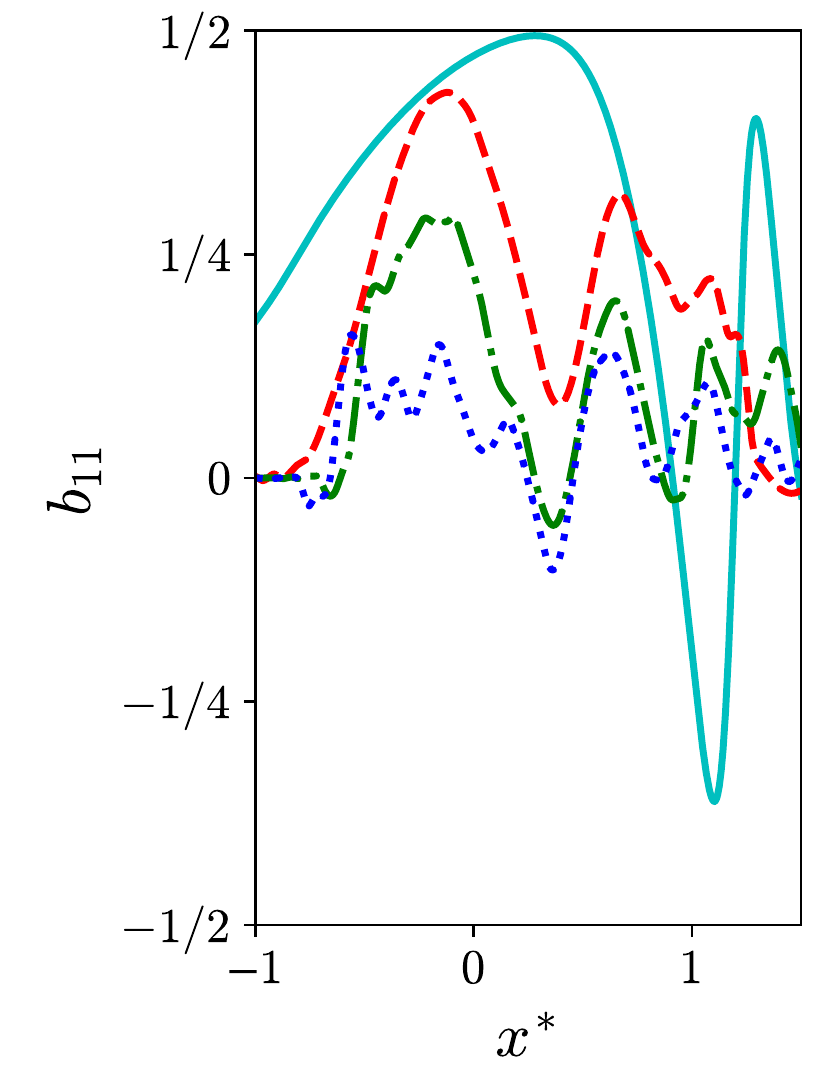}}
\subfigure[$\ $Before re-shock, 3D]{%
\includegraphics[height = 0.33\textwidth]{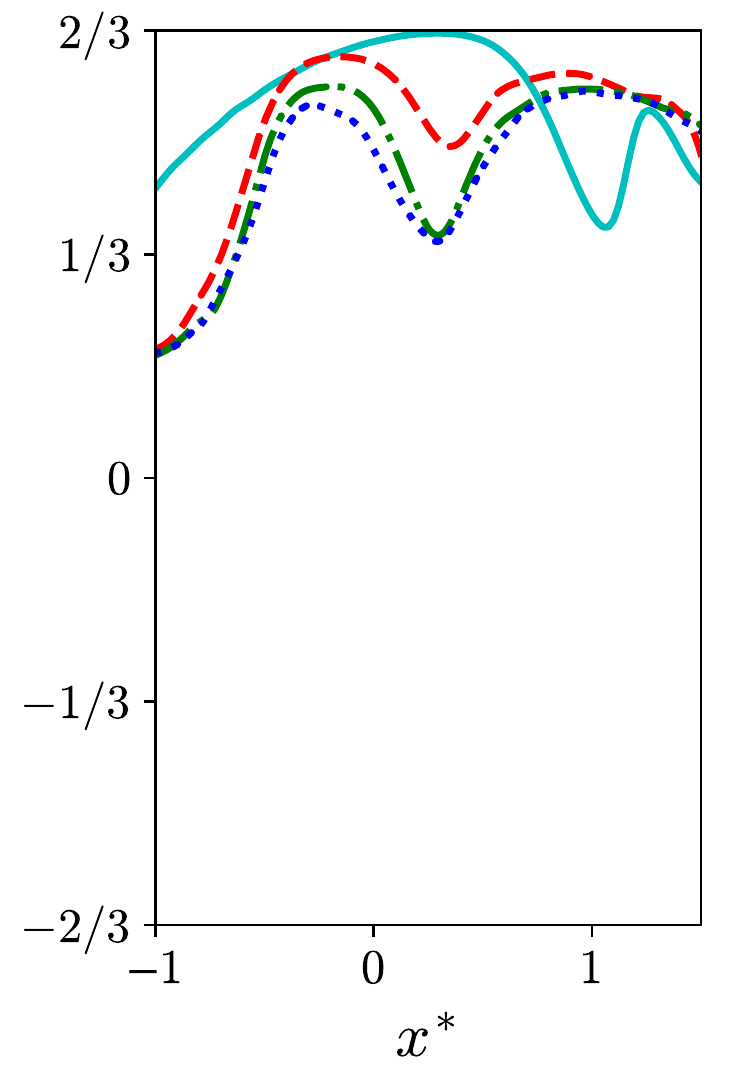}}
\subfigure[$\ $After re-shock, 2D]{%
\includegraphics[height = 0.33\textwidth]{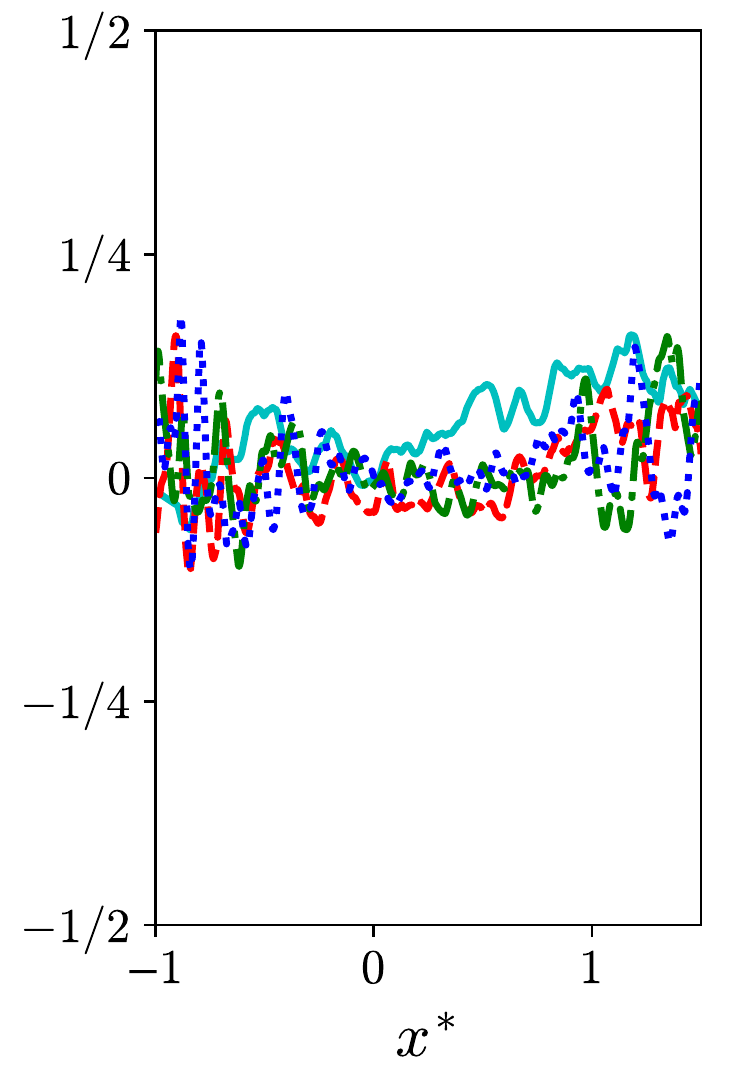}}
\subfigure[$\ $After re-shock, 3D]{%
\includegraphics[height = 0.33\textwidth]{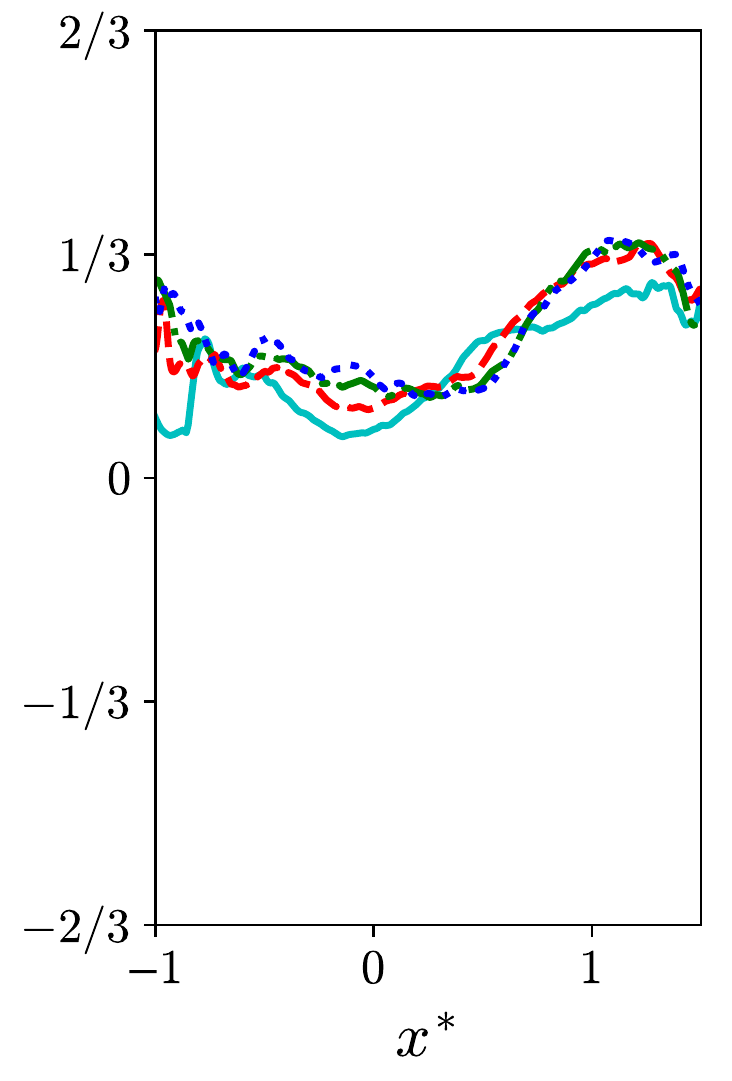}}
\caption{Profiles of Reynolds stress anisotropy component, $b_{11}$, at different times for the 2D and 3D problems with physical transport coefficients. Before re-shock: $t^{*}=0.91$ (cyan solid line); $t^{*}=7.25$ (red dashed line); $t^{*}=13.6$ (green dash-dotted line); $t^{*}=19.9$ (blue dotted line). After re-shock: $t^{*}=21.8$ (cyan solid line); $t^{*}=25.4$ (red dashed line); $t^{*}=29.0$ (green dash-dotted line); $t^{*}=31.7$ (blue dotted line).}
\label{fig:2D_vs_3D_anisotropy_profiles}
\end{figure*}

Figure~\ref{fig:2D_vs_3D_anisotropy_profiles} compares the mean profiles of $b_{11}$ at different times for the 2D and 3D cases with physical transport coefficients. After first shock, as the flow remains transitional, $\left<b_{11}\right>$ varies a lot inside the mixing layer for both 2D and 3D cases. However, consistent with figure~\ref{fig:2D_3D_transport_coeffs_anisotropy}, the overall profile for the 2D case is approaching zero at faster rate. After re-shock, $b_{11}$ has become more uniform inside the mixing layer for both 2D and 3D cases. However, for the 3D case, there is a slight asymmetry between the light and heavy fluid sides, with larger anisotropy on the light fluid side ($x^*>0$). As seen above, the kinetic energy is larger on the light fluid side, which corresponds to higher Reynolds numbers. Consistent with the time variation of $\left<b_{11}\right>$, its local values also remain larger at higher Reynolds numbers.

%%%%%%%%%%%%%%%%%%%%%%%%%%%%%%%%%%%%%%%%%%%%%%%%%%%%%%%%%%%%%%%%%%%%%%%%%%%%%%%%
%%%%%%%%%%%%%%%%%%%%%%%%%%%%%%%%%%%%%%%%%%%%%%%%%%%%%%%%%%%%%%%%%%%%%%%%%%%%%%%%
%%%%%%%%%%%%%%%%%%%%%%%%%%%%%%%%%%%%%%%%%%%%%%%%%%%%%%%%%%%%%%%%%%%%%%%%%%%%%%%%

\subsection{\label{sec:spectra} Spectra}

\begin{figure*}[!ht]
    \centering
    \subfigure[$\ $2D]{%
    \includegraphics[width=0.45\textwidth]{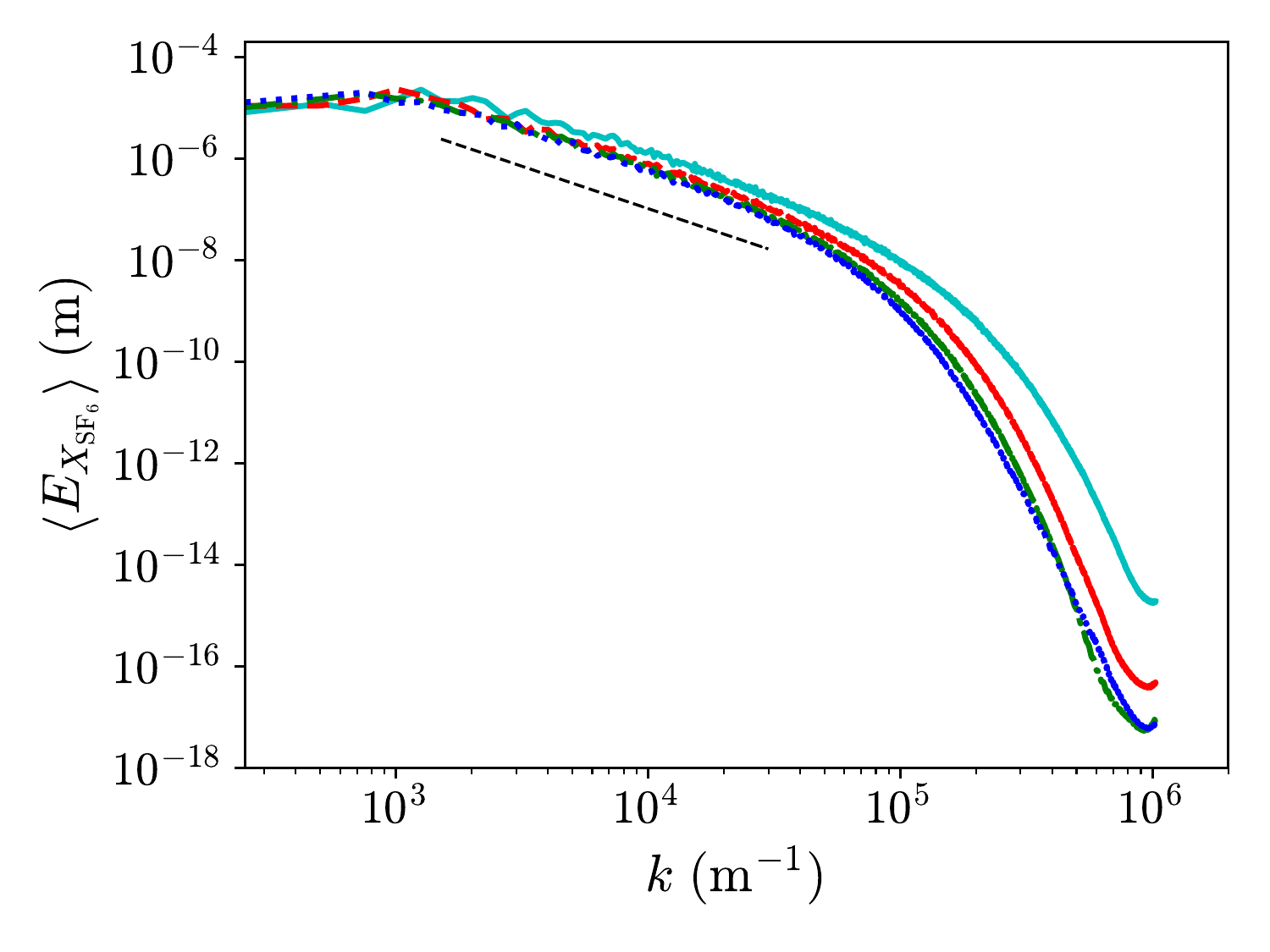}}
    \subfigure[$\ $3D]{%
    \includegraphics[width=0.45\textwidth]{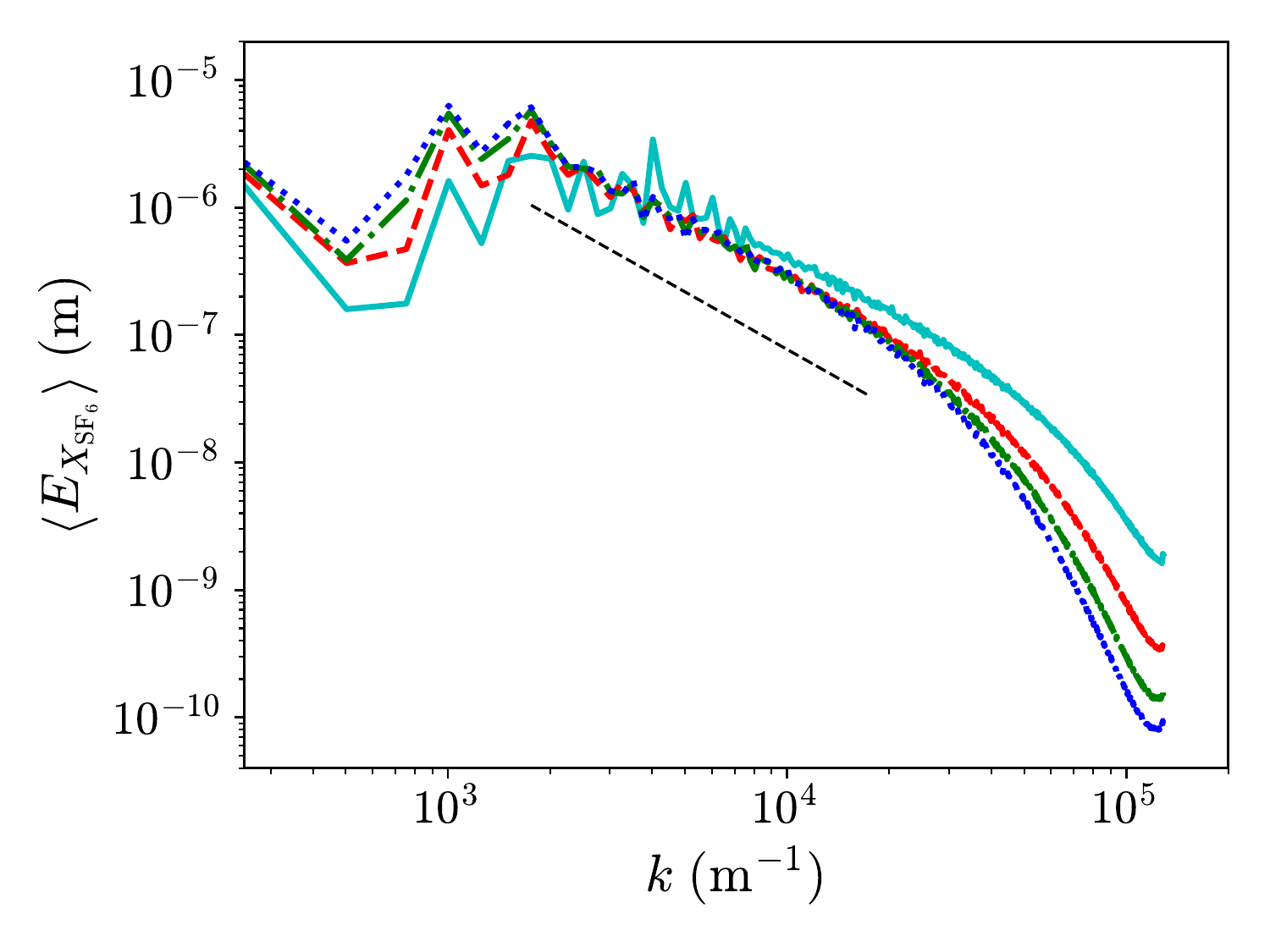}}
    \caption{Spectra of $\mathrm{SF_6}$ mole fraction, $X_{\mathrm{SF_6}}$, at different times after re-shock for the 2D and 3D problems with physical transport coefficients in the central part of mixing layer. Thin black dashed line: $k^{-5/3}$ or $k^{-3/2}$ (2D or 3D); cyan solid line: $t^{*}=21.8$; red dashed line: $t^{*}=25.4$; green dash-dotted line: $t^{*}=29.0$; blue dotted line: $t^{*}=31.7$.}
    \label{fig:mole_fraction_spectra_after_reshock}
\end{figure*}

Figure~\ref{fig:mole_fraction_spectra_after_reshock} shows the spectra of $\mathrm{SF6}$ mole fraction, $\left< E_{X_{\mathrm{SF6}}} \right>$, at different times after re-shock computed within the central part of mixing layer for both 2D and 3D cases with physical transport coefficients. At the end time of the simulations, inertial ranges are emerging for both cases. The scaling in the inertial range is close to $\left< E_{X_{\mathrm{SF6}}} \right> \propto k^{-5/3}$ for the 2D case and close to $\left< E_{X_{\mathrm{SF6}}} \right> \propto k^{-3/2}$ for the 3D case. The scaling of mole fraction spectrum at the end of 3D simulation is close to that of density spectra observed by \citet{tritschler2014richtmyer} in their 3D RMI simulations at early times after re-shock, although long after re-shock they found a smaller gradient scaling for the spectra. \citet{weber2014experimental} reported a -5/3 scaling for the mole fraction spectra from their 3D RMI experimental results at late times but the experiments did not have re-shock.

To study the energy spectrum in variable-density or compressible turbulence, it is common to decompose the energy as $|\sqrt{\rho} \bm{u}|^2$ or $|\sqrt{\rho} \bm{u}^{\prime}|^2$~\cite{cook2002energy,livescu2008variable,towery2016spectral,grete2017energy}. However, it is pointed out by \citet{zhao2018inviscid} that one cannot prove spectral locality using energy definitions of the form $|\sqrt{\rho} \bm{u}|^2$, because the corresponding transport equations contain terms that are divided by $\sqrt{\rho}$ and can no longer be arranged in divergence form. As a result, viscous effects may not decay at large scales when density variations are large. This lack of locality prevents the corresponding energy to develop an inertial range. Instead, they suggested that kinetic energy defined using Favre (density-weighted) filtering, $\hat{\bar{\rho}} |\hat{\tilde{\bm{u}}}|^2$, where $\hat{\bar{ \cdot }}$ and $\hat{\tilde{ \cdot }}$ represent the simple and Favre filtering respectively, can develop an inertial range. While useful for proving locality of energy transfer in the context of filtering, this quantity is not in quadratic form and, thus, not the proper way to form the energy spectrum. The energy spectrum based on $|\rho \bm{u}^{\prime\prime} / \sqrt{\bar{\rho}}|^2$ energy definition, satisfies
the requirement that terms in the corresponding transport equations remain in divergence form, while we still have a quadratic form for the energy.

\begin{figure*}[!ht]
    \centering
    \subfigure[$\ $2D]{%
    \includegraphics[width=0.45\textwidth]{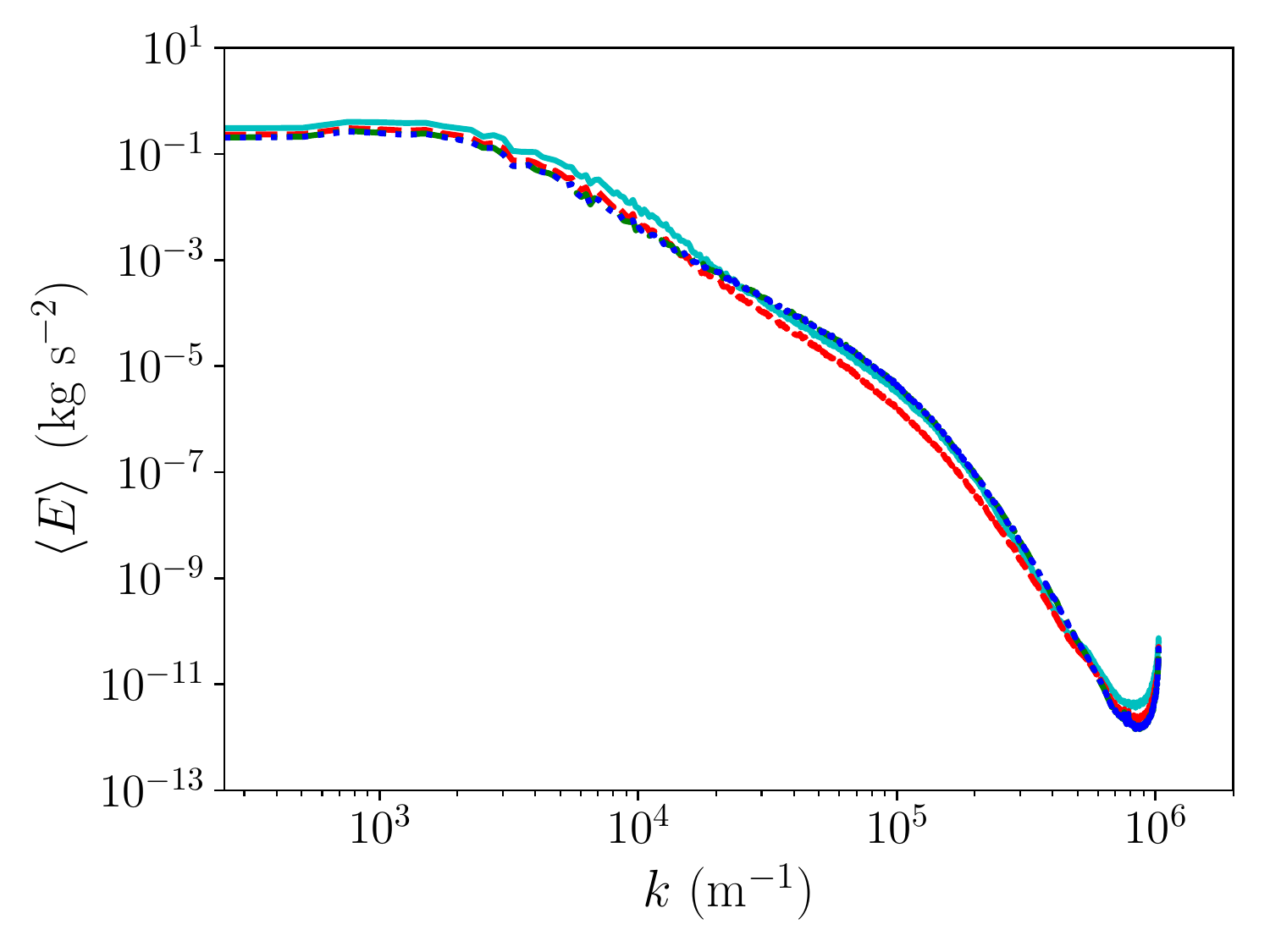}}
    \subfigure[$\ $3D]{%
    \includegraphics[width=0.45\textwidth]{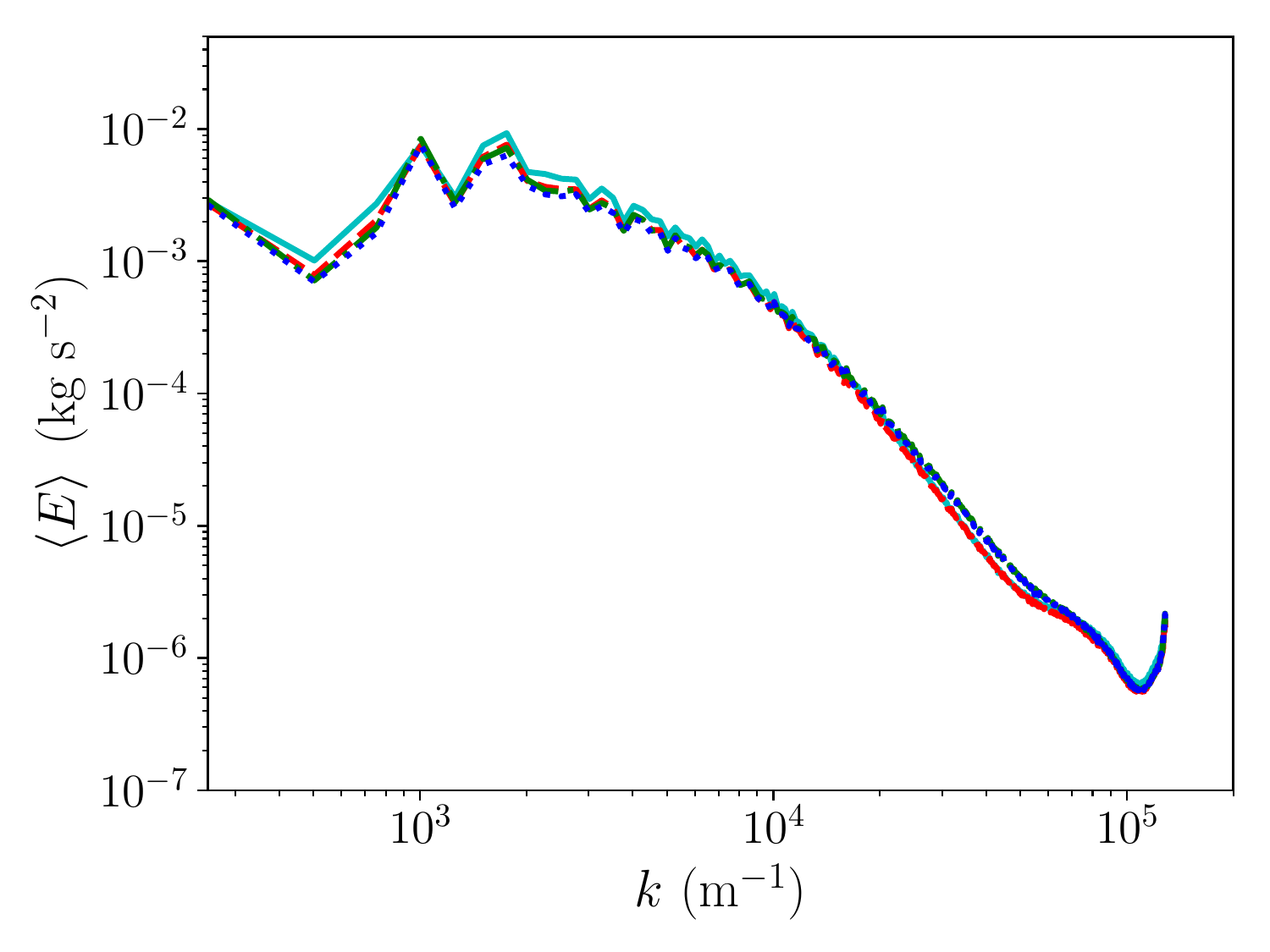}}
    \caption{Energy spectra computed with different energy definitions at the end of simulation ($t^{*}=31.7$) for the 2D and 3D problems with physical transport coefficients in the central part of mixing layer. Cyan solid line: $(\sqrt{\bar{\rho}} \bm{u}^{\prime})^2$; red dashed line: $(\sqrt{\rho} \bm{u}^{\prime})^2$; green dash-dotted line: $(\rho \bm{u}^{\prime} / \sqrt{\bar{\rho}})^2$; blue dotted line: $(\rho \bm{u}^{\prime\prime} / \sqrt{\bar{\rho}})^2$.}
    \label{fig:energy_spectra_comparison_end}
\end{figure*}

Figure~\ref{fig:energy_spectra_comparison_end} compares the energy spectra computed within the central part of mixing layer with four different energy definitions: (1) $|\sqrt{\bar{\rho}} \bm{u}^{\prime}|^2$, (2) $|\sqrt{\rho} \bm{u}^{\prime}|^2$, (3) $|\rho \bm{u}^{\prime} / \sqrt{\bar{\rho}}|^2$, and (4) $|\rho \bm{u}^{\prime\prime} / \sqrt{\bar{\rho}}|^2$, for both 2D and 3D cases with physical transport coefficients, at the end of simulations. Energy definitions using $|\rho \bm{u}^{\prime} / \sqrt{\bar{\rho}}|^2$ and $|\rho \bm{u}^{\prime\prime} / \sqrt{\bar{\rho}}|^2$ yield spectra that are almost indistinguishable for the flow in each case considered here. Also, switching $\bm{u}^{\prime}$ and $\bm{u}^{\prime\prime}$ has unnoticeable effect on the spectra for other two energy definitions. Although the spectra for the 3D case are close with the choices above, the spectra computed with $|\rho \bm{u}^{\prime} / \sqrt{\bar{\rho}}|^2$ or $|\rho \bm{u}^{\prime\prime} / \sqrt{\bar{\rho}}|^2$ for the 2D case seem to have clearer inertial range than spectra computed with the other two energy definitions, confirming the discussion above.

\begin{figure*}[!ht]
    \centering
    \subfigure[$\ $2D]{%
    \includegraphics[width=0.45\textwidth]{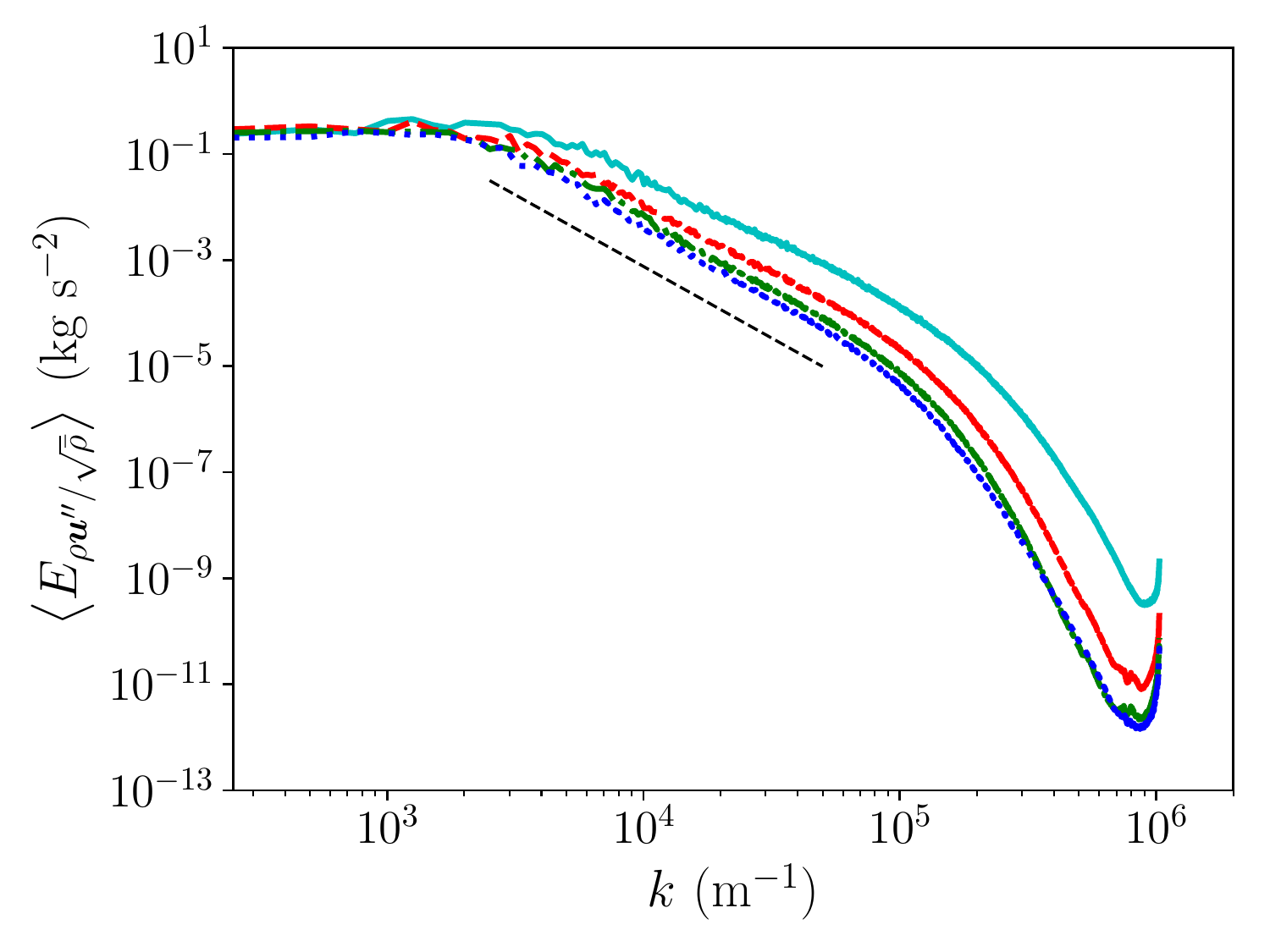}}
    \subfigure[$\ $3D]{%
    \includegraphics[width=0.45\textwidth]{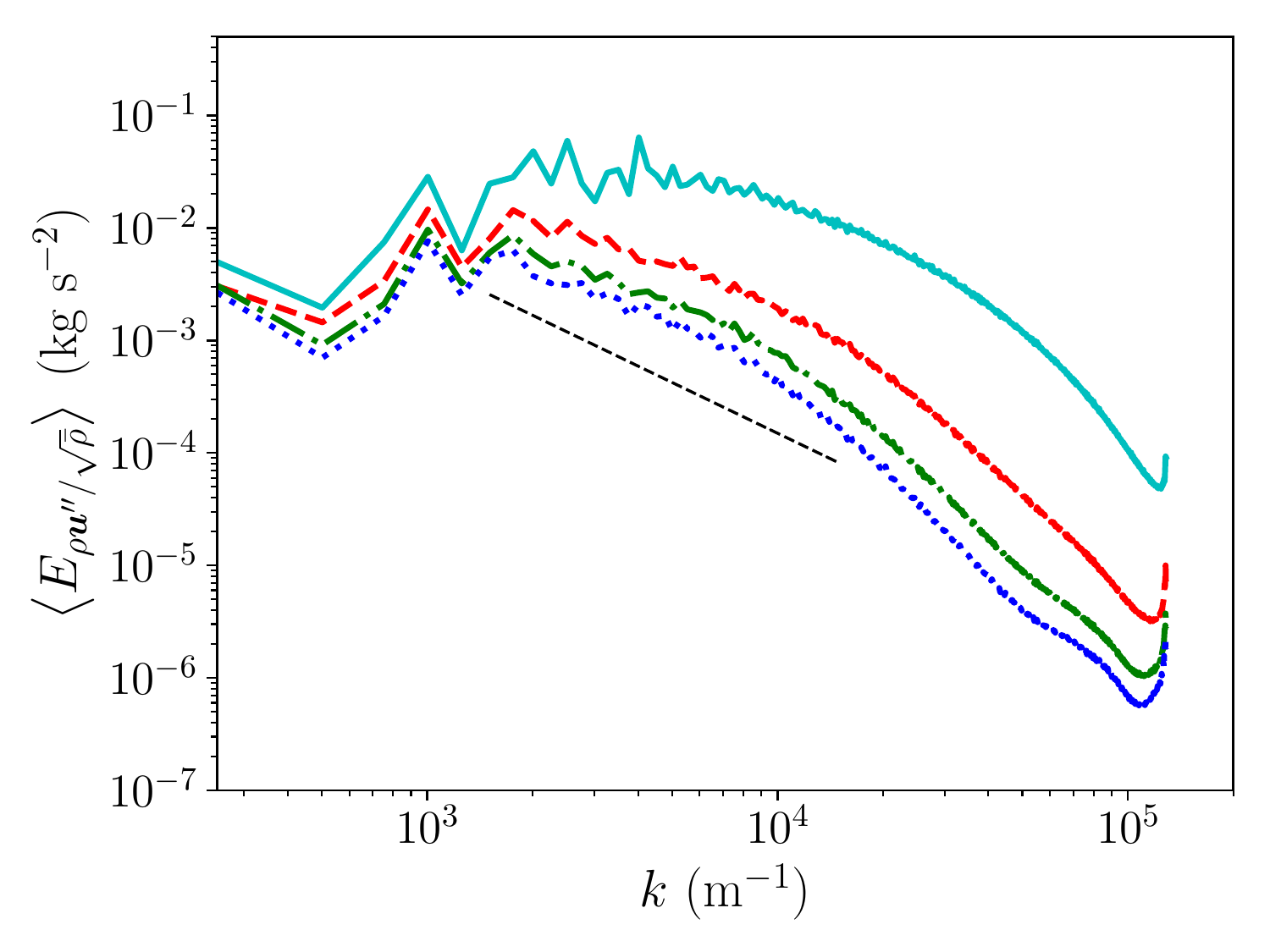}}
    \caption{Energy spectra at different times after re-shock for the 2D and 3D problems with physical transport coefficients in the central part of mixing layer. Thin black dashed line: $k^{-2.7}$ or $k^{-3/2}$ (2D or 3D); cyan solid line: $t^{*}=21.8$; red dashed line: $t^{*}=25.4$; green dash-dotted line: $t^{*}=29.0$; blue dotted line: $t^{*}=31.7$.}
    \label{fig:energy_spectra_after_reshock}
\end{figure*}

Figure~\ref{fig:energy_spectra_after_reshock} shows the energy spectra computed with $|\rho \bm{u}^{\prime\prime} / \sqrt{\bar{\rho}}|^2$, $\left< E_{\rho \bm{u}^{\prime\prime} / \sqrt{\bar{\rho}}} \right>$, at different times for both 2D and 3D cases. As seen from the figure, an inertial range is developing for each case. Comparing the 2D and 3D problems, there is a huge difference in scaling for the spectra in the inertial ranges. The scaling is close to $\left< E_{\rho \bm{u}^{\prime\prime} / \sqrt{\bar{\rho}}} \right> \propto k^{-2.7}$ for the 2D case but close to $\left< E_{\rho \bm{u}^{\prime\prime} / \sqrt{\bar{\rho}}} \right> \propto k^{-3/2}$ for the 3D case. \citet{thornber2015numerical} found the scaling of spectra of kinetic energy defined as $|\sqrt{\rho} \bm{u}|^2/2$ close to $E \propto k^{-3}$ at high wavenumbers at late times in their inviscid 2D RMI simulations with different initial perturbations. Similar to our results for the 3D case, \citet{tritschler2014richtmyer} also found scaling of spectra of turbulent kinetic energy defined as $|\sqrt{\rho} \bm{u^{\prime\prime}}|^2/2$ close to -3/2 scaling after re-shock from their 3D RMI simulations, although the post-re-shock 3D simulation results by \citet{lombardini2012transition} show that a -5/3 power law seems to be more appropriate for energy contributed from the streamwise velocity component.

%%%%%%%%%%%%%%%%%%%%%%%%%%%%%%%%%%%%%%%%%%%%%%%%%%%%%%%%%%%%%%%%%%%%%%%%%%%%%%%%
%%%%%%%%%%%%%%%%%%%%%%%%%%%%%%%%%%%%%%%%%%%%%%%%%%%%%%%%%%%%%%%%%%%%%%%%%%%%%%%%
%%%%%%%%%%%%%%%%%%%%%%%%%%%%%%%%%%%%%%%%%%%%%%%%%%%%%%%%%%%%%%%%%%%%%%%%%%%%%%%%

\section{Conclusions}

We have conducted high-resolution 2D and 3D Navier--Stokes simulations of multi-species mixing driven by RMI. Through grid sensitivity analysis of statistical quantities that are dependent on features of different scales, we have confirmed that the highest grid resolution used for 2D simulations was sufficient to provide accurate statistics that depend on smallest scale features in the flows. Although the highest grid resolution used for 3D simulations is not sufficient to capture statistics that are associated with finest scale features, such as scalar dissipation rate and enstrophy, the grid resolution is high enough for converged statistics that depend on large scale features, such as mixing width and TKE. It was found that, for the chosen initial configuration, the flows inside the mixing layers were only weakly compressible for most of the time under both 2D and 3D configurations. However, the effective Atwood number within the mixing layer remains high throughout the simulations, leading to noticeable variable-density effects. The importance of these effects could be inferred from the asymmetry in the spatial profiles of mole fraction variance and TKE, and also the mole fraction PDF's.

The mole fraction profiles become self-similar for both 2D and 3D problems after a short transient regime, following both first shock and re-shock. However, the mixing layer width grows faster for the 2D case after first shock and re-shock. The 2D and 3D scalings of mixing widths at late times after re-shock were quantified and compared with previous studies. The results show that the scalings converge, when the thickness of the layer at re-shock and time of re-shock are included in the scaling law. The scaling exponents for the 2D and 3D problems have been evaluated to be around 0.55 and 0.44, respectively, using different fitting techniques. Similar values have been obtained when calculated directly from the temporal derivatives of the mixing layer widths. Consistent with the differences in the scaling laws, there also appears to be asymptotic but different mixedness values for 2D and 3D problems, where that of the former is smaller. Overall, the temporal evolutions of mixedness and PDF of mole fraction at different times indicate that the fluids are more difficult to mix under the 2D configuration than the 3D configuration. 

We have also compared the time evolution of TKE and anisotropy of the Reynolds normal stresses for both configurations. It is clear that without the effect of vortex stretching to break large scale features into smaller scales, the mean TKE decays slower under the 2D configuration. On the other hand, the Reynolds normal stresses become isotropic quickly after re-shock for the 2D case, while they remain quite anisotropic under 3D configuration even at late times. Similar to mixing widths, the 2D and 3D scalings of TKE at late times after re-shock have been evaluated and the scaling for the 3D problem has been found to be similar to that for Batchelar-type decaying turbulence. This finding is consistent with some of the previous studies on RMI. The 2D and 3D cases have also been compared by considering the evolution of the mole fraction and energy spectra after re-shock. Different scalings in inertial ranges for the 2D and 3D spectra can be seen clearly at late times after re-shock. The energy spectra are based on a new definition of energy, which is consistent with the inviscid criterion for decomposing scales~\cite{zhao2018inviscid}. While the effect to isolate the viscous effects can not be seen in the 3D problem due to insufficient grid resolution, there is clearly a longer inertial range for the 2D case with the new definition of energy. This indicates that the new definition may be a more appropriate metric for the spectral analysis of turbulence energy for variable-density flows.

The effects of Reynolds number on 3D RMI have also been analyzed by varying the physical transport coefficients. Through artificially increasing the values of physical transport coefficients given by diffusivity/viscosity models, the Reynolds number based on mixing width or integral length scale is effectively reduced over time. The effect of reduced Reynolds number is small on the growth of mixing layer width, mixedness, and decay rate of TKE at late times after re-shock, but greatly affects the development of many statistical quantities between first shock and re-shock. During this time, the growth rate of mixing layer width decreases, while the decay rate of TKE increases for cases with smaller Reynolds numbers. At the same time, Reynolds normal stresses become isotropic at faster rate with decreased Reynolds number values. The changes of flow field variables due to Reynolds number effects before the occurrence of mixing transition address the importance of including viscous and diffusive effects in variable-density simulations for accurate predictions of the growth of instabilities.

\section{Acknowledgments}
This work was performed under the auspices of U.S. Department of Energy. M. L. Wong and S. K. Lele were supported by Los Alamos National Laboratory, under Grant No. 431679. Los Alamos National Laboratory is operated by Triad National Security, LLC, for the National Nuclear Security Administration of U.S. Department of Energy (Contract No. 89233218CNA000001). Computational resources were provided by the Los Alamos National Laboratory Institutional Computing Program.

\appendix

\section{\label{sec:2D_stat_convergence} Statistical sensitivity analysis of two-dimensional simulations}

\begin{figure*}[!ht]
\centering
\subfigure[$\ $Mixing width]{%
\includegraphics[width=0.42\textwidth]{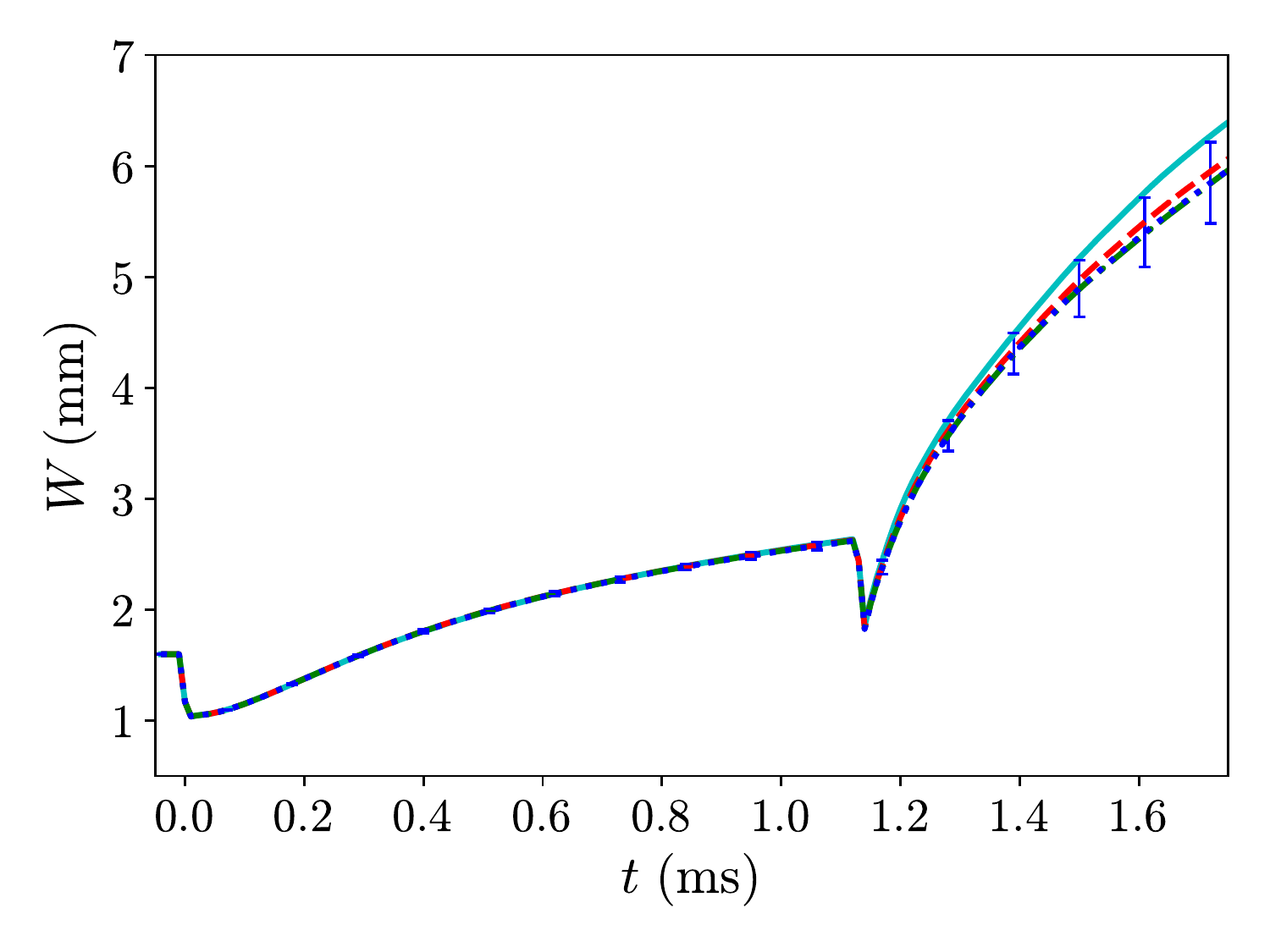}}
\subfigure[$\ $Mixedness]{%
\includegraphics[width=0.42\textwidth]{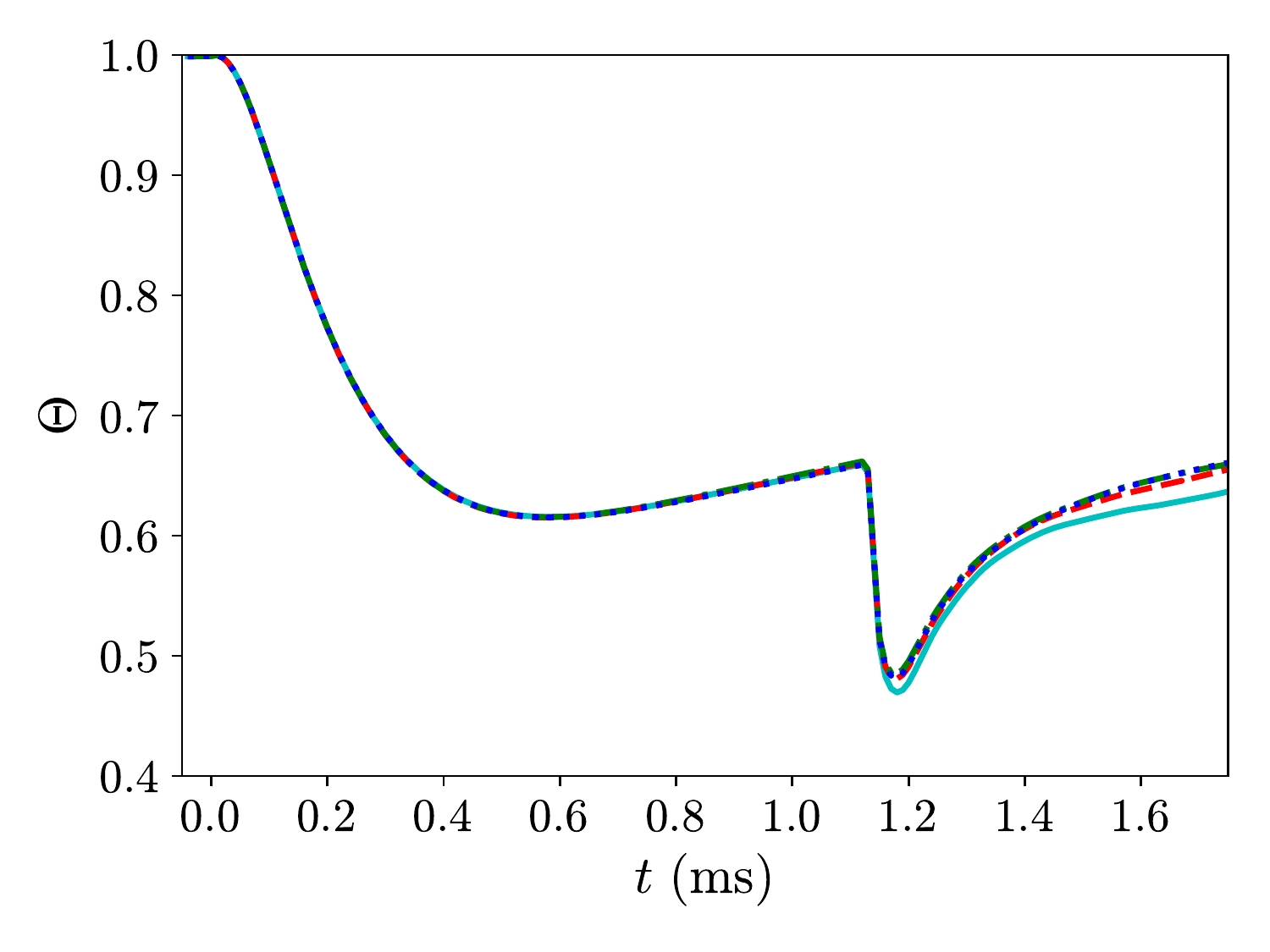}}
\subfigure[$\ $Integrated $\mathrm{TKE}$]{%
\includegraphics[width=0.42\textwidth]{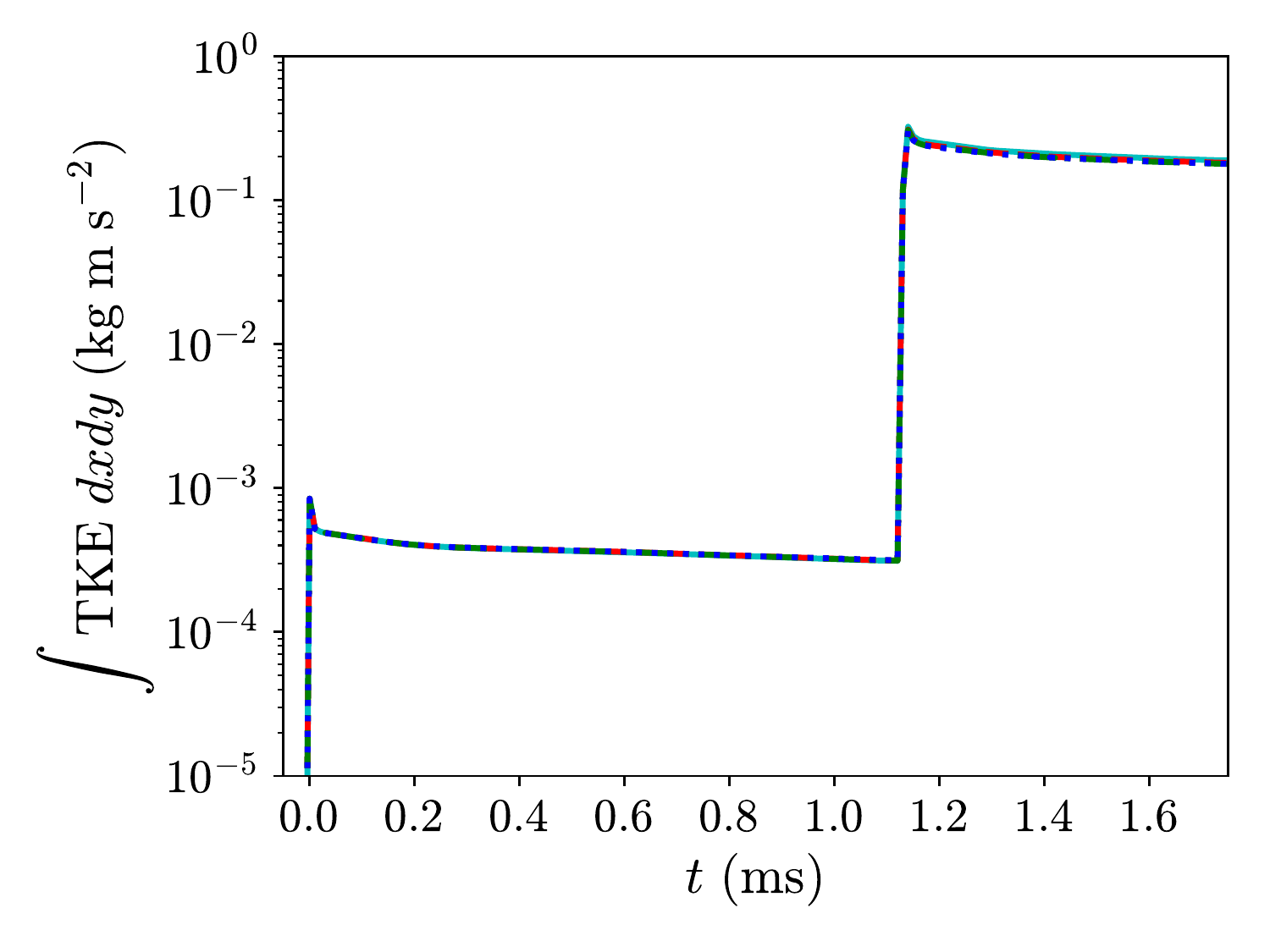}}
\subfigure[$\ $Integrated $\mathrm{TKE}$ (after re-shock)]{%
\includegraphics[width=0.42\textwidth]{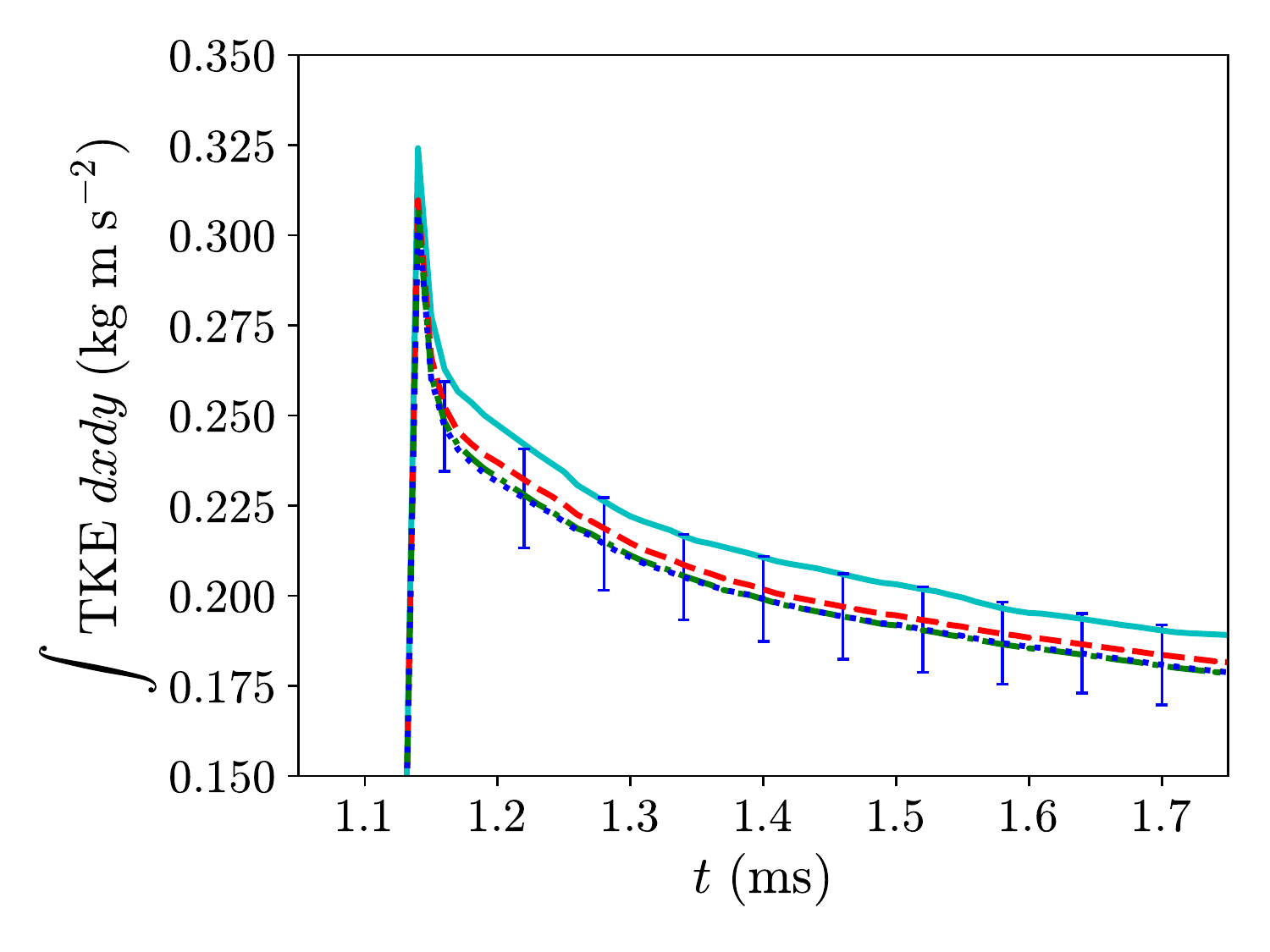}}
\subfigure[$\ $Integrated scalar dissipation rate]{%
\includegraphics[width=0.42\textwidth]{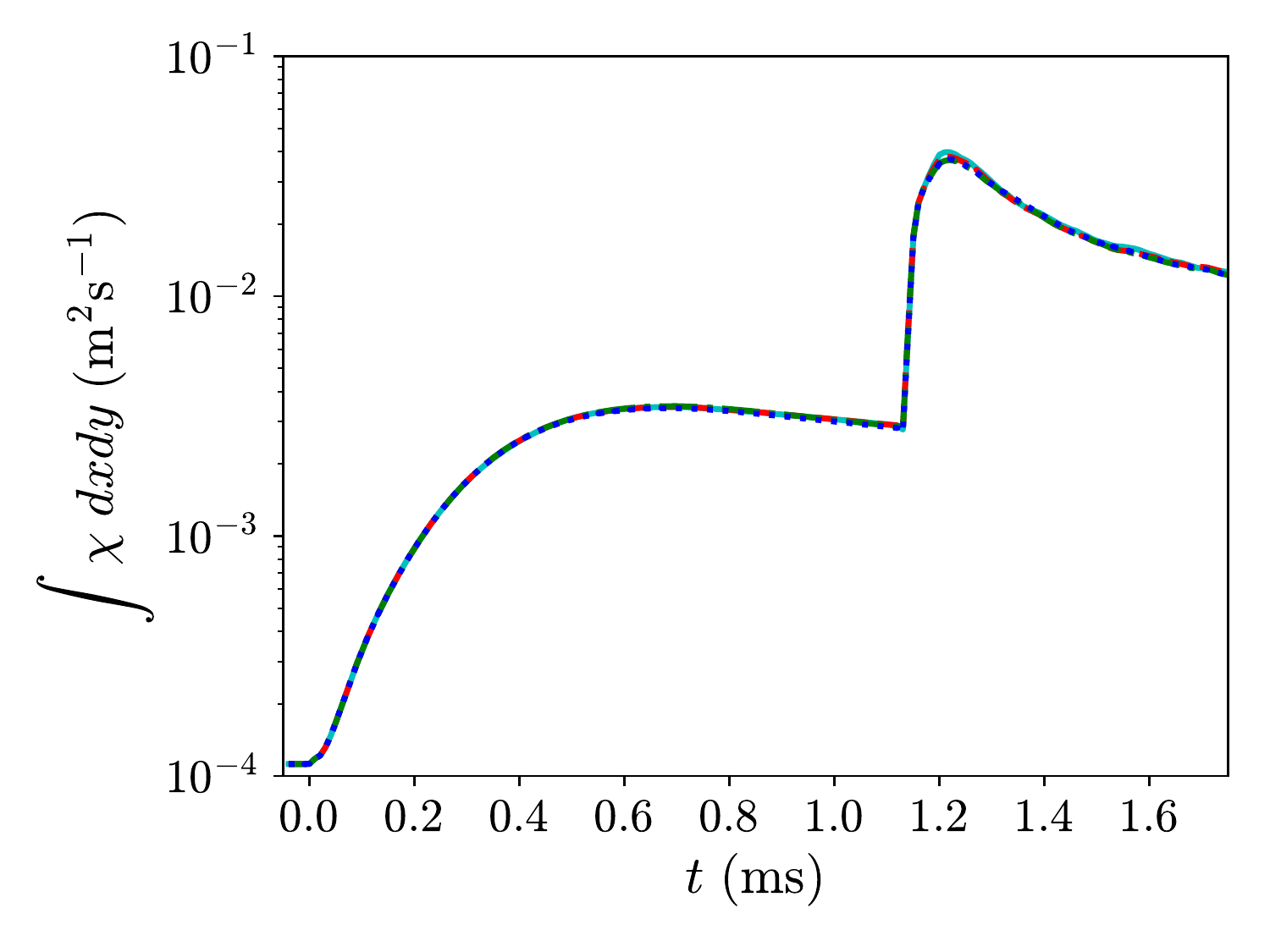}}
\subfigure[$\ $Integrated enstrophy]{%
\includegraphics[width=0.42\textwidth]{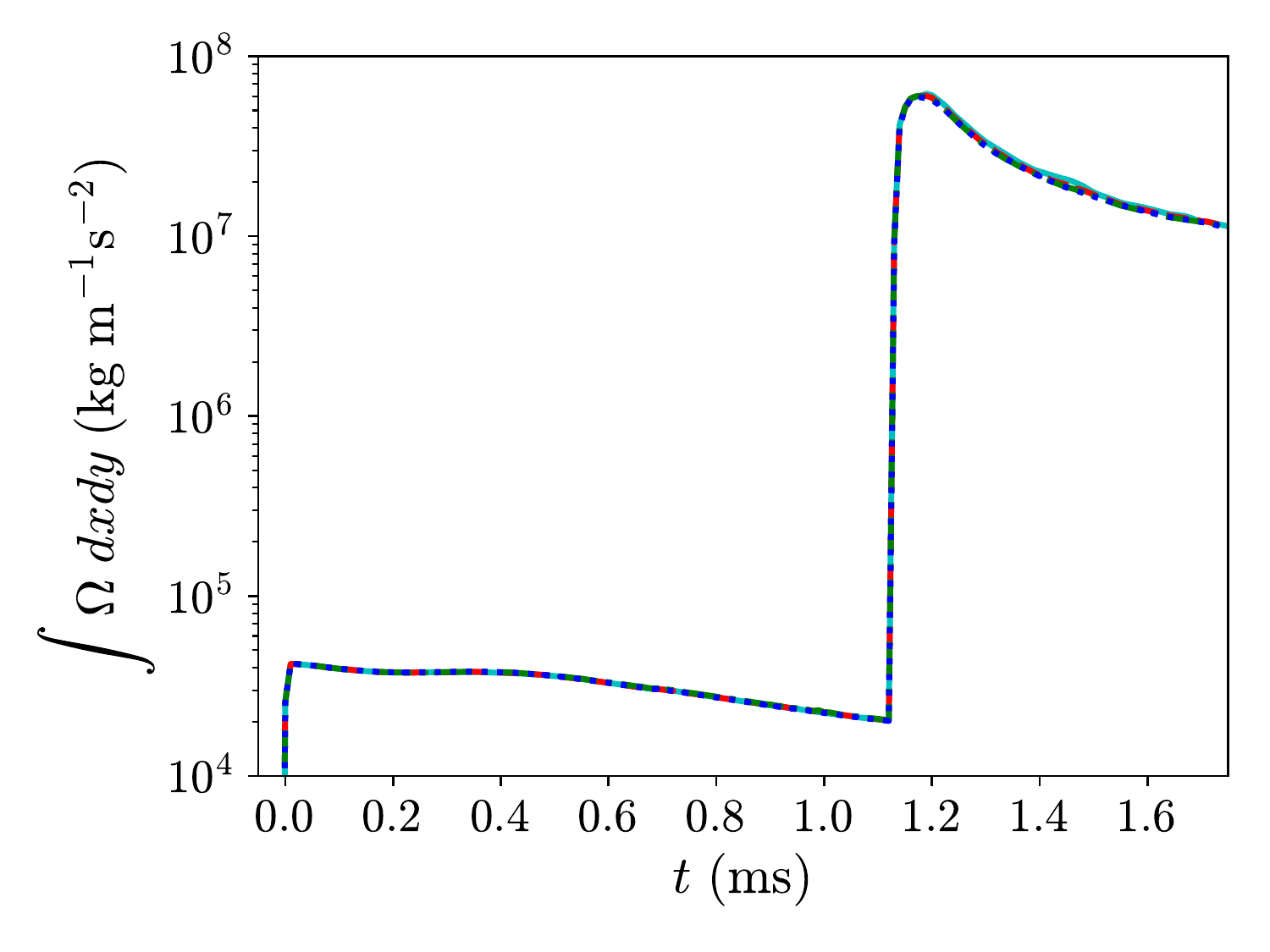}}
\caption{Statistical sensitivities of different statistical quantities over time for the 2D problem with grid G. Cyan solid line: 6 realizations; red dashed line: 12 realizations; green dash-dotted line: 18 realizations; blue dotted line: 24 realizations. The error bars in some of the sub-figures show the $\pm$ one standard deviation of the corresponding quantities over 24 realizations.}
\label{fig:2D_stat_sensitivity}
\end{figure*}

Figure~\ref{fig:2D_stat_sensitivity} shows the time evolution of various statistical quantities computed with different number of realizations and the grid G settings for the 2D problem. It can be seen from the figure that statistical convergence has already been obtained for the quantities considered with 24 realizations for all times. The standard deviations of mixing width and TKE (only after re-shock) over 24 realizations are also shown in the same figure. The standard deviations of mixing width and TKE are insignificant before re-shock. However, after re-shock, they have become large and hence statistical convergence can only be obtained for the quantities with a large number of realizations.

\section{\label{sec:p_over_T} Profiles of ratio of pressure to temperature}

\begin{figure*}[!ht]
\centering
\subfigure[$\ $Before re-shock, 2D]{%
\includegraphics[height = 0.33\textwidth]{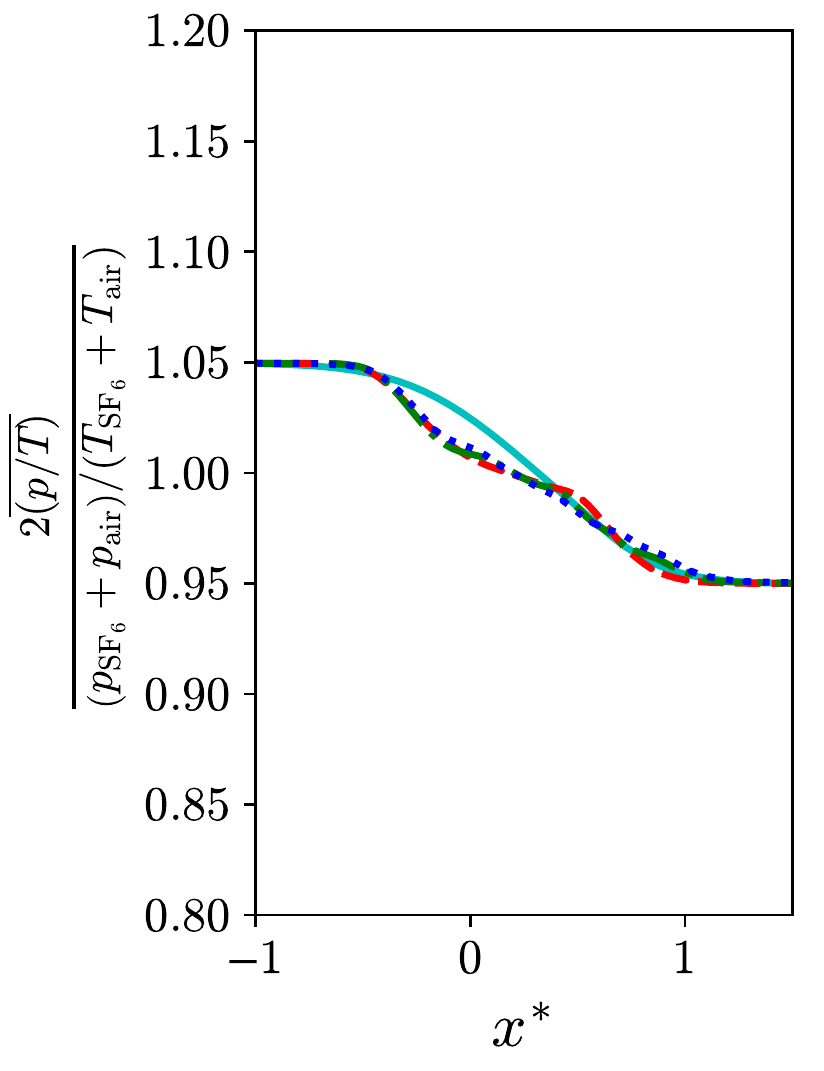}}
\subfigure[$\ $Before re-shock, 3D]{%
\includegraphics[height = 0.33\textwidth]{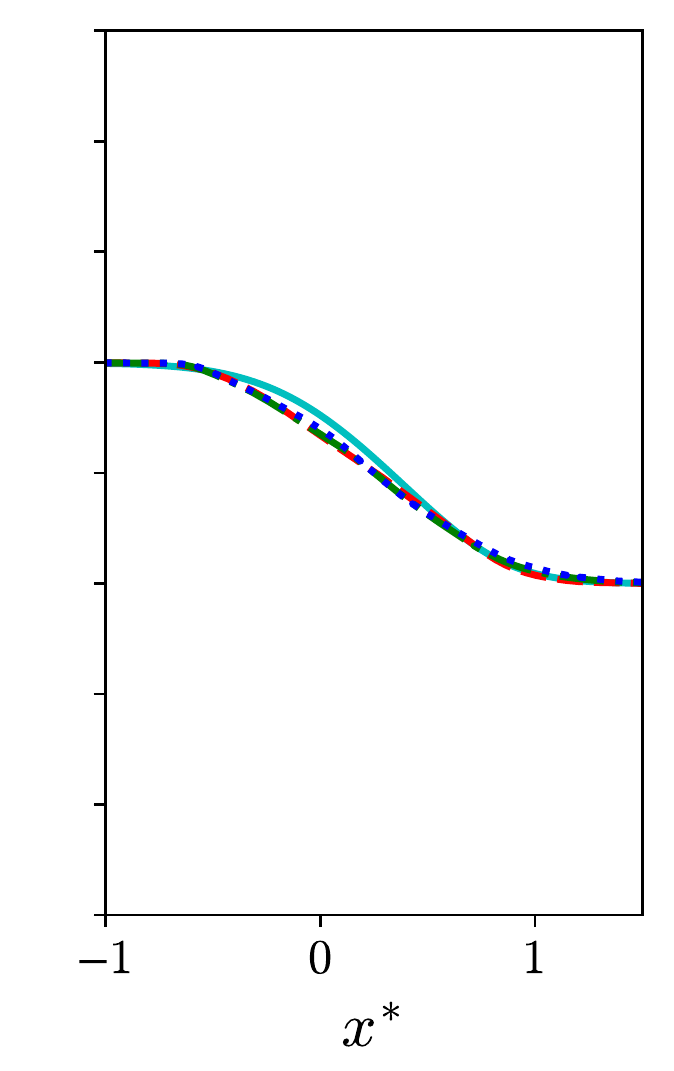}}
\subfigure[$\ $After re-shock, 2D]{%
\includegraphics[height = 0.33\textwidth]{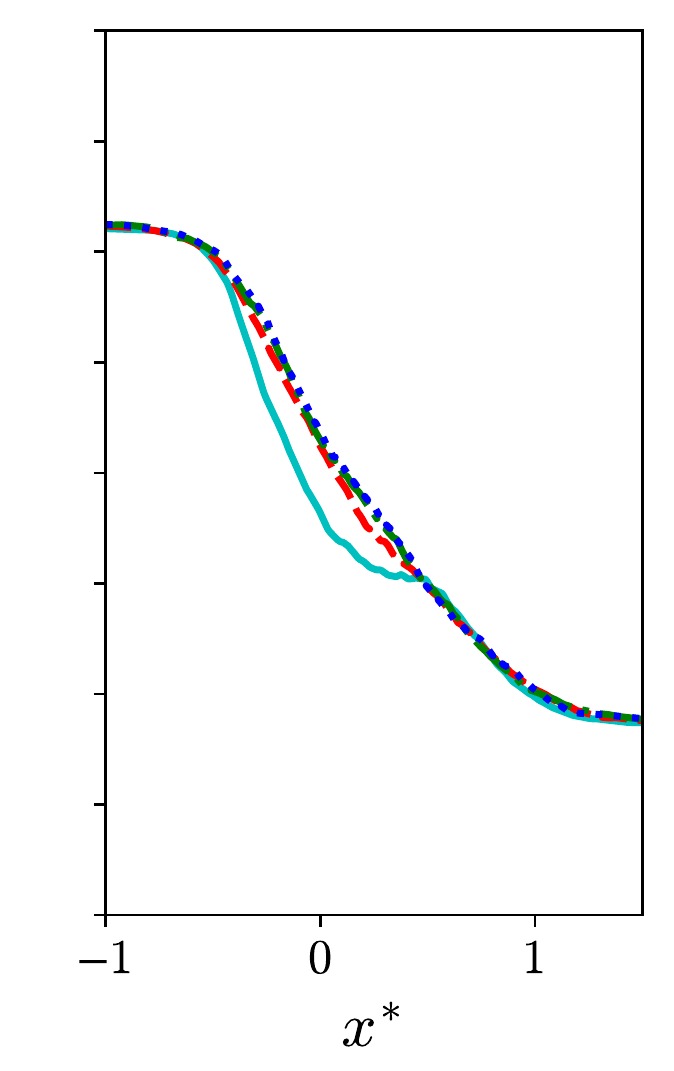}}
\subfigure[$\ $After re-shock, 3D]{%
\includegraphics[height = 0.33\textwidth]{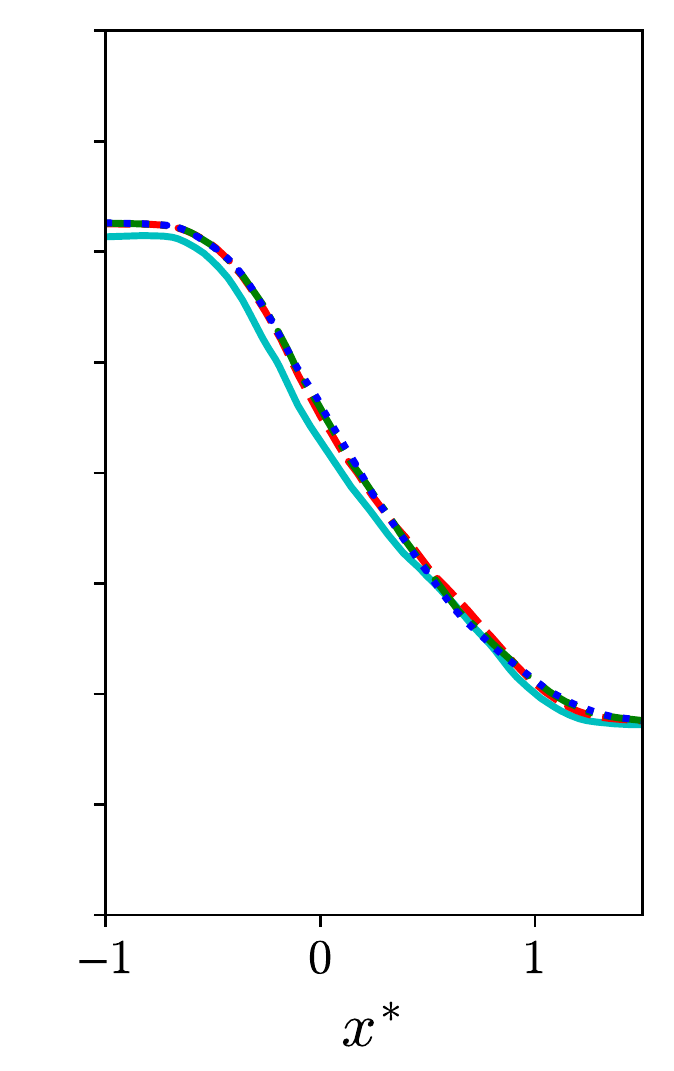}}
\caption{Profiles of normalized ratio of pressure to temperature at different times for the 2D and 3D problems with physical transport coefficients. Before re-shock: $t^{*}=0.9$ (cyan solid line); $t^{*}=7.5$ (red dashed line); $t^{*}=14.1$ (green dash-dotted line); $t^{*}=20.7$ (blue dotted line). After re-shock: $t^{*}=22.6$ (cyan solid line); $t^{*}=26.3$ (red dashed line); $t^{*}=30.1$ (green dash-dotted line); $t^{*}=32.9$ (blue dotted line).}
\label{fig:2D_vs_3D_mean_p_over_T_profiles}
\end{figure*}

\begin{figure*}[!ht]
\centering
\subfigure[$\ $Before re-shock, 2D]{%
\includegraphics[height = 0.33\textwidth]{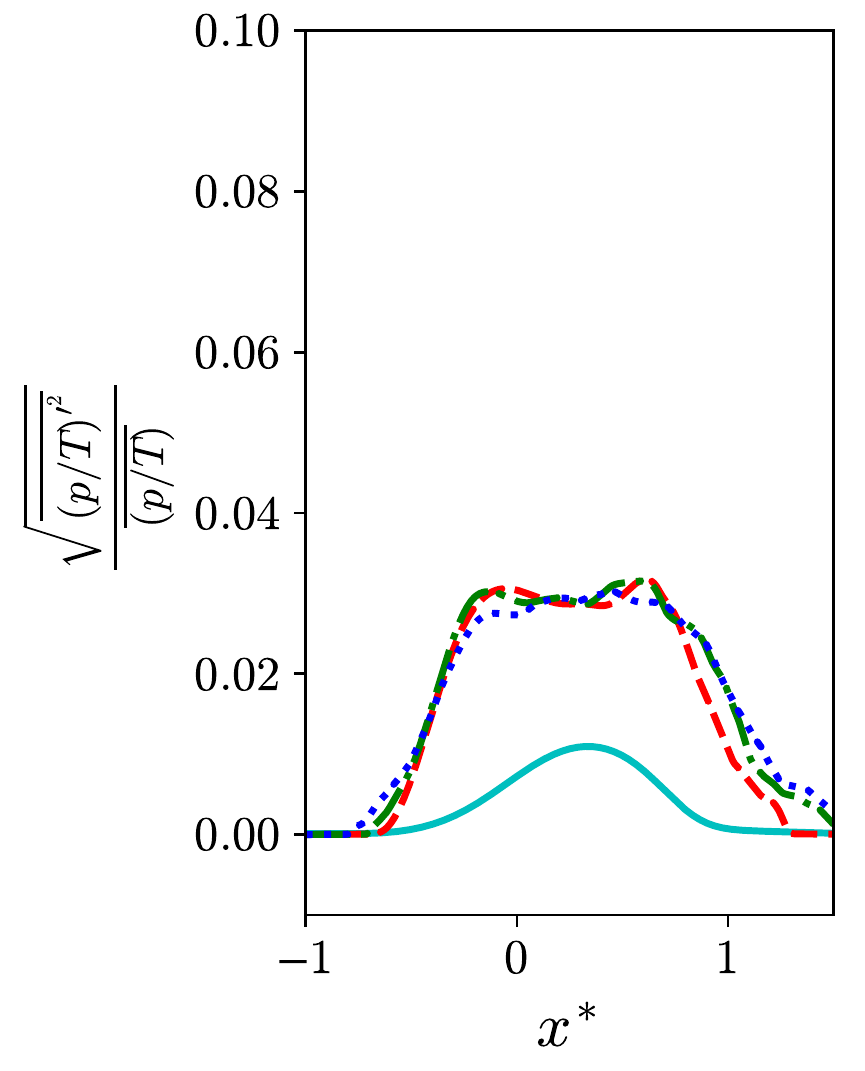}}
\subfigure[$\ $Before re-shock, 3D]{%
\includegraphics[height = 0.33\textwidth]{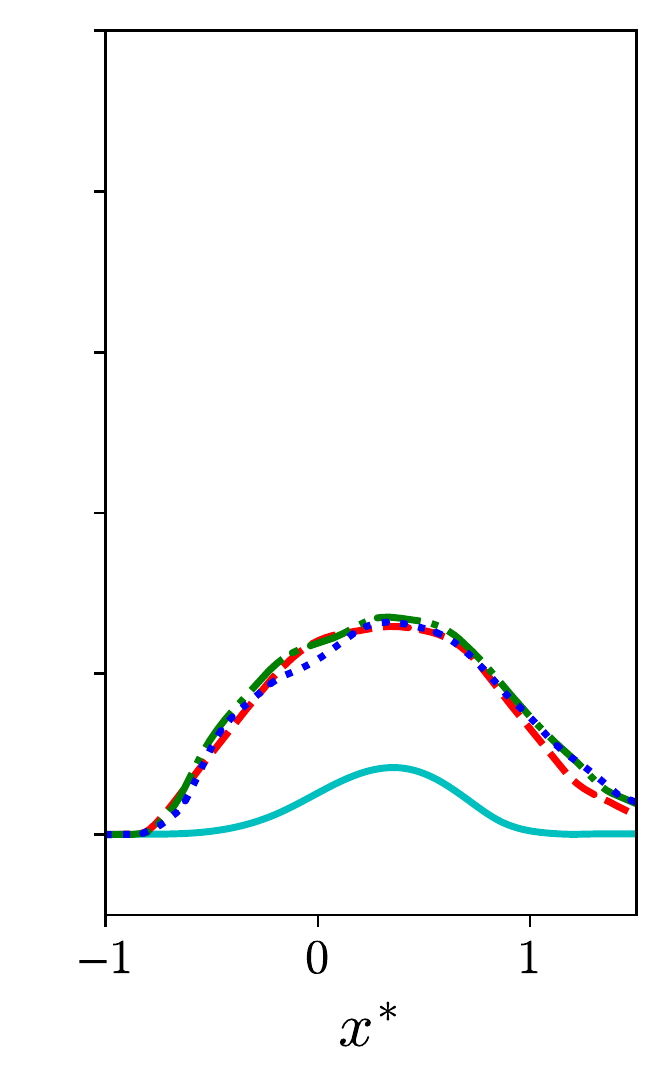}}
\subfigure[$\ $After re-shock, 2D]{%
\includegraphics[height = 0.33\textwidth]{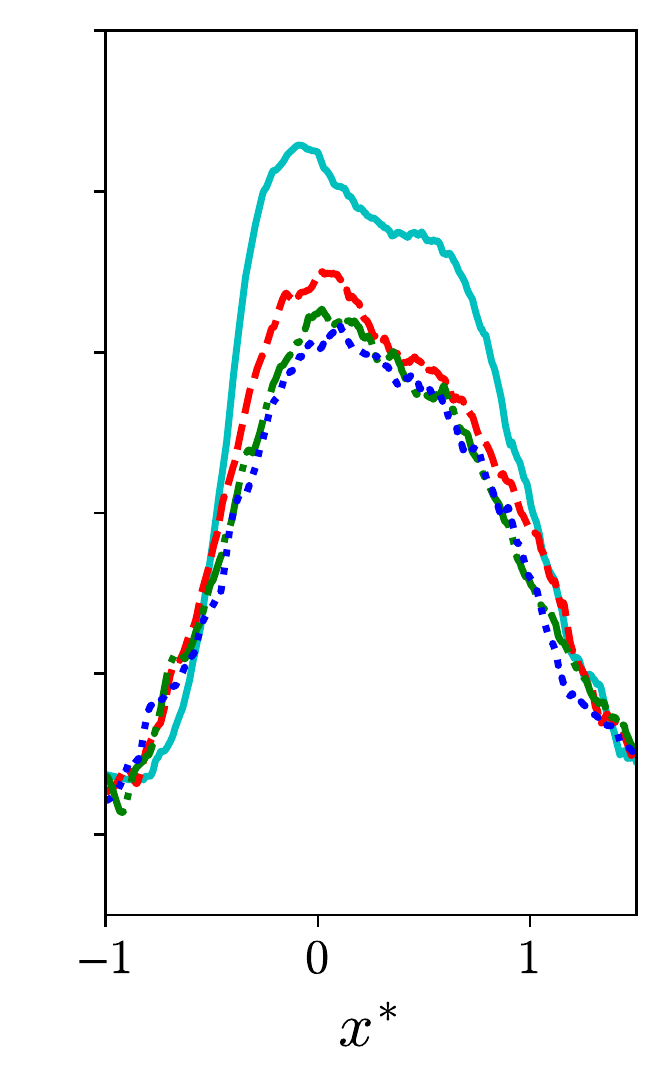}}
\subfigure[$\ $After re-shock, 3D]{%
\includegraphics[height = 0.33\textwidth]{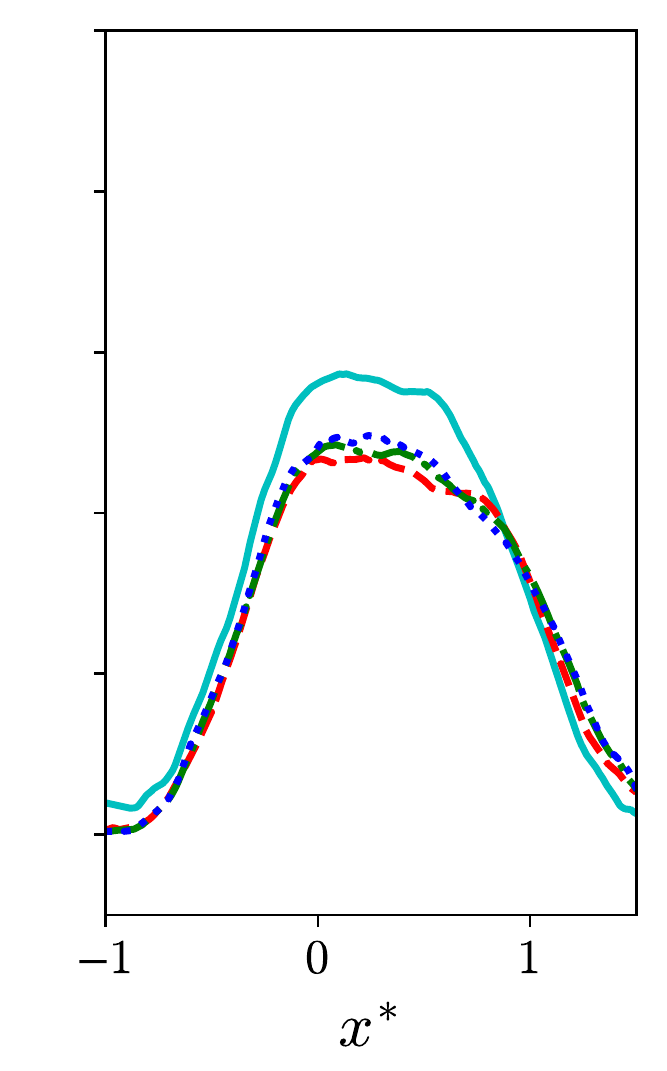}}
\caption{Profiles of normalized standard deviation of ratio of pressure to temperature at different times for the 2D and 3D problems with physical transport coefficients. Before re-shock: $t^{*}=0.9$ (cyan solid line); $t^{*}=7.5$ (red dashed line); $t^{*}=14.1$ (green dash-dotted line); $t^{*}=20.7$ (blue dotted line). After re-shock: $t^{*}=22.6$ (cyan solid line); $t^{*}=26.3$ (red dashed line); $t^{*}=30.1$ (green dash-dotted line); $t^{*}=32.9$ (blue dotted line).}
\label{fig:2D_vs_3D_stdev_p_over_T_profiles}
\end{figure*}

Figures~\ref{fig:2D_vs_3D_mean_p_over_T_profiles} and \ref{fig:2D_vs_3D_stdev_p_over_T_profiles} show the normalized mean and standard deviation profiles for the ratio $p/T$ across the mixing layers at different times for both 2D and 3D problems with physical transport coefficients. The mean profiles are normalized with the pressure ($p_{\mathrm{SF_6}}$ and $p_{\mathrm{air}}$) and temperature ($T_{\mathrm{SF_6}}$ and $T_{\mathrm{air}}$) on either side of the material interface from the solutions of the 1D flow representation. The standard deviation profiles are normalized by the mean profiles. It can be seen that the mean profiles at most only have around $\pm 10\%$ variations across the mixing layers and the normalized standard deviation profiles have magnitudes smaller than $10\%$ at all times. Therefore, the ratios $p/T$ inside the mixing layers are quasi-uniform for both 2D and 3D cases.

\section{\label{sec:length scales} Time evolution of length scales after re-shock}

In this section, the time evolution of different length scales after re-shock for the highest Reynolds number 3D case (with physical transport coefficients) are estimated and briefly discussed. According to~\citet{pope2000turbulent}, the Reynolds number associated with the outer-scale (largest scale) eddies, $Re_{l_o}$, can be related to the Taylor-scale Reynolds number, $Re_{\lambda_{T}}$ with a few assumptions. The Reynolds numbers are given by:
\begin{align}
    Re_{l_o}         &= \sqrt{\frac{3}{2}} \frac{\rho u_{\mathrm{turb}} l_o}{\mu}, \\
    Re_{\lambda_{T}} &= \frac{\rho u_{\mathrm{turb}} \lambda_{T}}{\mu},
\end{align}
where $l_o$ and $\lambda_{T}$ are the length scale of the largest eddies and Taylor microscale, respectively. The characteristic velocity scale for the turbulence, $u_{\mathrm{turb}}$, is the root mean square of the velocity fluctuation, $\bm{u}^{\prime}$:
\begin{equation}
    u_{\mathrm{turb}} = \sqrt{\frac{ \overline{ u_{i}^{\prime}u_{i}^{\prime}} }{3}}.
\end{equation}
Under the assumptions that the turbulence is isotropic and $l_o = ( \sqrt{3/2} u_{\mathrm{turb}} )^{3}/\epsilon$, where $\epsilon$ is an energy dissipation rate, a relation between $Re_{l_o}$ and $Re_{\lambda_{T}}$ can be obtained as~\cite{pope2000turbulent}:
\begin{equation}
    Re_{\lambda_{T}} = \sqrt{\frac{20}{3} Re_{l_o}}.
\end{equation}
Similar to section~\ref{sec:reduced_reynolds_number_simulations}, we approximate $u_{\mathrm{turb}} \approx u_{\mathrm{rms}}=\sqrt{ \overline{ u^{\prime\prime}_i u^{\prime\prime}_i } /3}$, since it can be seen from section~\ref{sec:spectra} that switching $\bm{u}^{\prime}$ and $\bm{u}^{\prime\prime}$ has unnoticeable effect on the energy contents. The length scale of the largest eddies is approximated by the mean of the integral length scales of velocity components over the central part of the mixing layer. Therefore, $Re_{l_o}$ and $Re_{\lambda_{T}}$ within the central part of the mixing layer and $\lambda_{T}$ can be estimated as:
\begin{align}
    l_o &\approx \frac{\left< l_{u} + l_{v} + l_{w} \right>}{3}, \\
    \left< Re_{l_o} \right> &\approx \sqrt{\frac{3}{2}} \left< \frac{\bar{\rho} u_{\mathrm{rms}} l_o }{\bar{\mu}} \right>, \\
    \left< Re_{\lambda_{T}} \right> &\approx \sqrt{\frac{20}{3} \left< Re_{l_o} \right>}, \\
    \lambda_{T} &\approx \left< \frac{\bar{\mu}}{\bar{\rho} u_{\mathrm{rms}}} \right> \left< Re_{\lambda_{T}} \right>.
\end{align}
The Kolmogorov microscale, $\eta$, which characterizes the smallest length scale in a turbulent flow, can be estimated as~\cite{pope2000turbulent}:
\begin{equation}
    \frac{\eta}{l_o} \approx \left< Re_{l_o} \right> ^{-3/4}.
\end{equation}
In the classical Kolmogorov theory~\cite{batchelor1953theory, pope2000turbulent}, the emergence of the inertial range in spectral space requires that there is an intermediate scale, $\lambda$, which is not influenced by the outer scale features and also eddies at the Kolmogorov microscale. In other words, $\eta \ll \lambda \ll l_o$ is necessary. \citet{dimotakis2000mixing} proposed a more refined and stricter condition for mixing transition to occur if there exists an effective range of scales that is bounded by:
\begin{equation}
    \lambda_{\nu} \leq \lambda \leq \lambda_L,
\end{equation}
where $\lambda_L$ is the Liepmann--Taylor scale and $\lambda_{\nu}$ is the inner viscous scale. $\lambda_L$ is the smallest scale that can be generated directly by the outer scale $l_o$ and $\lambda_{\nu}$ is the the upper limit of the viscous range of the energy spectrum. Observed from experimental data, \citet{dimotakis2000mixing} suggested that $\lambda_L$ is proportional to $\lambda_{T}$:
\begin{equation}
    \lambda_L \approx c_L \lambda_{T},
\end{equation}
where $c_L$ is a flow-dependent parameter. Following previous studies by~\citet{dimotakis2000mixing}, \citet{zhou2003onset}, and \citet{tritschler2014evolution}, we use $c_L=5$. Based on the experimental data on turbulent boundary layer flow by~\citet{saddoughi1994local} and results from RMI simulations by~\citet{tritschler2014evolution}, it is reasonable to estimate $\lambda_{\nu}$ as:
\begin{equation}
    \lambda_{\nu} \approx 50 \eta.
\end{equation}

\begin{figure*}[!ht]
\centering
\includegraphics[width = 0.5\textwidth]{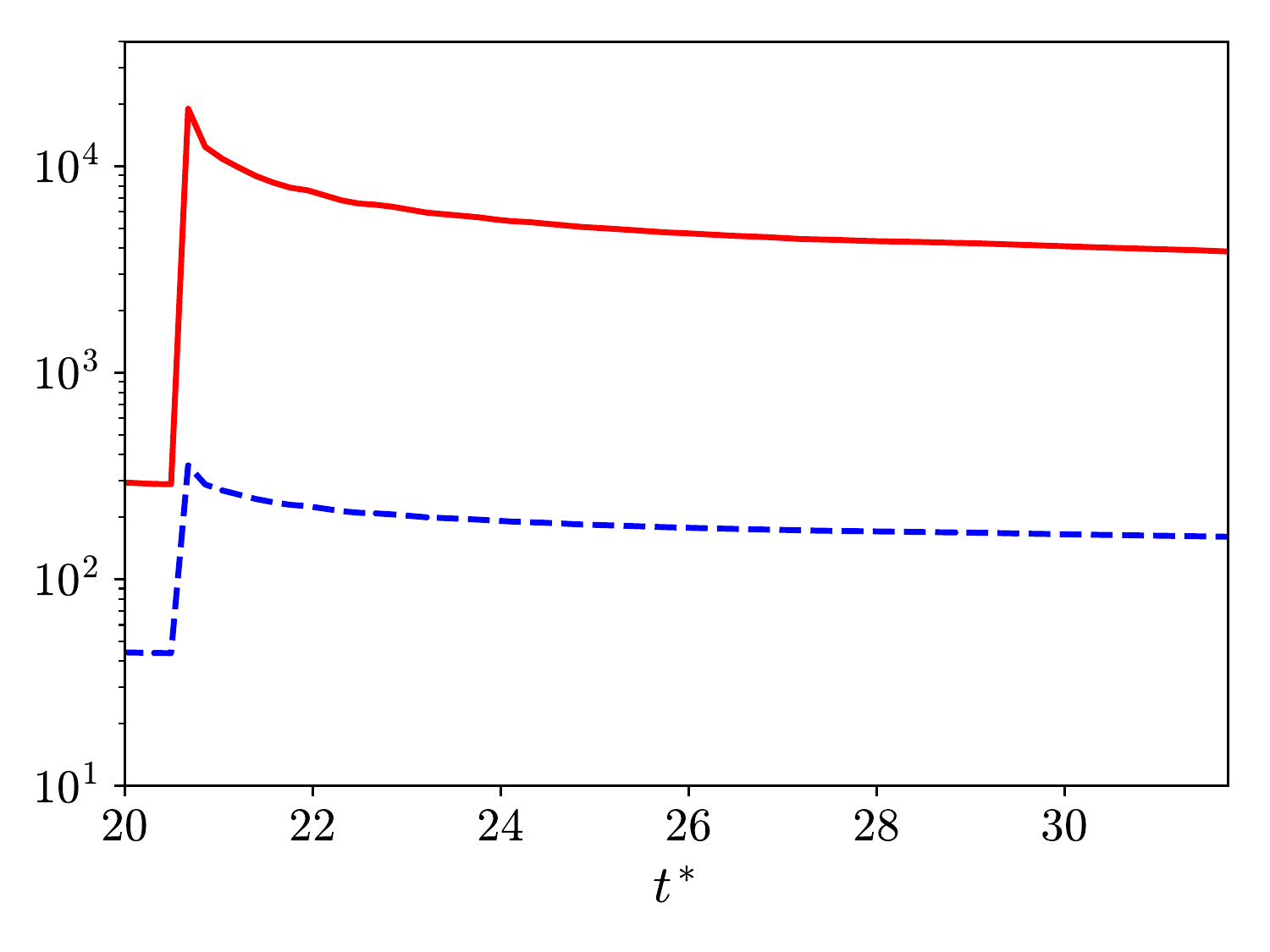}
\caption{Time evolution of the Reynolds numbers for the 3D problem with physical transport coefficients after re-shock. Red solid line: $\left< Re_{l_o} \right>$; blue dashed line: $\left< Re_{\lambda_T} \right>$.}
\label{fig:comparison_Reynold_numbers}
\end{figure*}

\begin{figure*}[!ht]
\centering
\includegraphics[width = 0.5\textwidth]{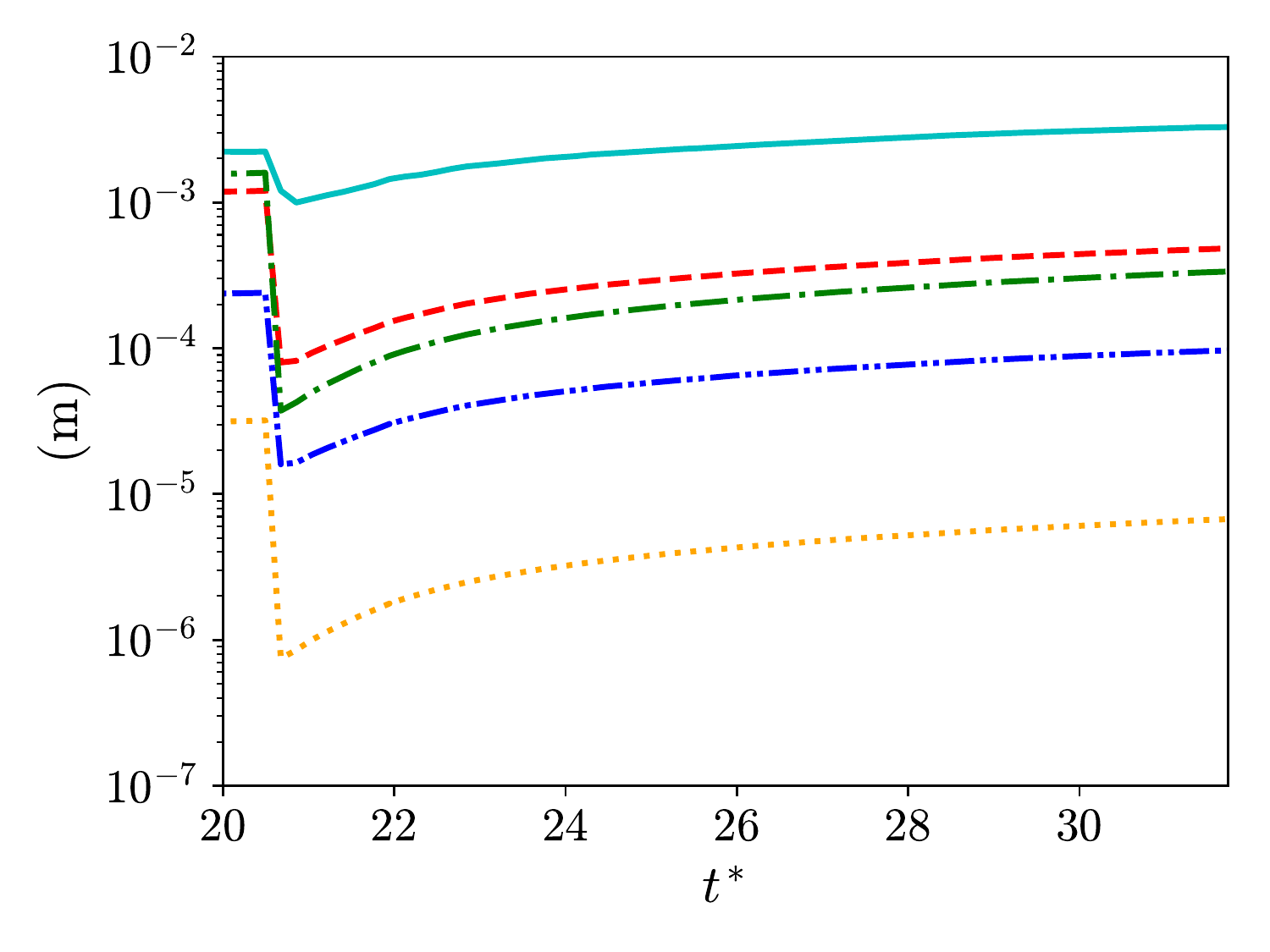}
\caption{Comparison of the length scales for the 3D problem with physical transport coefficients after re-shock. Cyan solid line: $l_o$; red dashed line: $\lambda_L$; green dash-dotted line: $\lambda_{\nu}$; blue dash-dot-dotted line: $\lambda_T$; orange dotted line: $\eta$.}
\label{fig:comparison_length_scales}
\end{figure*}

Figure~\ref{fig:comparison_Reynold_numbers} shows the time evolution of $\left< Re_{l_o} \right>$ and $\left< Re_{\lambda_T} \right>$ for the highest Reynolds number 3D case (with physical transport coefficients) after re-shock. \citet{dimotakis2000mixing} proposed that $\left< Re_{l_o} \right> \geq 10^{4}$ or $\left< Re_{\lambda_T} \right> \geq 100$ for mixing transition to occur. Although it is shown that $\left< Re_{l_o} \right>$ is only greater than $10^{4}$ for a short duration after re-shock, $\left< Re_{\lambda_T} \right>$ is greater than 100 throughout most of the simulation. This is an indication for mixing transition to occur in the corresponding 3D case. In figure~\ref{fig:comparison_length_scales}, the comparison of the time evolution of different estimated length scales discussed earlier is shown for the time duration after re-shock. It can be seen that $\lambda_L$ is always greater than $\lambda_{\nu}$ after re-shock. This means an uncoupled range of scales exists and this is consistent with the observation of inertial range emergence from the scale decomposition of energy in section~\ref{sec:spectra}. The coincidence of mixing transition with the appearance of the inertial range was also noted by~\citet{dimotakis2000mixing}. 

\section{\label{sec:spatial_numerics} Spatial discretizations}

The Navier--Stokes equations given by equations \eqref{eq:species_continuity_eqn}-\eqref{eq:mixture_energy_eqn} can be rewritten in vector form. For example, the vector form for a 3D problem can be represented as:
\begin{equation}
	\frac{\partial{\bm{Q}}}{\partial{t}} + \frac{\partial{\bm{F}}}{\partial{x}} + \frac{\partial{\bm{G}}}{\partial{y}} + \frac{\partial{\bm{H}}}{\partial{z}} 
	 = - \frac{\partial{\bm{F_v}}}{\partial{x}} - \frac{\partial{\bm{G_v}}}{\partial{y}} - \frac{\partial{\bm{H_v}}}{\partial{z}}, \label{eq:NS_vector_form}
\end{equation}
where $\bm{Q}$ is the vectors of conservative variables. $\bm{F}$, $\bm{G}$, and $\bm{H}$ are the convective flux vectors in the $x$, $y$, and $z$ directions. $\bm{F_v}$, $\bm{G_v}$, and $\bm{H_v}$ are the diffusive or viscous flux vectors in the $x$, $y$, and $z$ directions.

The explicit form of a sixth order weighted compact nonlinear scheme (WCNS) with nonlinear interpolation and approximate Riemann solver~\cite{wong2017high} is used to approximate the derivatives of the convective fluxes. The scheme is based on the sixth order accurate explicit midpoint-and-node-to-node differencing (MND) by \citet{nonomura2013robust}. For example, the first derivative of the convective flux in the $x$ direction, $\bm{F}$, at grid node $(i,j,k)$ is approximated by $\widehat{ \partial \bm{F} / \partial x } \big| _{i,j,k}$ as:
\begin{equation}
	\frac{\partial{\bm{F}}}{\partial{x}} \bigg|_{i,j,k} \approx \widehat{ \frac{\partial \bm{F}}{\partial x} } \bigg|_{i,j,k} = 
    \frac{1}{\Delta x}  \left[
      \frac{3}{2} \left(\tilde{\bm{F}}_{i+\frac{1}{2},j,k} - \tilde{\bm{F}}_{i-\frac{1}{2},j,k} \right)
      - \frac{3}{10} \left(\bm{F}_{i+1,j,k} - \bm{F}_{i-1,j,k} \right) + \frac{1}{30} \left(\tilde{\bm{F}}_{i+\frac{3}{2},j,k} - \tilde{\bm{F}}_{i-\frac{3}{2},j,k} \right)
    \right], \label{eq:MND_3D}
\end{equation}
where $\tilde{\bm{F}}_{i+\frac{1}{2},j,k}$ are fluxes approximated at midpoints between cell nodes by the sixth order accurate nonlinear localized dissipation interpolation and hybrid Riemann solver given by \citet{wong2017high}. $\bm{F}_{i,j,k}$ are the fluxes at cell nodes and $\Delta x$ is the uniform grid spacing in the $x$ direction. Note that equation~\eqref{eq:MND_3D} uses $\bm{F}_{i-1,j,k}$ and $\bm{F}_{i+1,j,k}$ instead of $\tilde{\bm{F}}_{i-1,j,k}$ and $\tilde{\bm{F}}_{i+1,j,k}$ since the fluxes at nodes can be directly evaluated from the conservative variables at nodes and require no interpolation. The MND scheme given by equation~\eqref{eq:MND_3D} can be rewritten in flux-difference form, which is given by:
\begin{equation}
	\left. \widehat{ \frac{\partial \bm{F}}{\partial x} } \right|_{i,j,k} = \frac{1}{\Delta x} \left( \widehat{\bm{F}}_{i+\frac{1}{2},j,k} - \widehat{\bm{F}}_{i-\frac{1}{2},j,k} \right), \label{eq:flux_derivative_definition}
\end{equation}
where
\begin{equation}
    \widehat{\bm{F}}_{i+\frac{1}{2},j,k} = \frac{1}{30} \tilde{\bm{F}}_{i-\frac{1}{2},j,k} - \frac{3}{10} {\bm{F}}_{i,j,k} + \frac{23}{15} \tilde{\bm{F}}_{i+\frac{1}{2},j,k} - \frac{3}{10} {\bm{F}}_{i+1,j,k} + \frac{1}{30} \tilde{\bm{F}}_{i+\frac{3}{2},j,k}.
\end{equation}

Any central explicit or compact (implicit) finite difference scheme can be rewritten into the flux-difference form and it is derived in~\citet*{subramaniam2019high}. The MND scheme in flux-difference form is implemented in HAMeRS to ensure that the discretizations of the convective fluxes are conservative at the coarse-fine AMR grid boundaries with the patch-based AMR method by~\citet{berger1989local}. Conservative linear spatial interpolation or prolongation is adopted for projecting solutions from coarse grids to fine grids in the simulations. Conservative averaging is also used to update coarse cells covered by finer cells. The conservative spatial discretizations ensure that shock waves are moving at the correct speeds. Note that explicit sixth order finite differences are used to compute the derivatives of diffusive and viscous fluxes in non-conservative form but the effects on the shock waves are negligible compared to those from the discretizations of the convective fluxes as the shocks are not well-resolved in the simulations.

\section{\label{sec:AMR_sensors} Sensors for adaptive mesh refinement}

Three different types of sensors were used in the AMR simulations to identify regions for refinement. They include the gradient sensor on pressure field, the multiresolution wavelet sensor on density field, and the value sensor on the mass fraction fields. In this section, the first two types of sensors are discussed.

\subsection{Gradient sensor}

The gradient sensor on pressure field, $p$, by \citet{jameson1981numerical} is chosen to detect shock waves. The sensor for a 3D problem is given by:
\begin{equation}
\begin{aligned}
 \left( {w_x} \right) _{i,j,k} &= \left| p_{i+1,j,k} - 2p_{i,j,k} + p_{i-1,j,k} \right|, \\
 \left( {w_y} \right) _{i,j,k} &= \left| p_{i,j+1,k} - 2p_{i,j,k} + p_{i,j-1,k} \right|, \\
 \left( {w_z} \right) _{i,j,k} &= \left| p_{i,j,k+1} - 2p_{i,j,k} + p_{i,j,k-1} \right|, \\
 {\tilde{w}}_{i,j,k} &= \frac{\sqrt{{ \left\{ {\left( {w_x} \right)}_{i,j,k} \right\} }^2 + { \left\{ {\left( {w_y} \right)}_{i,j,k} \right\} }^2 + { \left\{ {\left( {w_z} \right)}_{i,j,k} \right\} }^2 }}{\underset{(i,j,k)\in S}{\mathrm{mean}} \left(p_{i,j,k} \right) + \epsilon},
\end{aligned}
\end{equation}
\noindent where $S$ is the combined stencil of the three second order accurate central differences ($\left( {w_x} \right) _{i,j,k}$, $\left( {w_y} \right) _{i,j,k}$, and $\left( {w_z} \right) _{i,j,k}$) and $\epsilon$ is a small value to prevent division by zero. The local ``mean", ${\underset{(i,j)\in S}{\mathrm{mean}} \left(p_{i,j,k} \right)}$, is defined as:
\begin{equation}
 \underset{(i,j,k)\in S}{\mathrm{mean}} \left(p_{i,j,k} \right) = 
  \sqrt{ \left( p_{i+1,j,k} + 2p_{i,j,k} + p_{i-1,j,k} \right)^2 + \left( p_{i,j+1,k} + 2p_{i,j,k} + p_{i,j-1,k}  \right)^2 + \left( p_{i,j,k+1} + 2p_{i,j,k} + p_{i,j,k-1}  \right)^2}.
\end{equation}
Grid cells are tagged for refinement if ${\tilde{w}}_{i,j,k}$ is greater than a tolerance $tol_{\mathrm{local}}$. It should be noted that the normalized gradient sensor always has values between zero and one. $tol_{\mathrm{local}}=0.002$ is chosen for the gradient sensor for all simulations in this paper.

\subsection{Multiresolution wavelet sensor}

A multiresolution wavelet sensor on the density field, $\rho$, is used to detect mixing and chaotic regions. In wavelet decomposition, the amount of features at location $x=x_j$ having a characteristic length of $2^m$ or at scale level $m$ is measured through the wavelet coefficient $w^{(m)}_{j}$. The wavelet coefficients in 1D space are evaluated from the inner product of a quantitiy of interest ($\rho$ is used here) with some wavelet functions $\psi^{(m)}_{j}$:
\begin{equation}
 w^{(m)}_{j} = \left<\rho, \psi^{(m)}_{j} \right> = \int_{-\infty}^{\infty} \rho(x) \psi^{(m)}_{j} (x) dx, \quad \forall m \in \{1, \: 2, \: \dots, \: m_{\mathrm{max}} \}, \label{eq:wavelet_coeffs}
\end{equation}
where $m_{\mathrm{max}}$ is the largest scale level chosen by user. In order to obtain the wavelet coefficients at different scale levels at each grid point, we follow \citet{sjogreen2004multiresolution} to use the redundant wavelets. The wavelet function at scale level $m$ and location $x=x_j$ is defined as:
\begin{equation}
 \psi^{(m)}_{j} (x) = \frac{1}{2^m} \psi \left( \frac{x-x_j}{2^m} \right),
\end{equation}
\noindent where $\psi(x)$ is the mother wavelet. The Harten wavelet modified by \citet{sjogreen2004multiresolution} is used and the wavelet coefficients can be computed by:
\begin{align}
    w^{(m)}_{j} &= -\frac{1}{2} \left( \rho^{(m-1)}_{j+2^{m-1}} - 2\rho^{(m-1)}_j + \rho^{(m-1)}_{j-2^{m-1}} \right), \\
    \rho^{(m)}_{j} &= \frac{1}{2} \left( \rho^{(m-1)}_{j+2^{m-1}} + \rho^{(m-1)}_{j-2^{m-1}}\right),
\end{align}
where $\rho^{(m)}_{j}$ is the scaling coefficient at the same scale level and location as the wavelet coefficient, $w^{(m)}_{j}$. The wavelet and scaling coefficients at different levels and locations can be computed recursively from lower levels, where $\rho^{(0)}_j = \rho_j$ at the lowest level.

Wavelet coefficients in a 3D space are estimated from the 1D wavelet coefficients in different directions. The 1D wavelet coefficients in a 3D space are given by:
\begin{align}
 \left( {w_x} \right)^{(m)}_{i,j,k} &= \int_{-\infty}^{\infty} \rho(x,y,z) \psi^{(m)}_{i} (x) dx, \\
 \left( {w_y} \right)^{(m)}_{i,j,k} &= \int_{-\infty}^{\infty} \rho(x,y,z) \psi^{(m)}_{j} (y) dy, \\
 \left( {w_z} \right)^{(m)}_{i,j,k} &= \int_{-\infty}^{\infty} \rho(x,y,z) \psi^{(m)}_{k} (z) dz.
\end{align}
\noindent The 3D wavelet coefficients are then estimated from the 1D wavelet coefficients:
\begin{equation}
	w^{(m)}_{i,j,k} = \sqrt{ \left\{ {\left( {w_x} \right)^{(m)}_{i,j,k}} \right\}^2 + { \left\{ \left( {w_y} \right)^{(m)}_{i,j,k} \right\} }^2 + { \left\{ \left( {w_z} \right)^{(m)}_{i,j,k} \right\} }^2}.
\end{equation}

We define the wavelet sensors as the wavelet coefficients normalized by the local ``mean" of the scaling coefficients at one lower level:
\begin{equation}
	{\tilde{w}}^{(m)}_{i,j,k} = \frac{w^{(m)}_{i,j,k}}{\underset{(i,j,k)\in S}{\mathrm{mean}} \left( \rho^{(m-1)}_{i,j,k} \right) + \epsilon}.
\end{equation}
\noindent The local ``mean" of the scaling coefficients of Harten wavelet, $\underset{(i,j,k)\in S}{\mathrm{mean}} \left( \rho^{(m-1)}_{i,j,k} \right)$, is defined as:
\begin{equation}
\begin{split}
 {\underset{(i,j,k)\in S}{\mathrm{mean}} \left( \rho^{(m-1)}_{i,j,k} \right) } =
  \frac{1}{2} \left[ \left\{ { \rho }^{(m-1)}_{i+2^{m-1},j,k} + 2{ \rho }^{(m-1)}_{i,j,k} + { \rho }^{(m-1)}_{i-2^{m-1},j,k} \right\} ^2 
  + \left\{ { \rho }^{(m-1)}_{i,j+2^{m-1},k} + 2{ \rho }^{(m-1)}_{i,j,k} + { \rho }^{(m-1)}_{i,j-2^{m-1},k}  \right\}^2 \right. \\
  \left. + \left\{ { \rho }^{(m-1)}_{i,j,k+2^{m-1}} + 2{ \rho }^{(m-1)}_{i,j,k} + { \rho }^{(m-1)}_{i,j,k-2^{m-1}}  \right\}^2 \right]^{\frac{1}{2}}.
\end{split}
\end{equation}
If ${\tilde{w}}^{(m)}_{i,j}$ at any scale level is greater than a user-defined tolerance $tol_{\mathrm{local}}$, the corresponding grid cell is tagged for refinement. It should be noted that the local ``mean" is designed in a way that the sensor applied to the finest level of scale is the gradient sensor discussed in previous section. $m_{\mathrm{max}}=3$ and $tol_{\mathrm{local}}=0.004$ for all wavelet sensors are chosen for all simulations in this paper.

\section{\label{sec:appendix_TC} Transport coefficients}

The shear viscosity, $\mu_i$, of a species $i$ is given by the Chapman--Enskog's model~\cite{chapman1991mathematical}:
\begin{equation}
	\mu_i = 2.6693 \times 10^{-6} \frac{\sqrt{M_i T}}{\Omega_{\mu,i} \sigma^2_i},
\end{equation}
\noindent where $\sigma_i$ is the collision diameter and $\Omega_{\mu, i}$ is the collision integral of the species given by:
\begin{equation}
	\Omega_{\mu, i} = A \left( T^*_{i} \right)^B + C \textnormal{exp} \left( D T^*_{i} \right) + E \textnormal{exp} \left( F T^*_{i} \right),
\end{equation}
\noindent where $T^*_{i} = T/( \epsilon/k )_i$, $A = 1.16145$, $B = -0.14874$, $C = 0.52487$, $D = -0.7732$, $E = 2.16178$, and $F = -2.43787$. $T$ is the temperature of the species. $( \epsilon/k )_i$ is the Lennard--Jones energy parameter and $M_i$ is the molecular weight of the species. The values of $M_i$, $( \epsilon/k )_i$, and $\sigma_i$ of different species are given in table~\ref{tab:fluid_properties}.

The bulk viscosity, $\mu_v$, of air is given by the linear model by \citet{gu2014systematic}:
\begin{equation}
	\mu_{v} = A + B T,
\end{equation}
\noindent where $A=-3.15\times10^{-5}\ {\mathrm{kg \ m^{-1} s^{-1}}}$ and $B = 1.58\times10^{-7}\ {\mathrm{kg \ m^{-1} s^{-1} K^{-1}}}$.

The bulk viscosity, $\mu_v$, of $\mathrm{SF_6}$ is given by the Cramer's model~\cite{cramer2012numerical}:
\begin{align}
	\mu_{v} &= \left( \gamma - 1 \right)^2 \left. c_v \right|_{v}  (p \tau_v), \\
	\left. c_v \right|_{v} &= \left( \frac{c_{v}}{R} - \frac{f_{r} + 3}{2} \right), \\
    (p \tau_v) &= A \textnormal{exp} \left( \frac{B}{T^{\frac{1}{3}}} + \frac{C}{T^{\frac{2}{3}}} \right),
\end{align}
\noindent where $f_r = 3$, $A=0.2064\times10^{-5}\ {\mathrm{kg\ m^{-1}s^{-1}}}$, $B=121\ {\mathrm{K^{1/3}}}$, and $C=-339\ {\mathrm{K^{2/3}}}$ for $\mathrm{SF_6}$.

The thermal conductivity of species $i$, $\kappa_i$, is defined by:
\begin{equation}
	\kappa_i = c_{p,i} \frac{\mu_i}{Pr_i},
\end{equation}
\noindent where $Pr_i$ and $c_{p,i}$ are the species-specific Prandtl number and specific heat at constant pressure respectively.

Mass diffusion coefficient of a binary mixture, $D_{ij}$, is given by~\cite{poling2001properties}:
\begin{equation}
	D_{ij} = D_{i} = D_{j} = \frac{0.0266}{\Omega_{D,ij}} \frac{T^{3/2}}{p \sqrt{M_{ij}} \sigma_{ij}^2},
\end{equation}
\noindent where $p$ and $T$ are the pressure and temperature of the mixture. $\Omega_{D,ij}$ is the collision integral for diffusion given by:
\begin{equation}
	\Omega_{D,ij} = A \left( T^*_{ij} \right)^B + C \textnormal{exp} \left( D T^*_{ij} \right) + E \textnormal{exp} \left( F T^*_{ij} \right)	+ G \textnormal{exp} \left( H T^*_{ij} \right),
\end{equation}
\noindent where $T^*_{ij} = T/T_{\epsilon_{ij}}$, $A = 1.06036$, $B = -0.1561$, $C = 0.19300$, $D = -0.47635$, $E = 1.03587$, $F = -1.52996$, $G = 1.76474$, and $H = -3.89411$. $M_{ij}$, $\sigma_{ij}$, and $T_{\epsilon_{ij}}$ are the effective molecular weight, collision diameter, and Lennard--Jones energy parameter respectively for the mixture:
\begin{align}
	M_{ij} &= \frac{2}{\frac{1}{M_i} + \frac{1}{M_j}}, \\
    \sigma_{ij} &= \frac{\sigma_i + \sigma_j}{2}, \\
    T_{\epsilon_{ij}} &= \sqrt{\left( \frac{\epsilon}{k} \right)_i \left( \frac{\epsilon}{k} \right)_j }.
\end{align}
\noindent The values of $M_i$, $( \epsilon/k )_i$, and $\sigma_i$ of different species are given in table~\ref{tab:fluid_properties}.

\begin{table}[!ht]
\caption{\label{tab:fluid_properties}%
Fluid properties.}
\begin{ruledtabular}
\begin{tabular}{ c c c c c c c c c }
Gas & $\gamma_i$ & $c_{p,i}\ (\mathrm{J\ kg^{-1} K^{-1}})$ &
$c_{v,i}\ (\mathrm{J\ kg^{-1} K^{-1}})$ & $M_i\ (\mathrm{g\ mol^{-1}})$ &
$R_i\ (\mathrm{J\ kg^{-1} K^{-1}})$ & $\left( \epsilon/k \right)_i\ (\mathrm{K})$ &
$\sigma_i\ (\mbox{\normalfont\AA})$ & $Pr_i$ \\
\hline
$\mathrm{SF_6}$ & 1.09312 & 668.286 & 611.359 & 146.055 & 56.9269 & 222.1 & 5.128 & 0.79 \\
Air & 1.39909 & 1040.50 & 743.697 & 28.0135 & 296.802 & 78.6 & 3.711 & 0.71
\end{tabular}
\end{ruledtabular}
\end{table}

\section{\label{sec:appendix_mixing_rules} Mixing rules}
With the assumption that all species are at pressure and temperature equilibria, the ratio of specific heats of the mixture follows as:
\begin{equation}
    \gamma = \frac{c_p}{c_v}, \quad c_{p} = \sum_{i=1}^{N} Y_i c_{p,i}, \quad c_{v} = \sum_{i=1}^{N} Y_i c_{v,i}.
\end{equation}
The molecular weight of the mixture is given by:
\begin{equation}
    M = \left( \sum_{i=1}^{N} \frac{Y_i}{M_i} \right) ^{-1}.
\end{equation}

The mixture shear viscosity, bulk viscosity, and thermal conductivity are given by:
\begin{align}
	\mu    &= \frac{\sum^{N}_{i=1} \mu_i Y_i/\sqrt{M_i}}{\sum^{N}_{i=1} Y_i/\sqrt{M_i}}, \\
	\mu_v  &= \frac{\sum^{N}_{i=1} \mu_{v, i} Y_i/\sqrt{M_i}}{\sum^{N}_{i=1} Y_i/\sqrt{M_i}}, \\
	\kappa &= \frac{\sum^{N}_{i=1} \kappa_i Y_i/\sqrt{M_i}}{\sum^{N}_{i=1} Y_i/\sqrt{M_i}}.
\end{align}

\bibliography{references} 

\end{document}